\documentclass[12pt]{article}
\pdfoutput=1

\usepackage[height=8.85in,width=6.25in]{geometry}

\usepackage[utf8]{inputenc}
\usepackage{amsmath}
\usepackage{amssymb}
\usepackage{mathtools}
\numberwithin{equation}{section}
\usepackage{slashed}
\usepackage{braket}
\usepackage[svgnames]{xcolor}
\usepackage[colorlinks,citecolor=DarkGreen,linkcolor=FireBrick,urlcolor=FireBrick,linktocpage,breaklinks=true]{hyperref}
\usepackage{cite}
\usepackage[pdftex]{graphicx}
\usepackage{tikz}
\usepackage{tikz-cd}
\usepackage{times}
\usepackage{courier}
\usepackage{bm}
\usepackage{subfig}
\usepackage{dsfont}

\usepackage{xcolor}
\usepackage{mdframed}

\usetikzlibrary{decorations.pathreplacing,decorations.markings,calc}

\usepackage{mathrsfs} 

\definecolor{dgreen}{rgb}{0, 0.55, 0}

\usetikzlibrary{arrows.meta}
\newcommand{\midarrow}{\tikz[baseline] \draw[-{Stealth[scale=1.5]}] (0,0) -- +(.1,0);}

\usetikzlibrary{arrows,shapes,positioning,calc}
\usetikzlibrary{decorations.markings}
\usepackage[rightcaption]{sidecap}
\tikzstyle arrowstyle=[scale=1]
\tikzstyle directed=[postaction={decorate,decoration={markings,
    mark=at position .65 with {\arrow[arrowstyle]{stealth}}}}]

\usetikzlibrary{positioning}

\usepackage{colortbl}
\definecolor{llightyellow}{rgb}{1.0, 0.95, 0.7}
\definecolor{llightblue}{rgb}{0.7, 0.9, 1.0}
\definecolor{llightpink}{rgb}{1.0, 0.85, 0.95}
\definecolor{llightgreen}{rgb}{0.7, 1.0, 0.4}
\colorlet{lightyellow}{llightyellow!50!white}
\colorlet{lightblue}{llightblue!50!white}
\colorlet{lightgreen}{llightgreen!50!white}
\colorlet{lightpink}{llightpink!50!white}

\pgfdeclareradialshading{ring}{\pgfpoint{0cm}{0cm}}%
{rgb(0cm)=(1,1,1);
rgb(0.7cm)=(1,1,1);
rgb(0.719cm)=(1,1,1);
rgb(0.72cm)=(0.975,0,0);
rgb(0.9cm)=(1,1,1)}

\makeatletter
\newcommand{\neutralize}[1]{\expandafter\let\csname c@#1\endcsname\count@}
\makeatother

\newtheorem{thm}{Theorem}

\renewcommand{\thanks}[1]{\footnote{#1}}

\newcommand{\bea}{\begin{eqnarray}}
\newcommand{\eea}{\end{eqnarray}}
\newcommand{\be}{\begin{eqnarray}}
\newcommand{\ee}{\end{eqnarray}}

\def\cA{{\cal A}}

\def\cC{{\cal C}}
\def\cD{{\cal D}}

\def\cH{{\cal H}}

\def\cL{{\cal L}}
\def\cM{{\cal M}}
\def\cN{{\cal N}}
\def\cO{{\cal O}}
\def\cP{{\cal P}}

\def\cX{{\cal T}}

\def\cX{{\cal X}}

\def\cZ{{\cal Z}}

\def\ZZ{{\mathbb Z}}
\def\RR{{\mathbb R}}

\def\CC{{\mathbb C}}

\def\QQ{{\mathbb{Q}}}

\def\half{{1\over 2}}
\def\p{\partial}

\def\g{\gamma}

\def\eps{\epsilon}

\definecolor{grey}{rgb}{0.9,0.9,0.9}
\definecolor{dgrey}{rgb}{0.3,0.3,0.3}

\usetikzlibrary{decorations.pathmorphing}
\tikzset{snake it/.style={decorate, decoration=snake}}

\usetikzlibrary{decorations.pathreplacing,calligraphy}

\tikzset{
  on each segment/.style={
    decorate,
    decoration={
      show path construction,
      moveto code={},
      lineto code={
        \path [#1]
        (\tikzinputsegmentfirst) -- (\tikzinputsegmentlast);
      },
      curveto code={
        \path [#1] (\tikzinputsegmentfirst)
        .. controls
        (\tikzinputsegmentsupporta) and (\tikzinputsegmentsupportb)
        ..
        (\tikzinputsegmentlast);
      },
      closepath code={
        \path [#1]
        (\tikzinputsegmentfirst) -- (\tikzinputsegmentlast);
      },
    },
  },
  mid arrow/.style={postaction={decorate,decoration={
        markings,
        mark=at position .5 with {\arrow[#1]{stealth}}
      }}},
}

\tikzset{line/.style={line width=0.25mm},
curve/.style={line,smooth,tension=1},
->-/.style={decoration={
  markings,
  mark=at position #1 with {\arrow[>={Stealth[scale=1]}]{>}}},postaction={decorate}},
-<-/.style={decoration={
  markings,
  mark=at position #1 with {\arrow[>={Stealth[scale=1]}]{<}}},postaction={decorate}},
}

\def\m{\mu}
\def\n{\nu}

\def\){\right)}
\def\({\left( }
\def\]{\right] }
\def\[{\left[ }

\def\no{\nonumber}

\usetikzlibrary{intersections,shapes.arrows}

\usepackage{pgfplots}
\usepackage{braids}

\usepackage[most]{tcolorbox}

\makeatletter\def\l@subsubsection#1#2{}%
\makeatother

\begin{document}

\begin{titlepage}
\thispagestyle{empty}

\begin{flushright}
KYUSHU-HET-354
\end{flushright}

\bigskip

\begin{center}
\noindent{{\Large \textbf{
Introduction to Generalized Symmetries
}}}\\
\vspace{2cm}
Justin Kaidi${}^{1,2,3}$
\vspace{1cm}

\date{Last updated: \today}

${}^{1}${\small \sl 
Institute for Advanced Study, Kyushu University, Fukuoka 819-0395, Japan
}

${}^{2}${\small \sl 
Department of Physics, Kyushu University, Fukuoka 819-0395, Japan
}

$^3${ \small \sl  Quantum and Spacetime Research Institute (QuaSR), Kyushu University, Fukuoka 819-0395, Japan}

\vskip 2em
\end{center}

\begin{abstract}
These notes were prepared for a series of intensive lectures delivered at Hokkaido University, Nagoya University, Kyoto University, and Kyushu University.

We begin with a brief review of higher-form symmetries, anomalies, and discrete gauge theories, before introducing non-invertible symmetries in $(1+1)$-dimensional systems. 
The basic structure of fusion categories is then discussed, including a discussion of categorical analogs of discrete gauging and representation theory. We subsequently turn to $(3+1)$-dimensional theories, where several physical applications of non-invertible symmetries are discussed.
These notes are intended to be largely self-contained, and require no prior familiarity with subjects such as conformal field theory or lattice models.

 \end{abstract}

\end{titlepage}

\newpage
\tableofcontents
\newpage

\section{Introduction}
These lectures give a broad overview of \textit{generalized symmetries}, with a particular focus on the emerging notion of \textit{non-invertible symmetries}. Symmetries have long been central to our understanding of quantum field theory (QFT): they constrain dynamics, classify phases of matter, and often dictate the structure of observables. Though they have traditionally been understood as transformations forming a group, developments over the last decade have revealed that this group-theoretic framework is too narrow, and that many quantum systems exhibit richer symmetry structures which fall outside the group paradigm. A hallmark of such structures is the lack of inverse for some transformations, giving rise to the name non-invertible symmetry.


The study of non-invertible symmetries has seen rapid progress in recent years, with connections to conformal field theory, lattice models, and topological quantum field theory, as well as implications for dualities and dynamics in strongly-coupled systems. In these lectures we will introduce the basic ideas and provide examples, aiming for a self-contained exposition. We will assume only a basic familiarity with QFT and undergraduate mathematics. The goal is to build intuition for how generalized and non-invertible symmetries appear, how they are characterized, and why they play an increasingly important role in modern theoretical physics.

We should emphasize that there is essentially nothing new in these lecture notes, and in various places we will borrow from previous reviews of generalized symmetries, which include \cite{McGreevy:2022oyu,Freed:2022iao,Shao:2023gho, Brennan:2023mmt, Bhardwaj:2023kri, Gomes:2023ahz,Schafer-Nameki:2023jdn,Luo:2023ive}. The reader is encouraged to consult those references as well, which in some cases may be much clearer than the current reference. One potential ``novelty'' of the current notes is that, because some of the audiences to which these lectures were given were unfamiliar with conformal field theory (CFT), the presentation here makes no reference to CFT techniques---even the Ising model is treated only as a lattice model. This may make these notes more palatable to some readers,  though less palatable to others.

Because these lectures were given over the course of three (long and arduous) days, the notes are correspondingly split into three sections. Each section ends with a short list of relevant references. Though we have tried to include a decent number of citations, it would be impossible to cite every paper on the subject, since the literature is vast and ever-growing. Apologies in advance to anyone whose work was not properly accounted for. I would be happy to add references to any relevant papers upon request.

Finally, an unfortunate fact about the literature is that the distinction between Euclidean and Lorentzian signature is often blurred. For example, one often refers to Hilbert spaces and operators acting on them, which are intrinsically Lorentzian notions, but at the same time discusses e.g. linkings of the operators in Euclidean space. The Euclidean conventions are also common when determining where to distribute factors of $i$ throughout equations, though we will write most equations in a Lorentzian fashion here, in order to minimize the factors of $i$ appearing in the action/Lagrangian/currents. For the most part we will be rather cavalier about this issue, commenting on it only when it seems relevant. 

\subsection*{Acknowledgements}

These notes were prepared for a series of intensive lectures given at Hokkaido University, Kyoto University, and Nagoya University.
I would like to thank the organizers of those lectures---Tatsuo Kobayashi, Yu Nakayama, and Tadakatsu Sakai---as well as all of the participants for their many enlightening comments and questions. 
Almost all of the content of these lectures consists of things which I learned from my many excellent friends and collaborators, including Jan Albert, Vladimir Bashmakov, Yichul Choi, Michele Del Zotto, Thomas Dumitrescu, Wataru Harada, Azeem Hasan, Knight Hirasaki, Yamato Honda, Zohar Komargodski, Ying-Hsuan Lin, Yuefeng Liu, Julio Parra-Martinez, Emily Nardoni, Kantaro Ohmori, Sahand Seifnashri, Shu-Heng Shao, Xiaoyi Shi, Soichiro Shimamori, Zhengdi Sun, Yuji Tachikawa, Gabi Zafrir, and Yunqin Zheng. I look forward to learning more from them in the future.
Finally, I thank Kota Miki for catching typos in the draft.
\newpage 
\section{Topological Defects and Higher-Form Symmetries}

We begin our discussion by reviewing various well-known properties of standard continuous symmetries, before generalizing to continuous higher-form global symmetries. This material can be found in any review on the topic, and we will follow \cite{Brennan:2023mmt} particularly closely. Having done so, we will then move on to a discussion of discrete symmetries (both zero- and higher-form), their gauging, their anomalies, and their mixing into higher-group symmetries. This material is again fairly standard (see e.g. \cite{Brennan:2023mmt,Gomes:2023ahz,Schafer-Nameki:2023jdn,Bhardwaj:2023kri,Luo:2023ive}), but since all of the results discussed in this section will appear in some shape or form in the subsequent sections, we nevertheless include it for completeness.

\subsection{Noether's theorem}

We begin by considering the simplest case of a $U(1)$ symmetry. Familiar examples of theories with such a symmetry include complex scalar field theory,
\bea
S = \half \int d^d x\, \left( \p_\m \phi \p^\m \phi^* + m^2 |\phi|^2 \right)~, 
\eea
which is invariant under
\bea
\phi(x) \rightarrow e^{i \alpha} \phi(x)~, \hspace{0.5 in}\phi^*(x) \rightarrow e^{-i \alpha} \phi^*(x)~,
\eea
with $\alpha$ taking values in $\RR/2\pi \ZZ \cong U(1)$, and the theory of a Dirac fermion,
\bea
S = \int d^d x\, \overline \Psi (i \g^\m \p_\m - m ) \Psi~, 
\eea
which is invariant under 
\bea
\Psi(x) \rightarrow e^{i \alpha} \Psi(x)~, \hspace{0.5 in}\overline \Psi(x) \rightarrow e^{-i \alpha} \overline \Psi(x)~. 
\eea
For reasons that will become clear later, we will refer to this as a $U(1)$ \textit{zero-form} symmetry. 

There is a famous theorem due to Noether stating the following, 
\begin{thm}[Noether's theorem]

For every continuous symmetry there is an associated (equivalence class of) conserved current(s).
\end{thm} 

The proof is as follows. Denote our fields (not necessarily scalars) by $\phi^a(x)$ and our Lagrangian density by $\cL[\phi^a, \p_\m \phi^a]$. The action is given by 
\bea
S = \int d^d x\, \cL[\phi^a, \p_\m \phi^a]~,
\eea
and the equations of motion are
\bea
{\p \cL \over \p \phi^a} - \p_\m \left({\p \cL \over \p( \p_\m \phi^a)}\right) = 0~. 
\eea
We assume that our symmetry is continuous, so that we may consider transformations infinitesimally close to the identity, i.e. $\alpha \ll 1$ in the  case of $U(1)$. Say that under such an infinitesimal transformation we have 
\bea
\phi^a(x) \rightarrow \phi^a (x) +  \alpha\, \delta\phi^a(x)~. 
\eea
In the case in which $\phi^a$ has charge $q_a$ under a $U(1)$ symmetry, i.e. $\phi^a \rightarrow e^{i q_a \alpha} \phi^a $, we have simply $\delta \phi^a(x) = i q_a \phi^a(x)$. 

For the above transformation to be a symmetry of the classical action, we must have \bea
\delta_\alpha S = \int d^d x \, \delta_\alpha  \cL =0~. 
\eea
This requires that $\delta_\alpha \cL$ be a total derivative, and we furthermore assume that it is exactly zero.\footnote{It is also possible to consider a Lagrangian for which the variation is a non-trivial total derivative, in which case we refer to the symmetry as a \textit{quasi-symmetry} of the Lagrangian.  In this case Noether's theorem still holds, but the expression for the current must be modified.    } 
The variation can be written as 
\bea
(\alpha)^{-1}\delta_\alpha  \cL &=&  {\p \cL \over \p \phi^a} \delta \phi^a + {\p \cL \over \p(\p_\m \phi^a)} \delta (\p_\m \phi^a) = {\p \cL \over \p \phi^a} \delta \phi^a + {\p \cL \over \p(\p_\m \phi^a)} \p_\m (\delta \phi^a) 
\no\\
&=& \left[ {\p \cL \over \p \phi^a} - \p_\m \left({\p \cL \over \p( \p_\m \phi^a)}\right) \right] \delta \phi^a + \p_\m \left[ {\p \cL \over \p (\p_\m \phi^a)} \delta \phi^a \right] ~,
\eea
and on-shell, i.e. assuming that the equations of motion are satisfied, the first term vanishes, so we are left with 
\bea
\p_\m \left[ {\p \cL \over \p (\p_\m \phi^a)} \delta \phi^a \right]  = 0~. 
\eea 
Thus we find that 
\bea
\label{eq:Noethercurrentdef}
j_\m := {\p \cL \over \p (\p^\m \phi^a)} \delta \phi^a
\eea
satisfies the divergence-free equation $\p^\m j_\m = 0$. This $j_\m$ is called a \textit{conserved current}. In non-relativistic notation, the equation reads 
\bea
\p_0 j_0 - \p_i j_i = 0~, \hspace{0.5 in} i = 1, \dots, d-1~,
\eea
where the minus sign appears because we have lowered the $0$ index using $\eta_{\m\n} = \mathrm{diag}(-1, 1, \dots, 1)$. It is also useful to write this current in differential form notation, in which $ j = j_\m dx^\m$ is a one-form. In this notation the conservation equation reads $d * j = 0$.\footnote{Here $*$ is the Hodge star operation, which acts on differential $k$-forms in $d$ dimensions via $*: \,\omega_{\mu_1, \dots, \mu_k} dx^{\mu_1} \wedge \dots \wedge  dx^{\mu_k} \mapsto {1\over (d-k)!} {\eps^{\m_1 \dots \mu_k}}_{\mu_{k+1}\dots \mu_d} \omega_{\mu_1, \dots,\mu_k}  dx^{\mu_{k+1}} \wedge \dots \wedge dx^{\mu_d}$ and satisfies $**\omega = \pm(-1)^{k (d-k)} \omega$, with $+$ for Euclidean signature and $-$ for Lorentzian signature.  }

From the conserved current we may define the \textit{conserved charge}, 
\bea
Q:= -\int_{\mathrm{space}} d^{d-1} x\,\, j_0(x)~. 
\eea
This is conserved since we have 
\bea
\p_0 Q = -\int d^{d-1}x \,\, \p_0 j_0 = -\int d^{d-1} x \sum_{i=1}^{{d-1}} \p_i j_i (x) = 0~,
\eea
where in the final step we have used that the integrand is a total derivative, and space is assumed to have no boundary.

Note that the current is subject to an ambiguity, allowing for shifts by so-called \textit{improvement terms},
\bea
\label{eq:improvementterms1}
j_0 \rightarrow j_0 + \p_i X_i ~, \hspace{0.5 in} j_i \rightarrow j_i + \p_0 X_i + \p_j Z^{[ij]}~. 
\eea
The shift in $j_0$ is allowed since it does not change the charge, but must be accompanied by the first shift in $j_i$ so as to maintain the conservation equation. The second shift in $j_i$, namely the one involving $Z^{[ij]}$, is allowed since it changes neither the charge nor the conservation equation. 
These improvement terms are related to non-minimal couplings to background gauge fields, as we mention below. 

\subsection{Ward-Takahashi identities}

At the quantum level, the conservation of the Noether current can be violated at the location of charged operators. 
Indeed, imagine that we have a charge $q$ operator $\cO_q(x)$, whose correlation functions are given by the following,
\bea
\langle \cO_q(x) \dots \rangle = {1\over Z} \int \cD \phi^a\, \cO_q(x) \dots e^{i S [\phi^a]}~, \hspace{0.5 in} Z = \int \cD \phi^a e^{i S [\phi^a]}~;
\eea
 the $\dots$ represent potential operator insertions at separated points. 
Assuming an invariant path-integral measure, a change of integration variables with a smooth, spacetime-dependent parameter $\alpha$ leaves the expectation value unchanged:
\bea
\delta_\alpha \langle \cO_q(x) \dots \rangle= 0~.
\eea
On the other hand, note that 
\bea
\delta_\alpha \langle \cO_q(x) \dots \rangle &=& {1\over Z} \int \cD \phi^a \left[\left( \delta_\alpha \cO_q(x) + i \cO_q(x) \delta_\alpha S[\phi^a] \right) \dots + \cO_q(x) \delta_\alpha(\dots)\right] e^{i S [\phi^a]}
\no\\
&\vphantom{.}&\hspace{-1 in} = {1\over Z} \int \cD \phi^a \left[ \left( i q \alpha(x) \cO_q(x) - {i \cO_q(x) \int d^d y\, \alpha(y)\, \p^\m j_\m (y)} \right) \dots +  \cO_q(x) \delta_\alpha(\dots)\right] e^{i S [\phi^a]}~.
\no\\
\eea
{Choose $\alpha$ to vanish near the spectator insertions represented by $\dots$. Since it is otherwise arbitrary, we obtain the following local} \textit{{Ward-Takahashi} identity},
\bea
\label{eq:WTidentity}
\langle \p^\m j_\m (x) \cO_q(y) \dots \rangle = q\, \delta^{(d)}(x-y) \langle \cO_q(y) \dots \rangle~.
\eea
Whereas the classical conservation law would have naively led us to believe that the left-hand side is zero, we  see that we now get a non-zero value at coincident points. There are analogous contact terms at charged spectator insertions; in Lorentzian signature the correlators are time-ordered.

\subsection{Conserved charges and topological defects}
\label{sec:topdefintro}

Now say, as above, that we have a classical current satisfying $\p^\m j_\m = 0$. As we just said, the corresponding Noether charge at time $t$ is given by the integral of $-j_0$ over space. 
We can also define a slightly more general charge, living on any codimension-1 (i.e. $(d-1)$-dimensional) submanifold $M_{d-1}$ of spacetime, 
\bea
Q(M_{d-1}) := \int_{M_{d-1}} ds^\m \, j_\m ~, 
\eea
Here $ds^\m$ is the oriented surface element, with $ds^\m j_\m=*j$ on $M_{d-1}$. On a constant-time slice oriented by $d^{d-1}x$, this gives $Q=-\int d^{d-1}x\,j_0$.

In this case, the analog of conservation in time is the fact that the charge $Q(M_{d-1}) $ depends only on the homology class of $M_{d-1}$ in spacetime with the charged insertions removed. 
Indeed, suppose these manifolds bound an oriented region $N_d$ in spacetime that avoids charged insertions, such that $\p N_d =M_{d-1}\sqcup {\overline M}'_{d-1}$. Then by Stokes' theorem, 
\bea
\label{eq:Qtopological}
Q(M_{d-1}) - Q(M'_{d-1}) = \int_{N_d} d^d x \, \p^\m j_\m = 0~. 
\eea
Thus homologous codimension-1 supports have the same charge, provided the interpolation avoids charged insertions.

We may now define the \textit{symmetry generators}, 
\bea
\label{eq:symgendef}
U_\alpha(M_{d-1}) := e^{i \alpha Q(M_{d-1})}~. 
\eea
These again depend only topologically on $M_{d-1}$. 
In a quantum mechanical theory, how should we interpret the $U_\alpha(M_{d-1}) $? When $M_{d-1}$ is the whole space at a fixed time, then $U_\alpha(M_{d-1})$ should be interpreted as the conserved, unitary operator that acts on the Hilbert space $\cH$. On the other hand, when $M_{d-1}$ is localized in one of the spatial directions and extends along time, then the effect of $U_\alpha(M_{d-1}) $ is to modify the boundary conditions along that spatial direction. Quantizing in the presence of these modified boundary conditions gives a new Hilbert space of states, which we denote by $\cH_\alpha$. These are known as \textit{twisted sectors}. For a more general $M_{d-1}$ which is neither localized nor wrapping the entire time direction, the interpretation is less clear, but what is always true is that moving an operator past these defects leads to the action of the $U(1)$ symmetry on the operator. 

To see this, we may use the Ward-Takahashi identity (\ref{eq:WTidentity}) to compute
\bea
\label{eq:zeroformchargecomputation}
U_\alpha(M_{d-1}) \cO_q(x) U_{-\alpha}(M_{d-1}') &=& e^{i \alpha \int_{N_d} d^dy\, \p^\m j_\m(y) } \cO_q(x)
\no\\
&=& \sum_{n=0}^\infty {(i \alpha)^n \over n!} \left(\int_{N_d} d^dy\, \p^\m j_\m(y) \right)^n \cO_q(x)
\no\\
&=& \sum_{n=0}^\infty {(i \alpha q)^n \over n!} \left(\int_{N_d} d^dy\, \delta^{(d)}(x-y) \right)^n \cO_q(x)
\no\\
&=& \sum_{n=0}^\infty {(i \alpha q)^n \over n!} \cO_q(x)
\no\\
&=& e^{i q \alpha}\, \cO_q(x)~,
\eea
where $M_{d-1}$ is a manifold enclosing $x$, and $M_{d-1}'$ is a manifold which is continuously deformable to $M_{d-1}$, but which does not enclose $x$.
Pictorially, this result can be drawn as 
\bea
\begin{tikzpicture}[baseline=0]
  \shade[ball color = gray, opacity = 0.4] (0,0) circle (1cm);
 \draw[black] (0,0) circle (1cm);
 \draw[black] (-1,0) arc (180:360:1 and 0.2);
\draw[black,dashed] (1,0) arc (0:180:1 and 0.2);

\draw[black, fill=black](0,0) circle (.5ex); 
\node at (0,0.3) {$\cO_q$};
\node[left] at (-0.8,-0.6) {$U_\alpha$};

\end{tikzpicture}
\hspace{0.3 in}= \hspace{0.2 in} e^{i q \alpha}  \,\,\,
\begin{tikzpicture}[baseline=0]
\draw[black, fill=black](0,0) circle (.5ex); 
\node at (0,0.4) {$ \cO_q$};
\end{tikzpicture}
\eea
The symmetry generators $U_\alpha(M_{d-1})$ are our first examples of topological defects. 
Since everything is Abelian, it is easy to see that we have the following multiplication structure, 
\bea
U_{\alpha_1} (M_{d-1}) \times U_{\alpha_2} (M_{d-1}) = U_{\alpha_1+ \alpha_2} (M_{d-1}) ~, 
\eea
which is the usual group structure for $U(1)$. Since each element has an inverse, these defects are referred to as \textit{invertible}.

\paragraph{Example:} As a concrete example, consider a massless complex scalar field in $(1+1)$-dimensions. As we mentioned before,  there is a $U(1)$ global symmetry acting as $\phi \rightarrow e^{i \alpha} \phi$, and the Noether current is computed via (\ref{eq:Noethercurrentdef}) to be
\bea
j_\m = {{i\over 2}} \left[ \phi \p_\m \phi^* - \phi^* \p_\m \phi\right] ~. 
\eea
If we take the space  to be $S^1$ and quantize using the boundary conditions 
\bea
\phi(t, x + 2 \pi ) = \phi(t,x)~,
\eea
then we get a Hilbert space $\cH(S^1)$, and $U_\alpha(S^1) = e^{-i \alpha \int_{S^1} j_0}$ acts as a unitary operator on this Hilbert space. 
On the other hand, if we insert  $U_\alpha$ along the time direction, then the boundary conditions for the scalar field are modified to 
\bea
\phi(t, x + 2 \pi ) = e^{i \alpha}\phi(t,x)~,
\eea
and quantizing using these boundary conditions gives a new Hilbert space $\cH_\alpha(S^1)$.

\subsection{Coupling a current to a gauge field}

Given a conserved current, we can couple it to a background gauge field by adding the following term, to first order in the background, to the action\footnote{\label{footnote:gauge} Note that $A$ is strictly speaking not a 1-form---it is a \textit{1-form gauge field}, which means that it is locally a 1-form subject to a redundancy $A \sim A + d \lambda$ such that $\lambda \sim \lambda + 2 \pi$, and that it must have properly quantized fluxes $\oint {dA \over 2 \pi} \in \ZZ$.  }
\bea
S[A] = \int d^d x \,  A_\m(x) j^\m(x)  = \int A \wedge * j~. 
\eea
At finite background field, further terms can be required for gauge invariance; for a charged scalar these include the term quadratic in $A$ from the covariant kinetic term. Note that this is the \textit{minimal} coupling of the gauge field to the current. We may also consider non-minimal couplings such as $F_{\m\n}Z^{[\m\n]}$, but these can be removed by changing the current via the improvement terms in (\ref{eq:improvementterms1}).

The gauge field has the property that we may shift $A_\m \rightarrow A_\m + \p_\m \lambda(x)$, with $\lambda(x)$ a zero-form gauge field (i.e. a circle-valued field). For the linear coupling, current conservation gives
\bea
\delta_\lambda S[A] = \int d^d x \, \p_\m \lambda(x)\, j^\m(x) = - \int d^d x \, \lambda(x) \p_\m j^\m(x) = 0~. 
\eea
Off-shell gauge invariance of the full action also requires transforming the charged fields, with phase $e^{-i q_a\lambda}$ for $A\to A+d\lambda$ in the present convention. Background gauge fields are an extremely useful tool in the study of symmetries, especially in the quantum setting. Indeed, the question of whether or not a symmetry is present (i.e. whether or not a current $j_\mu$ is conserved) can be translated into the question of whether or not the path-integral is invariant under gauge transformations of the background gauge field. This will be important for understanding anomalies, as will be discussed later. 

Note that by ``background'' gauge field, we mean that the gauge field is non-dynamical, which means that it is not integrated over in the path integral. Coupling a theory to a background gauge field is \textit{not} the same as gauging the symmetry.\footnote{In condensed matter, it is sometimes referred to as \textit{weakly gauging}, but this terminology is not common in high energy theory. } To gauge a symmetry, we must couple to a background gauge field, and then make the gauge field \textit{dynamical}. We will adopt the convention that background gauge fields are written with upper case letters $A,B, \dots$, while dynamical gauge fields are written with lower case letters $a,b,\dots$. 

In fact, the background gauge field $A_\m(x)$ is related to the symmetry defects discussed before in the following way. 
Note that inserting the defect $U_{\alpha} (M_{d-1}) = e^{i \alpha \int_{M_{d-1}} ds^\m j_\m}$ into the path integral is equivalent to coupling a background gauge field of the form 
\bea
\label{eq:AmPD}
A = \alpha \, \omega_{M_{d-1}}~, 
\eea
where $\omega_{M_{d-1}}$ is the Poincar{\'e} dual of $M_{d-1}$, i.e. it is a 1-form such that $\int_{M_d} \omega_{M_{d-1}} \wedge \bullet = \int_{M_{d-1}} \bullet$. 
This is a singular gauge configuration, since $\omega_{M_{d-1}}$ is basically a delta-function localizing on $M_{d-1}$. 

For continuous symmetries such as the $U(1)$ case that we are currently considering, one may in general consider smooth gauge fields, which can be thought of as some sort of smearing out of symmetry generators. Specifying the gauge field requires more than just the data of the symmetry generators though---one also needs the smearing function. In contrast, when we discuss discrete symmetries later on, we will see that there are no smooth gauge fields in the first place, and specification of the background gauge field is completely equivalent to specifying the symmetry generators inserted on the manifold. 

Let us now use the background gauge field to reproduce the action of the symmetry generators on charged operators. We insert a symmetry generator $U_{\alpha} (M_{d-1}) $ and consider moving the operator $\cO_q$ from a point $x_i$ on one side to a point $x_f$ on the other along a path $\gamma$. Doing so leads to an Aharonov-Bohm phase, 
\bea
\cO_q(x_i) \,\rightarrow\, e^{i q \int_\gamma dx^\m A_\m} \cO_q(x_f)~. 
\eea
Plugging in (\ref{eq:AmPD}) and using the fact that $\gamma$ intersects $M_{d-1}$ at a single point, we then have 
\bea
\cO_q(x_i) \,\rightarrow\, e^{i q \alpha} \cO_q(x_f)~,
\eea
which is the expected symmetry action.

\subsection{One-form symmetries}
\label{sec:oneform}

Thus far, our current $j_\m$ had one tensor index. Next let us imagine that we had an anti-symmetric 2-tensor $j_{\m \n}$. It may not be obvious that such a thing is sensible, but we will see that it is, and will give various concrete examples later on. 
In this case the conservation law would look like 
\bea
\p^\m j_{\m\n} = 0~,
\eea 
or in non-relativistic notation,\footnote{Note that from the non-relativistic point of view, the first of these looks like a conservation equation, while the second looks like an arbitrary constraint. If we are willing to get rid of Lorentz invariance, we could indeed relax the second constraint, which gives rise to so-called \textit{vector symmetry} \cite{Seiberg:2019vrp}. In these lectures we will restrict ourselves to relativistic systems. Let us also mention that we could relax the condition that the indices of $j_{\m\n}$ be anti-symmetric, which would give a more general notion of \textit{higher-spin} symmetries. An example would be the energy-momentum tensor $T_{\m\n}$, whose corresponding charge is the 4-momentum $p_\m$.}
\bea
\p_0 j_{0i} - \p_j j_{ji} &=& 0~, \hspace{0.5 in} \n = i
\no\\
\p_i j_{i0} &=& 0~, \hspace{0.5 in} \n = 0~. 
\eea
In differential form notation, it is simply $d * j = 0$, where now $j$ is a 2-form. 
As before, we are free to introduce improvement terms, 
\bea
j_{0j} \rightarrow j_{0j} + \p_i X_{[ij]}~, \hspace{0.3 in} j_{ij} \rightarrow j_{ij} + \p_0 X_{[ij]} + \p_k Z^{[kij]}~.
\eea
In terms of these currents, we may now try to define two notions of charge, 
\begin{enumerate}
\item \textbf{Vector charge:} We begin by defining charge as a vector,
\bea
Q_i := \int_{\mathrm{space}} d^{d-1}x \, j_{0i}(x)~. 
\eea
This is conserved since 
\bea
\p_0 Q_i= \int_{\mathrm{space}} d^{d-1}x \, \p_0 j_{0i} = \int_{\mathrm{space}} d^{d-1}x \, \p_j j_{ji} = 0~. 
\eea
More generally, we can define $Q_i(M_{d-1})$, which depends topologically on $M_{d-1}$. 

\item \textbf{Scalar charge:} Alternatively, we can define a scalar charge 
\bea
Q(M_{d-2}) := \int_{M_{d-2}} ds^{\m\n} j_{\m\n}~.
\eea
Here $M_{d-2}$ is a codimension-2 manifold, and $ds^{\m\n} = \widehat{n}_1^\m \widehat{n}_2^\nu\, d^{d-2}x$ is its normal element.
The scalar charge is again topological, i.e.
\bea
Q(M_{d-2}) - Q(M'_{d-2}) = \int_{N_{d-1}} ds^\n \, \p_\m j_{\m\n} = 0~
\eea
where $M_{d-2}$ and $M'_{d-2}$ are bordant within spacetime away from charged insertions and $\p N_{d-1} = M_{d-2} \sqcup \overline{M}'_{d-2}$. 
\end{enumerate}

In these lectures, we will be concerned primarily with the scalar charges. Unlike the charges for $U(1)$ zero-form symmetries discussed in the previous sections, which were defined on codimension-1 submanifolds $M_{d-1}$, the scalar charges $Q(M_{d-2})$ appearing here are defined on codimension-2 submanifolds. The corresponding symmetry generators are defined as before, 
\bea
U_\alpha(M_{d-2}) := e^{i \alpha Q(M_{d-2})}~. 
\eea
However, note that we can no longer define the action of $U_\alpha(M_{d-2}) $ on local operators, since a codimension-2 operator cannot surround a point in a topologically non-trivial manner. For example, in three dimensions $M_{d-2}$ is a line, but a line cannot non-trivially wrap a point.
Instead, the things which can link non-trivially with codimension-2 surfaces are \textit{lines}, and hence the objects which are charged under these symmetries are line operators. 

Let us make this more explicit. By similar methods as before, one can obtain a Ward identity involving $j_{\m\n}$ and a charge $q$ line operator $L_q(\gamma)$ supported on a line $\gamma$, 
\bea
\langle \p^\m j_{\m \n}(x) L_q(\gamma) \dots \rangle = q \,\delta^{(d-1)}(x \in \gamma) \,\,\widehat v_\nu \,\langle L_q(\gamma) \dots \rangle ~,
\eea
where $\widehat v$ is a tangent vector to $\gamma$. The analog of (\ref{eq:zeroformchargecomputation}) is then to consider two manifolds $M_{d-2}$ and $M_{d-2}'$, the former of which links with $\gamma$ and the second of which does not, and then compute 
\bea
 U_\alpha(M_{d-2}) L_q(\gamma) U_\alpha(M'_{d-2})^{-1}= e^{i \alpha q}  L_q(\gamma)  ~.
\eea 
The computation is identical to what was done before. There is also a more direct path-integral derivation of this result, which we will give in an explicit example below. 

\paragraph{Example:} Consider $(3+1)$-dimensional Maxwell theory,\footnote{As an aside, note that the coupling $e$ is physically meaningful and cannot just be absorbed into redefinition of the integration variable $a$ because $a$ is constrained by the flux quantization condition $\oint {f \over 2 \pi}\in \ZZ$; rescaling $a$ would move $e$ into the quantization condition, but will not get rid of it completely. In this sense, $e$ is similar to the radius of the compact boson.}
\bea
S ={1\over 2 e^2} \int  f \wedge * f= {1 \over 4 e^2} \int d^4 x \, f_{\m\n} f^{\m\n} ~, 
\eea
where $f=da$ is the field strength. We write everything in lower case since they are dynamical here. We can define two currents 
\bea
j_{\m\n}^E = {1\over e^2} f_{\m\n}~, \hspace{0.5in} j_{\m\n}^M = {1\over 4 \pi} \eps_{\m\n\rho\sigma} f^{\rho \sigma} ~,
\eea
or in differential form notation 
\bea
\label{eq:Maxwellcurrents}
j^E = {1\over e^2} f~, \hspace{0.5in} j^M = {1\over 2 \pi} * f~.
\eea
Both are conserved---indeed, we have 
\bea
d* j^E = {1\over e^2}  d * f = 0~, \hspace{0.5 in} d* j^M = -{1\over 2 \pi} d f = 0~,
\eea
where the first follows from the equations of motion and the second follows from the Bianchi identity. The symmetry operators live on codimension-2 surfaces $M_2$, with
\bea
U_\alpha^E(M_2) = e^{i \alpha \int_{M_2}* j^E}~, \hspace{0.5 in} U_\alpha^M(M_2) = e^{i \alpha \int_{M_2} * j^M}~. 
\eea
We denote the corresponding symmetries by $U(1)_E^{(1)}$ and $U(1)_M^{(1)}$ and refer to them as the ``electric'' and ``magnetic'' one-form symmetries.  These symmetries act on Wilson and `t Hooft lines of the theory via linking, pictorially 
\bea
 \begin{tikzpicture}[baseline = -5]
\draw[blue,thick] (0,-1.2)--(0,0);
\draw [white!30, line width=3pt, distance = 0.3 in] (0,0) ellipse (1cm and 0.4cm);
\draw[red,thick] (0,0) ellipse (1cm and 0.4cm);
\draw [white!30, line width=3pt, distance = 0.3 in] (0,0)-- (0,1.2);
\draw[blue,thick] (0,0)-- (0,1.2);

\node[below] at (0,-1.2) {$W_q$};
\node[left] at (-1,-0.2) {$U^E_\alpha$};
    \end{tikzpicture} 
     \hspace{0.1 in}= e^{iq \alpha}     \hspace{0.1 in}\begin{tikzpicture}[baseline = -5]
\draw[blue,thick] (0,-1.2)--(0,0);
\draw[blue,thick] (0,0)-- (0,1.2);

\node[below] at (0,-1.2) {$W_q$};
    \end{tikzpicture} ~, \hspace{0.5 in} 
 \begin{tikzpicture}[baseline = -5]
\draw[dgreen,thick] (0,-1.2)--(0,0);
\draw [white!30, line width=3pt, distance = 0.3 in] (0,0) ellipse (1cm and 0.4cm);
\draw[red,thick] (0,0) ellipse (1cm and 0.4cm);
\draw [white!30, line width=3pt, distance = 0.3 in] (0,0)-- (0,1.2);
\draw[dgreen,thick] (0,0)-- (0,1.2);

\node[below] at (0,-1.2) {$T_m$};
\node[left] at (-1,-0.2) {$U^M_\alpha$};
    \end{tikzpicture} 
     \hspace{0.1 in}= e^{im \alpha}     \hspace{0.1 in}\begin{tikzpicture}[baseline = -5]
\draw[dgreen,thick] (0,-1.2)--(0,0);
\draw[dgreen,thick] (0,0)-- (0,1.2);

\node[below] at (0,-1.2) {$T_m$};
    \end{tikzpicture} ~. 
\eea

Let us analyze $U(1)_E^{(1)}$ in more detail. Define the following Wilson line 
\bea
W_q(\gamma) := e^{i q \oint_\gamma a}~.
\eea
Its charge is measured by its linking with $U_\alpha^E(M_2) $, such that we have 
\bea
U_\alpha^E(M_{2}) W_q(\gamma) U_\alpha^E(M'_{2})^{-1} = e^{i \alpha q\, \mathrm{Link}(M_{2},\gamma)} W_q(\gamma)~,
\eea
when $M_{2}$ links $\mathrm{Link}(M_{2},\gamma)$ times with $\gamma$, and $M'_{2}$  links none.
We may verify this action from the path integral as follows. Inserting the Wilson line in the path integral gives
\bea
\int \cD a \, W_q(\gamma)\, e^{{i \over 2 e^2 }\int_{M_4} f \wedge * f} = \int \cD a \, e^{ \int_{M_4} i  q \omega_\gamma \wedge a + {i \over 2 e^2 }f \wedge * f} ~, 
\eea
from which we see that the equations of motion are modified to
\bea
d * f =  q e^2 \omega_\gamma~. 
\eea
Then plugging this into 
\bea
U_\alpha^E(M_{2})  W_q(\gamma)  U_\alpha^E(M'_{2})^{-1}  = e^{{i\alpha} \int_{N_3} d* j^E}  W_q(\gamma) ~,
\eea
for $\p N_3 = M_2 \sqcup \overline M_2'$, the right-hand side becomes
\bea
e^{{i \alpha \over e^2} \int_{N_3} d* f} W_q(\gamma) = e^{{i q \alpha} \int_{N_3} \omega_\gamma}  W_q(\gamma) = e^{i q \alpha\, \mathrm{Link}(M_{2}, \gamma)} W_q(\gamma) ~,
\eea
as expected.

On the other hand, the magnetic one-form symmetry $U(1)_M^{(1)}$ is more difficult to understand in the current presentation. This is because the `t Hooft lines $T_m$ that are charged under it are not easily expressible in terms of $a$. Instead, here it is useful to pass to the electromagnetically dual description by defining the dual photon $\widehat a$ by $* da \propto d \widehat a$, in terms of which we can write the charge $m$ `t Hooft line as 
\bea
T_m(\gamma) = e^{i m \oint_\gamma \widehat a}~. 
\eea
In this dual description, effectively the same results as for $U(1)_E^{(1)}$ hold for $U(1)_M^{(1)}$. 
\newline

\begin{tcolorbox}
\textbf{Exercise:}  Thus far, we have imposed the Bianchi identity $df=0$ locally via $f=da$, while flux quantization follows from taking $a$ to be a compact $U(1)$ gauge field.
An alternative approach is to take $f$ to be an unconstrained 2-form, and to impose the closedness and quantization via a Lagrange multiplier. 
In other words, we introduce a new dynamical field $\widehat a$ which couples as 
\bea
S = \int  {1\over 2 e^2} f \wedge * f + {1\over 2 \pi} f \wedge d \widehat a~. \no
\eea
Integrating over local fluctuations of $\widehat a$ imposes $d f = 0$, which gives the desired Bianchi identity.
To furthermore get quantization of $f$, we take $\widehat a$ to be not a  1-form, but rather a \textit{1-form gauge field}; see footnote \ref{footnote:gauge}. 
Explain why this leads to quantization of $f$. (\textit{Hint: sum over fluxes of $\widehat a$}). 
Next, integrate out $f$ to obtain a theory involving only $\widehat a$. What happens to the coupling constant when we do so? 
This is known as \textit{electromagnetic-} or \textit{S-duality}.
\end{tcolorbox}

\subsection{Higher-form symmetries} 
\label{sec:higherform}

The one-form symmetries discussed above are the simplest example of so-called \textit{higher-form symmetries}. 
In particular, a continuous \textit{$p$-form global symmetry} has a conserved current which is a $(p+1)$-index antisymmetric tensor $j_{\m_1\dots \m_{p+1}}$ satisfying a conservation equation 
\bea
\p^{\m_1} j_{\m_1\dots \m_{p+1}} = 0 \hspace{0.3 in} \Leftrightarrow \hspace{0.3 in} d * j = 0~. 
\eea
In the same way as before, we define a $(d-p-1)$-dimensional topological symmetry generator 
\bea
U_\alpha(M_{d-p-1}) := e^{i \alpha \int_{M_{d-p-1}} * j }~, 
\eea
and the fact that $*j$ is closed implies that the symmetry generator is topological. These symmetry generators can link with $p$-dimensional objects $W_q(\Gamma_p)$, and from the Ward identity 
\bea
\langle d * j_{p+1}(x) W_q(\Gamma_p)  \dots \rangle= q \,\delta^{(d-p)}(x \in \Gamma_p) \, \langle W_q(\Gamma_p)\dots \rangle~
\eea
we obtain the following action
\bea
U_\alpha(M_{d-p-1}) W_q(\Gamma_p) U_{-\alpha}(M'_{d-p-1}) = e^{i q \alpha\,  \mathrm{Link}(M,\Gamma)} W_q(\Gamma_p)~. 
\eea
The dimensions of the various ingredients are summarized in Table \ref{tab:dimensions}.

\begin{table}[tp]
\begin{center}
\begin{tabular}{c|c|c|c}
& degree of current & dimension of symmetry generator & dimension of charged object
\\\hline
$0$-form & $1$ & $d-1$ & $0$
\\
$1$-form & $2$ & $d-2$ & $1$
\\
$p$-form & $p+1$ & $d-p-1$ & $p$
\end{tabular}
\end{center}
\caption{The degrees and dimensions of objects involved in higher-form symmetries. }
\label{tab:dimensions}
\end{table}%

Note that, as for standard currents, we may also couple higher-form currents to background gauge fields. Indeed, a $p$-form global symmetry can be coupled to a $(p+1)$-form background gauge field $A_{p+1}$ via
\bea
S_p = \int A_{p+1} \wedge * j_{p+1}~.
\eea
In this case the gauge transformation parameter of $A_{p+1}$ is a $p$-form gauge field,\footnote{A $p$-form gauge field $\Lambda_p$ is defined recursively as being locally a $p$-form, subject to the gauge redundancy $\Lambda_p \sim \Lambda_p + d \lambda_{p-1}$ with $\lambda_{p-1}$ a $(p-1)$-form gauge field, and having properly quantized fluxes $\oint {d \Lambda_p \over 2 \pi } \in \ZZ$.  } 
\bea
A_{p+1} \rightarrow A_{p+1} + d \Lambda_p~. 
\eea
Conservation of the current implies invariance of the path integral under these background gauge transformations. 

Before moving on to an example, let us make an important comment about $p$-form symmetries for $p \geq 1$. Whereas 0-form symmetries can be either Abelian or non-Abelian, higher-form symmetries \textit{must} be Abelian. The reason is that since the symmetry generators have codimension greater than one, there is no sense of ordering---the configurations $\langle U_{g_1}(M_{d-p-1}) U_{g_2}(M'_{d-p-1})\dots \rangle$ and $\langle U_{g_2}(M'_{d-p-1}) U_{g_1}(M_{d-p-1})\dots \rangle$ are topologically identical. 

\paragraph{Example:} Let us now give a concrete example of a $p$-form symmetry for $p>1$. 
Consider a compact boson in $(3+1)$-dimensions,
\bea
S = {R^2 \over 4 \pi} \int d\phi \wedge * d \phi~, 
\eea
where $\phi \sim \phi + 2 \pi$. In this theory, we may define the following currents 
\bea
j_s = {R^2\over 2 \pi} d \phi ~, \hspace{0.5 in}  j_{w} = {1\over 2 \pi }* d \phi~.
\eea
The first of these is a usual 1-form current, which is associated to the zero-form shift symmetry of $\phi$, i.e. $\phi \rightarrow \phi + \alpha$ for any constant $\alpha$. The fact that it is conserved follows from the equations of motion $d * d \phi = 0$. On the other hand, $j_w$ is conserved by the Bianchi identity, and is a 3-form current. As such, the corresponding symmetry is a 2-form symmetry, whose interpretation is as the \textit{winding} symmetry for the compact boson. 

The symmetry generators corresponding to the two symmetries are 
\bea
U(1)_s^{(0)}: \hspace{0.3 in} U^s_\alpha(M_3) &:=& e^{{i R^2 \alpha \over 2 \pi}\int_{M_3} {*d \phi} }~,
\no\\
U(1)_w^{(2)}: \hspace{0.3 in} U^w_\alpha(M_1) &:=& e^{{i \alpha\over 2 \pi} \int_{M_1} {d \phi } }~,
\eea
and the charged objects with respect to these symmetries are the following 
\bea
&U(1)_s^{(0)}:& \hspace{0.3 in}\cO_q(x) = e^{i q \phi(x)}~,
\no\\
&U(1)_w^{(2)}:& \hspace{0.3 in} W_n(\Gamma_2)~.
\eea
Here $\cO_q(x)$ is a local operator, while $W_n(\Gamma_2)$ is a vortex operator living on the 2-cycle $\Gamma_2$.\footnote{The vortex operator is defined by cutting out a tubular neighborhood of $\Gamma_2$ and imposing boundary conditions such that $\phi$ winds $n$ times around the excised disc, i.e. $\phi(x + 2 \pi) = \phi(x) + 2 \pi n $. } 
\newline

\begin{tcolorbox}
\textbf{Exercise:}
Unlike in $(3+1)$-dimensions, in $(1+1)$-dimensions the momentum and winding symmetries of the compact boson are both zero-form symmetries. Rewrite the action of the $(1+1)$-dimensional massless compact boson in terms of $f = d \phi$, and then treat $f$ as an unconstrained 1-form, enforcing the closedness and quantization via a Lagrange multiplier field $\widehat \phi$. Show that upon integrating out $f$, one obtains a theory of only $\widehat \phi$, which is a compact boson at the inverse radius $1/R$. This is known as \textit{T-duality}. It is the analog of electromagnetic duality mentioned above. 
\newline 

 More generally, consider a theory of $d$ massless compact bosons $\phi^\mu$ for $\mu = 1, \dots, d$ each of period $2\pi$, with action  
\bea
\label{eq:dcompactbos}
S = {1\over 4\pi}\int G_{\m\n}  d \phi^\m  \wedge * d\phi^\n - i B_{\m\n} d \phi^\m \wedge d \phi^\n ~. 
\eea
Repeat the above steps to obtain a dual theory 
\bea
S = {1\over 4\pi}\int \widehat G_{\m\n} d\widehat  \phi^\m \wedge * d \widehat \phi^\n - i \widehat B_{\m\n} d \widehat \phi^\m \wedge d \widehat \phi^\n~,\no
\eea
 where 
\bea
\widehat G = (G- B G^{-1} B)^{-1} ~, \hspace{0.5 in} \widehat B = - \widehat G B G^{-1}~. \no
\eea
This is a more general version of T-duality.
\end{tcolorbox}

\subsection{Anomalies}
\label{sec:anomaly}

We next briefly review the notion of an anomaly. Given a theory with a $p$-form global symmetry, we may couple the theory to a background gauge field $A_{p+1}$, in which case the partition function becomes a functional of the background gauge field, 
\bea
Z[A_{p+1}] = \int \cD \phi^a\, e^{i S [\phi^a, A_{p+1}]}~. 
\eea
If the current corresponding to the $p$-form symmetry is classically conserved, but the partition function is not invariant under gauge transformations of $A_{p+1}$, and this non-invariance cannot be removed by a local counterterm, then we say that the symmetry is \textit{anomalous}. In particular, we have 
\bea
Z[A_{p+1} + d \lambda_p] = e^{-i \int \cA[A_{p+1}, \lambda_p]} \times Z[A_{p+1}]~. 
\eea
for some $\cA[A_{p+1}, \lambda_p]$ such that $\cA[A_{p+1}, 0]=0$. 

One way to phrase this phase ambiguity is to say that $Z[A_{p+1}]$ is a section of a complex line bundle, or a vector in a one-dimensional complex vector space without canonical basis. Recall that for a one-dimensional complex vector space $V$, each element $v \in V$ is not just a number---to extract a number from $v$, we need to choose a basis vector $b$, at which point we can define $v/b \in \CC$. But a different choice of basis vector will give a different choice of number, and the two are related by ${v\over b} =\left( {b' \over b}\right)  {v \over b'}$ where $b'/b$ is the analog of our anomalous phase.
As such, instead of thinking of the partition function as a number $Z[A_{p+1}] \in \CC$, we should think of it as an element $Z[A_{p+1}] \in V(M_d)$, where $V(M_d)$ is a one-dimensional vector space associated with our spacetime $M_d$ and its background fields. 

This point of view gives rise to the following useful construction (see e.g. \cite{TachikawaTASI,Freed:2014iua,Freed:2016rqq,Monnier:2019ytc}). We would like to associate a one-dimensional vector space to any closed $d$-dimensional manifold. Recall that a $d$-dimensional QFT associates a vector space (i.e. the Hilbert space) to each $(d-1)$-dimensional manifold, and hence, when a compatible filling exists, we can consider $M_d$ as the boundary of a $(d+1)$-dimensional QFT $A$ on $N_{d+1}$, where $\p N_{d+1} = M_d$. The $(d+1)$-dimensional QFT $A$ must have the special property that its Hilbert space $\cH_A$ is one-dimensional---such an $A$ is called an \textit{invertible} theory.\footnote{If the Hilbert space is multi-dimensional, the theory on the boundary is not just anomalous, but \textit{relative} \cite{Freed:2012bs}.} We then have that $Z[A_{p+1}] \in \cH_A(M_d)$. The different choices of anomaly are captured by different choices of invertible field theory $A$ in $(d+1)$ dimensions. This picture of anomalies as being captured by invertible theories in one higher dimension is known as \textit{anomaly inflow} \cite{Faddeev:1984ung,Callan:1984sa}. Note that the coupled bulk-boundary system has no phase ambiguities, i.e. computing the partition function of the $(d+1)$-dimensional system on the manifold $N_{d+1}$ with boundary leads to a well-defined number. 

Slightly more explicitly, if we can write the total derivative of the anomalous phase $\cA[A_{p+1}, \lambda_p]$ as the variation of some $(d+1)$-form depending on the background gauge field $A_{p+1}$, 
\bea
d \cA[A_{p+1}, \lambda_p] = \delta_{\lambda_{p}} \widehat \cA[A_{p+1}]~,
\eea
then we can consider the following modified partition function 
\bea
\widehat Z[A_{p+1}] = Z[A_{p+1}] \times e^{i \int_{N_{d+1} }\widehat \cA [A_{p+1}]}~, 
\eea
and this is now gauge invariant. We can thus think of $\widehat \cA [A_{p+1}]$ as the action for the invertible field theory $A$ in $(d+1)$-dimensions.

We caution that there are two, very distinct notions of anomalies in the literature: ones that vanish when the background fields are turned off (so-called \textit{`t Hooft anomalies}), and ones which do not (ABJ anomalies of global currents involving dynamical gauge fields). In the former case, the anomaly is a useful property of the symmetry, which has various nice features like anomaly matching and anomaly inflow. In the latter case, the anomaly means that the symmetry is simply {broken}.\footnote{Though, as we will see later, the symmetry can sometimes be resurrected as a non-invertible symmetry.} In the modern parlance, when one refers to an anomaly without any qualifiers, one is almost always referring to the former, i.e. to an `t Hooft anomaly. However, both concepts are very important, and will play an important role in these lectures. 

Note that it is also possible that we can have anomalies that appear only when {multiple different} background gauge fields are turned on. In this case we say we have a \textit{mixed `t Hooft anomaly} between the corresponding symmetries. It is easiest to illustrate this by means of an example. 

\paragraph{Example 1:} In Section \ref{sec:oneform}, we saw that $(3+1)$-dimensional Maxwell theory had two one-form symmetries, with currents as below, 
\bea
j^E = {1 \over e^2} f ~, \hspace{0.5 in} j^M = {1\over 2 \pi} * f~. 
 \eea
 Let us now consider coupling these currents to background gauge fields by adding, 
 \bea
 S_\mathrm{coupling}[B_E, B_M] &=&  \int {1\over e^2} B_E \wedge * f + {1\over 2 \pi} B_M \wedge f
 \no\\
 &=& \int {1\over 2e^2} (B_E \wedge * f + f \wedge * B_E ) + {1\over 2 \pi} B_M \wedge f~.
 \eea
 We are also free to add any \textit{local counterterms} (i.e. terms built out of only background fields) to the action---for example, we can add 
 \bea
 S_\mathrm{l.c.} [B_E]  = {1\over 2e^2}  \int  B_E \wedge * B_E~.
 \eea
 In total then, our action becomes 
 \bea
 \label{eq:Maxwellcoupledtobkg}
 S[B_E, B_M] = \int {1\over 2e^2 }(f + B_E) \wedge * (f+B_E) + {1 \over 2 \pi} f \wedge B_M~. 
 \eea
 This action is clearly invariant under background gauge transformations  $B_M \rightarrow B_M + d \lambda_M$, since both $a$ and $\lambda_M$ have quantized fluxes, and hence ${1\over 2\pi}\int f \wedge d \lambda_M \in 2 \pi \ZZ$. Indeed, $f/(2\pi)$ and $d\lambda_M/(2\pi)$ represent integral cohomology classes, whose cup product has an integer pairing with the closed oriented spacetime.
 On the other hand, consider background gauge transformations $B_E \rightarrow B_E + d \lambda_E$. 
 In the first term, we can absorb this transformation by redefining $f \rightarrow f' = f + d \lambda_E$.\footnote{Note that as long as $\lambda_E$ is a proper 1-form gauge field, this does not spoil the quantization of $f'$.} However, due to the second term, doing so gives rise to 
 \bea
 \label{eq:BEvariation}
 \delta_{\lambda_E} S[B_E, B_M]  = - {1 \over 2 \pi}\int d \lambda_E \wedge B_M~.
 \eea
This does not in general take values in $2 \pi \ZZ$, since the periods of $B_M$ need not be quantized. Thus we see that when $B_M$ is turned on, we lose gauge invariance of $B_E$. Of course, we could always try to fix this non-invariance by adding in different local counterterms. For example, we could modify the action to 
  \bea
 S'[B_E, B_M] = \int {1\over 2e^2 }(f + B_E) \wedge * (f+B_E) + {1 \over 2 \pi} (f+B_E) \wedge B_M~. 
 \eea
 Doing so leads to invariance under background gauge transformations of $B_E$, but now we lose invariance under $B_M \rightarrow B_M + d \lambda_M$,
 \bea
\delta_{\lambda_M} S'[B_E, B_M] = {1\over 2 \pi} \int B_E \wedge d \lambda_M~. 
 \eea
 This is the hallmark of a mixed `t Hooft anomaly---we can use local counterterms to shift the anomaly from one symmetry to another, but we can never get rid of it completely. The conclusion is that $U(1)_E^{(1)}$ and $U(1)_M^{(1)}$ have a mixed `t Hooft anomaly. 
 
While we cannot use counterterms to get rid of the anomaly, note that we \textit{can} get rid of it by introducing the 5d inflow term 
\bea
\label{eq:Maxwellinflowact}
\widehat \cA = {1 \over 2 \pi } \int_{M_5} B_E \wedge d B_M~,\hspace{0.5 in} \p M_5 = M_4~.
\eea
Indeed, under $B_E \rightarrow B_E + d \lambda_E$ we have 
 \bea
 \delta_{\lambda_E}\widehat \cA  = {1 \over 2 \pi } \int_{M_5} d \lambda_E \wedge d B_M =  {1 \over 2 \pi } \int_{M_5} d (d \lambda_E \wedge B_M) = {1 \over 2 \pi } \int_{M_4} d \lambda_E \wedge  B_M~, 
 \eea
 which precisely cancels the variation in (\ref{eq:BEvariation}), while the gauge invariance under gauge transformations of $B_M$ is also preserved. The catch is that the coupled theory now depends on the auxiliary manifold $M_5$.   
 
 \paragraph{Example 2:} We next give the most familiar example of an anomaly of a global current, namely the \textit{ABJ anomaly} of massless QED. By massless QED, we mean Maxwell theory coupled to a charge-1 massless Dirac fermion, 
\bea
\cL = {1\over 4 e^2 } f_{\m\n} f^{\m\n} + i \overline \Psi (\p_\m - i a_\m) \g^\m \Psi~.
\eea
Classically, there is a $U(1)_A$ axial global symmetry that acts on the fermion as\footnote{The normalization in the exponent is chosen such that $\alpha \sim \alpha+2 \pi$. Indeed, for $\alpha = 2 \pi$ we have $\Psi \rightarrow e^{i \pi \gamma_5} \Psi = - \Psi$, which is already part of the $U(1)$ gauge group. \label{footnote:axial}}
\bea
\Psi \rightarrow e^{i \alpha \gamma_5 /2}\Psi~,
\eea
with the conventional axial current, twice the current normalized to generate the parameter $\alpha$ above,
\bea
j_\mu^A = \overline \Psi \gamma_5 \gamma_\mu \Psi~. 
\eea
By using e.g. the Fujikawa method described in any standard textbook on QFT, one can show that this current is not conserved, and we instead have 
\bea
d * j^A = {1\over 4 \pi^2} f \wedge f~. 
\eea
Clearly the right-hand side does not vanish even when background fields are set to zero---it depends on the dynamical field $f$, and hence indicates that the $U(1)_A$ axial symmetry is broken by a local nonconservation equation. In Section \ref{sec:masslessQED}, we will see that part of this broken symmetry can actually be resurrected as a non-invertible symmetry.

Note that ABJ anomalies can be thought of as originating from a mixed `t Hooft anomaly in the ungauged theory (i.e. the theory of just a massless Dirac fermion). 
Gauging of $U(1)_V$ turns the corresponding background gauge field into a dynamical gauge field, and leads to breaking of the $U(1)_A$ symmetry. In general, when we have a mixed `t Hooft anomaly, we are allowed to gauge one of the symmetries 
(as long as there are no self-anomalies), and this will lead to breaking of the other symmetry.

\subsection{Discrete gauge theories}
\label{sec:discrete}

Thus far we have been focusing on continuous symmetries, but discrete symmetries also play an important role in the study of generalized symmetries. 
In particular, it will be important for us to understand how to couple such symmetries to gauge fields. 

To do so, let us first recall some facts about the description of continuous gauge fields. We proceed as follows,
\begin{enumerate}
\item Choose an open cover $\{U_i\}$ of the spacetime $M_d$.
\item Choose a trivialization (i.e. a local section) in each patch.
\item On overlaps, the trivializations are related by transition functions $g_{ij}: U_i \cap U_j \rightarrow G$.
\item On triple overlaps $U_i \cap U_j \cap U_k$, the $g_{ij}$ are required to satisfy $g_{ij} g_{jk} g_{ki} = 1$.
\end{enumerate}
To specify a gauge field, on each patch we choose a  Lie-algebra-valued connection $A^{(i)}$, and the connections on adjacent patches are related via 
\bea
A^{(j)} = g_{ij}^{-1} A^{(i)} g_{ij} + g_{ij}^{-1} d g_{ij}~. 
\eea
The gauge field contains important geometric data about the bundle, captured by the curvature $F=dA+A\wedge A$. It also determines holonomies around loops, which depend only on their homotopy classes when the connection is flat. 

For a discrete group $G$ on the other hand, there is no nonzero Lie algebra, and hence no local connection degrees of freedom or curvature. 
Indeed, discrete gauge fields must always be (locally) flat. One way to understand this is to note that the holonomy should depend continuously on the path $\gamma$, but since the holonomy is valued in a discrete group it cannot change continuously, and hence is expected to be constant upon continuous deformation of $\gamma$. This means that the holonomy of any contractible cycle is trivial, which expresses local flatness.

Thus in the discrete case, all of the data of the gauge field is contained in the holonomies around non-contractible cycles, and this data is in turn contained in the transition functions described above. As such, specifying the gauge field basically just amounts to specifying the transition functions, each of which we may imagine as a symmetry generator $U_{g_{ij}}(M_{d-1})$ implementing the corresponding $g_{ij}$.\footnote{Here we are considering a zero-form symmetry, so that the symmetry generators live on codimension-1 loci. For higher-form symmetries, analogous statements hold, see e.g. \cite{Kapustin:2014gua}. } 
In other words, in the discrete case, a background gauge field is equivalent to a fine mesh of topological defects. By ``fine mesh,'' we mean that we first choose a triangulation of the manifold, and then insert codimension-1 defects on the corresponding cells of the dual decomposition. Different triangulations give rise to different meshes, each of which are related by gauge transformations. 

\subsubsection{BF theory description}

In the case of an Abelian discrete group such as $\ZZ_n$, we may write the corresponding discrete gauge theory as follows,
\bea
\label{eq:BFaction}
S_{BF} = {n \over 2 \pi} \int_{M_d} a_{p+1} \wedge d b_{d-p-2}
\eea
with $a$ and $b$ respectively the compact $U(1)$ gauge fields for $p$- and $(d-p-3)$-form symmetries, whose field strengths have periods in $2\pi\ZZ$. 
If $b_{d-p-2}$ is dynamical, we claim that the above action makes $a_{p+1} $ into a  $\ZZ_n$-valued $(p+1)$-form gauge field.\footnote{We can take $a_{p+1} $ to be either background or dynamical. When it is dynamical, this action also turns $b_{d-p-2}$  into a  $\ZZ_n$-valued gauge field. }

First, let us check that $S_{BF}$ is invariant under the following gauge transformations, 
\bea
a_{p+1}  \rightarrow a_{p+1}  + d \lambda_{p}~, \hspace{0.5 in} b_{d-p-2}  \rightarrow  b_{d-p-2}  + d \widehat \lambda_{d-p-3}~. 
\eea
For example, the first gives 
\bea
\delta_\lambda S_{BF} = {n \over 2 \pi} \int_{M_d} d \lambda_{p} \wedge d b_{d-p-2}~,
\eea
and on closed, oriented $M_d$, the integral below is the pairing of integral cohomology classes,
\bea
\delta_\lambda S_{BF} = 2 \pi n \int_{M_d} {d \lambda_{p} \over 2 \pi} \wedge {d b_{d-p-2} \over 2 \pi} \in 2 \pi n \ZZ~.
\eea
Hence as long as $n \in \ZZ$, the theory is gauge invariant.

It is clear that the equation of motion of $ b_{d-p-2}$ imposes that $a_{p+1}$ is locally flat, i.e. $d a_{p+1} = 0$, as expected for a discrete gauge field. But we can actually say more. To illustrate the global constraint, consider one independent flux through $\Sigma_{d-p-1}$ and the holonomy on its dual cycle $\Sigma_{p+1}$. Note that the gauge field $ b_{d-p-2}$ may be split into different topological sectors labelled by $k : = \oint_{\Sigma_{d-p-1}} {d b_{d-p-2} \over 2 \pi } \in \ZZ$. 
In each topological sector, we may then write the gauge field as a sum over two pieces---a certain representative in that sector, together with an element in the trivial topological sector. 
Hence when we perform the path integral in this theory, we may split the path integral over the gauge field $ b_{d-p-2}$ into a sum over $k$, as well as a path integral over the fields in the zero-flux sector. 
Integrating over the zero-flux fluctuations imposes $d a_{p+1}=0$. The sum over the remaining integer flux then gives
\bea
\sum_{k\in\ZZ} e^{i n k \int_{\Sigma_{p+1}} a_{p+1}}
&=& 2\pi\sum_{k\in\ZZ}\delta\left(n \int_{\Sigma_{p+1}} a_{p+1}-2\pi k\right)~,
\eea
with an analogous constraint for each independent holonomy. 
The periodic delta function means that Wilson $(p+1)$-surfaces are forced to satisfy $W^n = 1$, which is exactly as expected for a $\ZZ_n$ gauge theory.

Note that in the special case of $p=1$ in $d=4$, it is conventional to write the 2-form gauge field $a_{2}$ as $b$ and the 2-form field strength $d b_{1}$ as $f$, in which case the action looks like 
\bea
S_{BF} = {n \over 2 \pi} \int_{M_4} b \wedge f~. 
\eea
For this reason, discrete gauge theories written in this way are known as \textit{BF theories}. They will play a role in several places below. 

\subsubsection{Topological boundary conditions for BF theory}
\label{sec:topbcs}
It will also be useful to understand the topological boundary conditions of BF theory. 
There are two obvious boundary conditions that one can consider,\footnote{More generally, one could fix the right-hand side to a non-trivial flat background field, but for simplicity we just set it to be zero.}
\bea
&\mathrm{Dirichlet:}& \,\,\, a_{p+1}|_\p = 0~, 
\no\\
&\mathrm{Neumann:}& \,\,\, b_{d-p-2}|_\p = 0~, 
\eea
where the notions of ``Dirichlet'' and ``Neumann'' depend on whether we are considering the theory as an $a_{p+1}$ gauge theory or a $ b_{d-p-2}$ gauge theory---here we are taking the former perspective.
That these are legitimate boundary conditions follows from the fact that
\bea
\delta S&=&  {n \over 2 \pi} \int_{M_d^{\geq 0} } \left[ \delta a_{p+1} \wedge d b_{d-p-2} +   a_{p+1} \wedge d (\delta b_{d-p-2} )  \right] 
\no\\
&=&  {n \over 2 \pi} \int_{M_d^{> 0}} \left[ \delta a_{p+1} \wedge d b_{d-p-2} - (-1)^{p+1} d a_{p+1}  \wedge \delta b_{d-p-2} \right] 
\no\\
&\vphantom{.}& \hspace{0.9 in}+ (-1)^{p+1} {n \over 2 \pi} \int_{M_{d-1}|_0} a_{p+1}  \wedge \delta  b_{d-p-2}~,
\eea
where the boundary term vanishes in both cases.

An important property of these boundary conditions is that they are \textit{topological}, in the sense that any infinitesimal change in the location of the boundary does not affect correlation functions. This may be shown by an argument similar to the proof of topologicalness of the symmetry generators in Section \ref{sec:topdefintro}. For example, for the Dirichlet boundary condition we imagine changing from boundary $M_{d-1}$ to some infinitesimally different $M'_{d-1}$ continuously along a direction $r$. Choosing a gauge in which the normal components of $a_{p+1}$ vanish near the boundary gives
\bea
a_{p+1}|_{M_{d-1}} - a_{p+1}|_{M'_{d-1}} = \int_{r} d a_{p+1} = 0~,
\eea
where we have used that $d a_{p+1} = 0$ locally by the equations of motion.

 In higher dimensions there are many additional topological boundary conditions.\footnote{In $d \geq 3+1$, there are an infinite number of simple topological boundary conditions that can be obtained by stacking with decoupled $(d-1)$-dimensional TQFTs. In $d = 2+1$ the situation is better, since stacking with non-trivial $(1+1)$-dimensional TQFTs gives non-simple boundary conditions. } Later, we will use the topological nature of the Dirichlet boundary condition to construct non-invertible defects by gauging a discrete symmetry in half of space. 

\subsubsection{Discrete cocycle description}

An alternative description of discrete gauge theory is to work directly with discrete $p$-cochains $a \in C^p(M_d, \ZZ_n)$, which satisfy $\oint a \in \ZZ_n$ by definition. 
As opposed to the case of continuous gauge fields, for which the Wilson $p$-surfaces were written as $e^{i \oint a}$, in this case it is more natural to normalize the Wilson $p$-surfaces as 
\bea
W(\Gamma_p) = e^{i {2 \pi \over n} \oint_{\Gamma_p}a}~;
\eea
this ensures that the $a \sim a + n$ redundancy in $\ZZ_n$ $p$-cochains does not lead to an ambiguity in $W$, i.e. $W(\Gamma_p) = e^{i {2 \pi \over n} \oint_{\Gamma_p}a} \cong e^{i {2 \pi \over n} \oint_{\Gamma_p}(a+n)} = e^{ 2 \pi i}e^{i {2 \pi \over n} \oint_{\Gamma_p}a} = W(\Gamma_p)$. 
More generally, given an expression written in terms of continuous gauge fields, we may obtain one in terms of discrete gauge fields by making the replacement 
\bea
a^\mathrm{cont.} \,\,\longrightarrow\,\, {2\pi \over n} a^\mathrm{disc.}~. 
\eea
Additionally, in the discrete notation the wedge product and exterior derivative are replaced by analogous operations,
\bea
\wedge \,\,\longrightarrow\,\, \cup~, \hspace{0.6in} d \,\,\longrightarrow\,\, \delta~.
\eea 
Thus, for example, in discrete notation the BF action in (\ref{eq:BFaction}) would take the following form,
\bea
S_{BF} = {2 \pi \over n} \int_{M_d} a_{p+1} \cup \delta b_{d-p-2}~.
\eea
One might think that the BF term is somewhat meaningless now that $a$ and $b$ are already assumed to be $\ZZ_n$-valued, but it still tells us that $\delta a_{p+1} = 0$, so that $a_{p+1}$ is not just a $\ZZ_n$-cochain, but a $\ZZ_n$-cocycle. 

Finally, let us mention that because the gauge fields are now discrete, all path integrals involving $a^\mathrm{cont.}$ are to be replaced by sums over $a^\mathrm{disc.}$. In particular, we would like to replace the integral over $a^\mathrm{cont.}$ with a sum over flat $a^\mathrm{disc.} \in Z^{p+1}(M_d, \ZZ_n)$, with $Z^{p+1}$ being the set of $(p+1)$-cocycles. However, note that $a$ is a cocycle subject to a gauge redundancy, $a_{p+1} \sim a_{p+1} + \delta \lambda_p$, and we must divide out by this redundancy. At first glance, $\lambda_p$ is an element of $C^p(M_d, \ZZ_n)$, i.e. it is a $p$-cochain, but note that $\lambda_p$ is itself subject to gauge transformations $\lambda_p \rightarrow \lambda_p + \delta \lambda_{p-1}$, with $\lambda_{p-1}$ an element of $C^{p-1}(M_d, \ZZ_n)$. Likewise, $ \lambda_{p-1}$ is subject to a gauge transformation in terms of an element of $C^{p-2}(M_d, \ZZ_n)$, and so on and so forth. The correct replacement is thus as follows, 
\bea
\int \cD a \hspace{0.2in} \longrightarrow \hspace{0.2 in} \prod_{i=1}^{p+1} |C^{p+1-i}(M_d, \ZZ_n)|^{(-)^i} \sum_{a \in Z^{p+1}(M_d, \ZZ_n)}  ~.
\eea
In fact, it is more common to rewrite the sum in a slightly different way. To do so, recall that we may define cohomology groups as 
\bea
H^{p}(M_d, \ZZ_n) = {Z^p (M_d, \ZZ_n) \over B^p(M_d, \ZZ_n)}~, 
\eea
where $B^p$ is the group of exact $p$-cochains. In particular, we have $|B^p| = |C^{p-1}| / |Z^{p-1}|$ and $|B^0|=1$, where for simplicity we dropped the arguments of the various groups involved. 
These then give 
\bea
|B^p| = {|C^{p-1}| \over |Z^{p-1}| }= {|C^{p-1}| \over |H^{p-1}| |B^{p-1}|} = {|C^{p-1}| |Z^{p-2}| \over |H^{p-1}| |C^{p-2}|} = {|C^{p-1}| |H^{p-2}| \over |H^{p-1}| |C^{p-2}|} |B^{p-2}|~,
\eea
and repeating this sort of manipulation eventually gives
\bea
|B^p| = \prod_{i=1}^p \left({|H^{p-i}| \over |C^{p-i}|}\right)^{(-)^i}~.
\eea
 Using this, we conclude that as long as the summand is gauge invariant, we may rewrite \bea
\prod_{i=1}^{p+1} |C^{p+1-i}|^{(-)^i} \sum_{a \in Z^{p+1}} &=& |B^{p+1}| \prod_{i=1}^{p+1} |C^{p+1-i}|^{(-)^i} \sum_{a \in H^{p+1}} 
\no\\
&=& \prod_{i=1}^{p+1} |H^{p+1-i}|^{(-)^i} \sum_{a \in H^{p+1}} ~. 
\eea
Thus, for example, for a zero-form symmetry we replace 
\bea
\label{eq:zeroformcoeff}
\int \cD a \hspace{0.2in} \longrightarrow \hspace{0.2 in} {1\over |H^0(M_d, \ZZ_n)|} \sum_{a \in H^1(M_d, \ZZ_n)}~,
\eea
while for a one-form symmetry we replace 
\bea
\int \cD a \hspace{0.2in} \longrightarrow \hspace{0.2 in} {|H^0(M_d, \ZZ_n)|\over |H^1(M_d, \ZZ_n)|} \sum_{a \in H^2(M_d, \ZZ_n)}~.
\eea

\paragraph{Bocksteins:} Though it will not play much of a role below, let us briefly mention a bit of math that appears throughout the literature on this subject; this material can be skipped on a first reading. Associated to any short exact sequence of Abelian groups
\bea
0 \rightarrow K \rightarrow G \rightarrow H \rightarrow 0
\eea
is a long exact sequence of cohomology groups
\bea
\dots \rightarrow H^p(M_d, G) \rightarrow H^p (M_d,H) \xrightarrow{\beta} H^{p+1}(M_d,K) \rightarrow \dots
\eea
We will be interested in the map $\beta: \,H^p (M_d,H)\rightarrow H^{p+1}(M_d,K)$, which can be described as follows. 
Begin with a cocycle $a \in C^p(M_d, H)$ with $\delta a = 0$, and then lift it to a cochain $\tilde a \in  C^p(M_d, G)$, such that the projection to $H$ reproduces $a$. 
Note that $\delta \tilde a \in C^{p+1}(M_d, G)$ is not necessarily $0$, but since $a$ was closed it must be in the image of the map from $K$ to $G$. We may then use this to define a $[\delta \tilde a] \in H^{p+1}(M_d,K)$, identifying $K$ with its image in $G$, which we denote by $\beta([a]):=[\delta \tilde a]$, referred to as the \textit{Bockstein} of $a$. Morally, it tells us the obstruction to $a$ being flat when lifted to $G$.

As a warm-up example, consider the short-exact sequence\footnote{In this context, the Bockstein operation is equivalent to the first Steenrod square.}  
\bea
0 \rightarrow \ZZ_2 \xrightarrow{\imath} \ZZ_4 \xrightarrow{\pi} \ZZ_2 \rightarrow 0~
\eea
with $\imath$ the inclusion map and $\pi$ a projection. Given a cocycle $a \in Z^p(M_d, \ZZ_2)$, we choose a lift $\tilde a \in C^p(M_d, \ZZ_4)$ such that $\pi(\tilde a) = a$, and in particular $\pi(\delta \tilde a) = \delta a = 0$. The latter means that $\delta \tilde a$ lies in the image of $\imath: \ZZ_2 \rightarrow \ZZ_4$, so there exists a unique $x \in C^{p+1}(M_d, \ZZ_2)$ such that $\delta \tilde a = \imath(x)$, and from our above general discussion $\beta(a) = x$. 
Concretely, the inclusion map $\imath$ is given simply by multiplying by $2$, so we have $\delta \tilde a = 2 \beta(a)$, or
\bea
\beta(a) = \half \delta \tilde a\,\,\,\mathrm{mod}\,\,2~, 
\eea
where the division is done in the even subgroup of $\ZZ_4$. Note that we can change the choice of $\ZZ_4$ lift $\tilde a$ by an element of the kernel of $\pi$, i.e. $\tilde a \rightarrow \tilde a + 2\lambda$ for $\lambda \in C^p(M_d, \ZZ_2)$, but this only gives rise to a change $\beta(a) \rightarrow \beta(a) + \delta \lambda$, so that the cohomology class of $\beta(a)$ is independent of the particular lift. 

We now move on to the example of particular interest to us, namely the short-exact sequence
\bea
0 \rightarrow \ZZ \xrightarrow{\times n} \ZZ \rightarrow \ZZ_n \rightarrow 0~. 
\eea
If $a \in Z^p(M_d, \ZZ_n)$ is a discrete cocycle, let $\tilde a \in C^p(M_d, \ZZ)$ be an integer lift. 
Because $\delta a = 0$ mod $n$, we have that $\delta \tilde a = n x $ for some integer $(p+1)$-cochain $x$, and the Bockstein is given by
\bea
\beta(a) = {1\over n} \delta \tilde a ~.
\eea
We can also write the discrete BF action using integer lifts. First, starting with $a,b \in C^*(M_d, \ZZ_n)$, we choose integer lifts $\tilde a, \tilde b \in C^*(M_d, \ZZ)$, 
and then write 
\bea
S = {2 \pi \over n} \int a \cup \delta b =  {2 \pi \over n} \int \tilde a \cup \delta \tilde b  
\eea
where the choice of integer lift does not matter since the action is only defined modulo $2 \pi \ZZ$. 
Now if we further take $b$ to be flat, then we may make use of the Bockstein from above, which gives its value on flat $b$
\bea
S = 2 \pi \int \tilde a \cup \beta(b)\;=0\quad\mathrm{mod}\;2\pi~.
\eea
This vanishing on flat $b$ does not replace the off-shell BF action, whose sum over $b$ imposes flatness of $a$.

\subsection{Dual Symmetries and Twisted Sectors}
\label{sec:dualandtwist}

We next discuss the gauging of non-anomalous $\ZZ_n$ $p$-form symmetries at the level of the partition function/spectrum of the theory. To do this, we couple the theory $\cX$ to a background gauge field $A_{p+1}$ for $\ZZ_n$, and then make it dynamical by performing a sum to get 
\bea
Z_{\cX/\ZZ_n} \propto \sum_{a_{p+1}} Z_\cX[a_{p+1}]~.
\eea
When we perform the sum, we are also free to add a coupling to another background field $\widehat A_{d-p-1}$ to give 
\bea
Z_{\cX/\ZZ_n}[\widehat A_{d-p-1}] \propto \sum_{a_{p+1}}  e^{{2 \pi i \over n} \int a_{p+1} \cup \widehat A_{d-p-1}}Z_\cX[a_{p+1}]~.
\eea
This field $\widehat A_{d-p-1}$ can be thought of as the background field for a $\ZZ_n$ $(d-p-2)$-form symmetry, which is known as the \textit{dual} or \textit{quantum symmetry} \cite{Vafa:1989ih}. 

A special class of examples are those for which $p = d-p-2$, i.e. $p = {d-2 \over 2}$. This includes zero-form symmetries in $d=2$ or one-form symmetries in $d=4$. In this case the original symmetry and the dual symmetry are the same, and thus it is sometimes possible that the theory after gauging is the \textit{same} as the theory before gauging. As will be discussed in Sections \ref{sec:daytwo} and \ref{sec:daythree}, this leads to one of the simplest constructions of non-invertible defects, known as ``half-space gauging.''

Let us now focus for simplicity on the case of zero-form symmetries in two-dimensions. In this case, we will write simply $a = a_{1}$, and we have
\bea
\label{eq:gaugedpartfunc}
Z_{\cX/\ZZ_n}[\widehat A] ={1\over n} \sum_{a}  e^{{2 \pi i \over n} \int a \cup \widehat A}Z_\cX[a]~,
\eea
with overall coefficient as given in (\ref{eq:zeroformcoeff}); note that $|H^0(M_2, \ZZ_n)| = n$ for connected $M_2$. 
Now let us translate this into the notation of topological defects. For simplicity, take the spacetime to be a torus, with $\mathfrak{A}$- and $\mathfrak{B}$-cycles as follows, 
\begin{equation}
   \begin{tikzpicture}[baseline={([yshift=+.5ex]current bounding box.center)},vertex/.style={anchor=base,
    circle,fill=black!25,minimum size=18pt,inner sep=2pt},scale=0.5]
    \filldraw[grey] (-2,-2) rectangle ++(4,4);
    \draw[thick, dgrey] (-2,-2) -- (-2,+2);
    \draw[thick, dgrey] (-2,-2) -- (+2,-2);
    \draw[thick, dgrey] (+2,+2) -- (+2,-2);
    \draw[thick, dgrey] (+2,+2) -- (-2,+2);
    \draw[very thick, blue, ->-=0.3 ] (0,-2) -- (0,2);
       \draw[very thick, red, ->-=0.3 ] (-2,0) -- (2,0);
   
    \node[blue, below] at (0,-2) { $\mathfrak{B}$};
    \node[red, right] at (2,0) { $\mathfrak{A}$};
\end{tikzpicture} ~,
\end{equation}
and with intersection pairing  $\langle \mathfrak{A}, \mathfrak{B} \rangle = - \langle \mathfrak{B}, \mathfrak{A} \rangle = 1$.
A generic 1-cocycle can be written, up to a coboundary, as a linear combination of the Poincar{\'e} duals of $ \mathfrak{A}$ and $\mathfrak{B}$, 
and we choose to write 
\bea
a = - h\, \omega_\mathfrak{A} + g\, \omega_\mathfrak{B}~, \hspace{0.5 in}\widehat A  = -  \widehat h\, \omega_\mathfrak{A} + \widehat g\, \omega_\mathfrak{B}~
\eea
such that 
\bea
\int_{\mathfrak{A} } a =  g~, \hspace{0.4 in} \int_{\mathfrak{B} } a =  h~, \hspace{0.4in}\int_{\mathfrak{A} } \widehat A =  \widehat g~, \hspace{0.4 in} \int_{\mathfrak{B} } \widehat A = \widehat h~,
\eea
with $g,h \in \ZZ_n$.
In other words, the gauge field $a$ on the torus corresponds to the following mesh of defects 
\bea
   \begin{tikzpicture}[baseline={([yshift=+.5ex]current bounding box.center)},vertex/.style={anchor=base,
    circle,fill=black!25,minimum size=18pt,inner sep=2pt},scale=0.5]
    \filldraw[grey] (-2,-2) rectangle ++(4,4);
    \draw[thick, dgrey] (-2,-2) -- (-2,+2);
    \draw[thick, dgrey] (-2,-2) -- (+2,-2);
    \draw[thick, dgrey] (+2,+2) -- (+2,-2);
    \draw[thick, dgrey] (+2,+2) -- (-2,+2);
    \draw[ thick, ->-=0.3 ] (0,-2) -- (0,2);
       \draw[ thick, -<-=0.3 ] (-2,0) -- (2,0);
   
    \node[ below] at (0,-2) { $U_g$};
    \node[ right] at (2,0) { $U_h$};
\end{tikzpicture} ~,
\eea
or, more precisely, since only trivalent junctions of defects are defined,\footnote{There is an obvious ambiguity in how we resolve the 4-valent junction into trivalent vertices, but the two possibilities should give the same result as long as the symmetry is non-anomalous. This will be discussed more in Section \ref{sec:grouplike}. }
\bea
   \begin{tikzpicture}[baseline={([yshift=+.5ex]current bounding box.center)},vertex/.style={anchor=base,
    circle,fill=black!25,minimum size=18pt,inner sep=2pt},scale=0.5]
    \filldraw[grey] (-2,-2) rectangle ++(4,4);
    \draw[thick, dgrey] (-2,-2) -- (-2,+2);
    \draw[thick, dgrey] (-2,-2) -- (+2,-2);
    \draw[thick, dgrey] (+2,+2) -- (+2,-2);
    \draw[thick, dgrey] (+2,+2) -- (-2,+2);
    \draw[ thick, ->-=0.3 ] (0,-2) -- (0,2);
       \draw[ thick, -<-=0.3 ] (-2,0) -- (2,0);
   
    \node[ below] at (0,-2) { $U_g$};
    \node[ right] at (2,0) { $U_h$};
\end{tikzpicture} \hspace{0.4 in}\longrightarrow  \hspace{0.4 in}
 \begin{tikzpicture}[baseline={([yshift=+.5ex]current bounding box.center)},vertex/.style={anchor=base,
    circle,fill=black!25,minimum size=18pt,inner sep=2pt},scale=0.5]
    \filldraw[grey] (-2,-2) rectangle ++(4,4);
    \draw[thick, dgrey] (-2,-2) -- (-2,+2);
    \draw[thick, dgrey] (-2,-2) -- (+2,-2);
    \draw[thick, dgrey] (+2,+2) -- (+2,-2);
    \draw[thick, dgrey] (+2,+2) -- (-2,+2);
    \draw[thick, black, -stealth] (0,-2) -- (0.354,-1.354);
    \draw[thick, black] (0,-2) -- (0.707,-0.707);
    \draw[thick, black, -stealth] (2,0) -- (1.354,-0.354);
    \draw[thick, black] (2,0) -- (0.707,-0.707);
    \draw[thick, black, -stealth] (-0.707,0.707) -- (-0.354,1.354);
    \draw[thick, black] (0,2) -- (-0.707,0.707);
    \draw[thick, black, -stealth] (-0.707,0.707) -- (-1.354,0.354);
    \draw[thick, black] (-2,0) -- (-0.707,0.707);
    \draw[thick, black, -stealth] (0.707,-0.707) -- (0,0);
    \draw[thick, black] (0.707,-0.707) -- (-0.707,0.707);

    \node[black, below] at (0,-2) {\scriptsize $U_g$};
    \node[black, right] at (2,0) {\scriptsize $U_h$};
    \node[black, above] at (0.6,-0.2) {\scriptsize $ U_{g+h}$};
\end{tikzpicture}~. 
\eea
The sum over $a$ then becomes simply a sum over $g,h$. As for the coupling to the dual gauge field, we have 
\bea
\int a \cup \widehat A &=& \int ( - h\, \omega_\mathfrak{A} + g\, \omega_\mathfrak{B}) \cup ( - \widehat h\, \omega_\mathfrak{A} + \widehat g\, \omega_\mathfrak{B}) 
\no\\
&=& -h \widehat g \int \omega_\mathfrak{A}  \cup \omega_\mathfrak{B} - g \widehat h \int \omega_\mathfrak{B} \cup \omega_\mathfrak{A} 
\no\\
&=& g \widehat h - h \widehat g~.
\eea
We thus conclude that the expression for the gauged partition function (\ref{eq:gaugedpartfunc}) can be written in terms of defects as follows,
\bea
Z_{\cX/\ZZ_n} \left( \begin{tikzpicture}[baseline={([yshift=+.5ex]current bounding box.center)},vertex/.style={anchor=base,
    circle,fill=black!25,minimum size=18pt,inner sep=2pt},scale=0.5]
    \filldraw[grey] (-2,-2) rectangle ++(4,4);
    \draw[thick, dgrey] (-2,-2) -- (-2,+2);
    \draw[thick, dgrey] (-2,-2) -- (+2,-2);
    \draw[thick, dgrey] (+2,+2) -- (+2,-2);
    \draw[thick, dgrey] (+2,+2) -- (-2,+2);
    \draw[thick, black, -stealth] (0,-2) -- (0.354,-1.354);
    \draw[thick, black] (0,-2) -- (0.707,-0.707);
    \draw[thick, black, -stealth] (2,0) -- (1.354,-0.354);
    \draw[thick, black] (2,0) -- (0.707,-0.707);
    \draw[thick, black, -stealth] (-0.707,0.707) -- (-0.354,1.354);
    \draw[thick, black] (0,2) -- (-0.707,0.707);
    \draw[thick, black, -stealth] (-0.707,0.707) -- (-1.354,0.354);
    \draw[thick, black] (-2,0) -- (-0.707,0.707);
    \draw[thick, black, -stealth] (0.707,-0.707) -- (0,0);
    \draw[thick, black] (0.707,-0.707) -- (-0.707,0.707);

    \node[black, below] at (0,-2) {\scriptsize $U_{\widehat g}$};
    \node[black, right] at (2,0) {\scriptsize $U_{\widehat h}$};
    \node[black, above] at (0.6,-0.2) {\scriptsize $U_{\widehat g+\widehat h}$};
\end{tikzpicture}  \right)
= \,\,\,
{1\over n} \sum_{g,h \in \ZZ_n} e^{{2\pi i \over n} (g \widehat h - h \widehat g)}\,\,\,
Z_{\cX}\left( \begin{tikzpicture}[baseline={([yshift=+.5ex]current bounding box.center)},vertex/.style={anchor=base,
    circle,fill=black!25,minimum size=18pt,inner sep=2pt},scale=0.5]
    \filldraw[grey] (-2,-2) rectangle ++(4,4);
    \draw[thick, dgrey] (-2,-2) -- (-2,+2);
    \draw[thick, dgrey] (-2,-2) -- (+2,-2);
    \draw[thick, dgrey] (+2,+2) -- (+2,-2);
    \draw[thick, dgrey] (+2,+2) -- (-2,+2);
    \draw[thick, black, -stealth] (0,-2) -- (0.354,-1.354);
    \draw[thick, black] (0,-2) -- (0.707,-0.707);
    \draw[thick, black, -stealth] (2,0) -- (1.354,-0.354);
    \draw[thick, black] (2,0) -- (0.707,-0.707);
    \draw[thick, black, -stealth] (-0.707,0.707) -- (-0.354,1.354);
    \draw[thick, black] (0,2) -- (-0.707,0.707);
    \draw[thick, black, -stealth] (-0.707,0.707) -- (-1.354,0.354);
    \draw[thick, black] (-2,0) -- (-0.707,0.707);
    \draw[thick, black, -stealth] (0.707,-0.707) -- (0,0);
    \draw[thick, black] (0.707,-0.707) -- (-0.707,0.707);

    \node[black, below] at (0,-2) {\scriptsize $ U_g$};
    \node[black, right] at (2,0) {\scriptsize $ U_h$};
    \node[black, above] at (0.6,-0.2) {\scriptsize $ U_{g+h}$};
\end{tikzpicture}\right) ~,
\eea
where $g,h$ capture the dynamical gauge field $a$ and are thus summed over, whereas $\widehat g, \widehat h$ specify the background gauge field $\widehat A$.  In particular, if we are only interested in the case in which $\widehat A$ is turned off, then we have the simpler expression 
\bea
Z_{\cX/\ZZ_n} \left(\,\, \begin{tikzpicture}[baseline=0,scale=0.5]
    \filldraw[grey] (-2,-2) rectangle ++(4,4);
    \draw[thick, dgrey] (-2,-2) -- (-2,+2);
    \draw[thick, dgrey] (-2,-2) -- (+2,-2);
    \draw[thick, dgrey] (+2,+2) -- (+2,-2);
    \draw[thick, dgrey] (+2,+2) -- (-2,+2);
\end{tikzpicture} \,\,\right)
= \,\,\, {1\over n} \sum_{g,h \in \ZZ_n }
Z_{\cX}\left( \begin{tikzpicture}[baseline=0,scale=0.5]
    \filldraw[grey] (-2,-2) rectangle ++(4,4);
    \draw[thick, dgrey] (-2,-2) -- (-2,+2);
    \draw[thick, dgrey] (-2,-2) -- (+2,-2);
    \draw[thick, dgrey] (+2,+2) -- (+2,-2);
    \draw[thick, dgrey] (+2,+2) -- (-2,+2);
    \draw[thick, black, -stealth] (0,-2) -- (0.354,-1.354);
    \draw[thick, black] (0,-2) -- (0.707,-0.707);
    \draw[thick, black, -stealth] (2,0) -- (1.354,-0.354);
    \draw[thick, black] (2,0) -- (0.707,-0.707);
    \draw[thick, black, -stealth] (-0.707,0.707) -- (-0.354,1.354);
    \draw[thick, black] (0,2) -- (-0.707,0.707);
    \draw[thick, black, -stealth] (-0.707,0.707) -- (-1.354,0.354);
    \draw[thick, black] (-2,0) -- (-0.707,0.707);
    \draw[thick, black, -stealth] (0.707,-0.707) -- (0,0);
    \draw[thick, black] (0.707,-0.707) -- (-0.707,0.707);

    \node[black, below] at (0,-2) {\scriptsize $ U_g$};
    \node[black, right] at (2,0) {\scriptsize $ U_h$};
    \node[black, above] at (0.6,-0.2) {\scriptsize $ U_{g+h}$};
\end{tikzpicture}\right) ~.
\eea
Physically, this may be understood as follows. First, if we focus on the terms with $g=0$ on the right-hand side, we see that we are simply inserting ${1\over n} \sum_{h\in \ZZ_n} U_h$ on the horizontal cycle. If we take our foliation to be in the vertical direction, this corresponds to inserting the operator ${1\over n} \sum_{h\in \ZZ_n} U_h$ into the path integral, which projects onto $\ZZ_n$-invariant states. Thus the terms with $g=0$ just capture the states in the original theory which are invariant under $\ZZ_n$, and hence which survive the gauging. 
On the other hand, the states with $g$ inserted vertically correspond to \textit{twisted sector} states, as we have already discussed below (\ref{eq:symgendef}). For each fixed $g$, we again have the projector ${1\over n} \sum_{h\in \ZZ_n} U_h$ inserted, so we are projecting onto invariant states in the twisted sectors as well.

To be even more explicit, let us focus on the case of $n=2$, in which case we have 
\bea
\label{eq:toruspartfuncsum}
Z_{\cX/\ZZ_2} 
= \,\,\, {1\over 2} \left( \begin{tikzpicture}[baseline=0,scale=0.4]
    \filldraw[grey] (-2,-2) rectangle ++(4,4);
    \draw[thick, dgrey] (-2,-2) -- (-2,+2);
    \draw[thick, dgrey] (-2,-2) -- (+2,-2);
    \draw[thick, dgrey] (+2,+2) -- (+2,-2);
    \draw[thick, dgrey] (+2,+2) -- (-2,+2);
    \end{tikzpicture} 
+
 \begin{tikzpicture}[baseline=0,scale=0.4]
    \filldraw[grey] (-2,-2) rectangle ++(4,4);
    \draw[thick, dgrey] (-2,-2) -- (-2,+2);
    \draw[thick, dgrey] (-2,-2) -- (+2,-2);
    \draw[thick, dgrey] (+2,+2) -- (+2,-2);
    \draw[thick, dgrey] (+2,+2) -- (-2,+2);
       \draw[ thick, -<-=0.3 ] (-2,0) -- (2,0);

    \node[ right] at (2,0) {\scriptsize $U_g$};

   \end{tikzpicture} 
+ \begin{tikzpicture}[baseline=0,scale=0.4]
    \filldraw[grey] (-2,-2) rectangle ++(4,4);
    \draw[thick, dgrey] (-2,-2) -- (-2,+2);
    \draw[thick, dgrey] (-2,-2) -- (+2,-2);
    \draw[thick, dgrey] (+2,+2) -- (+2,-2);
    \draw[thick, dgrey] (+2,+2) -- (-2,+2);
    
        \draw[ thick, ->-=0.3 ] (0,-2) -- (0,2);
        \node[ below] at (0,-2) {\scriptsize $U_g$};
    
\end{tikzpicture}
+\begin{tikzpicture}[baseline=0,scale=0.4]
    \filldraw[grey] (-2,-2) rectangle ++(4,4);
    \draw[thick, dgrey] (-2,-2) -- (-2,+2);
    \draw[thick, dgrey] (-2,-2) -- (+2,-2);
    \draw[thick, dgrey] (+2,+2) -- (+2,-2);
    \draw[thick, dgrey] (+2,+2) -- (-2,+2);
    \draw[thick, black, -stealth] (0,-2) -- (0.354,-1.354);
    \draw[thick, black] (0,-2) -- (0.707,-0.707);
    \draw[thick, black, -stealth] (2,0) -- (1.354,-0.354);
    \draw[thick, black] (2,0) -- (0.707,-0.707);
    \draw[thick, black, -stealth] (-0.707,0.707) -- (-0.354,1.354);
    \draw[thick, black] (0,2) -- (-0.707,0.707);
    \draw[thick, black, -stealth] (-0.707,0.707) -- (-1.354,0.354);
    \draw[thick, black] (-2,0) -- (-0.707,0.707);
    \draw[thick, black,dashed] (0.707,-0.707) -- (-0.707,0.707);

    \node[black, below] at (0,-2) {\scriptsize $ U_g$};
    \node[black, right] at (2,0) {\scriptsize $ U_g$};
\end{tikzpicture}  
\right)~
\eea 
in terms of the single  generator $g$. Note that we have dropped the $Z_\cX$ on the right-hand side and represented the partition function simply by the picture of the torus itself, as is standard in the literature.
The first two terms give the $\ZZ_2$-invariant states of the original theory, while the second two terms correspond to the even twisted sector states of the original theory.

An extremely confusing point is that in the older literature, the first two terms are referred to as the untwisted sector of the gauged theory, while the second two terms are referred to as the twisted sector of the gauged theory. On the other hand, in the modern terminology, one would call both of these sectors untwisted in the gauged theory. Indeed, in modern parlance, any state which can be prepared using a local operator (and hence is counted by the partition function) is untwisted. In the modern language, the original theory contains both an untwisted sector and twisted sector, and only the former contributes to the original partition function---states in the latter are created by twist operators attached to a $\ZZ_2$ line, and hence do not contribute to the partition function. Then when we gauge the $\ZZ_2$ symmetry, part of the untwisted sector is projected out into the twisted sector, and conversely part of the twisted sector becomes untwisted. The gauged partition function ultimately only counts untwisted sector states in the gauged theory, but the origin of these untwisted sectors can be either the untwisted or twisted sectors in the ungauged theory. 

More explicitly, let us separate states of the ungauged theory $\cX$ into four sectors, depending on if they are $\ZZ_2$ even/odd, and if they are untwisted/twisted (i.e. if the operator creating them is a local operator, or an operator attached to a $\ZZ_2$ line),
\bea
\label{tab:table1}
\begin{tabular}{c|cc}
$\cX$ & \text{untwisted} & \text{twisted} \\
\hline
\text{even} & \cellcolor{lightblue} $S$ & \cellcolor{lightpink} $U$ \\
\text{odd} & \cellcolor{lightgreen} $T$ &  \cellcolor{lightyellow} $V$
\end{tabular}
\eea
Upon gauging, we lose the $\ZZ_2$ symmetry (all surviving states are invariant under it) but we gain a dual $\widehat \ZZ_2$ symmetry, and can again separate states of the gauged theory $\cX/\ZZ_2$ into four sectors, depending on if they are $\widehat \ZZ_2$ even/odd, and if they are untwisted/twisted (i.e. if the operator creating them is a local operator, or an operator attached to a $\widehat \ZZ_2$ line),
\bea
\label{tab:table2}
\begin{tabular}{c|cc}
$\cX/\ZZ_2$ & \text{untwisted} & \text{twisted} \\
\hline
\text{even} & \cellcolor{lightblue} $S$ & \cellcolor{lightgreen} $T$ \\
\text{odd} &  \cellcolor{lightpink} $U$ &  \cellcolor{lightyellow} $V$
\end{tabular}
\eea
We see that upon discrete gauging, we never really lose or gain states---they just get reorganized. In the current case of $\ZZ_2$ gauging,  the $\ZZ_2$-even twisted sector states in $\cX$ become $\widehat \ZZ_2$-odd untwisted sector states in $\cX/\ZZ_2$ and vice versa. The partition function only ever contains contributions from the first column.

\subsection{Higher-group symmetries} 
\label{sec:highergroup}

Above we have considered theories with both zero-form and higher-form symmetries---for example, the compact boson in $(3+1)$ dimensions.
An interesting phenomenon is that sometimes the zero- and higher-form symmetries can combine in a non-trivial extension to form a \textit{higher-group symmetry}, see e.g. \cite{Baez:2010ya,Cordova:2018cvg,Benini:2018reh}. Concretely, consider a theory with a collection of $p_i$-form symmetries with $p_1 < p_2< \dots< p_n$. If we had a simple direct product of these symmetries, then each would have independent gauge transformations, 
\bea
\delta_i A_{p_i+1} = d \lambda_{p_i}~. 
\eea
But we can also consider the case in which the gauge transformations are given by 
\bea
\delta A_{p_i+1}  = d \lambda_{p_i} + \sum_{j \leq i} \lambda_{p_j} \wedge \alpha_j^{(i)} (\{A_{p_j+1}\}) + \dots 
\eea
where $\alpha_j^{(i)}$ are $(p_i - p_j + 1)$-forms that depend on $A_{p_j+1}$ for $j < i$, and the dots represent terms which are non-linear in gauge transformation parameters. This corresponds to a non-trivial extension of the $p_1$-form symmetry by the higher-form symmetries, and we call the result a \textit{$(p_n+1)$-group} symmetry. 

It is useful to jump directly into an example here. Say that we have a $(3+1)$-dimensional theory on $M_4$ with a zero-form symmetry $\ZZ_{2}^{(0)}$ and a one-form symmetry $\ZZ_2^{(1)}$. Let us denote the background gauge fields for these two symmetries by $A_1$ and $B_2$, and further assume that they have a mixed `t Hooft anomaly given by the inflow action\bea
\label{eq:anomalyhighergroup}
\widehat \cA[A_1, B_2] = \pi \int_{M_5} A_1 \cup B_2 \cup B_2~, \hspace{0.5 in} \p M_5 = M_4~. 
\eea
We now gauge $\ZZ_2^{(0)}$ by promoting the background field $A_1$ to a dynamical field $a_1$, 
\bea
Z_{\cX/\ZZ_2}[\widehat A_3, B_2] = \half \sum_{a_1 \in H^1(M_4, \ZZ_2)} (-1)^{\int a_1 \cup \widehat A_3}  Z_{\cX}[a_1, B_2] 
\eea
where $\widehat A_3$ is the background gauge field for the dual $\ZZ_2^{(2)}$ symmetry. 

When we perform this gauging, the anomaly (\ref{eq:anomalyhighergroup}) naively gives us two choices:  we either remain in $(3+1)$ dimensions and break the $\ZZ_2^{(1)}$ symmetry, or we couple the $(3+1)$-dimensional theory to a $(4+1)$-dimensional bulk theory cancelling the anomaly, in which case we preserve the $\ZZ_2^{(1)}$ symmetry. However, there is actually another option. Let us begin by adding in the inflow term, which gives the gauged partition function,
\bea
Z_{\cX/\ZZ_2}[\widehat A_3, B_2] = \half \sum_{a_1 }  Z_{\cX}[a_1, B_2] \, e^{i \pi \int_{M_5}a_1 \cup (B_2)^2 } e^{i \pi \int_{M_4} a_1 \cup \widehat A_3}~.
\eea
From this expression, we see that we can actually remove the $M_5$ dependence by modifying the standard cocycle condition $\delta \widehat A_3=0$ of the dual background gauge field to the twisted cocycle condition $\delta \widehat A_3 = - B_2 \cup B_2$. One effect of this is that the gauge transformations of $B_2$ and $\widehat A_3$ can no longer be separated---indeed, transformations of $B_2$ must be accompanied by transformations of $\widehat A_3$ in order to maintain the twisted cocycle condition.  
The fact that the $\ZZ_2^{(1)}$ and $\widehat \ZZ_2^{(2)}$ gauge transformations are related in this way means that the full group is a non-trivial extension of the former by the latter, 
\bea
0 \rightarrow \widehat \ZZ_2^{(2)} \rightarrow G^{(3)} \rightarrow \ZZ_2^{(1)} \rightarrow 0~. 
\eea
This is a concrete example of a \textit{3-group}. 

\subsection*{Further References}

The realization of symmetries as topological defects and the subsequent generalization to higher-form symmetries is due to \cite{Frohlich:2009gb,Kapustin:2014gua,Gaiotto:2014kfa}, with some earlier results pointing to the existence of higher-form symmetries in \cite{Batista:2004sc,Nussinov:2006iva,Nussinov:2009zz}. Perhaps the earliest ``killer app'' of higher-form symmetry was that of \cite{Gaiotto:2017yup}, and by now the techniques have become absolutely ubiquitous, see e.g. \cite{Cobanera:2011wn,DelZotto:2015isa,Eckhard:2019jgg,Bergman:2020ifi,Morrison:2020ool,DelZotto:2020sop,Albertini:2020mdx,Bah:2020uev,DelZotto:2020esg,Bhardwaj:2020phs,Apruzzi:2020zot,BenettiGenolini:2020doj,Gukov:2020btk,Closset:2020scj,Closset:2020afy,Apruzzi:2021phx,Apruzzi:2021vcu,Hosseini:2021ged,Cvetic:2021sxm,Buican:2021xhs,Iqbal:2021rkn,Braun:2021sex,Cvetic:2021maf,Closset:2021lhd,Lee:2021obi,Lee:2021crt,Bah:2021brs,Closset:2021lwy,DelZotto:2022fnw,Cvetic:2022uuu,Beratto:2021xmn,Cvetic:2023pgm,Anber:2024gis,Balasubramanian:2024nei,Berean-Dutcher:2025ohp}. The generalization to higher-group symmetries and the exploration of their applications can be found in \cite{Baez:2010ya,Cordova:2018cvg,Benini:2018reh,Cordova:2020tij,DeWolfe:2020uzb,Iqbal:2020lrt,Hidaka:2020izy,Brennan:2020ehu,Bhardwaj:2021wif,Hidaka:2021mml,Hidaka:2021kkf,Apruzzi:2021mlh,DelZotto:2022joo}. 
The anomaly inflow formalism originated in the work of \cite{Faddeev:1984ung,Callan:1984sa}, with the modern formulation being given in \cite{Freed:2014iua,Freed:2016rqq,Monnier:2019ytc}. A useful summary of many of the techniques involved in the study of discrete gauge theories can be found in \cite{Kapustin:2014gua}. 

\newpage
\section{Topological Defect Lines in Two Dimensions}
\label{sec:daytwo}

Thus far, all of the symmetries that we have discussed have had the structure of a (higher-)group. In particular, the fusion of two symmetry generators always gave a unique third element, labelled by the composition of group elements, 
\bea
U_{g_1} \times U_{g_2} = U_{g_1 g_2}~,
\eea
and every symmetry defect $U_g$ had an inverse $U_{g^{-1}}$ labelled by the inverse element. 

More generally, it can be possible for the fusion of two topological defects to produce a \textit{sum} of topological defects, 
\bea
U_{g_1} \times U_{g_2} = U_{g_3} + U_{g_4} + \dots~.
\eea
The meaning of the sum here is that, when inserted into a correlation function, we have 
\bea
\langle U_{g_1} \times U_{g_2} \dots \rangle = \langle U_{g_3} \dots \rangle +\langle  U_{g_4} \dots \rangle + \dots~.
\eea
One can likewise define the Hilbert space associated with $U_{g_3} + U_{g_4}+ \dots$ to be $\cH_{g_3} \oplus \cH_{g_4} \oplus \dots$. 
Requiring that there be a Hilbert space interpretation tells us that the coefficients in these linear combinations must be non-negative integers. 

Because topological defects with the above fusion rules do not have an inverse, we say that they generate \textit{non-invertible symmetries}. The reason that they are still referred to as symmetries in the first place is that, as we will see later, just like standard symmetries they are invariants of symmetry-preserving RG flows, and they can be used to obtain a variety of selection rules. It is also possible to label operators in the theory by representations of these non-invertible symmetries. 

By far the most well-understood examples of non-invertible symmetries are in two-dimensions, where the mathematical structure takes the form of a \textit{fusion category}. Standard physics references on this topic include \cite{Kitaev:2005hzj,Bhardwaj:2017xup,Chang:2018iay}. Below, we will begin by rephrasing the structure of familiar group symmetry in the categorical language, before moving on to the more general case. Our treatment will be far from mathematically rigorous, and will only highlight the salient features needed for practicing physicists. Readers looking for a more rigorous treatment should consult one of the works listed above.

\subsection{Group symmetry in categorical language}
\label{sec:grouplike}
Let us begin by focusing on the case of an invertible finite zero-form symmetry $G$ in $(1+1)$-dimensions. 
We may describe this symmetry with the following data,
\begin{itemize}
\item \textbf{Objects:} topological lines $U_g$ for each $g \in G$. Each line is endowed with an orientation, and every line $U_g$ has a unique orientation reversal $\overline U_g$ given by the inverse element $\overline U_g = U_{g^{-1}}$.
\item \textbf{Morphisms:} topological local operators $\cO_g^h$ that act as homomorphisms between two topological lines, i.e. $\cO_g^h \in \mathrm{Hom}(U_g, U_h)$. Pictorially,
\bea
   \begin{tikzpicture}[baseline=0,scale=0.8]
   \draw[->-=0.5,thick] (0,0) -- (2,0);
   \draw[->-=0.5,thick] (2,0) -- (4,0);
   \filldraw (2,0) circle (0.4ex);
          \node[left] at (0,0) {$U_g$};
     \node[right] at (4,0) {$U_h$};
     \node[below] at (2,0) {$\cO_g^h$};
    \end{tikzpicture} ~.
\eea
By definition, for \textit{simple} lines $g \neq h$, the morphism space $\mathrm{Hom}(U_g, U_h)$ is trivial, i.e. there does not exist any topological local operator between the two lines, whereas if $g = h$ the morphism space is one-dimensional, corresponding to the trivial operator on the line $U_g$. For non-simple lines, the morphism space contains projection operators onto simple constituents. 
\end{itemize}
\noindent

The data of objects and morphisms furnishes the structure of a category, but for categories describing symmetries, there is an extra necessary ingredient known as the \textit{fusion} of line operators. 
Globally, this fusion can be thought of as follows,
\bea
\begin{tikzpicture}[baseline=15, scale=0.4]
   \draw[->-=0.5,thick] (0,0) -- (0,4);
   \draw[->-=0.5,thick] (2,0) -- (2,4);
     \node[below] at (0,0) {$U_g$};
     \node[below] at (2,0) {$U_h$};
    \end{tikzpicture} 
  \hspace{0.1 in}  =  \hspace{0.1 in} 
    \begin{tikzpicture}[baseline=15, scale=0.4]
        \draw[->-=0.5,thick] (1,0) -- (1,4);
         \node[below] at (1,0) {$U_{gh}$};
       \end{tikzpicture}~.
\eea
Locally, two lines can be fused to obtain a third line by using an operator in $\mathrm{Hom}(U_g \otimes U_h, U_{gh})$.
A similar thing is done for the splitting junction, and each of these can be represented by a trivalent vertex,
\bea
 \begin{tikzpicture}[baseline={([yshift=-1ex]current bounding box.center)},vertex/.style={anchor=base,
    circle,fill=black!25,minimum size=18pt,inner sep=2pt},scale=0.4]
   \draw[->-=0.5,thick] (-2,-2) -- (0,0);
   \draw[->-=0.5,thick] (2,-2) -- (0,0);
   \draw[->-=0.7,thick] (0,0)--(0,2) ;
     \node[below] at (-2,-2.2) {$U_g$};
     \node[below] at (2,-2) {$U_h$};
     \node[above] at (0,2) {$U_{gh}$};
     \node[above] at (0,-0.2) {\footnotesize$\times$};
    \end{tikzpicture} \in\, \mathrm{Hom}(U_g \otimes U_h, U_{gh})~, \hspace{0.4 in}
     \begin{tikzpicture}[baseline={([yshift=-1ex]current bounding box.center)},vertex/.style={anchor=base,
    circle,fill=black!25,minimum size=18pt,inner sep=2pt},scale=0.4]
   \draw[->-=0.6,thick] (0,-2) -- (0,0);
   \draw[->-=0.6,thick]  (0,0) -- (2,2);
   \draw[->-=0.6,thick] (0,0)--(-2,2) ;
     \node[above] at (2,2) {$U_h$};
     \node[above] at (-2,2) {$U_g$};
     \node[below] at (0,-2) {$U_{gh}$};
      \node[] at (0,-0.4) {\footnotesize$\times$};
    \end{tikzpicture}
    \in\, \mathrm{Hom}(U_{gh},U_g \otimes U_h)~. 
\eea
We mark one of the lines by an $\times$ to make it explicit which hom-space the junction belongs to, although it is already implicit in the direction of the arrows. 
As we mentioned above, in the group case these hom-spaces are one-dimensional. 
As such, we have the following completeness and orthogonality relations,
\bea
\label{eq:groupbasisconventions}
\begin{tikzpicture}[baseline=20,scale=0.4]
   \draw[->-=0.5,thick] (0,0) -- (0,4);
   \draw[->-=0.5,thick] (2,0) -- (2,4);
     \node[below] at (0,0) {$U_g$};
     \node[below] at (2,0) {$U_h$};
    \end{tikzpicture} 
   \hspace{0.1 in} =\hspace{0.1 in} 
    \begin{tikzpicture}[baseline=20,scale=0.4]
   \draw[->-=0.7,thick] (0,0) -- (1,1);
   \draw[->-=0.7,thick] (2,0) -- (1,1);
    \draw[->-=0.6,thick] (1,1) -- (1,3);
     \draw[->-=0.7,thick] (1,3) -- (0,4);
       \draw[->-=0.7,thick] (1,3) -- (2,4);
     \node[below] at (0,0) {$U_g$};
     \node[below] at (2,0) {$U_h$};
      \node[above] at (0,4) {$U_g$};
     \node[above] at (2,4) {$U_h$};
      \node[right] at (1,2) {$U_{gh}$};
      \node[above] at (1,0.6) {\footnotesize$\times$};
      \node[below] at (1,3.4) {\footnotesize$\times$};
      
    \end{tikzpicture} ~, 
    \hspace{0.7 in}
\begin{tikzpicture}[baseline=5,scale=0.4]
  
    \draw[thick] (0,0) to [out = 180, in = 180,distance = 1.2 cm]  node[rotate=90]{\midarrow} (0,2);
     \draw[thick] (0,0) to [out = 0, in = 0,distance=1.2 cm]  node[rotate=90]{\midarrow}(0,2);
   \draw[thick, ->-=0.5] (0,-1.5) -- (0,0); 
   \draw[thick, ->-=0.9] (0,2) -- (0,3.5);

    \node[left] at (-0.9,1) {\footnotesize $U_g$};
        \node[right] at (0.9,1) {\footnotesize $ U_h$};
    \node[below] at (0,-1.5) {\footnotesize $U_{gh}$};
    \node[above] at (0,3.5) {\footnotesize $U_{gh}$};
      \node[below] at (0,0.3) {\footnotesize$\times$};
       \node[above] at (0,1.7) {\footnotesize$\times$};
       
      
\end{tikzpicture}
\quad= \,\, 
\begin{tikzpicture}[baseline=5,scale=0.4]
  
   \draw[thick, ->-=0.5] (0,-1.5) -- (0,3.5); 
    \node[below] at (0,-1.5) {\footnotesize $U_{gh}$};
    
\end{tikzpicture}~.
\eea
In the dual picture of open patches on a manifold (see the discussion in Section \ref{sec:discrete}),  each line represents a transition function between two open patches, and the above relations correspond to the requirement that changing our open cover does not change any physical results.

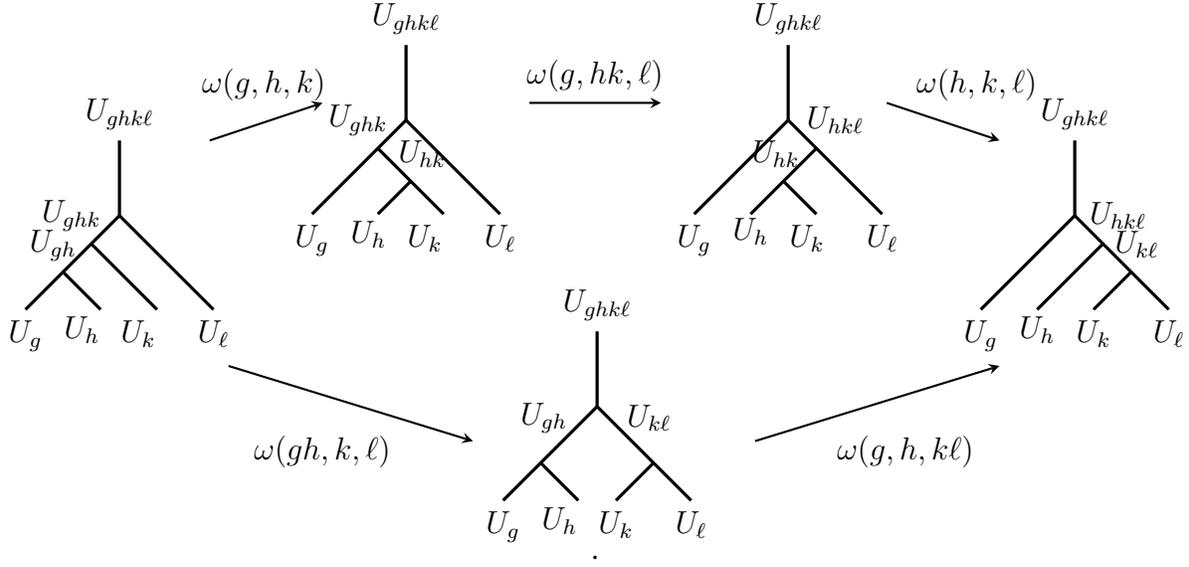
\begin{figure}[tbp]
\begin{center}
\begin{tikzpicture}[baseline=0,scale = 0.5, baseline=-10,rotate=180]
\draw [very thick] (0,0) to (0,-2);
\draw [very thick] (0,0) to (2.5,2.5);
\draw [very thick] (0,0) to (-2.5,2.5);
\draw [very thick] (-0.75,0.75) to (1.75-0.75,1.75+0.75);
\draw [very thick] (-1.5,1.5) to (1-1.5,1+1.5);
\node[above] at (0,-2) {$U_{ghk\ell}$};
\node[below] at (-2.5,2.5) {$U_\ell$};
\node[below] at (1.75-0.75,1.75+0.65) {$U_h$};
\node[below] at (1-1.5,1+1.5) {$U_k$};
\node[below] at (2.5,2.5) {$U_g$};
\node[right] at (-0.1,0) {$U_{hk\ell}$};
\node[right] at (-0.77,0.76) {$U_{k \ell}$};

\begin{scope}[xshift=5in, yshift = 2in]
\draw [very thick] (0,0) to (0,-2);
\draw [very thick] (0,0) to (2.5,2.5);
\draw [very thick] (0,0) to (-2.5,2.5);
\draw [very thick] (1.5,1.5) to (-1+1.5,1+1.5);
\draw [very thick] (-1.5,1.5) to (1-1.5,1+1.5);
\node[above] at (0,-2) {$U_{g hk \ell}$};
\node[below] at (-2.5,2.5) {$U_\ell$};
\node[below] at (1.75-0.75,1.75+0.65) {$U_h$};
\node[below] at (1-1.5,1+1.5) {$U_k$};
\node[below] at (2.5,2.5) {$U_g$};
\node[right] at (-0.5,0.3) {$U_{k \ell}$};
\node[left] at (0.5,0.3) {$U_{gh}$};
\end{scope}

\begin{scope}[xshift=10in, yshift = 0in]
\draw [very thick] (0,0) to (0,-2);
\draw [very thick] (0,0) to (2.5,2.5);
\draw [very thick] (0,0) to (-2.5,2.5);
\draw [very thick] (0.75,0.75) to (-1.75+0.75,1.75+0.75);
\draw [very thick] (1.5,1.5) to (-1+1.5,1+1.5);
\node[above] at (0,-2) {$U_{ghk\ell}$};
\node[below] at (-2.5,2.5) {$U_\ell$};
\node[below] at (1.75-0.75,1.75+0.65) {$U_h$};
\node[below] at (1-1.5,1+1.5) {$U_k$};
\node[below] at (2.5,2.5) {$U_g$};
\node[left] at (0.2,0) {$U_{ghk}$};
\node[left] at (0.85,0.78) {$U_{gh}$};
\end{scope}

\begin{scope}[xshift=7in, yshift = -1in]
\draw [very thick] (0,0) to (0,-2);
\draw [very thick] (0,0) to (2.5,2.5);
\draw [very thick] (0,0) to (-2.5,2.5);
\draw [very thick] (0.75,0.75) to (-1.75+0.75,1.75+0.75);
\draw [very thick] (-0.125,1.625) to (0.875-0.125,0.875+1.625);
\node[above] at (0,-2) {$U_{ghk\ell}$};
\node[below] at (-2.5,2.5) {$U_\ell$};
\node[below] at (1.75-0.75,1.75+0.65) {$U_h$};
\node[below] at (1-1.5,1+1.5) {$U_k$};
\node[below] at (2.5,2.5) {$U_g$};
\node[left] at (0.2,0) {$U_{ghk}$};
\node[right] at (0.5,0.9) {$U_{hk}$};
\end{scope}

\begin{scope}[xshift=3in, yshift = -1in]
\draw [very thick] (0,0) to (0,-2);
\draw [very thick] (0,0) to (2.5,2.5);
\draw [very thick] (0,0) to (-2.5,2.5);
\draw [very thick] (-0.75,0.75) to (1.75-0.75,1.75+0.75);
\draw [very thick] (0.125,1.625) to (-0.875+0.125,0.875+1.625);
\node[above] at (0,-2) {$U_{ghk\ell}$};
\node[below] at (-2.5,2.5) {$U_\ell$};
\node[below] at (1.75-0.75,1.75+0.56) {$U_h$};
\node[below] at (1-1.5,1+1.5) {$U_k$};
\node[below] at (2.5,2.5) {$U_g$};
\node[right] at (-0.2,0) {$U_{hk\ell}$};
\node[left] at (-0.6,0.9) {$U_{hk}$};
\end{scope}

\draw[thick,stealth-] (2,-2) -- (5,-3);
\draw[thick,stealth-] (11,-3) -- (14.5,-3);
\draw[thick,stealth-] (20,-3) -- (23,-2);
\draw[thick,stealth-] (2,4) -- (8.5,6);
\draw[thick,stealth-] (16,6) -- (6.5+16,4);

\node[right] at (4.5, -3.5) {$\omega(h,k,\ell)$};
\node[above] at (12.75, -3) {$\omega(g,hk,\ell)$};
\node[right] at (23.5, -3.5) {$\omega(g,h,k)$};
\node[above] at (4.5, 7) {$\omega(g,h,k\ell)$};
\node[above] at (20, 7) {$\omega(gh,k,\ell)$};
\end{tikzpicture} ~.
\caption{The pentagon identity for invertible defects. This condition translates to the closedness of $\omega$ in group cohomology.}
\label{fig:pentagon}
\end{center}
\end{figure}

An important property of the fusion is that it is not necessarily strictly associative (even though the group multiplication is always associative). Indeed, there can exist a non-trivial \textit{associator homomorphism},
\bea
\omega \in \mathrm{Hom}\left( (U_g \otimes U_h) \otimes U_k, U_g \otimes (U_h \otimes U_k) \right) ~. 
\eea
Pictorially, this is captured by the following, 
\bea
\label{eq:grouplikeFsymbol}
   \begin{tikzpicture}[baseline={([yshift=-1ex]current bounding box.center)},vertex/.style={anchor=base,
    circle,fill=black!25,minimum size=18pt,inner sep=2pt},scale=0.4]
   \draw[->-=0.2,->-=0.6,->-=0.95,thick] (0,0) -- (4,4);
   \draw[->-=0.7,thick] (2.6,0) -- (1.3,1.3);
    \draw[->-=0.6,thick] (6,0) -- (3,3);
          \node[below] at (0,0) {$U_g$};
     \node[below] at (3,0) {$U_h$};
      \node[below] at (6,0) {$U_k$};
     \node[above] at (4,4) {$U_{ghk}$};
      \node[right] at (2,1.6) {$U_{gh}$};
     \node at (1.65,1.65) {\footnotesize$+$};
      \node at (3.3,3.3) {\footnotesize$+$};
    \end{tikzpicture} 
    = \,\,
\omega(g,h,k)\,\,\,
       \begin{tikzpicture}[baseline={([yshift=-1ex]current bounding box.center)},vertex/.style={anchor=base,
    circle,fill=black!25,minimum size=18pt,inner sep=2pt},scale=0.4]
   \draw[->-=0.45,->-=0.95,thick] (0,0) -- (4,4);
   \draw[->-=0.6,thick] (3,0) -- (4.5,1.5);
    \draw[->-=0.3,->-=0.8,thick] (6,0) -- (3,3);
          \node[below] at (0,0) {$U_g$};
     \node[below] at (3,0) {$U_h$};
      \node[below] at (6,0) {$U_k$};
     \node[above] at (4,4) {$U_{ghk}$};
      \node[right] at (2,1.8) {$U_{hk}$};
     \node at (4.2,1.8) {\footnotesize$+$};
      \node at (3.3,3.3) {\footnotesize$+$};
    \end{tikzpicture} ~.
\eea
Note that one is always free to make a redefinition of the hom-space basis vectors via factors $\beta, \gamma$,
\bea
\label{eq:groupjunctiongaugetrans}
 \begin{tikzpicture}[baseline={([yshift=-1ex]current bounding box.center)},vertex/.style={anchor=base,
    circle,fill=black!25,minimum size=18pt,inner sep=2pt},scale=0.4]
   \draw[->-=0.5,thick] (-2,-2) -- (0,0);
   \draw[->-=0.5,thick] (2,-2) -- (0,0);
   \draw[->-=0.7,thick] (0,0)--(0,2) ;
     \node[below] at (-2,-2.2) {$U_g$};
     \node[below] at (2,-2) {$U_h$};
     \node[above] at (0,2) {$U_{gh}$};
     \node[above] at (0,-0.2) {\footnotesize$\times$};
    \end{tikzpicture} \longrightarrow \beta(g,h)
     \begin{tikzpicture}[baseline={([yshift=-1ex]current bounding box.center)},vertex/.style={anchor=base,
    circle,fill=black!25,minimum size=18pt,inner sep=2pt},scale=0.4]
   \draw[->-=0.5,thick] (-2,-2) -- (0,0);
   \draw[->-=0.5,thick] (2,-2) -- (0,0);
   \draw[->-=0.7,thick] (0,0)--(0,2) ;
     \node[below] at (-2,-2.2) {$U_g$};
     \node[below] at (2,-2) {$U_h$};
     \node[above] at (0,2) {$U_{gh}$};
     \node[above] at (0,-0.2) {\footnotesize$\times$};
    \end{tikzpicture}, \hspace{0.4 in}
     \begin{tikzpicture}[baseline={([yshift=-1ex]current bounding box.center)},vertex/.style={anchor=base,
    circle,fill=black!25,minimum size=18pt,inner sep=2pt},scale=0.4]
   \draw[->-=0.6,thick] (0,-2) -- (0,0);
   \draw[->-=0.6,thick]  (0,0) -- (2,2);
   \draw[->-=0.6,thick] (0,0)--(-2,2) ;
     \node[above] at (2,2) {$U_h$};
     \node[above] at (-2,2) {$U_g$};
     \node[below] at (0,-2) {$U_{gh}$};
            \node[] at (0,-0.4) {\footnotesize$\times$};
    \end{tikzpicture}
   \longrightarrow \gamma(g,h)
     \begin{tikzpicture}[baseline={([yshift=-1ex]current bounding box.center)},vertex/.style={anchor=base,
    circle,fill=black!25,minimum size=18pt,inner sep=2pt},scale=0.4]
   \draw[->-=0.6,thick] (0,-2) -- (0,0);
   \draw[->-=0.6,thick]  (0,0) -- (2,2);
   \draw[->-=0.6,thick] (0,0)--(-2,2) ;
     \node[above] at (2,2) {$U_h$};
     \node[above] at (-2,2) {$U_g$};
     \node[below] at (0,-2) {$U_{gh}$};
      \node[] at (0,-0.4) {\footnotesize$\times$};
    \end{tikzpicture}~,\hspace{0.2 in}
\eea
which is referred to as a \textit{gauge transformation}.\footnote{This should not be confused with the background gauge transformations of $G$, which correspond to changing our mesh of topological defects. }
We take $\gamma(g,h) = \beta(g,h)^{-1}$ such that the conventions in (\ref{eq:groupbasisconventions}) remain unchanged. 
Upon such a gauge transformation, the associator is modified via 
\bea
\label{eq:Fgaugeredun1}
  \omega(g,h,k)\rightarrow   {\beta(g,hk) \beta(h,k) \over \beta(g,h) \beta(gh,k) } \omega(g,h,k)~,
\eea
and hence the associators themselves are not, in general, gauge invariant.
At the same time, the associators are subject to the so-called \textit{pentagon identity}, illustrated in Figure \ref{fig:pentagon}, which in equation form is 
\bea
\label{eq:pentagonid}
\omega(g,h,k) \omega(g, hk, \ell) \omega(h,k,\ell) = \omega(gh,k,\ell) \omega(g,h,k\ell)~.
\eea
Now let us define an operation 
\bea
\label{eq:deltadefinition}
\delta f(g_1, \dots, g_{n+1}) &:=& f(g_2, \dots, g_{n+1}) f(g_1, g_2, \dots, g_n)^{(-1)^{n+1}} 
\\
&\vphantom{.}& \hspace{0.5 in}\times\prod_{i=1}^n f(g_1, g_2, \dots, g_{i-1}, g_i g_{i+1}, g_{i+2}, \dots, g_{n+1})^{(-1)^i} ~. 
\no
\eea
In terms of this operation, the condition in (\ref{eq:pentagonid}) becomes 
\bea
\label{eq:omegaconst1}
\delta \omega(g,h,k,\ell) = 1~, 
\eea
while the redundancy in (\ref{eq:Fgaugeredun1}) becomes 
\bea
\label{eq:omegaconst2}
\omega(g,h,k) \sim \delta \beta(g,h,k) \, \omega(g,h,k)~. 
\eea

\begin{tcolorbox}
\textbf{Exercise:} Verify equations (\ref{eq:omegaconst1}) and (\ref{eq:omegaconst2}).
\end{tcolorbox}
\noindent
This should be reminiscent of the definition of a cohomology class---namely closed forms modulo exact forms. Indeed, an element $\omega(g,h,k)$ which satisfies (\ref{eq:pentagonid}) is known as a \textit{3-cocycle}; its equivalence class under the redundancy (\ref{eq:omegaconst2}) is an element of \textit{group cohomology} $H^3(G,U(1))$. This is the same space in which potential anomalies of $G$ take values, so the physical interpretation of the associator $\omega(g,h,k)$ is as the anomaly of $G$. Physically, changing the mesh from the left-hand side of (\ref{eq:grouplikeFsymbol}) to the right-hand side is like performing a background gauge transformation for $G$, and $\omega(g,h,k)$ captures non-invariance of the partition function under this transformation. 
The fusion category describing a discrete symmetry $G$ with anomaly $\omega$ is denoted by $\mathrm{Vec}_G^\omega$.

\subsection{General fusion categories}

We now generalize the discussion above to more general fusion categories $\cC$. As before, the basic data of the category is 
\begin{itemize}
\item \textbf{Objects:} topological lines $a, b, c, \dots \in \cC$.\footnote{These are not to be confused with dynamical gauge fields $a,b,c,\dots$ appearing in other sections. In the current subsection we will not discuss dynamical gauge fields, so this should not lead to any confusion. }  Every element $a$ has a unique orientation-reversal, or \textit{dual}, element $\bar a$. Note that $\bar a$ is not necessarily the inverse of $a$; indeed, $a$ will not in general  admit an inverse.
\item \textbf{Morphisms:} topological local operators $\cO_a^b$ that act as homomorphisms between two topological lines, i.e. $\cO_a^b \in \mathrm{Hom}(a,b)$. A \textit{simple object} is one such that $\dim\mathrm{Hom}(a,a)=1$. 
\end{itemize}
We also need a notion of fusion, with $a,b,c$ now labelling simple objects,
\bea
\label{eq:fusionrulesdef}
a \otimes b = \bigoplus_c N_{ab}^c\, c~,
\eea
where the fusion coefficients $N_{ab}^c$ are non-negative integers. We assume that the fusion rules are commutative, so that $N_{ab}^c = N_{ba}^c$. We also have $N_{a \bar a}^1 = N_{\bar a a}^1 = 1$ for any simple $a$.
Corresponding to these are trivalent junctions representing elements of the hom-spaces, 
\bea
\label{eq:noninvHomspaces}
 \begin{tikzpicture}[baseline={([yshift=-1ex]current bounding box.center)},vertex/.style={anchor=base,
    circle,fill=black!25,minimum size=18pt,inner sep=2pt},scale=0.4]
   \draw[->-=0.5,thick] (-2,-2) -- (0,0);
   \draw[->-=0.5,thick] (2,-2) -- (0,0);
   \draw[->-=0.7,thick] (0,0)--(0,2) ;
     \node[below] at (-2,-2.2) {$a$};
     \node[below] at (2,-2) {$b$};
     \node[above] at (0,2) {$c$};
     \node[above] at (0,-0.2) {\footnotesize$\times$};
     \node[right] at (0,0) {\scriptsize$\mu$};
    \end{tikzpicture} \in\, \mathrm{Hom}(a\otimes b, c)~, \hspace{0.4 in}
     \begin{tikzpicture}[baseline={([yshift=-1ex]current bounding box.center)},vertex/.style={anchor=base,
    circle,fill=black!25,minimum size=18pt,inner sep=2pt},scale=0.4]
   \draw[->-=0.6,thick] (0,-2) -- (0,0);
   \draw[->-=0.6,thick]  (0,0) -- (2,2);
   \draw[->-=0.6,thick] (0,0)--(-2,2) ;
     \node[above] at (2,2) {$b$};
     \node[above] at (-2,2) {$a$};
     \node[below] at (0,-2) {$c$};
       \node[right] at (0,-0.2) {\scriptsize$\overline \mu$};
      \node[] at (0,-0.4) {\footnotesize$\times$};
    \end{tikzpicture}
    \in\, \mathrm{Hom}(c,a\otimes b)~,
\eea
where $\mu = 1, \dots, N_{ab}^c$.  Similar to before, the basis vectors of the hom-spaces are chosen such that they satisfy completeness and orthogonality relations, which in the current case take the following form,
\bea
\label{eq:basisconventions}
\begin{tikzpicture}[baseline={([yshift=-1ex]current bounding box.center)},vertex/.style={anchor=base,
    circle,fill=black!25,minimum size=18pt,inner sep=2pt},scale=0.4]
   \draw[->-=0.5,thick] (0,0) -- (0,4);
   \draw[->-=0.5,thick] (2,0) -- (2,4);
     \node[below] at (0,-0.2) {$a$};
     \node[below] at (2,0) {$b$};
    \end{tikzpicture} 
    =\sum_{c} \sum_{\mu=1}^{N^c_{ab}} \sqrt{d_c \over d_a d_b} \,\,\,
    \begin{tikzpicture}[baseline={([yshift=-1ex]current bounding box.center)},vertex/.style={anchor=base,
    circle,fill=black!25,minimum size=18pt,inner sep=2pt},scale=0.4]
   \draw[->-=0.7,thick] (0,0) -- (1,1);
   \draw[->-=0.7,thick] (2,0) -- (1,1);
    \draw[->-=0.6,thick] (1,1) -- (1,3);
     \draw[->-=0.7,thick] (1,3) -- (0,4);
       \draw[->-=0.7,thick] (1,3) -- (2,4);
     \node[below] at (0,-0.2) {$a$};
     \node[below] at (2,0) {$b$};
      \node[above] at (0,4) {$a$};
     \node[above] at (2,4) {$b$};
      \node[right] at (1,2) {$c$};
      \node[above] at (1,0.6) {\footnotesize$\times$};
      \node[below] at (1,3.4) {\footnotesize$\times$};
      
      \node[left] at (1,1.1) {\scriptsize$\mu$};
       \node[left] at (1,2.8) {\scriptsize$\overline\mu$};
    \end{tikzpicture} ~, 
    \hspace{0.5 in}
\begin{tikzpicture}[baseline={([yshift=-2ex]current bounding box.center)},vertex/.style={anchor=base,
    circle,fill=black!25,minimum size=18pt,inner sep=2pt},scale=0.4]
  
    \draw[thick] (0,0) to [out = 180, in = 180,distance = 1.2 cm]  node[rotate=90]{\midarrow} (0,2);
     \draw[thick] (0,0) to [out = 0, in = 0,distance=1.2 cm]  node[rotate=90]{\midarrow}(0,2);
   \draw[thick, ->-=0.5] (0,-1.5) -- (0,0); 
   \draw[thick, ->-=0.9] (0,2) -- (0,3.5);

    \node[left] at (-0.9,1) {\footnotesize $a$};
        \node[right] at (0.9,1) {\footnotesize $ b$};
    \node[below] at (0,-1.5) {\footnotesize $c$};
    \node[above] at (0,3.5) {\footnotesize $d$};
      \node[below] at (0,0.3) {\footnotesize$\times$};
       \node[above] at (0,1.7) {\footnotesize$\times$};
       
       \node[left] at (0,2.2) {\scriptsize$\mu$};
       \node[left] at (0,-0.3) {\scriptsize$\overline\nu$};
      
\end{tikzpicture}
\quad=\delta_{cd} \delta_{\mu \nu} \sqrt{{d_a d_b \over d_c}} \,\, 
\begin{tikzpicture}[baseline={([yshift=-1ex]current bounding box.center)},vertex/.style={anchor=base,
    circle,fill=black!25,minimum size=18pt,inner sep=2pt},scale=0.4]
  
   \draw[thick, ->-=0.5] (0,-1.5) -- (0,3.5); 
    \node[below] at (0,-1.5) {\footnotesize $c$};
    
\end{tikzpicture}~. 
\eea
Here $d_a$ is the \textit{quantum dimension} of the line $a$, which is additive under direct sums and multiplicative under fusion, $d_a d_b=\sum_c N_{ab}^c d_c$, satisfying $d_a = d_{\bar a}$ and $d_a >0$ in a unitary theory. The quantum dimension can be defined in a number of ways, one of which is as the expectation value of a loop of $ a$ and $\bar a$,\bea
\label{eq:quantumdimdef}
\begin{tikzpicture}[baseline={([yshift=-0.5ex]current bounding box.center)},vertex/.style={anchor=base,
    circle,fill=black!25,minimum size=18pt,inner sep=2pt},scale=0.4]
  
    \draw[thick] (0,0) to [out = 180, in = 180,distance = 1.25 cm] node[rotate=90]{\midarrow} (0,2);
     \draw[thick] (0,0) to [out = 0, in = 0,distance=1.25 cm] node[rotate=90]{\midarrow} (0,2);
   \draw[thick, dashed] (0,-1) -- (0,0); 
   \draw[thick, dashed] (0,2) -- (0,3); 

  \filldraw  (0,0) circle (2pt);
  \filldraw  (0,2) circle (2pt);
  
    \node[left] at (-1,1) {\footnotesize $  a$};
        \node[right] at (1,1.1) {\footnotesize $\bar a$};
       
\end{tikzpicture} = \,\,\, d_a ~;
\eea
in other words, we just define it in terms of the second equation of (\ref{eq:basisconventions}). Another definition is as the normalized vacuum expectation value of a loop of $a$ wrapping the non-trivial cycle on a cylinder. Note that for invertible elements we have $d_a = 1$ in a unitary theory, as follows from multiplicativity and $d_a >0$. 

As before, the fusion rules can have a non-trivial associator, whose matrix elements in the current context are known as the \textit{$F$-symbols} of the category. Pictorially, they may be represented as follows, 
\bea
   \begin{tikzpicture}[baseline={([yshift=-1ex]current bounding box.center)},vertex/.style={anchor=base,
    circle,fill=black!25,minimum size=18pt,inner sep=2pt},scale=0.4]
   \draw[->-=0.2,->-=0.6,->-=0.95,thick] (0,0) -- (4,4);
   \draw[->-=0.7,thick] (2.6,0) -- (1.3,1.3);
    \draw[->-=0.6,thick] (6,0) -- (3,3);
          \node[below] at (0,-0.2) {$a$};
     \node[below] at (3,0) {$b$};
      \node[below] at (6,-0.2) {$c$};
     \node[above] at (4,4) {$d$};
      \node[right] at (2,2) {$e$};
     \node at (1.65,1.65) {\footnotesize$+$};
      \node at (3.3,3.3) {\footnotesize$+$};
      \node[left] at (1.3,1.3) {\scriptsize$\mu$};
       \node[left] at (3,3) {\scriptsize$\nu$};
    \end{tikzpicture} 
    = 
    \sum_{f} \sum_{\sigma = 1}^{N_{bc}^f} \sum_{\rho = 1}^{N_{af}^d} (F_{abc}^d)_{(e\mu\nu)(f \rho\sigma)}\,\,\,
       \begin{tikzpicture}[baseline={([yshift=-1ex]current bounding box.center)},vertex/.style={anchor=base,
    circle,fill=black!25,minimum size=18pt,inner sep=2pt},scale=0.4]
   \draw[->-=0.45,->-=0.95,thick] (0,0) -- (4,4);
   \draw[->-=0.6,thick] (3,0) -- (4.5,1.5);
    \draw[->-=0.3,->-=0.8,thick] (6,0) -- (3,3);
          \node[below] at (0,-0.2) {$a$};
     \node[below] at (3,0) {$b$};
      \node[below] at (6,-0.2) {$c$};
     \node[above] at (4,4) {$d$};
      \node[right] at (2.5,1.8) {$f$};
     \node at (4.2,1.8) {\footnotesize$+$};
      \node at (3.3,3.3) {\footnotesize$+$};
      \node[left] at (3,3) {\scriptsize $\rho$};
      \node[right] at (4.4,1.5) {\scriptsize $\sigma$};
    \end{tikzpicture} ~.
\eea
The $F$-symbols play an analogous role as the anomaly $\omega(g,h,k)$ in (\ref{eq:grouplikeFsymbol}), but are more complicated due to the various sums and non-trivial junction vectors that appear in the non-invertible case. There are also so-called $G$-symbols, defined by 
\bea
  \begin{tikzpicture}[baseline={([yshift=-1ex]current bounding box.center)},vertex/.style={anchor=base,
    circle,fill=black!25,minimum size=18pt,inner sep=2pt},scale=0.5]
   \draw[->-=0.2,->-=0.6,->-=0.95,thick] (4,0) -- (0,4);
   \draw[->-=0.7,thick] (2.5,1.5) -- (5,4);
   \draw[->-=0.7,thick] (1,3) -- (2,4);
   \node[below] at (4,0) {$d$};
    \node[above] at (0,4) {$a$};
    \node[above] at (2,4) {$b$};
    \node[above] at (5,4) {$c$};
      \node at (2.8,1.2) {\footnotesize$+$};
      \node at (1.3,2.7) {\footnotesize$+$};
        \node[left] at (2.5,1.3) {\scriptsize $\overline\nu$};
      \node[left] at (1,2.8) {\scriptsize $\overline\mu$};
       \node[right] at (1.7,2.6) {$e$};
           \end{tikzpicture} 
    = 
    \sum_{f} \sum_{\sigma = 1}^{N_{bc}^f} \sum_{\rho = 1}^{N_{af}^d} (G^{abc}_d)_{(e\overline\mu\overline\nu)(f \overline\rho\overline\sigma)}\,\,\,
       \begin{tikzpicture}[baseline={([yshift=-1ex]current bounding box.center)},vertex/.style={anchor=base,
    circle,fill=black!25,minimum size=18pt,inner sep=2pt},scale=0.5]
   \draw[->-=0.2,->-=0.6,->-=0.95,thick] (4,0) -- (0,4);
   \draw[->-=0.8,->-=0.2,thick] (2.5,1.5) -- (5,4);
   \draw[->-=0.7,,thick] (3.5,2.5) -- (2,4);
   \node[below] at (4,0) {$d$};
    \node[above] at (0,4) {$a$};
    \node[above] at (2,4) {$b$};
    \node[above] at (5,4) {$c$};
      \node at (2.8,1.2) {\footnotesize$+$};
      \node at (3.3,2.3) {\footnotesize$+$};
        \node[left] at (2.5,1.3) {\scriptsize $\overline\rho$};
      \node[right] at (3.5,2.5) {\scriptsize $\overline\sigma$};
       \node[right] at (2,2.4) {$f$};
           \end{tikzpicture} ~,
\eea
which are related to the $F$-symbols by 
\bea
(G^{abc}_d)_{(e\overline\mu\overline\nu)(f \overline\rho\overline\sigma)} = (F_{abc}^d)_{(f \rho\sigma)(e\mu\nu)}^{-1}~;
\eea
this relation is derived by considering two ways of going between the following two diagrams,
\bea
 \begin{tikzpicture}[baseline={([yshift=-1ex]current bounding box.center)},vertex/.style={anchor=base,
    circle,fill=black!25,minimum size=18pt,inner sep=2pt},scale=0.5]
   \draw[thick] (0,0) -- (2,2);
    \draw[thick] (2,2) -- (0,4);
     \draw[thick] (1,1) -- (0,2);
      \draw[thick] (0,2) -- (1,3);
       \draw[thick] (0.5,1.5) -- (1.5,2.5);
 
           \end{tikzpicture} \hspace{0.2 in} \longrightarrow \hspace{0.2 in} 
         \begin{tikzpicture}[baseline={([yshift=-1ex]current bounding box.center)},vertex/.style={anchor=base,
    circle,fill=black!25,minimum size=18pt,inner sep=2pt},scale=0.5]
   \draw[thick] (0,0) -- (0,4);
           \end{tikzpicture}   ~,
\eea
namely using an $F$-move together with two applications of the orthogonality relation of (\ref{eq:basisconventions}),  or a $G$-move with two applications of the orthogonality relation.

As before, one is always free to make a redefinition of the hom-space basis vectors via unitary matrices, 
\bea
\label{eq:junctiongaugetrans}
 \begin{tikzpicture}[baseline={([yshift=-1ex]current bounding box.center)},vertex/.style={anchor=base,
    circle,fill=black!25,minimum size=18pt,inner sep=2pt},scale=0.4]
   \draw[->-=0.5,thick] (-2,-2) -- (0,0);
   \draw[->-=0.5,thick] (2,-2) -- (0,0);
   \draw[->-=0.7,thick] (0,0)--(0,2) ;
     \node[below] at (-2,-2.2) {$a$};
     \node[below] at (2,-2) {$b$};
     \node[above] at (0,2) {$c$};
     \node[above] at (0,-0.2) {\footnotesize$\times$};
     \node[right] at (0,0) {\scriptsize$\mu$};
    \end{tikzpicture} \longrightarrow \sum_{\nu=1}^{N_{ab}^c}\,\, (U_{ab}^c)_{\mu \nu}
     \begin{tikzpicture}[baseline={([yshift=-1ex]current bounding box.center)},vertex/.style={anchor=base,
    circle,fill=black!25,minimum size=18pt,inner sep=2pt},scale=0.4]
   \draw[->-=0.5,thick] (-2,-2) -- (0,0);
   \draw[->-=0.5,thick] (2,-2) -- (0,0);
   \draw[->-=0.7,thick] (0,0)--(0,2) ;
     \node[below] at (-2,-2.2) {$a$};
     \node[below] at (2,-2) {$b$};
     \node[above] at (0,2) {$c$};
     \node[above] at (0,-0.2) {\footnotesize$\times$};
     \node[right] at (0,0) {\scriptsize$\nu$};
    \end{tikzpicture}, \hspace{0.4 in}
     \begin{tikzpicture}[baseline={([yshift=-1ex]current bounding box.center)},vertex/.style={anchor=base,
    circle,fill=black!25,minimum size=18pt,inner sep=2pt},scale=0.4]
   \draw[->-=0.6,thick] (0,-2) -- (0,0);
   \draw[->-=0.6,thick]  (0,0) -- (2,2);
   \draw[->-=0.6,thick] (0,0)--(-2,2) ;
     \node[above] at (2,2) {$b$};
     \node[above] at (-2,2) {$a$};
     \node[below] at (0,-2) {$c$};
       \node[right] at (0,-0.2) {\scriptsize$\overline \mu$};
      \node[] at (0,-0.4) {\footnotesize$\times$};
    \end{tikzpicture}
   \longrightarrow \sum_{\nu=1}^{N_{ab}^c}\,\, (V_c^{ab})_{\overline\mu \overline\nu}
     \begin{tikzpicture}[baseline={([yshift=-1ex]current bounding box.center)},vertex/.style={anchor=base,
    circle,fill=black!25,minimum size=18pt,inner sep=2pt},scale=0.4]
   \draw[->-=0.6,thick] (0,-2) -- (0,0);
   \draw[->-=0.6,thick]  (0,0) -- (2,2);
   \draw[->-=0.6,thick] (0,0)--(-2,2) ;
     \node[above] at (2,2) {$b$};
     \node[above] at (-2,2) {$a$};
     \node[below] at (0,-2) {$c$};
       \node[right] at (0,-0.2) {\scriptsize$\overline \nu$};
      \node[] at (0,-0.4) {\footnotesize$\times$};
    \end{tikzpicture}.\hspace{0.2 in}
\eea
We take $(V_c^{ab})_{\overline\mu \overline\nu} = (U_{ab}^c)^{-1}_{\nu \mu}$ such that the conventions in (\ref{eq:basisconventions}) remain unchanged. 
Upon such a gauge transformation, the $F$-symbols are modified via 
\bea
\label{eq:Fgaugeredun}
 (F_{abc}^d)_{(e\mu\nu)(f \rho\sigma)}\rightarrow  \sum_{\alpha = 1}^{N_{ab}^e} \sum_{\beta = 1}^{N_{ec}^d} \sum_{\gamma=1}^{N_{af}^d} \sum_{\delta = 1}^{N_{bc}^f}  (U_{ab}^e)^{-1}_{\mu \alpha} (U_{ec}^d)^{-1}_{\nu \beta}  (F_{abc}^d)_{(e\alpha\beta)(f \gamma\delta)} (U_{af}^d)_{\gamma \rho} (U_{bc}^f)_{\delta \sigma} ~,
\eea
and hence the $F$-symbols themselves are not, in general, gauge invariant. 
At the same time, the $F$-symbols are subject to a pentagon identity of exactly the same form as in Figure \ref{fig:pentagon}.\\

\begin{tcolorbox}
\textbf{Exercise:} 
For a self-dual simple object $a=\bar a$, using the definition of the quantum dimension and $F$-symbols, show that we have 
\bea
\label{eq:FSisotopydef}
\begin{tikzpicture}[scale=0.8,baseline=20]
    \begin{scope}[ thick, every node/.style={sloped,allow upside down}]
         \draw[ postaction={decorate}, decoration={markings,
      mark=at position 0.4 with {\arrow[scale=0.9]{Stealth}}, mark=at position 0.99 with {\arrow[scale=0.9]{Stealth[reversed]}} }] (-1,-0.5+0.5) .. controls (-1,1.8+0.5) and (0,1.4+0.5) .. (0,0.5+0.5);
      \draw[ postaction={decorate}, decoration={markings,
      mark=at position 0.7 with {\arrow[scale=0.9]{Stealth}} }] (0,0.5+0.5) .. controls (0,-0.3+0.5) and (1,-0.8+0.5) .. (1,1.6+0.5);
      \filldraw  (-0.45,1.21+0.5) circle (1pt);
      \filldraw  (0.45,-0.15+0.5) circle (1pt);
        \node[ below] at (-1,-0.4+0.5) {$a$};
        \node[ right] at (-0.1,0.9+0.5) {$\bar{a}$};
        \node[right] at (1,1.5+0.5) {$a$};
        
       \draw[dashed] (0.46,0.2)   -- (0.46,0.2-0.5)  ;
       \draw[dashed] (-0.46,1.8)   -- (-0.46,1.8+0.5)  ;
    \end{scope}
\end{tikzpicture}
\hspace{0.1 in}= \quad \kappa_a \hspace{0.1 in} 
\begin{tikzpicture}[scale=0.8,baseline=20]
\begin{scope}[ thick, every node/.style={sloped,allow upside down}]
   \draw(0,-0.7+0.5) -- node{\midarrow} (0,1.5+0.5);
   \node[right] at (0,0) {$a$};
\end{scope}
\end{tikzpicture}~\no
\eea
where
\bea
\kappa_a = d_a \, (F_{a \bar a a}^a)_{11}~.\no
\eea
(\textit{Hint: Take the upper identity line to end on the upwards-going $a$ line, and then perform an $F$-move with this as the internal line}.)
Then check that this quantity is gauge invariant with the unit vertices held fixed. 
In fact, this quantity can be shown to take values $\pm1$ and is known as the \textit{Frobenius-Schur (FS) indicator} of $a$. 
A non-trivial FS indicator is a sign of a mixed anomaly between isotopy and orientation-reversal invariances \cite{Chang:2018iay,Kitaev:2005hzj,Simon:2022ohj}.
\end{tcolorbox}

For $a\ne\bar a$, the bending phase $\kappa_a$ is gauge dependent; below we choose it to be $+1$. Note that using the phase $\kappa_b$, we may re-express the 
fusion trivalent junctions in (\ref{eq:noninvHomspaces}) as 
\begin{equation}
     \begin{tikzpicture}[baseline={([yshift=-1ex]current bounding box.center)},vertex/.style={anchor=base,
    circle,fill=black!25,minimum size=18pt,inner sep=2pt},scale=0.4]
   \draw[->-=0.5,thick] (-2,-2) -- (0,0);
   \draw[->-=0.5,thick] (2,-2) -- (0,0);
   \draw[->-=0.7,thick] (0,0)--(0,2) ;
     \node[below] at (-2,-2.2) {$a$};
     \node[below] at (2,-2) {$b$};
     \node[above] at (0,2) {$c$};
     \node[above] at (0,-0.2) {\footnotesize$\times$};
     \node[right] at (0,0) {\scriptsize$\mu$};
    \end{tikzpicture}
    = \kappa_b
   \begin{tikzpicture}[baseline={([yshift=-1ex]current bounding box.center)},vertex/.style={anchor=base,
    circle,fill=black!25,minimum size=18pt,inner sep=2pt},scale=0.4]
   \draw[->-=0.5,thick] (-2,-2) -- (0,0);
   \draw[->-=0.2,-<-=0.5,->-=0.85,thick] (5,-2) -- (0,0);
   \draw[->-=0.7,thick] (0,0)--(0,2) ;
     \node[below] at (-2,-2.2) {$a$};
     \node[below] at (1.5,1) {$b$};
     \node[below] at (2.5,-1.2) {${\overline b}$};
     \node[below] at (4.5,-2) {$b$};
     \node[above] at (0,2) {$c$};
     \node[above] at (0,-0.2) {\footnotesize$\times$};
     \node[right] at (0,0.3 ) {\scriptsize$\mu$};
     \filldraw[black] (2,-0.8) circle (1mm); 
     \filldraw[black] (3.5,-1.4) circle (1mm); 
     \draw[dashed] (2,-0.8) to [out = -90, in = -90,distance = 0.8 cm]   (-0.8,-0.8);

    \end{tikzpicture}
    \label{eq:Bsymbollike}~.
\end{equation}
Then performing a $G$-move with internal line being the identity line shown above, and shrinking the resulting bubble using (\ref{eq:basisconventions}), we obtain
\bea
   \begin{tikzpicture}[baseline={([yshift=-1ex]current bounding box.center)},vertex/.style={anchor=base,
    circle,fill=black!25,minimum size=18pt,inner sep=2pt},scale=0.4]
   \draw[->-=0.5,thick] (-2,-2) -- (0,0);
   \draw[->-=0.5,thick] (2,-2) -- (0,0);
   \draw[->-=0.7,thick] (0,0)--(0,2) ;
     \node[below] at (-2,-2.2) {$a$};
     \node[below] at (2,-2) {$b$};
     \node[above] at (0,2) {$c$};
     \node[above] at (0,-0.2) {\footnotesize$\times$};
     \node[right] at (0,0) {\scriptsize$\mu$};
    \end{tikzpicture} = \,\, \kappa_b \sqrt{\frac{d_a d_b}{d_c}} \sum_{\nu=1}^{N_{{\overline b}c}^a}\,\, (F_{ab{\overline b}}^a)_{(c {\mu\nu})1}
     \begin{tikzpicture}[baseline={([yshift=-1ex]current bounding box.center)},vertex/.style={anchor=base,
    circle,fill=black!25,minimum size=18pt,inner sep=2pt},scale=0.4]
   \draw[->-=0.5,thick] (-2,-2) -- (0,0);
   \draw[-<-=0.8,->-=0.3,thick] (2,-2) -- (0,0);
   \draw[->-=0.7,thick] (0,0)--(0,2) ;
     \node[below] at (-2,-2.2) {$a$};
     \node[below] at (2,-2) {$b$};
     \node[below] at (0.35,-0.35) {$\overline b$};
     \node[above] at (0,2) {$c$};
     \node[above] at (-0.5,-1.05) {\footnotesize$+$};
     \filldraw[black] (1.1,-1.1) circle (1mm);
     \node[left] at (0,0) {\scriptsize$\bar\nu$};
    \end{tikzpicture}
        \label{eq:definitionofBsymbol1}
    \eea
  and likewise for the splitting junctions,
    \bea
    \begin{tikzpicture}[baseline={([yshift=-1ex]current bounding box.center)},vertex/.style={anchor=base,
    circle,fill=black!25,minimum size=18pt,inner sep=2pt},scale=0.35]
   \draw[->-=0.6,thick] (0,-2) -- (0,0);
   \draw[->-=0.6,thick]  (0,0) -- (2,2);
   \draw[->-=0.6,thick] (0,0)--(-2,2) ;
     \node[above] at (2,2) {$b$};
     \node[above] at (-2,2) {$a$};
     \node[below] at (0,-2) {$c$}; 
       \node[right] at (0,-0.2) {\scriptsize$\overline \mu$};
      \node[] at (0,-0.4) {\footnotesize$\times$};
    \end{tikzpicture} = \kappa_b \sqrt{\frac{d_a d_b}{d_c}}\sum_{\nu=1}^{N_{{\overline b}c}^a}\,\, (F^a_{ab{\overline b}})^{-1}_{1(c{\mu\nu})}
       \begin{tikzpicture}[baseline={([yshift=-1ex]current bounding box.center)},vertex/.style={anchor=base,
    circle,fill=black!25,minimum size=18pt,inner sep=2pt},scale=0.35]
   \draw[->-=0.6,thick] (0,-2) -- (0,0);
   \draw[-<-=0.4,->-=0.95,thick]  (0,0) -- (2,2);
   \draw[->-=0.6,thick] (0,0)--(-2,2) ;
     \node[above] at (0.5,0.6) {$\overline b$};
     \node[above] at (2,2) {$b$};
     \node[above] at (-2,2) {$a$};
     \node[below] at (0,-2) {$c$};
     \filldraw[black] (1.1,1.1) circle (1mm);
       \node[below] at (-0.4,0.2) {\scriptsize$ \nu$};
      \node[] at (-0.4,0.4) {\footnotesize$+$};
    \end{tikzpicture} ~.
    \label{eq:definitionofBsymbol2}
\eea
The coefficients appearing here are referred to as \textit{$B$-symbols} in some of the literature \cite{Kitaev:2005hzj,Choi:2024tri}. 

To give just one example of the use of these definitions in a non-trivial computation, let us derive the opposite orientation version of the completeness relation in (\ref{eq:basisconventions}). In particular, we have 
 \bea
\begin{tikzpicture}[baseline={([yshift=-1ex]current bounding box.center)},vertex/.style={anchor=base,
    circle,fill=black!25,minimum size=18pt,inner sep=2pt},scale=0.5]
   \draw[->-=0.5,thick] (0,0) -- (0,4);
   \draw[-<-=0.5,thick] (2,0) -- (2,4);
     \node[below] at (0,-0.2) {$a$};
     \node[below] at (2,0) {$b$};
    \end{tikzpicture} 
&=& \:\:\kappa_b\:\:
    \begin{tikzpicture}[baseline={([yshift=-1ex]current bounding box.center)},vertex/.style={anchor=base,
    circle,fill=black!25,minimum size=18pt,inner sep=2pt},scale=0.4]
   \draw[->-=0.5,thick] (0,0) -- (0,6);
   \draw[-<-=0.2,->-=0.55,-<-=0.9,thick] (2,0) -- (2,6);
     \node[left] at (0,-0.2) {$a$};
     \node[right] at (2,0) {$b$};
     \node[right] at (2,3) {${\overline b}$};
     \node[right] at (2,5.5) {$b$};
  \filldraw[black] (2,2) circle (1mm);   
  \filldraw[black] (2,4) circle (1mm); 
    \end{tikzpicture} 
    =\kappa_b \sum_{c} \sum_{\mu=1}^{N^c_{a{\overline b}}} \sqrt{d_c \over d_a d_b} \,\,\,
    \begin{tikzpicture}[baseline={([yshift=-1ex]current bounding box.center)},vertex/.style={anchor=base,
    circle,fill=black!25,minimum size=18pt,inner sep=2pt},scale=0.5]
   \draw[->-=0.7,thick] (0,0) -- (1,1);
   \draw[-<-=0.3,->-=0.8,thick] (3,-1) -- (1,1);
    \draw[->-=0.6,thick] (1,1) -- (1,3);
     \draw[->-=0.7,thick] (1,3) -- (0,4);
       \draw[->-=0.3,-<-=0.8,thick] (1,3) -- (3,5);
     \node[below] at (0,-0.2) {$a$};
     \node[below] at (1.4,0.5) {${\overline b}$};
     \node[below] at (2.5,-0.5) {$b$};
      \node[above] at (0,4) {$a$};
     \node[above] at (1.4,3.5) {${\overline b}$};
     \node[above] at (2.5,4.5) {$b$};
      \node[right] at (1,2) {$c$};
      \node[above] at (1,0.8) {\footnotesize$\times$};
      \node[below] at (1,3.2) {\footnotesize$\times$};
      \filldraw[black] (2,4) circle (1mm); 
      \filldraw[black] (2,0) circle (1mm); 
      \node[left] at (1,1.1) {\scriptsize$\mu$};
       \node[left] at (1,2.8) {\scriptsize$\overline\mu$};
    \end{tikzpicture}
    \no\\
  &=& \kappa_b \sum_{c} \sum_{\mu=1}^{N^c_{a{\overline b}}} \sum_{\nu, \rho=1}^{N^a_{bc}} \sqrt{d_c \over d_a d_b} {\frac{d_a d_b}{d_c}}  (F_{a{\overline b}b}^a)_{(c {\mu\nu})1} (F_{a{\overline b}b}^a)^{-1}_{1(c {\mu\rho})} \,\,\,
    \begin{tikzpicture}[baseline={([yshift=-1ex]current bounding box.center)},vertex/.style={anchor=base,
    circle,fill=black!25,minimum size=18pt,inner sep=2pt},scale=0.5]
   \draw[->-=0.7,thick] (0,0) -- (1,1);
   \draw[-<-=0.2,->-=0.55,-<-=0.9,thick] (4,-2) -- (1,1);
    \draw[->-=0.6,thick] (1,1) -- (1,3);
     \draw[->-=0.7,thick] (1,3) -- (0,4);
       \draw[-<-=0.2,->-=0.55,-<-=0.9,thick] (1,3) -- (4,6);
     \node[below] at (0,-0.2) {$a$};
     \node[below] at (1.5,0.3) {$b$};
     \node[below] at (2.5,-0.6) {${\overline b}$};
     \node[above] at (3.8,-1.2) {$b$};
      \node[above] at (0,4) {$a$};
     \node[above] at (1.5,3.7) {$b$};
     \node[above] at (2.5,4.5) {${\overline b}$};
     \node[below] at (3.8,5.5) {$b$};
      \node[right] at (1,2) {$c$};
      \node[above] at (0.8,0.3) {\footnotesize$+$};
      \node[below] at (0.8,3.7) {\footnotesize$+$};
      \filldraw[black] (2,4) circle (1mm); 
      \filldraw[black] (3,5) circle (1mm); 
      \filldraw[black] (2,0) circle (1mm); 
      \filldraw[black] (3,-1) circle (1mm); 
      \node[left] at (1,1.1) {\scriptsize$\overline\nu$};
       \node[left] at (1,2.8) {\scriptsize$\rho$};
    \end{tikzpicture}\\
&=&\kappa_b \sum_{c}  \sum_{\nu =1}^{N^a_{bc}}  \sqrt{d_c \over d_a d_b}
\begin{tikzpicture}[baseline={([yshift=-1ex]current bounding box.center)},vertex/.style={anchor=base,
    circle,fill=black!25,minimum size=18pt,inner sep=2pt},scale=0.6]
   \draw[->-=0.7,thick] (0,0) -- (1,1);
   \draw[-<-=0.7,thick] (2,0) -- (1,1);
    \draw[->-=0.6,thick] (1,1) -- (1,3);
     \draw[->-=0.7,thick] (1,3) -- (0,4);
       \draw[-<-=0.7,thick] (1,3) -- (2,4);
     \node[below] at (0,-0.2) {$a$};
     \node[below] at (2,0) {$b$};
      \node[above] at (0,4) {$a$};
     \node[above] at (2,4) {$b$};
      \node[right] at (1,2) {$c$};
      \node[above] at (0.85,0.4) {\footnotesize$+$};
      \node[below] at (0.85,3.6) {\footnotesize$+$};
      
      \node[left] at (1,1.1) {\scriptsize$\overline\nu$};
       \node[left] at (1,2.8) {\scriptsize$\nu$};
    \end{tikzpicture}
\eea
where in the first step we have used the definition of the FS indicator, in the second step we have used the usual completeness relation, in the third step we have used the results of (\ref{eq:definitionofBsymbol1}) and (\ref{eq:definitionofBsymbol2}), and in the final step we have again used the definition of the FS indicator.

Returning to the discussion of $F$-symbols and the pentagon identity, a result known as \textit{Ocneanu rigidity} in the math literature \cite{etingof2005fusion,gainutdinov2023davydov} states that the solutions to the pentagon identity are discrete up to gauge equivalence. Hence given a set of fusion rules, there do not exist continuous families of categories with different $F$-symbols---only a discrete set. Because there are only discrete sets of possibilities, we do not expect the category to change under continuous processes like symmetry-preserving RG flows, which is why it is natural to think of this structure as describing some generalization of symmetry.\footnote{Of course, it is still conceivable that there are discontinuous jumps, but this is also the case for standard symmetry. }

\subsection{Low-rank fusion categories}

The number of simple objects in a fusion category is known as its \textit{rank}. For low-rank pivotal fusion categories, one can actually classify the full set of solutions to the pentagon identities, as was done for ranks 2 and 3  in \cite{ostrik2002fusion} and \cite{ostrik2013pivotal}, respectively.\footnote{Even stronger results exist in the literature---for example, modular categories have been classified up to at least rank 5 \cite{bruillard2016classification}, while multiplicity-free pivotal fusion categories have been classified up to rank 7 \cite{vercleyen2024low}.} We will now review this classification.

\subsubsection{Rank 2}
\label{sec:rank2}

We begin with rank 2, for which there are two simple topological lines $1$ and $W$. The fusion algebra takes the general form $W^2 = 1 + n W$, where $n$ is \textit{a priori} any non-negative integer. However, it turns out that only $n= 0,1$ allow for solutions to the pentagon identity. 

\begin{itemize}
\item $n=0$: In this case the fusion rules are $W^2 = 1$, so $W$ is just the generator of a $\ZZ_2$ global symmetry. Thus we reduce to the group-like case studied in Section \ref{sec:grouplike}. The pentagon identities simplify in this case to a single  non-trivial identity $\omega(W,W,W)^2 =1$, from which we conclude that there are two solutions 
\bea
\omega(W,W,W) = \pm 1~. 
\eea
These solutions correspond to a $\ZZ_2$ global symmetry with and without `t Hooft anomaly. In particular, the anomalous case has the following crossing relation, 
\bea
   \begin{tikzpicture}[baseline={([yshift=-1ex]current bounding box.center)},vertex/.style={anchor=base,
    circle,fill=black!25,minimum size=18pt,inner sep=2pt},scale=0.4]
   \draw[->-=0.5,thick] (0,0) -- (1.3,1.3);
     \draw[dashed] (3,3) -- (1.3,1.3);
      \draw[->-=0.5,thick] (3,3) -- (4,4);
   \draw[->-=0.7,thick] (2.6,0) -- (1.3,1.3);
    \draw[->-=0.6,thick] (6,0) -- (3,3);
 \node[below] at (0,-0.2) {$W$};
     \node[below] at (3,0) {$W$};
      \node[below] at (6,-0.2) {$W$};
     \node[above] at (4,4) {$W$};
    \end{tikzpicture} 
   \hspace{0.2 in} =   \hspace{0.2 in} 
   -
       \begin{tikzpicture}[baseline={([yshift=-1ex]current bounding box.center)},vertex/.style={anchor=base,
    circle,fill=black!25,minimum size=18pt,inner sep=2pt},scale=0.4]
   \draw[->-=0.45,->-=0.95,thick] (0,0) -- (4,4);
   \draw[->-=0.6,thick] (3,0) -- (4.5,1.5);
   \draw[dashed] (4.5,1.5) -- (3,3);
    \draw[->-=0.5,thick] (6,0) -- (4.5,1.5);
          \node[below] at (0,-0.2) {$W$};
     \node[below] at (3,0) {$W$};
      \node[below] at (6,-0.2) {$W$};
     \node[above] at (4,4) {$W$};
    \end{tikzpicture} ~.
\eea

\item $n=1$: In this case the fusion algebra is $W^2 = 1+W$, which is known as the \textit{Fibonacci} or \textit{Lee-Yang algebra}. 
Since the quantum dimension is additive and multiplicative, we must have $d_W^2 = 1 + d_W$, which has solutions $d_W = \half ( 1\pm \sqrt{5})$. The choice of $d_W = \half(1+ \sqrt{5}) = \varphi$, i.e. the golden ratio, gives the so-called \textit{Fibonacci category}. The non-trivial $F$-symbols are uniquely determined up to gauge equivalence; in a convenient gauge they are
\bea
\label{eq:FibonacciF}
F_{WWW}^W = \left( \begin{matrix} \varphi^{-1} & \varphi^{-1/2} \\  \varphi^{-1/2} & - \varphi^{-1} \end{matrix}  \right)~. 
\eea
On the other hand, the choice of $d_W = \half( 1- \sqrt{5})$ gives the \textit{Lee-Yang category}. It cannot be realized in a unitary model, but underlies e.g. the non-unitary minimal model $\cM(2,5)$.
\end{itemize}

\subsubsection{Rank 3}

We next consider rank 3. In this case there are 3 simple topological lines $1, X, Y$. There are now five allowed fusion algebras, and various ways to complete each of these algebras into a fusion category,
\begin{enumerate}
\item $X^2 = Y, Y^2 = X, XY = 1$:  three solutions to the pentagon identity
\item $X^2 =1, Y^2 = 1+X, XY = Y$:  two solutions to the pentagon identity
\item $X^2 = 1, Y^2 = 1 + X + Y, XY = Y$:  three solutions to the pentagon identity
\item $X^2 = 1 + Y, Y^2 = 1 + X + Y, XY = X + Y$:  three solutions to the pentagon identity
\item $X^2 = 1, Y^2 = 1+ X+ 2Y, XY = Y$:  four solutions to the pentagon identity
\end{enumerate}
In the first case, we recognize the fusion rules as those for a $\ZZ_3$ global symmetry. The three solutions to the pentagon identity correspond to the three different values of the mod 3 `t Hooft anomaly. On the other hand, the remaining four cases are all non-invertible. 

Let us now focus on the second case. In this case, one often uses the notation $X = \eta$ and $Y = \cN$, so that the fusion rules read
\bea
\label{eq:Isingfusionrules}
\eta^2 = 1~, \hspace{0.4 in} \eta \times \cN = \cN~, \hspace{0.4 in} \cN\times \cN = 1 + \eta~.
\eea
These are known as the \textit{Ising fusion rules}, since they are realized in the critical Ising model, as we will discuss below. 
We see that $\eta$ generates a standard $\ZZ_2$ symmetry, but $\cN$ is a non-invertible object. One can check that the two solutions to the pentagon identities are given by 
\bea
\label{eq:IsingFsymbols}
F_{\eta\cN \eta}^\cN = F_{\cN \eta \cN}^\eta = -1~, \hspace{0.5 in} F_{\cN \cN \cN}^\cN = \pm {1 \over \sqrt{2}}\left( \begin{matrix} 1 & 1 \\  1 & -1\end{matrix}  \right) ~,
\eea
with all other $F$-symbols being trivial. The choice with the $+$ sign is often referred to as the \textit{Ising fusion category}. 

The Ising category is in some sense the simplest and most famous of all fusion categories. One of the reasons why it is so famous is that it is realized in the simplest non-trivial unitary conformal field theory (CFT), the critical Ising model, as we will now discuss.

\subsection{The Ising model}

The Ising model can be described as a lattice model in the following way.
We begin with a two-dimensional Euclidean square lattice, which for simplicity we take to be periodic in both directions.
We assign a spin $0,1$ to each lattice site, and then couple the sites together 
via nearest neighbor interactions such that the total energy is 
\bea
E[s] = - J \sum_{\langle i j \rangle} (-1)^{s_i + s_j}~,
\eea
where $J>0$ is a coupling constant, $s_i$ is the spin associated with site $i$, and $\langle ij\rangle$ labels nearest neighbor pairs. 
We choose $J$ to be positive so that the energy decreases when the spins are aligned---in this case the system acts as a ferromagnet. 
The partition function of the Ising model at inverse temperature $\beta$ is then given by 
\bea
Z_\mathrm{Ising}(\beta,J) = \sum_{\{s\}} e^{- \beta E[s]} =  \sum_{\{s\}} e^{\beta J \sum_{\langle i j \rangle} (-1)^{s_i + s_j}}~.
\eea

\subsubsection{Kramers-Wannier duality}
An important property of the Ising model is that it has a $\ZZ_2$ spin flip symmetry which flips all $s_i$ simultaneously, i.e. $s_i \rightarrow s_i + 1$ mod $2$. 
This symmetry is non-anomalous and can be gauged. Doing so gives the gauged theory $\mathrm{Ising}/\ZZ_2$ with partition function 
\bea
Z_{\mathrm{Ising}/\ZZ_2}(\beta,J) =\half \sum_{\mathrm{b.c.}}   \sum_{\{s\}} e^{\beta J \sum_{\langle i j \rangle} (-1)^{s_i + s_j}}~,
\eea
where the sum $\sum_{\mathrm{b.c.}} $ is a sum over boundary conditions along the two cycles of the lattice, which is analogous to (\ref{eq:toruspartfuncsum}) on the continuum torus. 

Since the dependence on $\beta$ and $J$ above comes only in the combination $K:=\beta J$, we will denote the partition functions as $Z_\mathrm{Ising}(K) $ and $Z_{\mathrm{Ising}/\ZZ_2}(K) $, and will refer to $K$ as the temperature. One of the most important features of the Ising model is so-called \textit{Kramers-Wannier (KW) duality} \cite{Kramers:1941kn,Kramers:1941zz}, which states that\footnote{The overall factor of $\mathrm{sinh}$ can be removed by addition of appropriate local counterterms, and hence is not particularly important for the statement of the duality.}
\bea
\label{eq:KWdualitystatement}
 Z_{\mathrm{Ising}/\ZZ_2}(K) = {(\mathrm{sinh}\,2 K)^{N }} Z_\mathrm{Ising}(\widetilde K) ~, 
\eea 
where $N$ is the number of lattice sites, and $\widetilde K$ is related to $K$ via 
\bea
\mathrm{sinh}\,2K \, \mathrm{sinh}\,2 \widetilde K = 1~.
\eea
Note that when $K$ is very large, $\widetilde K$ is very small, and vice versa, so this is roughly speaking a duality between the Ising model at high temperature and the $\ZZ_2$-gauged Ising model at low temperature, and vice versa. There is a special, \textit{critical temperature} $K_c = \widetilde K_c$ at which Kramers-Wannier is not just a duality between distinct theories, but a self-duality of a single theory. At this temperature, the continuum limit is a CFT and has a new non-invertible symmetry, as we will discuss momentarily. 

Before discussing the critical point, let us try to understand explicitly how KW duality arises. Here we follow closely the discussion in \cite{Ising}.
Our starting point will be the following function of $K$, 
\bea
\label{eq:genZK}
\cZ(K) =  {1\over 2^N } \sum_{\{a\}} \sum_{\{b\}} e^{K \sum_{\ell} (-1)^{a_\ell} + i \pi \sum_{p = \langle \ell_1 \ell_2 \ell_3 \ell_4 \rangle} b_p (a_{\ell_1} + a_{\ell_2} + a_{\ell_3} + a_{\ell_4})}~. 
\eea
Here the $a_\ell$ are $\ZZ_2$-variables labelled by links $\ell$, where each link $\ell$ is defined by a pair of nearest neighbor points $\langle i j\rangle$. On the other hand, the $b_p$ are $\ZZ_2$-variables labelled by plaquettes, each of which is specified by a set of four links $\langle\ell_1 \ell_2 \ell_3 \ell_4\rangle$ forming a closed loop. What we will now show is that one way of evaluating $\cZ(K)$ gives the right-hand side of (\ref{eq:KWdualitystatement}), while another way of evaluating it gives the left-hand side of (\ref{eq:KWdualitystatement}), and hence the two must be equal. 

Let us start by reproducing the right-hand side. To do so, we first perform the sum over $a_\ell$ in (\ref{eq:genZK}). Since every link $\ell$ is shared by two plaquettes, say $p_1$ and $p_2$, we may rewrite, 
\bea
\cZ(K) &=&  {1\over 2^N }\sum_{\{b\}}  \prod_\ell  \sum_{a_\ell = 0,1} e^{K (-1)^{a_\ell} + i \pi a_\ell (b_{p_1} + b_{p_2})}~
\no\\
&=&  {1\over 2^N }\sum_{\{b\}}  \prod_\ell \left( e^{K} + (-1)^{b_{p_1} + b_{p_2}}e^{-K} \right)
\no\\
&=& {1\over 2^N }\sum_{\{b\}}  \prod_\ell 2\, \mathrm{cosh}K\, (\mathrm{tanh}K)^{b_{p_1} + b_{p_2} \, \mathrm{mod}\,2}~.
\eea
Some additional trigonometric manipulations then allow one to rewrite this as 
\bea
\cZ(K) &=&  {1\over 2^N }\sum_{\{b\}}  \prod_\ell  \sqrt{2\, \mathrm{sinh}\,2K} e^{\widetilde K (-1)^{b_{p_1}+b_{p_2}}}~,
\eea
where $\widetilde K$ is such that 
\bea
\mathrm{tanh}\,K = e^{- 2 \widetilde K}\hspace{0.3 in} \Rightarrow \hspace{0.3 in}  \mathrm{sinh}\,2K \, \mathrm{sinh}\,2 \widetilde K = 1~.
\eea

\begin{tcolorbox}
\textbf{Exercise:} Check these trigonometric manipulations. 
\end{tcolorbox}
Finally, noting that we may equivalently label each link by the pair of plaquettes that share it (instead of the pair of vertices it connects), we have 
\bea
\label{eq:Isingplaquetteexp}
\cZ(K) &=&  ({ \mathrm{sinh}\,2K})^N  \sum_{\{b\}}  \prod_{\ell = \langle p_1 p_2\rangle}  e^{\widetilde K (-1)^{b_{p_1}+b_{p_2}}}~
\no\\
&=& ({ \mathrm{sinh}\,2K})^N Z_{\mathrm{Ising}}(\widetilde K)~.
\eea
In other words, the role of vertices and plaquettes were exchanged, but the result is still the Ising model partition function.

Next we reproduce the left-hand side of (\ref{eq:KWdualitystatement}). In this case, we first evaluate the sum over $b_p$. Then noting that 
\bea
\half \sum_{b_p=0,1} e^{i \pi a_\ell b_p } = \delta_{a_\ell,0}~,
\eea
where the delta function is defined modulo 2, we have 
\bea
\cZ(K) = \sum_{\{a\}} e^{K \sum_\ell (-1)^{a_\ell}} \prod_{p = \langle \ell_1 \ell_2 \ell_3 \ell_4\rangle} \delta_{a_{\ell_1} +a_{\ell_2}+ a_{\ell_3}+a_{\ell_4} ,0  }~. 
\eea
Note that if we impose $a_{\ell_1} +a_{\ell_2}+ a_{\ell_3}+a_{\ell_4} = 0 $ mod $2$ on any plaquette, then the sets $\{a_\ell\}$ and $\{s_i\}$ are actually in 1-to-2 correspondence. Indeed, 
\begin{itemize}
\item For every set of $\{s_i\}$, we can get  the set of $\{a_\ell\}$ by defining 
\bea
a_{\langle i j \rangle} = s_i + s_j \,\,\, \mathrm{mod} \,\, 2~. 
\eea
This choice satisfies the condition $a_{\ell_1} +a_{\ell_2}+ a_{\ell_3}+a_{\ell_4} = 0 $ mod $2$. 

\item Conversely, for every set of $\{a_\ell\}$, we may define a set of $\{s_i\}$ by starting at some arbitrary lattice site, assigning it spin $s_0$, and then defining all other spins via 
\bea
s_i := s_0 + a_{\ell_1} + \dots +a_{\ell_n}~,
\eea
where $\ell_1, \dots, \ell_n$ form a path from site 0 to site $i$. Modulo a subtlety that we will discuss in a moment, the choice of path does not matter as long as $a_{\ell_1} +a_{\ell_2}+ a_{\ell_3}+a_{\ell_4} = 0 $ mod $2$, but note that there are two choices of the initial $s_0$, so from the set $\{a_\ell\}$ there are two   sets $\{s_i\}$. 
\end{itemize}
Based on this 1-to-2 correspondence, we might then expect that 
\bea
\sum_{\{a\}} \prod_{p = \langle \ell_1 \ell_2 \ell_3 \ell_4\rangle} \delta_{a_{\ell_1} +a_{\ell_2}+ a_{\ell_3}+a_{\ell_4} ,0  } \hspace{0.1 in} \longrightarrow \hspace{0.1 in} \half \sum_{\{s\}}~~~.
\eea
This is almost correct, but on lattices with non-trivial topology (such as our current case with periodic boundary conditions) there is an additional subtlety. In particular, we assumed in the second bullet point that any choice of path $\ell_1, \dots, \ell_n$ from site 0 to site $i$ gave the same result. However, this can be false on non-contractible cycles. In particular, set $i=0$ and consider a path that goes from 0 to itself around a non-contractible cycle. If $a_{\ell_1} + \dots +a_{\ell_n}= 0$ mod 2 along this cycle, then $s_0 \rightarrow s_0$, and we have periodic boundary conditions, but if $a_{\ell_1} + \dots +a_{\ell_n}= 1$ mod 2 then we obtain anti-periodic boundary conditions. So what we really have is 
\bea
\sum_{\{a\}} \prod_{p = \langle \ell_1 \ell_2 \ell_3 \ell_4\rangle} \delta_{a_{\ell_1} +a_{\ell_2}+ a_{\ell_3}+a_{\ell_4} ,0  } \hspace{0.1 in} \longrightarrow \hspace{0.1 in} \half \sum_{\mathrm{b.c.}}\sum_{\{s\}}~~~.
\eea
From this, we conclude that 
\bea
\cZ(K) = Z_{\mathrm{Ising}/\ZZ_2}(K)~, 
\eea
thus proving the Kramers-Wannier duality relation.

\subsubsection{Operators}

Let us now briefly discuss the operators in the Ising model, and their behavior under duality.
The most basic operators are the spin operators---in the presentation of the Ising model in terms of plaquettes in (\ref{eq:Isingplaquetteexp}), these are written as 
\bea
\sigma_p = (-1)^{b_p}~,
\eea
which can be thought of as local operators living on the dual lattice. 
There is also an extended operator living on closed loops $\gamma$, 
\bea
\eta(\gamma) = (-1)^{\Sigma_{\ell \in \gamma} a_\ell}~.
\eea
From the expression of $\cZ(K)$ in (\ref{eq:genZK}), it is easy to see that inserting this into the path-integral leads to a shift $b_p \rightarrow b_p+1$ for plaquettes enclosed in the loop $\gamma$. In other words, this is the generator of the spin-flip $\ZZ_2$ symmetry that we had discussed before. 
It acts on the operators $\sigma_p$ with a minus sign, pictorially 
\bea
\begin{tikzpicture}[baseline=0]
 
\draw[black, fill=black](0,0) circle (.3ex); 
\draw[red, thick ](0,0) circle (4.4ex); 
\node at (0,0.3) {$\sigma$};
\node[left] at (-0.7,-0.6) {$\color{red}\eta$};

\end{tikzpicture}
\hspace{0.3 in}= \hspace{0.2 in}- \,\,\,
\begin{tikzpicture}[baseline=0]
\draw[black, fill=black](0,0) circle (.3ex); 
\node at (0,0.4) {$\sigma$};
\end{tikzpicture}~~~.
\eea

We could also try to define $\eta$ on curves which are \textit{not} closed, i.e. which have an endpoint. This gives a so-called \textit{twist defect}, i.e. an operator inserted at a lattice site $i$ which is then attached to a line of $\eta$. With its tail running to a boundary or infinity, this operator is denoted by\footnote{In flat space the $\eta$ tail could run off to infinity, but on a compact lattice such as the one we are considering here, the tail must end on a second $\mu$ operator. In other words,  $\mu$ always comes in pairs on a compact manifold.}
\bea
\mu_i = (-1)^{a_{\ell_1} + a_{\ell_2} + \dots}~. 
\eea
Here $\ell_1,\ell_2,\dots$ form the tail starting at $i$; on a compact lattice the same open-line insertion has a $\mu$ at each endpoint.

Note that when we gauge the $\ZZ_2$ generated by $\eta$, on the one hand $\ZZ_2$ odd operators are projected out, which means that $\sigma_p$ ceases to be a local operator, but on the other hand we make the $\eta$ line transparent, so that $\ZZ_2$ twisted operators can become local; this sort of interchange was discussed in detail in Section \ref{sec:dualandtwist}. We thus see that, upon Kramers-Wannier duality, the operators $\sigma$ and $\mu$ are exchanged, and we also exchange the lattice and dual lattices, as well as the temperatures $K$ and $\widetilde K$. 

Finally, let us introduce the energy operator 
\bea
\eps_\ell = (-1)^{b_{p_1} + b_{p_2}}~,
\eea
where $\ell$ is the link shared by $p_1$ and $p_2$. In the continuum limit, after subtracting its identity component, this flows to an operator whose role is to shift the temperature. In particular, if we start at the critical temperature $K_c = \widetilde K_c$ and modify the Hamiltonian to $H \rightarrow H - \delta K \sum_\ell \eps_\ell$, then we will change the coupling to $K_c \rightarrow K = K_c + \delta K$. Such a shift in $K$ corresponds to an opposite shift in $\widetilde K$, namely $\widetilde K = K_c - \delta K$ to first order in $\delta K$, and this implies that when we pass to the Kramers-Wannier dual description, we must have $\eps \rightarrow - \eps$. 
 
\subsubsection{Ising fusion rules}

Let us summarize the above discussion. The Ising model has a spin operator $\sigma$ and a $\ZZ_2$ symmetry generated by $\eta$ such that $\eta \sigma = - \sigma$. There also exists a twist operator $\mu$ which is attached to an $\eta$ line. Upon gauging $\eta$, the operators $\sigma$ and $\mu$ exchange roles. The ungauged theory at temperature $K$ and the gauged theory at temperature $\widetilde K$ are actually equivalent, i.e. $Z_{\mathrm{Ising}/\ZZ_2}(K) \cong Z_\mathrm{Ising}(\widetilde K) $. 

We would now like to connect this discussion to the Ising fusion rules given in (\ref{eq:Isingfusionrules}).  
To do so, let us consider gauging the $\ZZ_2$ symmetry in half of the space, e.g. the right-hand side. 
When we perform the gauging, we impose Dirichlet boundary conditions at the boundary, which are topological; see Section \ref{sec:topbcs}. 
This should give us a topological interface between the Ising model at temperature $K$ and the $\ZZ_2$-gauged Ising model at the same temperature, 
\bea
\begin{tikzpicture}[baseline={([yshift=-1ex]current bounding box.center)},vertex/.style={anchor=base,
    circle,fill=black!25,minimum size=18pt,inner sep=2pt},scale=0.5]
  
   \draw[thick] (0,-1.5) -- (0,3.5); 
   
   \shade[top color=blue, bottom color=white,opacity = 0.1] (0,-1.5) --(0,3.5) --(-5,3.5) -- (-5, -1.5) -- (0,-1.5) ;
      \shade[top color=red, bottom color=white,opacity = 0.1] (0,-1.5) --(0,3.5) --(7,3.5) -- (7, -1.5) -- (0,-1.5) ;
   
    \node[left] at (-1,1.3) {$\mathrm{Ising}[K]$}; 
     \node[right] at (1,1.3) {$\mathrm{Ising}/\ZZ_2[K]$}; 
\end{tikzpicture}~~~.
\eea
However, making use of Kramers-Wannier duality, we note that the right-hand side is also equivalent to the Ising model at temperature $\widetilde K$, 
\bea
\begin{tikzpicture}[baseline={([yshift=-1ex]current bounding box.center)},vertex/.style={anchor=base,
    circle,fill=black!25,minimum size=18pt,inner sep=2pt},scale=0.5]
  
   \draw[thick] (0,-1.5) -- (0,3.5); 
   
   \shade[top color=blue, bottom color=white,opacity = 0.1] (0,-1.5) --(0,3.5) --(-5,3.5) -- (-5, -1.5) -- (0,-1.5) ;
      \shade[top color=red, bottom color=white,opacity = 0.1] (0,-1.5) --(0,3.5) --(5,3.5) -- (5, -1.5) -- (0,-1.5) ;
   
    \node[left] at (-1,1.3) {$\mathrm{Ising}[K]$}; 
    \node[right] at (1,1.3) {$\mathrm{Ising}[\widetilde K]$}; 
\end{tikzpicture}~~~.
\eea
Thus we see that at the critical temperature $K_c = \widetilde K_c$, i.e. $K_c = \half \log(1+ \sqrt{2})$, the \textit{interface} between two theories becomes a \textit{defect} in a single theory---in particular, it gives rise to a new symmetry generator $\cN$. The conclusion is that the critical Ising model has an additional symmetry $\cN$, which captures its self-duality under gauging of the $\ZZ_2$ symmetry.

We may understand the fusion rules of $\cN$ as follows. First, recall that upon $\ZZ_2$ gauging, the roles of $\sigma$ and $\mu$ are interchanged. Thus moving $\sigma$ past the defect $\cN$ must turn it into $\mu$, and hence the defect $\cN$ must be able to absorb the line coming off of $\mu$,\footnote{The line coming off of $\mu$ should be anchored on $\cN$ and not run off to infinity, since local moves should not change the asymptotic state.}

\bea
\label{eq:sigmatomu}
\begin{tikzpicture}[baseline=20,scale=0.5]
  
   \draw[thick] (0,0) -- (0,4); 
   \draw[black, fill=black](-2,2)  circle (.5ex); 
    \node[below] at (0,0) {$\cN$}; 
    \node[left] at (-2,2)  {$\sigma$}; 
\end{tikzpicture}
\hspace{0.5 in}\Longrightarrow \hspace{0.5 in}
\begin{tikzpicture}[baseline=20,scale=0.5]
  
   \draw[thick] (0,0) -- (0,4); 
   \draw[red, thick](3,2)--(0,2);
   \draw[black, fill=black](3,2)  circle (.5ex); 
    \node[below] at (0,0) {$\cN$}; 
     \node[below] at (1.5,2) {$\eta$}; 
    \node[right] at (3,2)  {$\mu$}; 
\end{tikzpicture}~~~. 
\eea
From this we conclude that $\cN \times \eta = \cN$. 

Next, imagine inserting two copies of $\cN$ on e.g. the $\mathfrak{B}$-cycle of the torus. The first copy of $\cN$ gauges the $\ZZ_2$ symmetry, while the second $\cN$ ungauges it. This configuration thus corresponds to gauging the $\ZZ_2$ symmetry in the annulus between the two copies of $\cN$. Recall from Section \ref{sec:dualandtwist} that gauging a $\ZZ_2$ symmetry amounts to inserting a copy of $1+ \eta$ on all cycles (this will be discussed further in Section \ref{sec:non-invgauging}). If we are inserting  the defects $\cN$ along the $\mathfrak{B}$-cycle, then the only cycle in the strip is again the $\mathfrak{B}$-cycle, and hence this configuration amounts to simply inserting $1+ \eta$ along the $\mathfrak{B}$-cycle, i.e.
\bea
\begin{tikzpicture}[baseline=20,scale=0.5]
  
  \shade[top color=red, bottom color=white,opacity = 0.1] (0,0)--(2,0)--(2,4)--(0,4)--(0,0);

   \draw[thick] (0,0) -- (0,4); 
    \node[below] at (0,0) {$\cN$}; 
    
       \draw[thick] (2,0) -- (2,4); 
    \node[below] at (2,0) {$\cN$}; 
\end{tikzpicture}
\hspace{0.5 in}\Longrightarrow \hspace{0.5 in}
\begin{tikzpicture}[baseline=20,scale=0.5]
  
 \draw[red, thick] (1,0) -- (1,4); 
    \node[below] at (1,0) {$1+\eta$}; 
\end{tikzpicture}~~~. 
\eea
From this, we conclude that we have the fusion rules $\cN \times \cN = 1+ \eta$. 

In conclusion, the fusion rules of the topological defects of the critical Ising model are given by 
\bea
\eta^2=1~, \hspace{0.4 in} \eta \times \cN = \cN~, \hspace{0.4 in} \cN \times \cN = 1 + \eta~.
\eea
One may further study the $F$-symbols of this model, upon which one finds that they take the form of (\ref{eq:IsingFsymbols}), with the choice of overall sign of $F_{\cN\cN\cN}^\cN$ being $+1$. Let us finally note that the action of $\eta$ and $\cN$ on the operators $1, \eps,$ and $\sigma$ of the theory are as follows, 
\bea
\label{eq:Isingactions}
&\vphantom{.}& \hspace{0.08 in}\eta\, |1\rangle \,\,=\,\, |1\rangle ~, \hspace{0.8 in} \eta\, |\epsilon \rangle  \,\,=\,\, |\epsilon \rangle~, \hspace{0.95 in} \eta\, |\sigma\rangle \,\,=\,\, - |\sigma\rangle~,
\no\\
&\vphantom{.}& \cN\, |1\rangle  \,\,=\,\, \sqrt{2} \, |1\rangle ~, \hspace{0.5 in} \cN\, |\epsilon \rangle \,\,=\,\, - \sqrt{2} |\epsilon \rangle~, \hspace{0.5 in} \cN\, |\sigma\rangle \,\,=\,\, 0~.
\eea
The first line follows straightforwardly from our previous discussions. In the last equation of the second line, we are thinking of $\cN$ as a map from the untwisted Hilbert space to itself, and $\cN\eta=\cN$ together with $\eta|\sigma\rangle=-|\sigma\rangle$ implies $\cN|\sigma\rangle=-\cN|\sigma\rangle=0$.
The action of $\cN$ on the identity is simply its quantum dimension, which from the fusion algebra must be $\sqrt{2}$. 
Finally, the minus sign appearing in the action of $\cN$ on $\eps$ comes from the fact that $\eps$ changes sign upon Kramers-Wannier duality, as explained in the previous subsection.

\subsection{Tambara-Yamagami categories and self-duality}
\label{sec:TY}

The Ising category studied above is one of the simplest examples of a more general class of categories 
known as \textit{Tambara-Yamagami (TY) categories} \cite{tambara1998tensor}. These are categories which have invertible elements 
$U_g$ labelled by elements $g$ of a discrete Abelian group $G$, together with a single non-invertible element $\cN$ 
with the following fusion rules, 
\bea
U_{g_1} \times U_{g_2} = U_{g_1 g_2}~, \hspace{0.3 in} U_g \times \cN = \cN~, \hspace{0.3 in} \cN \times \cN = \sum_{g\in G} U_g~. 
\eea
The Ising category studied above corresponds to the case of $G= \ZZ_2$ (and a particular choice of $F$-symbols). 

Previously, we saw that the non-invertible defect $\cN$ captured self-duality under gauging the $\ZZ_2$ symmetry of the Ising model. 
More generally, it turns out that a theory is self-dual under gauging $G$ if and only if the symmetry $G$ admits a TY extension.
One direction of this follows via the half-space gauging construction.
Let us now discuss the other direction, at the level of the torus partition function. We begin with the schematic derivation, giving the more rigorous computation afterwards.  

First assume that we have a defect $\cN$, and insert a loop of it on a contractible cycle of the torus, 
which only changes the partition function by an overall factor that we divide out by,
\bea
 \begin{tikzpicture}[baseline=0,scale=0.5]
    \filldraw[grey] (-2,-2) rectangle ++(4,4);
    \draw[thick, dgrey] (-2,-2) -- (-2,+2);
    \draw[thick, dgrey] (-2,-2) -- (+2,-2);
    \draw[thick, dgrey] (+2,+2) -- (+2,-2);
    \draw[thick, dgrey] (+2,+2) -- (-2,+2);
\end{tikzpicture}
\hspace{0.1 in} \propto \hspace{0.1 in}  \begin{tikzpicture}[baseline=0,scale=0.5]
    \filldraw[grey] (-2,-2) rectangle ++(4,4);
    \draw[thick, dgrey] (-2,-2) -- (-2,+2);
    \draw[thick, dgrey] (-2,-2) -- (+2,-2);
    \draw[thick, dgrey] (+2,+2) -- (+2,-2);
    \draw[thick, dgrey] (+2,+2) -- (-2,+2);
    
    \draw[red,thick] (0,0) circle (1cm);
    \node[red] at (1.1,-1.1) {$\cN$};
\end{tikzpicture}~.
\eea
Then deforming $\cN$ topologically so that it wraps the sum of all cycles using the fusion rules gives 
\bea
\label{eq:TYselfdualinterm}
 \begin{tikzpicture}[baseline=0,scale=0.5]
    \filldraw[grey] (-2,-2) rectangle ++(4,4);
    \draw[thick, dgrey] (-2,-2) -- (-2,+2);
    \draw[thick, dgrey] (-2,-2) -- (+2,-2);
    \draw[thick, dgrey] (+2,+2) -- (+2,-2);
    \draw[thick, dgrey] (+2,+2) -- (-2,+2);
    
    \draw[red,thick] (0,0) circle (1cm);
    \node[red] at (1.1,-1.1) {$\cN$};
\end{tikzpicture}
\hspace{0.1 in} = \hspace{0.1 in}  \begin{tikzpicture}[baseline=0,scale=0.5]
    \filldraw[grey] (-2,-2) rectangle ++(4,4);
    \draw[thick, dgrey] (-2,-2) -- (-2,+2);
    \draw[thick, dgrey] (-2,-2) -- (+2,-2);
    \draw[thick, dgrey] (+2,+2) -- (+2,-2);
    \draw[thick, dgrey] (+2,+2) -- (-2,+2);
    
    \draw[thick,red] (-2,0.5) to [out = 0, in = -90] (-0.5,2); 
     \draw[thick,red] (2,0.5) to [out = 180, in = -90] (0.5,2); 
     
       \draw[thick,red] (2,-0.5) to [out = 180, in = 90] (0.5,-2); 
     \draw[thick,red] (-2,-0.5) to [out = 0, in = 90] (-0.5,-2); 
    
    \node[red] at (0.5,-0.5) {$\cN$};
\end{tikzpicture}
\hspace{0.1 in} \propto\sum_{g,h \in G} \hspace{0.1 in}  \begin{tikzpicture}[baseline=0,scale=0.5]
    \filldraw[grey] (-2,-2) rectangle ++(4,4);
    \draw[thick, dgrey] (-2,-2) -- (-2,+2);
    \draw[thick, dgrey] (-2,-2) -- (+2,-2);
    \draw[thick, dgrey] (+2,+2) -- (+2,-2);
    \draw[thick, dgrey] (+2,+2) -- (-2,+2);
    
    \draw[thick] (-2,0) -- (-0.5,0);
    \draw[thick] (2,0) -- (0.5,0);
    \draw[thick] (0,-2) -- (0,-0.5);
    \draw[thick] (0,2) -- (0,0.5);
     \draw[red,thick] (0,0) circle (0.5cm);
    \node[red] at (0.7,-0.7) {$\cN$};
     \node[right] at(2,0) {$h$};
      \node[below] at  (0,-2) {$g$};
    
   \end{tikzpicture}~,
\eea
where from now on, for simplicity, we label the lines corresponding to group-like symmetries $g \in G$ by the group element $g$ itself, instead of by $U_g$. 
Finally, shrinking the bubble gives 
\bea
 \begin{tikzpicture}[baseline=0,scale=0.5]
    \filldraw[grey] (-2,-2) rectangle ++(4,4);
    \draw[thick, dgrey] (-2,-2) -- (-2,+2);
    \draw[thick, dgrey] (-2,-2) -- (+2,-2);
    \draw[thick, dgrey] (+2,+2) -- (+2,-2);
    \draw[thick, dgrey] (+2,+2) -- (-2,+2);
\end{tikzpicture}
\hspace{0.2in}\propto \sum_{g,h \in G}\hspace{0.2in}
 \begin{tikzpicture}[baseline=0,scale=0.5]
    \filldraw[grey] (-2,-2) rectangle ++(4,4);
    \draw[thick, dgrey] (-2,-2) -- (-2,+2);
    \draw[thick, dgrey] (-2,-2) -- (+2,-2);
    \draw[thick, dgrey] (+2,+2) -- (+2,-2);
    \draw[thick, dgrey] (+2,+2) -- (-2,+2);
    
    \draw[thick] (-2,0) -- (2,0);
    \draw[thick] (0,-2) -- (0,2);
     \node[right] at(2,0) {$h$};
      \node[below] at  (0,-2) {$g$};
   
   \end{tikzpicture}~,
\eea
where the right-hand side is basically the $G$-gauged partition function. This is the statement of self-duality. 
Of course, in the computation above we have been very cavalier about issues such as the orientation of $\cN$  and overall normalization.  A more precise treatment should keep track of the orientation of $\cN$ and make use of the $F$-symbols to go from the second to the third expression in (\ref{eq:TYselfdualinterm}), as we do now.

The more precise derivation proceeds as follows. We begin by using our convention in (\ref{eq:quantumdimdef}) to write 
\bea
 \begin{tikzpicture}[baseline=0,scale=0.5]
    \filldraw[grey] (-2,-2) rectangle ++(4,4);
    \draw[thick, dgrey] (-2,-2) -- (-2,+2);
    \draw[thick, dgrey] (-2,-2) -- (+2,-2);
    \draw[thick, dgrey] (+2,+2) -- (+2,-2);
    \draw[thick, dgrey] (+2,+2) -- (-2,+2);
\end{tikzpicture}
\hspace{0.1 in} \,\,\,=\,\,\, {1\over d_\cN} \hspace{0.1 in}  \begin{tikzpicture}[baseline=0,scale=0.5]
    \filldraw[grey] (-2,-2) rectangle ++(4,4);
    \draw[thick, dgrey] (-2,-2) -- (-2,+2);
    \draw[thick, dgrey] (-2,-2) -- (+2,-2);
    \draw[thick, dgrey] (+2,+2) -- (+2,-2);
    \draw[thick, dgrey] (+2,+2) -- (-2,+2);
    
        \draw[red,thick] (0,-1) to [out = 180, in = 180,distance = 1.25 cm] node[rotate=90]{\midarrow} (0,1);
     \draw[red,thick] (0,-1) to [out = 0, in = 0,distance=1.25 cm] node[rotate=90]{\midarrow} (0,1);

  \filldraw[red]  (0,-1) circle (2pt);
  \filldraw[red]  (0,1) circle (2pt);
 
\end{tikzpicture}~~,
\eea
where all red lines represent $\cN$---note that we are using that $\overline \cN = \cN$ for TY categories. Now we deform this topologically to obtain

\bea
 \begin{tikzpicture}[baseline=0,scale=0.5]
    \filldraw[grey] (-2,-2) rectangle ++(4,4);
    \draw[thick, dgrey] (-2,-2) -- (-2,+2);
    \draw[thick, dgrey] (-2,-2) -- (+2,-2);
    \draw[thick, dgrey] (+2,+2) -- (+2,-2);
    \draw[thick, dgrey] (+2,+2) -- (-2,+2);
    
        \draw[red,thick] (0,-1) to [out = 180, in = 180,distance = 1.25 cm] node[rotate=90]{\midarrow} (0,1);
     \draw[red,thick] (0,-1) to [out = 0, in = 0,distance=1.25 cm] node[rotate=90]{\midarrow} (0,1);

  \filldraw[red]  (0,-1) circle (2pt);
  \filldraw[red]  (0,1) circle (2pt);
 
\end{tikzpicture}
\hspace{0.2 in} \,\,=\,\, \hspace{0.2 in}
\begin{tikzpicture}[baseline={([yshift=+.5ex]current bounding box.center)},vertex/.style={anchor=base,
    circle,fill=black!25,minimum size=18pt,inner sep=2pt},scale=0.5,rotate=90]
    \filldraw[grey] (-2,-2) rectangle ++(4,4);
    \draw[thick, dgrey] (-2,-2) -- (-2,+2);
    \draw[thick, dgrey] (-2,-2) -- (+2,-2);
    \draw[thick, dgrey] (+2,+2) -- (+2,-2);
    \draw[thick, dgrey] (+2,+2) -- (-2,+2);
    
     
   \draw[thick, red, ->-=0.6] (-2,0.5)  [out = 0, in = -90] to (-0.5,2); 
      \draw[thick, red, -<-=0.3, ->-=0.85] (0.5,2) to [out = -90, in = 180]  (2,0.5);
      \draw[thick, red, -<-=0.6] (2,-0.5)  [out = 180, in = 90] to (0.5,-2); 
    \draw[thick, red,->-=0.35, -<-=0.85] (-0.5,-2)  [out = 90, in = 0] to (-2,-0.5); 
   
     \filldraw[red]  (-0.95,-0.95) circle (2pt);
      \filldraw[red]  (0.95,0.95) circle (2pt);

    
\end{tikzpicture}~~. 
\eea
If we imagine an identity line as being stretched from the vertex in the bottom right to the middle of the line on the top right, then we may use an $F$-symbol to obtain\footnote{For graphical simplicity, we drop the $\times$ that we were using to dress trivalent junctions in previous subsections. We can always reinstate them by just checking the number of lines going into/out of the vertex. } 
\bea
\begin{tikzpicture}[baseline={([yshift=+.5ex]current bounding box.center)},vertex/.style={anchor=base,
    circle,fill=black!25,minimum size=18pt,inner sep=2pt},scale=0.6,rotate=90]
    \filldraw[grey] (-2,-2) rectangle ++(4,4);
    \draw[thick, dgrey] (-2,-2) -- (-2,+2);
    \draw[thick, dgrey] (-2,-2) -- (+2,-2);
    \draw[thick, dgrey] (+2,+2) -- (+2,-2);
    \draw[thick, dgrey] (+2,+2) -- (-2,+2);
    
     
   \draw[thick, red, ->-=0.6] (-2,0.5)  [out = 0, in = -90] to (-0.5,2); 
      \draw[thick, red, -<-=0.3, ->-=0.85] (0.5,2) to [out = -90, in = 180]  (2,0.5);
      \draw[thick, red, -<-=0.6] (2,-0.5)  [out = 180, in = 90] to (0.5,-2); 
    \draw[thick, red,->-=0.35, -<-=0.85] (-0.5,-2)  [out = 90, in = 0] to (-2,-0.5); 
   
     \filldraw[red]  (-0.95,-0.95) circle (2pt);
      \filldraw[red]  (0.95,0.95) circle (2pt);

    
\end{tikzpicture}
\hspace{0.3 in} &=& \,\,\,
\sum_h (F_{\cN\cN\cN}^\cN)_{1 h}  \hspace{0.3 in}
\begin{tikzpicture}[baseline={([yshift=+.5ex]current bounding box.center)},vertex/.style={anchor=base,
    circle,fill=black!25,minimum size=18pt,inner sep=2pt},scale=0.6,rotate=90]
    \filldraw[grey] (-2,-2) rectangle ++(4,4);
    \draw[thick, dgrey] (-2,-2) -- (-2,+2);
    \draw[thick, dgrey] (-2,-2) -- (+2,-2);
    \draw[thick, dgrey] (+2,+2) -- (+2,-2);
    \draw[thick, dgrey] (+2,+2) -- (-2,+2);
    
     
   \draw[thick, red, ->-=0.6] (-2,0.5)  [out = 0, in = -90] to (-0.5,2); 
      \draw[thick, red, -<-=0.3, ->-=0.85] (0.5,2) to [out = -90, in = 180]  (2,0.5);
   
   \draw[thick, red, -<-=0.35, -<-=0.85](2,-0.5)  -- (-2,-0.5); 
   
      \draw[thick, -<-=0.75](0,-0.5)  -- (0,-1.2); 
      \filldraw[red]  (0.95,0.95) circle (2pt);
      
      \draw[thick, red, -<-=0.6] (0,-1.2)  [out = 0, in = 90] to (0.5,-2); 
 \draw[thick, red, -<-=0.6] (0,-1.2)  [out = 180, in = 90] to (-0.5,-2);

     \node[below] at (0.9,-1) {\footnotesize $h$};
    
\end{tikzpicture}
\no\\
&=& \,\,\,
\sum_h (F_{\cN\cN\cN}^\cN)_{1 h}  \hspace{0.3 in}
 \begin{tikzpicture}[baseline=0,scale=0.6]
    \filldraw[grey] (-2,-2) rectangle ++(4,4);
    \draw[thick, dgrey] (-2,-2) -- (-2,+2);
    \draw[thick, dgrey] (-2,-2) -- (+2,-2);
    \draw[thick, dgrey] (+2,+2) -- (+2,-2);
    \draw[thick, dgrey] (+2,+2) -- (-2,+2);
    
  \draw[thick,red,->-=0.3,-<-=0.65,->-=0.95] (-1,-2) -- (-1,2);
    \draw[thick,red,->-=0.3,->-=0.8] (1,-2) -- (1,2);
    
    \draw[thick,->-=0.6] (-1,0) -- (-2,0);
      \draw[thick,-<-=0.6] (1,0) -- (2,0);
  
  \filldraw[red]  (-1,1) circle (2pt);

\end{tikzpicture}~~,
\eea
where in the second equality we used a topological move. Next we imagine an identity line stretched from the 
top left vertex to the middle of the top right line, do an $F$-move, and use a topological deformation to give
\bea
\label{eq:TYselfdualbubble}
\begin{tikzpicture}[baseline={([yshift=+.5ex]current bounding box.center)},vertex/.style={anchor=base,
    circle,fill=black!25,minimum size=18pt,inner sep=2pt},scale=0.6,rotate=90]
    \filldraw[grey] (-2,-2) rectangle ++(4,4);
    \draw[thick, dgrey] (-2,-2) -- (-2,+2);
    \draw[thick, dgrey] (-2,-2) -- (+2,-2);
    \draw[thick, dgrey] (+2,+2) -- (+2,-2);
    \draw[thick, dgrey] (+2,+2) -- (-2,+2);
    
     
   \draw[thick, red, ->-=0.6] (-2,0.5)  [out = 0, in = -90] to (-0.5,2); 
      \draw[thick, red, -<-=0.3, ->-=0.85] (0.5,2) to [out = -90, in = 180]  (2,0.5);
      \draw[thick, red, -<-=0.6] (2,-0.5)  [out = 180, in = 90] to (0.5,-2); 
    \draw[thick, red,->-=0.35, -<-=0.85] (-0.5,-2)  [out = 90, in = 0] to (-2,-0.5); 
   
     \filldraw[red]  (-0.95,-0.95) circle (2pt);
      \filldraw[red]  (0.95,0.95) circle (2pt);

    
\end{tikzpicture}
\hspace{0.2 in}= \,\,\,\sum_{g,h} (F_{\cN \cN \cN}^{\cN})_{1h} (F_{\cN\cN\cN}^\cN)_{g1}^{-1} \hspace{0.2 in}
\begin{tikzpicture}[baseline={([yshift=+.5ex]current bounding box.center)},vertex/.style={anchor=base,
    circle,fill=black!25,minimum size=18pt,inner sep=2pt},scale=0.6]
    \filldraw[grey] (-2,-2) rectangle ++(4,4);
    \draw[thick, dgrey] (-2,-2) -- (-2,+2);
    \draw[thick, dgrey] (-2,-2) -- (+2,-2);
    \draw[thick, dgrey] (+2,+2) -- (+2,-2);
    \draw[thick, dgrey] (+2,+2) -- (-2,+2);
    \draw[thick, black, ->-=0.5] (0,-2) -- (0,-0.5);
    \draw[thick, black, ->-=0.5] (0,0.5) -- (0,2);
    \draw[thick, black, ->-=0.5] (-0.5,0) -- (-2,0) ;
    \draw[thick, black, ->-=0.5] (2,0) -- (0.5,0) ;
    
 \draw[thick, red ] (0,0.5) to [out = 180, in = 90] node[rotate=225]{\midarrow} (-0.5, 0);
      \draw[thick, red] (0,0.5) to [out = 0, in = 90] node[rotate=135]{\midarrow}  (0.5, 0);
        \draw[thick, red] (0,-0.5) to [out = 180, in = -90] node[rotate=135]{\midarrow}  (-0.5, 0);
         \draw[thick, red ] (0.5, 0) to [out = -90, in = 0] node[rotate=45]{\midarrow} (0,-0.5);
         
    \node[black, below] at (0,-2) {\footnotesize $g$};
    \node[black, right] at (2,0) {\footnotesize $h$};
    
\end{tikzpicture}~. 
\eea
We must now shrink the internal bubble. In order to do so, let us consider the upper-left portion of the bubble. This may be manipulated as follows, 
\bea
\label{eq:intermediatethingbubble}
\begin{tikzpicture}[baseline={([yshift=+.5ex]current bounding box.center)},vertex/.style={anchor=base,
    circle,fill=black!25,minimum size=18pt,inner sep=2pt},scale=0.6]

    \draw[red, thick, ->- = 0.5] (-2,-2) --(-1,-1);; 
     \draw[red, thick, ->- = 0.5] (2,-2) --(1,-1);
     
      \draw[ thick,->-=0.5] (-1,-1)-- (-1,1); 
  \draw[ thick,->-=0.5]  (1,-1)  -- (1,1); 
 \draw[red, thick, -<-=0.5] (-1,-1) --(1,-1); 
 
 \node[above] at (-1,1) {$h$};
 \node[above] at (1,1) {$g$};
\end{tikzpicture}~
\hspace{0.3 in} = \hspace{0.3 in} 
\begin{tikzpicture}[baseline={([yshift=+.5ex]current bounding box.center)},vertex/.style={anchor=base,
    circle,fill=black!25,minimum size=18pt,inner sep=2pt},scale=0.6]

    \draw[red, thick, ->- = 0.5] (-2,-2) --(-1,-1); 
     \draw[red, thick, ->- = 0.5] (2,-2) --(1,-1);
     
 \draw[red, thick, -<-=0.5] (-1,-1) --(1,-1); 
 
 \draw[thick, ->-=0.7 ]  (-1,-1) to[out = 90, in = 180] (0,-0.5); 
  \draw[thick, ->-=0.7 ]  (1,-1) to[out = 90, in = 0] (0,-0.5); 
    \draw[thick, ->-=0.7 ]  (0,-0.5)-- (0,0.5); 
      \draw[thick, ->-=0.7 ]  (0,0.5) -- (-1,1); 
     \draw[thick, ->-=0.7 ]  (0,0.5) -- (1,1);

 \node[above] at (-1,1) {$h$};
 \node[above] at (1,1) {$g$};
  \node[left] at (-0.8,-0.7) {$h$};
 \node[right] at (0.8,-0.7) {$g$};
  \node[left] at (0,0) {$gh$};
 
\end{tikzpicture}
\hspace{0.3 in} = \,\,\,
F_{\cN \cN g}^{gh}\hspace{0.1 in} 
\begin{tikzpicture}[baseline={([yshift=+.5ex]current bounding box.center)},vertex/.style={anchor=base,
    circle,fill=black!25,minimum size=18pt,inner sep=2pt},scale=0.6]

    \draw[red, thick, ->- = 0.5] (-2,-2) --(0,-1);
     \draw[red, thick, ->- = 0.5] (2,-2) --(0,-1);
     
 \draw[ thick, ->- =0.7] (0,-1) -- (0,0); 
 
      \draw[thick, ->-=0.7 ]  (0,0) -- (-1,1); 
     \draw[thick, ->-=0.7 ]  (0,0) -- (1,1);

 \node[above] at (-1,1) {$h$};
 \node[above] at (1,1) {$g$};

  \node[left] at (0,-0.4) {$gh$};
 
\end{tikzpicture}~,
\eea
where in the first step we used the first equation of (\ref{eq:groupbasisconventions}), and in the second step we used an $F$-move and the second equation of (\ref{eq:basisconventions}). Doing the same thing on the bottom right of (\ref{eq:TYselfdualbubble}) then gives 
\bea
 \begin{tikzpicture}[baseline=0,scale=0.5]
    \filldraw[grey] (-2,-2) rectangle ++(4,4);
    \draw[thick, dgrey] (-2,-2) -- (-2,+2);
    \draw[thick, dgrey] (-2,-2) -- (+2,-2);
    \draw[thick, dgrey] (+2,+2) -- (+2,-2);
    \draw[thick, dgrey] (+2,+2) -- (-2,+2);
\end{tikzpicture}
\hspace{0.1 in} &=& {1\over d_\cN}
\sum_{g,h} (F_{\cN \cN \cN}^{\cN})_{1h} (F_{\cN\cN\cN}^\cN)_{g1}^{-1} F_{\cN \cN g}^{gh} (F_{\cN \cN h}^{gh})^{-1} \hspace{0.2 in}
\begin{tikzpicture}[baseline={([yshift=+.5ex]current bounding box.center)},vertex/.style={anchor=base,
    circle,fill=black!25,minimum size=18pt,inner sep=2pt},scale=0.5]
    \filldraw[grey] (-2,-2) rectangle ++(4,4);
    \draw[thick, dgrey] (-2,-2) -- (-2,+2);
    \draw[thick, dgrey] (-2,-2) -- (+2,-2);
    \draw[thick, dgrey] (+2,+2) -- (+2,-2);
    \draw[thick, dgrey] (+2,+2) -- (-2,+2);
    \draw[thick, black, ->-=0.5] (0,-2) -- (0.707,-0.707);
    \draw[thick, black, ->-=0.5] (2,0) -- (0.707,-0.707);
    \draw[thick, black, ->-=0.7] (-0.707,0.707) -- (0,2);
    \draw[thick, black, -<-=0.5] (-2,0) -- (-0.707,0.707);
    \draw[thick, black, ->-=0.9] (0.707,-0.707) -- (0.293,-0.293);
    \draw[thick, black, -<-=0.7] (-0.707,0.707) -- (-0.293,0.293);
    \draw[thick, red, ->-=0.6] (0.293,-0.293) arc(-45:135:0.414);
    \draw[thick, red, ->-=0.6] (0.293,-0.293) arc(-45:-225:0.414);
    \node[black, below] at (0,-2) {\footnotesize $ g$};
    \node[black, right] at (2,0) {\footnotesize $ h$};
  
\end{tikzpicture}
\no\\
&=& 
\sum_{g,h} (F_{\cN \cN \cN}^{\cN})_{1h} (F_{\cN\cN\cN}^\cN)_{g1}^{-1} F_{\cN \cN g}^{gh} (F_{\cN \cN h}^{gh})^{-1} \hspace{0.2 in}
\begin{tikzpicture}[baseline={([yshift=+.5ex]current bounding box.center)},vertex/.style={anchor=base,
    circle,fill=black!25,minimum size=18pt,inner sep=2pt},scale=0.5]
    \filldraw[grey] (-2,-2) rectangle ++(4,4);
    \draw[thick, dgrey] (-2,-2) -- (-2,+2);
    \draw[thick, dgrey] (-2,-2) -- (+2,-2);
    \draw[thick, dgrey] (+2,+2) -- (+2,-2);
    \draw[thick, dgrey] (+2,+2) -- (-2,+2);
    \draw[thick, black, -stealth] (0,-2) -- (0.354,-1.354);
    \draw[thick, black] (0,-2) -- (0.707,-0.707);
    \draw[thick, black, -stealth] (2,0) -- (1.354,-0.354);
    \draw[thick, black] (2,0) -- (0.707,-0.707);
    \draw[thick, black, -stealth] (-0.707,0.707) -- (-0.354,1.354);
    \draw[thick, black] (0,2) -- (-0.707,0.707);
    \draw[thick, black, -stealth] (-0.707,0.707) -- (-1.354,0.354);
    \draw[thick, black] (-2,0) -- (-0.707,0.707);
    \draw[thick, black, -stealth] (0.707,-0.707) -- (0,0);
    \draw[thick, black] (0.707,-0.707) -- (-0.707,0.707);

    \node[black, below] at (0,-2) {\scriptsize $ g$};
    \node[black, right] at (2,0) {\scriptsize $ h$};
    \node[black, above] at (0.2,0) {\scriptsize $ g  h$};
\end{tikzpicture}~,
\eea
where we again made use of (\ref{eq:basisconventions}). 
In fact, it is possible to simplify the product of the $F$-symbols above by using the pentagon identity, giving the final result 
\bea
 \begin{tikzpicture}[baseline=0,scale=0.5]
    \filldraw[grey] (-2,-2) rectangle ++(4,4);
    \draw[thick, dgrey] (-2,-2) -- (-2,+2);
    \draw[thick, dgrey] (-2,-2) -- (+2,-2);
    \draw[thick, dgrey] (+2,+2) -- (+2,-2);
    \draw[thick, dgrey] (+2,+2) -- (-2,+2);
\end{tikzpicture}
\hspace{0.1 in} &=& {1\over |G|} \sum_{g,h} {F_{hg\cN}^\cN \over F_{gh \cN}^\cN} \,\,\, \begin{tikzpicture}[baseline={([yshift=+.5ex]current bounding box.center)},vertex/.style={anchor=base,
    circle,fill=black!25,minimum size=18pt,inner sep=2pt},scale=0.5]
    \filldraw[grey] (-2,-2) rectangle ++(4,4);
    \draw[thick, dgrey] (-2,-2) -- (-2,+2);
    \draw[thick, dgrey] (-2,-2) -- (+2,-2);
    \draw[thick, dgrey] (+2,+2) -- (+2,-2);
    \draw[thick, dgrey] (+2,+2) -- (-2,+2);
    \draw[thick, black, -stealth] (0,-2) -- (0.354,-1.354);
    \draw[thick, black] (0,-2) -- (0.707,-0.707);
    \draw[thick, black, -stealth] (2,0) -- (1.354,-0.354);
    \draw[thick, black] (2,0) -- (0.707,-0.707);
    \draw[thick, black, -stealth] (-0.707,0.707) -- (-0.354,1.354);
    \draw[thick, black] (0,2) -- (-0.707,0.707);
    \draw[thick, black, -stealth] (-0.707,0.707) -- (-1.354,0.354);
    \draw[thick, black] (-2,0) -- (-0.707,0.707);
    \draw[thick, black, -stealth] (0.707,-0.707) -- (0,0);
    \draw[thick, black] (0.707,-0.707) -- (-0.707,0.707);

    \node[black, below] at (0,-2) {\scriptsize $ g$};
    \node[black, right] at (2,0) {\scriptsize $ h$};
    \node[black, above] at (0.2,0) {\scriptsize $ g  h$};
\end{tikzpicture}~. 
\eea
This is precisely the result expected for self-duality under gauging of $G$ with a discrete torsion specified by $F_{hg\cN}^\cN/F_{gh\cN}^\cN$.

\begin{tcolorbox}
\textbf{Exercise:} Verify that the pentagon identities actually give such a simplification. In particular, you will have to show that 
\bea
(F_{\cN\cN\cN}^\cN)_{gh} = (F_{\cN\cN\cN}^\cN)_{11} {F_{g \cN \cN}^g F_{g \cN h^{-1}}^\cN \over F_{\cN h h^{-1}}^\cN F_{\cN \cN h^{-1}}^1}~,
\eea 
and 
\bea
F_{xy\cN}^\cN F_{x \cN \cN}^z F_{y \cN \cN}^{x^{-1} z} = F_{xy, \cN, \cN}^z~,\hspace{0.5 in} x,y,z \in G~. 
\eea
You will also need to use that $(F_{\cN\cN\cN}^\cN)_{11}^2 = 1/d_\cN^2 =1/|G|$. 
\end{tcolorbox}

\noindent
A similar result can be obtained when a background for the dual $G$ symmetry is turned on \cite{Kaidi:2025hyr},
\bea
\begin{tikzpicture}[baseline={([yshift=+.5ex]current bounding box.center)},vertex/.style={anchor=base,
    circle,fill=black!25,minimum size=18pt,inner sep=2pt},scale=0.5]
    \filldraw[grey] (-2,-2) rectangle ++(4,4);
    \draw[thick, dgrey] (-2,-2) -- (-2,+2);
    \draw[thick, dgrey] (-2,-2) -- (+2,-2);
    \draw[thick, dgrey] (+2,+2) -- (+2,-2);
    \draw[thick, dgrey] (+2,+2) -- (-2,+2);
    \draw[thick, black, -stealth] (0,-2) -- (0.354,-1.354);
    \draw[thick, black] (0,-2) -- (0.707,-0.707);
    \draw[thick, black, -stealth] (2,0) -- (1.354,-0.354);
    \draw[thick, black] (2,0) -- (0.707,-0.707);
    \draw[thick, black, -stealth] (-0.707,0.707) -- (-0.354,1.354);
    \draw[thick, black] (0,2) -- (-0.707,0.707);
    \draw[thick, black, -stealth] (-0.707,0.707) -- (-1.354,0.354);
    \draw[thick, black] (-2,0) -- (-0.707,0.707);
    \draw[thick, black, -stealth] (0.707,-0.707) -- (0,0);
    \draw[thick, black] (0.707,-0.707) -- (-0.707,0.707);

    \node[black, below] at (0,-2) {\scriptsize $ \widehat g$};
    \node[black, right] at (2,0) {\scriptsize $ \widehat h$};
    \node[black, above] at (0.2,0) {\scriptsize $ \widehat g \widehat h$};
\end{tikzpicture}
\hspace{0.1 in} &=& {1\over |G|} {F_{\cN \widehat h \widehat g}^\cN \over F_{\cN \widehat g \widehat h}^\cN} \sum_{g,h} {F_{h\cN \widehat g^{-1}}^\cN \over F_{g \cN \widehat h^{-1}}^\cN}  {F_{hg\cN}^\cN \over F_{gh \cN}^\cN} \,\,\, \begin{tikzpicture}[baseline={([yshift=+.5ex]current bounding box.center)},vertex/.style={anchor=base,
    circle,fill=black!25,minimum size=18pt,inner sep=2pt},scale=0.5]
    \filldraw[grey] (-2,-2) rectangle ++(4,4);
    \draw[thick, dgrey] (-2,-2) -- (-2,+2);
    \draw[thick, dgrey] (-2,-2) -- (+2,-2);
    \draw[thick, dgrey] (+2,+2) -- (+2,-2);
    \draw[thick, dgrey] (+2,+2) -- (-2,+2);
    \draw[thick, black, -stealth] (0,-2) -- (0.354,-1.354);
    \draw[thick, black] (0,-2) -- (0.707,-0.707);
    \draw[thick, black, -stealth] (2,0) -- (1.354,-0.354);
    \draw[thick, black] (2,0) -- (0.707,-0.707);
    \draw[thick, black, -stealth] (-0.707,0.707) -- (-0.354,1.354);
    \draw[thick, black] (0,2) -- (-0.707,0.707);
    \draw[thick, black, -stealth] (-0.707,0.707) -- (-1.354,0.354);
    \draw[thick, black] (-2,0) -- (-0.707,0.707);
    \draw[thick, black, -stealth] (0.707,-0.707) -- (0,0);
    \draw[thick, black] (0.707,-0.707) -- (-0.707,0.707);

    \node[black, below] at (0,-2) {\scriptsize $ g$};
    \node[black, right] at (2,0) {\scriptsize $ h$};
    \node[black, above] at (0.2,0) {\scriptsize $ g  h$};
\end{tikzpicture}~. 
\eea

\subsection{Gauging non-invertible symmetries}
\label{sec:non-invgauging}
Just as invertible global symmetries can be promoted to gauge symmetries, it is natural to ask if there is a sense in which non-invertible symmetries can be gauged. Indeed, such a generalization exists (see e.g. \cite{Frohlich:2009gb,Diatlyk:2023fwf,Perez-Lona:2023djo,Perez-Lona:2024sds}). 

\subsubsection{Algebra objects }To begin, let us rephrase the gauging of an invertible symmetry as follows. 
As we have seen above, for discrete Abelian group $G$, the torus partition function of the gauged theory is given by,\footnote{A similar formula holds for non-Abelian $G$ as long as we restrict the sum to commuting pairs.}
\bea
Z_{\cX /G}
&=& {1\over |G|} \sum_{g,h\in G} \mu(g,h) \mu^\vee(h,g) \,\,\, \begin{tikzpicture}[baseline={([yshift=+.5ex]current bounding box.center)},vertex/.style={anchor=base,
    circle,fill=black!25,minimum size=18pt,inner sep=2pt},scale=0.5]
    \filldraw[grey] (-2,-2) rectangle ++(4,4);
    \draw[thick, dgrey] (-2,-2) -- (-2,+2);
    \draw[thick, dgrey] (-2,-2) -- (+2,-2);
    \draw[thick, dgrey] (+2,+2) -- (+2,-2);
    \draw[thick, dgrey] (+2,+2) -- (-2,+2);
    \draw[thick, black, -stealth] (0,-2) -- (0.354,-1.354);
    \draw[thick, black] (0,-2) -- (0.707,-0.707);
    \draw[thick, black, -stealth] (2,0) -- (1.354,-0.354);
    \draw[thick, black] (2,0) -- (0.707,-0.707);
    \draw[thick, black, -stealth] (-0.707,0.707) -- (-0.354,1.354);
    \draw[thick, black] (0,2) -- (-0.707,0.707);
    \draw[thick, black, -stealth] (-0.707,0.707) -- (-1.354,0.354);
    \draw[thick, black] (-2,0) -- (-0.707,0.707);
    \draw[thick, black, -stealth] (0.707,-0.707) -- (0,0);
    \draw[thick, black] (0.707,-0.707) -- (-0.707,0.707);

    \node[black, below] at (0,-2) {\scriptsize $ g$};
    \node[black, right] at (2,0) {\scriptsize $ h$};
    \node[black, above] at (0.2,0) {\scriptsize $ g  h$};
\end{tikzpicture}~,
\eea
where $\mu(g,h) \mu^\vee(h,g)$ captures the discrete torsion; this notation will be explained shortly. 
We may now rephrase this gauging prescription as follows. Begin by introducing the non-simple line 
\bea
\cA := \bigoplus_{g \in G} g~.
\eea
Because $\cA$ is composed of $g \in G$, there exist non-zero, and in particular one-dimensional, hom-spaces $\mathrm{Hom}(g,\cA)$ and $\mathrm{Hom}(\cA,g)$, whose basis vectors can be chosen as inclusion and projection maps, respectively, for the constituents of $\cA$. The basis vectors for these spaces are written graphically as follows,
\begin{equation}\label{eq:ALvbasis}
{\begin{tikzpicture}[baseline=0]
\draw[ thick, decoration={markings, mark=at position 0.5 with {\arrow{stealth}}}, postaction={decorate}] (0,-0.5) -- (0,0);
\draw[red, thick, decoration={markings, mark=at position 0.7 with {\arrow{stealth}}}, postaction={decorate}] (0,0) -- (0,0.5);
\filldraw[] (0,0) circle (1pt);

\node[ below] at (0,-0.5) {\footnotesize $g$};
\node[red, above] at (0,0.5) {\footnotesize $\cA$};
\end{tikzpicture}} \in \mathrm{Hom}(g, \cA)~, \hspace{0.5 in} {\begin{tikzpicture}[baseline=0]
\draw[red, thick, decoration={markings, mark=at position 0.5 with {\arrow{stealth}}}, postaction={decorate}] (0,-0.5) -- (0,0);
\draw[ thick,, decoration={markings, mark=at position 0.7 with {\arrow{stealth}}}, postaction={decorate}] (0,0) -- (0,0.5);
\filldraw[] (0,0) circle (1pt);

\node[red, below] at (0,-0.5) {\footnotesize $\cA$};
\node[ above] at (0,0.5) {\footnotesize $g$};
\end{tikzpicture}} \in \mathrm{Hom}(\cA, g)~.
\end{equation}
In particular, for any $G$, we will always have $\dim\mathrm{Hom}(1,\cA)=\dim\mathrm{Hom}(\cA,1)=1$, capturing the presence of a topological endpoint for $\cA$. 
We may choose a complete orthonormal basis such that the following conditions are satisfied, 
\begin{equation}\label{eq:grouporthonormalA}
 \sum_{g} {\begin{tikzpicture}[baseline=0,square/.style={regular polygon,regular polygon sides=4}]
\draw[ thick] (0,0) -- (0,0.5);
\draw[red, thick, decoration={markings, mark=at position 0.7 with {\arrow{stealth}}}, postaction={decorate}] (0,0.5) -- (0,1);
\filldraw[] (0,0.5) circle (1pt);
\draw[red, thick, decoration={markings, mark=at position 0.5 with {\arrow{stealth}}}, postaction={decorate}] (0,-1) -- (0,-0.5);
\draw[ thick, decoration={markings, mark=at position 1 with {\arrow{stealth}}}, postaction={decorate}] (0,-0.5) -- (0,0);
\filldraw[] (0,-0.5) circle (1pt);

\node[ right] at (0,0) {\footnotesize $g$};
\node[red, above] at (0,1) {\footnotesize $\cA$};
\node[red, below] at (0,-1) {\footnotesize $\cA$};
\end{tikzpicture}} \,\,=\,\, {\begin{tikzpicture}[baseline=0,square/.style={regular polygon,regular polygon sides=4}] 
\draw[red, thick, decoration={markings, mark=at position 0.6 with {\arrow{stealth}}}, postaction={decorate}] (0,-1) -- (0,1);

\node[red, above] at (0,1) {\footnotesize $\cA$};
\node[red, below] at (0,-1) {\footnotesize $\cA$};
\end{tikzpicture}}~, \hspace{0.7 in}
{\begin{tikzpicture}[baseline=0,square/.style={regular polygon,regular polygon sides=4}]
\draw[red, thick, decoration={markings, mark=at position 0.3 with {\arrow{stealth}}}, postaction={decorate}] (0,0) -- (0,0.5);
\draw[ thick, decoration={markings, mark=at position 0.7 with {\arrow{stealth}}}, postaction={decorate}] (0,0.5) -- (0,1);
\filldraw[] (0,0.5) circle (1pt);
\draw[ thick,  decoration={markings, mark=at position 0.5 with {\arrow{stealth}}}, postaction={decorate}] (0,-1) -- (0,-0.5);
\draw[red,thick] (0,-0.5) -- (0,0);
\filldraw[] (0,-0.5) circle (1pt);

\node[red, left] at (0,0) {\footnotesize $\cA$};
\node[ above] at (0,1) {\footnotesize $g$};
\node[ below] at (0,-1) {\footnotesize $h$};
\end{tikzpicture}} \,\,=\,\, \delta^{g,h}{\begin{tikzpicture}[baseline=0]
\draw[ thick, decoration={markings, mark=at position 0.6 with {\arrow{stealth}}}, postaction={decorate}] (0,-1) -- (0,1);

\node[ above] at (0,1) {\footnotesize $g$};
\node[ below] at (0,-1) {\footnotesize $g$};
\end{tikzpicture}}~,
\end{equation}
with the first condition capturing completeness and the second capturing orthonormality. 
We are always free to rescale these junction vectors by a phase, 
\bea
\label{eq:ALvgaugeredundancy}
{\begin{tikzpicture}[baseline=0]
\draw[ thick, decoration={markings, mark=at position 0.5 with {\arrow{stealth}}}, postaction={decorate}] (0,-0.5) -- (0,0);
\draw[red, thick, decoration={markings, mark=at position 0.7 with {\arrow{stealth}}}, postaction={decorate}] (0,0) -- (0,0.5);
\filldraw[] (0,0) circle (1pt);

\node[ below] at (0,-0.5) {\footnotesize $g$};
\node[red, above] at (0,0.5) {\footnotesize $\cA$};
\end{tikzpicture}} \rightarrow \chi(g) {\begin{tikzpicture}[baseline=0]
\draw[ thick, decoration={markings, mark=at position 0.5 with {\arrow{stealth}}}, postaction={decorate}] (0,-0.5) -- (0,0);
\draw[red, thick, decoration={markings, mark=at position 0.7 with {\arrow{stealth}}}, postaction={decorate}] (0,0) -- (0,0.5);
\filldraw[] (0,0) circle (1pt);

\node[ below] at (0,-0.5) {\footnotesize $g$};
\node[red, above] at (0,0.5) {\footnotesize $\cA$};
\end{tikzpicture}}~, \hspace{0.5 in} {\begin{tikzpicture}[baseline=0]
\draw[red, thick, decoration={markings, mark=at position 0.5 with {\arrow{stealth}}}, postaction={decorate}] (0,-0.5) -- (0,0);
\draw[ thick,, decoration={markings, mark=at position 0.7 with {\arrow{stealth}}}, postaction={decorate}] (0,0) -- (0,0.5);
\filldraw[] (0,0) circle (1pt);

\node[red, below] at (0,-0.5) {\footnotesize $\cA$};
\node[ above] at (0,0.5) {\footnotesize $g$};
\end{tikzpicture}} \rightarrow \chi(g)^{-1} {\begin{tikzpicture}[baseline=0]
\draw[red, thick, decoration={markings, mark=at position 0.5 with {\arrow{stealth}}}, postaction={decorate}] (0,-0.5) -- (0,0);
\draw[ thick,, decoration={markings, mark=at position 0.7 with {\arrow{stealth}}}, postaction={decorate}] (0,0) -- (0,0.5);
\filldraw[] (0,0) circle (1pt);

\node[red, below] at (0,-0.5) {\footnotesize $\cA$};
\node[ above] at (0,0.5) {\footnotesize $g$};
\end{tikzpicture}}
\eea
which preserves the two conditions. As in (\ref{eq:junctiongaugetrans}), this is referred to as a ``gauge redundancy.''

In addition to the line $\cA$ itself, we must also specify rules for local fusion of the line, known as a \textit{multiplication} junction $\mu$. By using the hom spaces above, it is possible to represent the multiplication rule in terms of that of the constituent simple lines as,
\bea
\label{eq:groupalgebramultdef}
{\begin{tikzpicture}[baseline=0,scale=0.9]
\draw [red, thick, decoration={markings, mark=at position 0.5 with {\arrow{stealth}}}, postaction={decorate}] (-1,-1) -- (0,0);
\draw [red, thick, decoration={markings, mark=at position 0.5 with {\arrow{stealth}}}, postaction={decorate}] (1,-1) -- (0,0);
\draw [red, thick, decoration={markings, mark=at position 0.6 with {\arrow{stealth}}}, postaction={decorate}] (0,0) -- (0,1);

\filldraw[red] (0,0) circle (1.5pt);
 \node[above,red] at (0,-0.1) {\footnotesize$\times$};
\node[red,right] at (0,0) {\scriptsize$\mu$};

\node[below] at (-1.2,-1) {\footnotesize \color{red}$\cA$};
\node[below] at (1.1,-1) {\footnotesize \color{red}$\cA$};
\node[above] at (0,1) {\footnotesize \color{red}$\cA$};
\end{tikzpicture}}
\displaystyle :={1\over \sqrt{|G|}}\sum_{g,h \in G} \,\,\mu(g,h)\hspace{-0.1 in}
{\begin{tikzpicture}[baseline=0,scale=0.9]

\draw [ thick, decoration={markings, mark=at position 0.5 with {\arrow{stealth}}}, postaction={decorate}] (0.7,-0.7) -- (0,0);
\draw [red, thick, decoration={markings, mark=at position 0.5 with {\arrow{stealth}}}, postaction={decorate}] (1,-1) -- (0.7,-0.7);

 \draw [red, thick, decoration={markings, mark=at position 0.5 with {\arrow{stealth}}}, postaction={decorate}] (-1,-1) -- (-0.7,-0.7);
\draw [ thick, decoration={markings, mark=at position 0.5 with {\arrow{stealth}}}, postaction={decorate}] (-0.7,-0.7) -- (0,0);
\draw [ thick, decoration={markings, mark=at position 0.5 with {\arrow{stealth}}}, postaction={decorate}] (0,0) -- (0,0.6);
\draw [red, thick, decoration={markings, mark=at position 0.8 with {\arrow{stealth}}}, postaction={decorate}] (0,0.6) -- (0,1);

\node[below] at (-1.2,-1) {\footnotesize \color{red}$\cA$};
\node[below] at (1.1,-1) {\footnotesize \color{red}$\cA$};
\node[above] at (0,1) {\footnotesize \color{red}$\cA$};

\node[below] at (-0.6,0.1) {\footnotesize $g$};
\node[below] at (0.6,0.1) {\footnotesize $h$};
\node[right] at (-0,0.4) {\footnotesize $g h$};

\filldraw[] (-0.7,-0.7) circle (1pt);
\filldraw[] (0.7,-0.7) circle (1pt);
\filldraw[] (0,0.6) circle (1pt);

 \node[above] at (0,0.1) {\footnotesize$\times$};
\end{tikzpicture}}~.
\hspace{0 in}
\eea
Note that for notational simplicity, one often drops the hom elements on the right-hand side and simply draws the simple internal lines, i.e.
\bea
\label{eq:simplifiedmults}
 \begin{tikzpicture}[baseline={([yshift=-1ex]current bounding box.center)},vertex/.style={anchor=base,
    circle,fill=black!25,minimum size=18pt,inner sep=2pt},scale=0.4]
   \draw[->-=0.5,thick,red] (-2,-2) -- (0,0);
   \draw[->-=0.5,thick,red] (2,-2) -- (0,0);
   \draw[->-=0.7,thick,red] (0,0)--(0,2) ;
     \node[below,red] at (-2,-2.2) {$\cA$};
     \node[below,red] at (2,-2) {$\cA$};
     \node[above,red] at (0,2) {$\cA$};
     \node[above,red] at (0,-0.2) {\footnotesize$\times$};
       \filldraw[red]  (0,0) circle (3pt);
    \node[red,right] at (0,0) {\scriptsize$\mu$};
    \end{tikzpicture} = {1\over \sqrt{|G|}}\sum_{g,h} \mu(g,h)
     \begin{tikzpicture}[baseline={([yshift=-1ex]current bounding box.center)},vertex/.style={anchor=base,
    circle,fill=black!25,minimum size=18pt,inner sep=2pt},scale=0.4]
   \draw[->-=0.5,thick] (-2,-2) -- (0,0);
   \draw[->-=0.5,thick] (2,-2) -- (0,0);
   \draw[->-=0.7,thick] (0,0)--(0,2) ;
     \node[below] at (-2,-2.2) {$g$};
     \node[below] at (2,-2) {$h$};
     \node[above] at (0,2) {$gh$};
     \node[above] at (0,-0.2) {\footnotesize$\times$};
    \end{tikzpicture} ~,
\eea
but what is really meant by this is what is in (\ref{eq:groupalgebramultdef}).  
The multiplication must be chosen such that the following associativity condition is satisfied, 
\bea
\label{eq:associativityforA}
{\begin{tikzpicture}[baseline=0,scale=0.6]
\draw [red, thick, decoration={markings, mark=at position 0.5 with {\arrow{stealth}}}, postaction={decorate}] (1,1) -- (2,2);
\draw [red, thick, decoration={markings, mark=at position 0.5 with {\arrow{stealth}}}, postaction={decorate}] (0,0) -- (1,1);
\draw [red, thick, decoration={markings, mark=at position 0.5 with {\arrow{stealth}}}, postaction={decorate}] (-1,-1) -- (0,0);
\draw [red, thick, decoration={markings, mark=at position 0.5 with {\arrow{stealth}}}, postaction={decorate}] (1,-1) -- (0,0);
\draw [red, thick, decoration={markings, mark=at position 0.5 with {\arrow{stealth}}}, postaction={decorate}] (3,-1) -- (1,1);

\filldraw[red] (0,0) circle (2pt);
\node[red,left] at (0,0) {\scriptsize$\mu$};
\filldraw[red] (1,1) circle (2pt);
\node[red,left] at (1,1) {\scriptsize$\mu$};

\node[below] at (-1.2,-1) {\footnotesize \color{red}$\cA$};
\node[below] at (1.1,-1) {\footnotesize \color{red}$\cA$};
\node[below] at (3.6,-1) {\footnotesize \color{red}$\cA$};

\node[above] at (2,2) {\footnotesize \color{red}$\cA$};
\end{tikzpicture}}
\hspace{0.05 in}= \hspace{0.1 in}
{\begin{tikzpicture}[baseline=0,square/.style={regular polygon,regular polygon sides=4},scale=0.6]
\draw [red, thick, decoration={markings, mark=at position 0.5 with {\arrow{stealth}}}, postaction={decorate}] (1,1) -- (2,2);
\draw [red, thick] (0,0) -- (1,1);
\draw [red, thick, ->-=0.5] (-1,-1) -- (0,0);
\draw [red, thick, decoration={markings, mark=at position 0.5 with {\arrow{stealth}}}, postaction={decorate}] (1,-1) -- (2,0);
\draw [red, thick, decoration={markings, mark=at position 0.5 with {\arrow{stealth}}}, postaction={decorate}] (3,-1) -- (2,0);
\draw [red, thick, decoration={markings, mark=at position 0.5 with {\arrow{stealth}}}, postaction={decorate}] (2,0) -- (1,1);

\filldraw[red] (2,0) circle (2pt);
\node[red,left] at (1,1) {\scriptsize$\mu$};
\filldraw[red] (1,1) circle (2pt);
\node[red,right] at (2,0) {\scriptsize$\mu$};

\node[below] at (-1.2,-1) {\footnotesize \color{red}$\cA$};
\node[below] at (1.1,-1) {\footnotesize \color{red}$\cA$};
\node[below] at (3.6,-1) {\footnotesize \color{red}$\cA$};

\node[above] at (2,2) {\footnotesize \color{red}$\cA$};
\end{tikzpicture}}~. 
\eea
This condition may be rewritten in terms of the coefficients $\mu(g,h)$ by expanding the junctions using  (\ref{eq:groupalgebramultdef}),
shrinking an internal $\cA$ line using (\ref{eq:grouporthonormalA}), using an $F$-symbol of the constituent simple lines on the left, and comparing to the result on the right. 
This gives rise to the constraint 
\bea
\mu(g,h) \mu(gh,k) \omega(g,h,k) = \mu(g,hk) \mu(h,k)~,
\eea
or, in terms of the operation $\delta$ defined in (\ref{eq:deltadefinition}),\footnote{Note that the gauge redundancy in (\ref{eq:ALvgaugeredundancy}), which shifts the multiplication by 
 \bea
 \mu(g,h) \sim  \mu(g,h)  \delta \chi(g,h)~,
 \eea
 does not modify this relation. 
}
\bea
 \label{eq:relomegadelmu}
\omega(g,h,k) = \delta \mu(g,h,k)~. 
\eea
This tells us that the associator cocycle $\omega(g,h,k) $ must be cohomologically trivial---this encodes the lack of anomaly for the group $G$ we will be gauging. 
The pair of $\cA$ together with a multiplication operation satisfying the above is referred to as an \textit{algebra}. 

Similar to an algebra, one may also define a \textit{coalgebra} by equipping $\cA$ with a \textit{comultiplication} operation, i.e. a splitting junction,
\bea
{\begin{tikzpicture}[baseline=0,scale=0.9]
\draw [red, thick, -<-=0.5] (-1,1) -- (0,0);
\draw [red, thick, -<-=0.5] (1,1) -- (0,0);
\draw [red, thick, -<-=0.6] (0,0) -- (0,-1);

\filldraw[red] (0,0) circle (1.5pt);
\node[red,right] at (-0,-0) {\scriptsize$\overline\mu^\vee$};
 \node[below,red] at (0,0) {\footnotesize$\times$};

\node[above] at (-1.2,1) {\footnotesize \color{red}$\cA$};
\node[above] at (1.1,1) {\footnotesize \color{red}$\cA$};
\node[below] at (0,-1) {\footnotesize \color{red}$\cA$};
\end{tikzpicture}}
  :={1\over \sqrt{|G|}} \sum_{g, h \in G} \,\,\mu^\vee({g,h})
  \hspace{-0.1 in}
{\begin{tikzpicture}[baseline=0,scale=0.9]
 \draw [red, thick, , -<-=0.5] (-1,1) -- (-0.7,0.7);
 \draw [ thick, , -<-=0.5] (-0.7,0.7) -- (0,0);
\draw [ thick, , -<-=0.5] (0.7,0.7) -- (0,0);
\draw [red, thick, , -<-=0.5] (1,1) -- (0.7,0.7);

\draw [ thick, , -<-=0.5] (0,0) -- (0,-0.6);
\draw [red, thick, , -<-=0.7] (0,-0.6) -- (0,-1);

\filldraw[] (-0.7,0.7) circle (1pt);
\filldraw[] (0.7,0.7) circle (1pt);
\filldraw[] (0,-0.6) circle (1pt);

\node[left] at (-0.3,0.2) {\footnotesize $g$};
\node[right] at (0.3,0.2) {\footnotesize$h$};
\node[right] at (0,-0.4) {\footnotesize$g h$};

\node[above] at (-1.2,1) {\footnotesize \color{red}$\cA$};
\node[above] at (1.1,1) {\footnotesize \color{red}$\cA$};
\node[below] at (0,-1) {\footnotesize \color{red}$\cA$};

 \node[below] at (0,-0.1) {\footnotesize$\times$};
 
\end{tikzpicture}}~,\eea
or in simplified notation 
\bea
\begin{tikzpicture}[baseline={([yshift=-1ex]current bounding box.center)},vertex/.style={anchor=base,
    circle,fill=black!25,minimum size=18pt,inner sep=2pt},scale=0.4]
   \draw[->-=0.6,thick,red] (0,-2) -- (0,0);
   \draw[->-=0.6,thick,red]  (0,0) -- (2,2);
   \draw[->-=0.6,thick,red] (0,0)--(-2,2) ;
     \node[above,red] at (2,2) {$\cA$};
     \node[above,red] at (-2,2) {$\cA$};
     \node[below,red] at (0,-2) {$\cA$};
       \filldraw[red]  (0,0) circle (3pt);
     \node[red,right] at (0,-0.2) {\scriptsize$\overline \mu^\vee$};
      \node[red] at (0,-0.4) {\footnotesize$\times$};
    \end{tikzpicture}
  ={1\over \sqrt{|G|}}\sum_{g,h \in G} \mu^\vee(g,h)
     \begin{tikzpicture}[baseline={([yshift=-1ex]current bounding box.center)},vertex/.style={anchor=base,
    circle,fill=black!25,minimum size=18pt,inner sep=2pt},scale=0.4]
   \draw[->-=0.6,thick] (0,-2) -- (0,0);
   \draw[->-=0.6,thick]  (0,0) -- (2,2);
   \draw[->-=0.6,thick] (0,0)--(-2,2) ;
     \node[above] at (2,2) {${h}$};
     \node[above] at (-2,2) {${g}$};
     \node[below] at (0,-2) {$gh$};
      \node[] at (0,-0.4) {\footnotesize$\times$};
    \end{tikzpicture}~. 
\eea
The coefficients $\mu^\vee(g,h)$ are constrained to satisfy the upside-down analog of (\ref{eq:associativityforA}). 

The algebra objects that we will be concerned with here will be equipped with \textit{both} multiplication and comultiplication operations. 
Given such an algebra object, we may rewrite the gauged torus partition function in terms of it as follows, 
\bea
\label{eq:gaugingintermsofalgebra}
Z_{\cX /G}
&=&  \begin{tikzpicture}[baseline={([yshift=+.5ex]current bounding box.center)},vertex/.style={anchor=base,
    circle,fill=black!25,minimum size=18pt,inner sep=2pt},scale=0.5]
    \filldraw[grey] (-2,-2) rectangle ++(4,4);
    \draw[thick, dgrey] (-2,-2) -- (-2,+2);
    \draw[thick, dgrey] (-2,-2) -- (+2,-2);
    \draw[thick, dgrey] (+2,+2) -- (+2,-2);
    \draw[thick, dgrey] (+2,+2) -- (-2,+2);
    \draw[thick, red, -stealth] (0,-2) -- (0.354,-1.354);
    \draw[thick, red] (0,-2) -- (0.707,-0.707);
    \draw[thick, red, -stealth] (2,0) -- (1.354,-0.354);
    \draw[thick, red] (2,0) -- (0.707,-0.707);
    \draw[thick, red, -stealth] (-0.707,0.707) -- (-0.354,1.354);
    \draw[thick, red] (0,2) -- (-0.707,0.707);
    \draw[thick, red, -stealth] (-0.707,0.707) -- (-1.354,0.354);
    \draw[thick, red] (-2,0) -- (-0.707,0.707);
    \draw[thick, red, -stealth] (0.707,-0.707) -- (0,0);
    \draw[thick, red] (0.707,-0.707) -- (-0.707,0.707);

     \filldraw[red]  (-0.707,0.707) circle (3pt);
          \filldraw[red]  (0.707,-0.707) circle (3pt);
            \node[right,red] at (0.5,-1) {\scriptsize$\mu$};
              \node[left,red] at (-0.5,1) {\scriptsize$\mu^\vee$};
          
    \node[red, below] at (0,-2) {\scriptsize $ \cA$};
    \node[red, right] at (2,0) {\scriptsize $ \cA$};
    \node[red, above] at (0.2,0) {\scriptsize $ \cA$};
\end{tikzpicture}~,
\eea
where the choice of discrete torsion is now captured by the choice of (co-)multiplication.
For the gauging process to be unambiguous (i.e. non-anomalous), it is clear that we must have the additional constraint 
\bea
\label{eq:Frobeniuscondition}
 \begin{tikzpicture}[baseline={([yshift=+.5ex]current bounding box.center)},vertex/.style={anchor=base,
    circle,fill=black!25,minimum size=18pt,inner sep=2pt},scale=0.5]

    \draw[thick, red, -stealth] (0,-2) -- (0.354,-1.354);
    \draw[thick, red] (0,-2) -- (0.707,-0.707);
    \draw[thick, red, -stealth] (2,0) -- (1.354,-0.354);
    \draw[thick, red] (2,0) -- (0.707,-0.707);
    \draw[thick, red, -stealth] (-0.707,0.707) -- (-0.354,1.354);
    \draw[thick, red] (0,2) -- (-0.707,0.707);
    \draw[thick, red, -stealth] (-0.707,0.707) -- (-1.354,0.354);
    \draw[thick, red] (-2,0) -- (-0.707,0.707);
    \draw[thick, red, -stealth] (0.707,-0.707) -- (0,0);
    \draw[thick, red] (0.707,-0.707) -- (-0.707,0.707);

     \filldraw[red]  (-0.707,0.707) circle (3pt);
          \filldraw[red]  (0.707,-0.707) circle (3pt);
            \node[right,red] at (0.5,-1) {\scriptsize$\mu$};
              \node[left,red] at (-0.5,1) {\scriptsize$\mu^\vee$};
          
\end{tikzpicture} 
\hspace{0.2 in}= \hspace{0.2 in}
\begin{tikzpicture}[baseline={([yshift=+.5ex]current bounding box.center)},vertex/.style={anchor=base,
    circle,fill=black!25,minimum size=18pt,inner sep=2pt},scale=0.5]
\begin{scope}[yscale=-1,xscale=1]
    \draw[thick, red, -stealth] (0.707,-0.707)-- (0.354,-1.354);
    \draw[thick, red] (0,-2) -- (0.707,-0.707);
    \draw[thick, red, -stealth] (2,0) -- (1.354,-0.354);
    \draw[thick, red] (2,0) -- (0.707,-0.707);
    \draw[thick, red, -stealth] (0,2) -- (-0.354,1.354);
    \draw[thick, red] (0,2) -- (-0.707,0.707);
    \draw[thick, red, -stealth] (-0.707,0.707) -- (-1.354,0.354);
    \draw[thick, red] (-2,0) -- (-0.707,0.707);
    \draw[thick, red, -stealth] (-0.707,0.707) -- (0,0);
    \draw[thick, red] (0.707,-0.707) -- (-0.707,0.707);

     \filldraw[red]  (-0.707,0.707) circle (3pt);
          \filldraw[red]  (0.707,-0.707) circle (3pt);
            \node[right,red] at (0.5,-1) {\scriptsize$\mu$};
              \node[left,red] at (-0.5,1) {\scriptsize$\mu^\vee$};
          
    \end{scope}
\end{tikzpicture} 
~,
\eea
which is known as the \textit{Frobenius} condition. Algebras which satisfy this constraint are referred to as \textit{Frobenius algebras}. 
One also requires the consistency conditions shown below, known as the separability condition,
\bea
{\begin{tikzpicture}[baseline=0,scale=0.5]
\draw [red, thick, decoration={markings, mark=at position 0.5 with {\arrow{stealth}}}, postaction={decorate}] (0,-2) -- (0,-0.6);
\draw [red, thick, decoration={markings, mark=at position 0.5 with {\arrow{stealth}}}, postaction={decorate}] (0,0.6) -- (0,2);
\draw [red,thick,decoration={markings, mark=at position 0.5 with {\arrow{stealth}}}, postaction={decorate}] (0,-0.6) arc [radius=.6, start angle=-90, end angle=90];
\draw [red,thick,decoration={markings, mark=at position 0.5 with {\arrow{stealth[reversed]}}}, postaction={decorate}] (0,0.6) arc [radius=.6, start angle=90, end angle=270];
\filldraw[red] (0,0.6)  circle (2pt);
\node[red,left] at (-0.1,0.8)  {\scriptsize$\mu$};
\filldraw[red] (0,-0.6) circle (2pt);
\node[red,right] at (0,-0.8)  {\scriptsize$\mu^\vee$};
\node[below] at (0,-2) {\footnotesize \color{red}$\cA$};
\end{tikzpicture}}
\hspace{0.1 in}= \hspace{0.2 in}
{\begin{tikzpicture}[baseline=0,scale=0.5]
\draw [red, thick, decoration={markings, mark=at position 0.5 with {\arrow{stealth}}}, postaction={decorate}] (0,-2) -- (0,2);
\node[below] at (0,-2) {\footnotesize \color{red}$\cA$};
\end{tikzpicture}}
\eea
which ensures that configurations of $\cA$ that are continuously deformable into one another encode the same gauging.\footnote{In addition to the conditions discussed here, there is also a symmetricity condition; see e.g. Figure 4.5 of \cite{Bhardwaj:2017xup}. }
The upshot is then that discrete gaugings of $G$ correspond to separable (symmetric) Frobenius algebras.

We may now generalize this to the non-invertible case. We proceed similarly to before by defining a certain object
\bea
\cA = \bigoplus_{a \in \cC} Z_a\, a~, 
\eea
with $Z_a$ being non-negative integer coefficients; once again, we require $Z_1 = 1$. 
The hom-spaces $\mathrm{Hom}(a,\cA)$ and $\mathrm{Hom}(\cA,a)$ are now $Z_a$-dimensional, and the basis vectors may be represented graphically as before. Now, however, we must add a label keeping track of which basis vector we are considering, 
\bea
{\begin{tikzpicture}[baseline=0]
\draw[ thick, decoration={markings, mark=at position 0.5 with {\arrow{stealth}}}, postaction={decorate}] (0,-0.5) -- (0,0);
\draw[red, thick, decoration={markings, mark=at position 0.7 with {\arrow{stealth}}}, postaction={decorate}] (0,0) -- (0,0.5);
\filldraw[] (0,0) circle (1pt);
\node[ left] at (0,0) {\footnotesize $i$};

\node[ below] at (0,-0.5) {\footnotesize $a$};
\node[red, above] at (0,0.5) {\footnotesize $\cA$};
\end{tikzpicture}} \in \mathrm{Hom}(a, \cA)~, \hspace{0.5 in} {\begin{tikzpicture}[baseline=0]
\draw[red, thick, decoration={markings, mark=at position 0.5 with {\arrow{stealth}}}, postaction={decorate}] (0,-0.5) -- (0,0);
\draw[ thick,, decoration={markings, mark=at position 0.7 with {\arrow{stealth}}}, postaction={decorate}] (0,0) -- (0,0.5);
\filldraw[] (0,0) circle (1pt);

\node[red, below] at (0,-0.5) {\footnotesize $\cA$};
\node[ above] at (0,0.5) {\footnotesize $a$};
\node[ left] at (0,0) {\footnotesize $\bar{\imath}$};
\end{tikzpicture}} \in \mathrm{Hom}(\cA, a)~, \hspace{0.3 in} i, \bar{\imath} = 1, \dots, Z_a~;
\eea
the completeness and  orthonormality conditions are similarly modified, 
\begin{equation}\label{eq:orthonormalA}
 \sum_{a} \sum_{i=1}^{Z_a} {\begin{tikzpicture}[baseline=0,square/.style={regular polygon,regular polygon sides=4}]
\draw[ thick] (0,0) -- (0,0.5);
\draw[red, thick, decoration={markings, mark=at position 0.7 with {\arrow{stealth}}}, postaction={decorate}] (0,0.5) -- (0,1);
\filldraw[] (0,0.5) circle (1pt);
\draw[red, thick, decoration={markings, mark=at position 0.5 with {\arrow{stealth}}}, postaction={decorate}] (0,-1) -- (0,-0.5);
\draw[ thick, decoration={markings, mark=at position 1 with {\arrow{stealth}}}, postaction={decorate}] (0,-0.5) -- (0,0);
\filldraw[] (0,-0.5) circle (1pt);

\node[ right] at (0,0) {\footnotesize $a$};
\node[left] at (0,0.5) {$\bar{\imath}$};
\node[left] at (0,-0.5) {$i$};
\node[red, above] at (0,1) {\footnotesize $\cA$};
\node[red, below] at (0,-1) {\footnotesize $\cA$};
\end{tikzpicture}} \,\,=\,\, {\begin{tikzpicture}[baseline=0,square/.style={regular polygon,regular polygon sides=4}] 
\draw[red, thick, decoration={markings, mark=at position 0.6 with {\arrow{stealth}}}, postaction={decorate}] (0,-1) -- (0,1);

\node[red, above] at (0,1) {\footnotesize $\cA$};
\node[red, below] at (0,-1) {\footnotesize $\cA$};
\end{tikzpicture}}~, \hspace{0.7 in}
{\begin{tikzpicture}[baseline=0,square/.style={regular polygon,regular polygon sides=4}]
\draw[red, thick, decoration={markings, mark=at position 0.3 with {\arrow{stealth}}}, postaction={decorate}] (0,0) -- (0,0.5);
\draw[ thick, decoration={markings, mark=at position 0.7 with {\arrow{stealth}}}, postaction={decorate}] (0,0.5) -- (0,1);
\filldraw[] (0,0.5) circle (1pt);
\draw[ thick,  decoration={markings, mark=at position 0.5 with {\arrow{stealth}}}, postaction={decorate}] (0,-1) -- (0,-0.5);
\draw[red,thick] (0,-0.5) -- (0,0);
\filldraw[] (0,-0.5) circle (1pt);

\node[left] at (0,0.5) {$\bar{\imath}$};
\node[left] at (0,-0.5) {$j$};

\node[red, left] at (0,0) {\footnotesize $\cA$};
\node[ above] at (0,1) {\footnotesize $a$};
\node[ below] at (0,-1) {\footnotesize $b$};
\end{tikzpicture}} \,\,=\,\, \delta^{i,j} \delta^{a,b}{\begin{tikzpicture}[baseline=0]
\draw[ thick, decoration={markings, mark=at position 0.6 with {\arrow{stealth}}}, postaction={decorate}] (0,-1) -- (0,1);

\node[ above] at (0,1) {\footnotesize $a$};
\node[ below] at (0,-1) {\footnotesize $a$};
\end{tikzpicture}}~.
\end{equation}
The multiplication and co-multiplication are defined in a similar manner to before,\footnote{The notation $ c \prec a b$ means that $c$ is contained in the fusion of $a$ and $b$.}
\bea
\label{eq:algebramultdef}
{\begin{tikzpicture}[baseline=0,scale=0.9]
\draw [red, thick, decoration={markings, mark=at position 0.5 with {\arrow{stealth}}}, postaction={decorate}] (-1,-1) -- (0,0);
\draw [red, thick, decoration={markings, mark=at position 0.5 with {\arrow{stealth}}}, postaction={decorate}] (1,-1) -- (0,0);
\draw [red, thick, decoration={markings, mark=at position 0.6 with {\arrow{stealth}}}, postaction={decorate}] (0,0) -- (0,1);

\filldraw[red] (0,0) circle (1.5pt);
 \node[above,red] at (0,-0.1) {\footnotesize$\times$};
\node[red,right] at (0,0) {\scriptsize$\mu$};

\node[below] at (-1.2,-1) {\footnotesize \color{red}$\cA$};
\node[below] at (1.1,-1) {\footnotesize \color{red}$\cA$};
\node[above] at (0,1) {\footnotesize \color{red}$\cA$};
\end{tikzpicture}}
\displaystyle &:=&{1\over \sqrt{d_\cA}}\sum_{\substack{a,b \\ c \prec a b}}\sum_{i,j,k}{\sum_{\alpha=1}^{N_{ab}^{c}}} \,\,(\mu_{a,b}^{c;\alpha})_{i,j,k}\hspace{-0.1 in}
{\begin{tikzpicture}[baseline=0,scale=0.9]

\draw [ thick, decoration={markings, mark=at position 0.5 with {\arrow{stealth}}}, postaction={decorate}] (0.7,-0.7) -- (0,0);
\draw [red, thick, decoration={markings, mark=at position 0.5 with {\arrow{stealth}}}, postaction={decorate}] (1,-1) -- (0.7,-0.7);

 \draw [red, thick, decoration={markings, mark=at position 0.5 with {\arrow{stealth}}}, postaction={decorate}] (-1,-1) -- (-0.7,-0.7);
\draw [ thick, decoration={markings, mark=at position 0.5 with {\arrow{stealth}}}, postaction={decorate}] (-0.7,-0.7) -- (0,0);
\draw [ thick, decoration={markings, mark=at position 0.5 with {\arrow{stealth}}}, postaction={decorate}] (0,0) -- (0,0.6);
\draw [red, thick, decoration={markings, mark=at position 0.8 with {\arrow{stealth}}}, postaction={decorate}] (0,0.6) -- (0,1);

\node[below] at (-1.2,-1) {\footnotesize \color{red}$\cA$};
\node[below] at (1.1,-1) {\footnotesize \color{red}$\cA$};
\node[above] at (0,1) {\footnotesize \color{red}$\cA$};

\node[left] at (-0.7,-0.6) {\scriptsize $\bar{\imath}$};
\node[right] at (0.7,-0.6) {\scriptsize $\bar{\jmath}$};
\node[left] at (0,0.6) {\scriptsize $k$};

\node[below] at (-0.6,0.1) {\footnotesize $a$};
\node[below] at (0.6,0.1) {\footnotesize $b$};
\node[right] at (-0,0.4) {\footnotesize $c$};
\node[below] at (0,0) {\scriptsize $\alpha$};

\filldraw[] (-0.7,-0.7) circle (1pt);
\filldraw[] (0.7,-0.7) circle (1pt);
\filldraw[] (0,0.6) circle (1pt);
\filldraw[] (0,0) circle (1pt);

 \node[above] at (0,0.1) {\footnotesize$\times$};
\end{tikzpicture}}~,
\no\\
{\begin{tikzpicture}[baseline=0,scale=0.9]
\draw [red, thick, -<-=0.5] (-1,1) -- (0,0);
\draw [red, thick, -<-=0.5] (1,1) -- (0,0);
\draw [red, thick, -<-=0.6] (0,0) -- (0,-1);

\filldraw[red] (0,0) circle (1.5pt);
\node[red,right] at (-0,-0) {\scriptsize$\overline\mu^\vee$};
 \node[below,red] at (0,0) {\footnotesize$\times$};

\node[above] at (-1.2,1) {\footnotesize \color{red}$\cA$};
\node[above] at (1.1,1) {\footnotesize \color{red}$\cA$};
\node[below] at (0,-1) {\footnotesize \color{red}$\cA$};
\end{tikzpicture}}
  &:=&{1\over \sqrt{d_\cA}} \sum_{\substack{a,b \\ c \prec ab}} \sum_{i,j,k}{\sum_{\bar\alpha=1}^{N_{ab}^{c}}} \,\,(\mu_{a,b}^{c;\bar \alpha})^\vee_{i,j,k}
  \hspace{-0.1 in}
{\begin{tikzpicture}[baseline=0,scale=0.9]
 \draw [red, thick, , -<-=0.5] (-1,1) -- (-0.7,0.7);
 \draw [ thick, , -<-=0.5] (-0.7,0.7) -- (0,0);
\draw [ thick, , -<-=0.5] (0.7,0.7) -- (0,0);
\draw [red, thick, , -<-=0.5] (1,1) -- (0.7,0.7);

\draw [ thick, , -<-=0.5] (0,0) -- (0,-0.6);
\draw [red, thick, , -<-=0.7] (0,-0.6) -- (0,-1);

\filldraw[] (-0.7,0.7) circle (1pt);
\filldraw[] (0.7,0.7) circle (1pt);
\filldraw[] (0,-0.6) circle (1pt);
\filldraw[] (0,0) circle (1pt);

\node[left] at (-0.7,0.6) {\scriptsize $i$};
\node[right] at (0.7,0.6) {\scriptsize $j$};
\node[left] at (0,-0.6) {\scriptsize $\bar k$};
\node[above] at (0,0) {\scriptsize $\bar \alpha$};

\node[left] at (-0.3,0.2) {\footnotesize $a$};
\node[right] at (0.3,0.2) {\footnotesize$b$};
\node[right] at (0,-0.4) {\footnotesize$c$};

\node[above] at (-1.2,1) {\footnotesize \color{red}$\cA$};
\node[above] at (1.1,1) {\footnotesize \color{red}$\cA$};
\node[below] at (0,-1) {\footnotesize \color{red}$\cA$};

 \node[below] at (0,-0.1) {\footnotesize$\times$};
 
\end{tikzpicture}}~,
\eea
and it is once again common to use a short-hand notation for the right-hand side, 
\bea
 \begin{tikzpicture}[baseline={([yshift=-1ex]current bounding box.center)},vertex/.style={anchor=base,
    circle,fill=black!25,minimum size=18pt,inner sep=2pt},scale=0.4]
   \draw[->-=0.5,thick,red] (-2,-2) -- (0,0);
   \draw[->-=0.5,thick,red] (2,-2) -- (0,0);
   \draw[->-=0.7,thick,red] (0,0)--(0,2) ;
     \node[below,red] at (-2,-2.2) {$\cA$};
     \node[below,red] at (2,-2) {$\cA$};
     \node[above,red] at (0,2) {$\cA$};
     \node[above,red] at (0,-0.2) {\footnotesize$\times$};
       \filldraw[red]  (0,0) circle (3pt);
    \node[red,right] at (0,0) {\scriptsize$\mu$};
    \end{tikzpicture} &=& {1\over \sqrt{d_\cA}}\sum_{\substack{a,b \\ c \prec a b } } \sum_{i,j,k}{\sum_{\alpha=1}^{N_{ab}^{c}}}\,\, (\mu_{a,b}^{c;\alpha})_{i,j,k}
     \begin{tikzpicture}[baseline={([yshift=-1ex]current bounding box.center)},vertex/.style={anchor=base,
    circle,fill=black!25,minimum size=18pt,inner sep=2pt},scale=0.4]
   \draw[->-=0.5,thick] (-2,-2) -- (0,0);
   \draw[->-=0.5,thick] (2,-2) -- (0,0);
   \draw[->-=0.7,thick] (0,0)--(0,2) ;
   \filldraw[] (0,0) circle (3pt);
     \node[below] at (-2,-2.2) {$a_i$};
     \node[below] at (2,-2) {$b_j$};
     \node[above] at (0,2) {$c_k$};
     \node[below] at (0,0) {\scriptsize $ \alpha$};
     \node[above] at (0,-0.2) {\footnotesize$\times$};
    \end{tikzpicture}~,
    \no\\
     \begin{tikzpicture}[baseline={([yshift=-1ex]current bounding box.center)},vertex/.style={anchor=base,
    circle,fill=black!25,minimum size=18pt,inner sep=2pt},scale=0.4]
   \draw[->-=0.6,thick,red] (0,-2) -- (0,0);
   \draw[->-=0.6,thick,red]  (0,0) -- (2,2);
   \draw[->-=0.6,thick,red] (0,0)--(-2,2) ;
     \node[above,red] at (2,2) {$\cA$};
     \node[above,red] at (-2,2) {$\cA$};
     \node[below,red] at (0,-2) {$\cA$};
       \filldraw[red]  (0,0) circle (3pt);
     \node[red,right] at (0,-0.2) {\scriptsize$\overline \mu^\vee$};
      \node[red] at (0,-0.4) {\footnotesize$\times$};
    \end{tikzpicture}
  &=&{1\over \sqrt{d_\cA}}\sum_{\substack{a,b \\ c \prec a b } } \sum_{i,j,k}{\sum_{\bar\alpha=1}^{N_{ab}^{c}}}\,\, (\mu_{a,b}^{c;\bar \alpha})^\vee_{i,j,k}
     \begin{tikzpicture}[baseline={([yshift=-1ex]current bounding box.center)},vertex/.style={anchor=base,
    circle,fill=black!25,minimum size=18pt,inner sep=2pt},scale=0.4]
   \draw[->-=0.6,thick] (0,-2) -- (0,0);
   \draw[->-=0.6,thick]  (0,0) -- (2,2);
   \draw[->-=0.6,thick] (0,0)--(-2,2) ;
   \filldraw[] (0,0) circle (3pt);
     \node[above] at (2,2) {$b_j$};
     \node[above] at (-2,2) {$a_i $};
     \node[below] at (0,-2) {$c_k$};
     \node[above] at (0,0) {\scriptsize $\bar \alpha$};
      \node[] at (0,-0.4) {\footnotesize$\times$};
    \end{tikzpicture}.\hspace{0.2 in}
\eea
For consistency, we must again impose the associativity and separability constraints, as well as the Frobenius condition in (\ref{eq:Frobeniuscondition}). 
These constrain not only $\mu$ and $\mu^\vee$, but also the allowed coefficients $Z_a$. 
Of course, there need not be a unique solution---each different solution corresponds to a different algebra. 
 One then gauges the algebra in exactly the same way as before  by inserting a fine mesh of $\cA$. 
 For example, on the torus we have the configuration shown in (\ref{eq:gaugingintermsofalgebra}). 
 
 \subsubsection{Morita equivalence}
Note that it is useful to introduce a notion of equivalence classes of algebra objects, known as \textit{Morita equivalence}.
In particular, $\cA_1$ is said to be Morita equivalent to $\cA_2$, i.e. $\cA_1 \sim \cA_2$, when their categories of modules are equivalent as $\cC$-module categories. A sufficient condition is that there exists a simple object $\cN$ such that 
\bea
\cA_2 = \cN \otimes \cA_1 \otimes \overline \cN~. 
\eea
Here the right-hand side carries the canonical algebra structure induced from $\cA_1$.
The reason for defining such equivalence classes is that gauging Morita equivalent objects gives the \textit{same} theory. 
This is rather easy to understand pictorially, where the argument is schematically as follows,
\bea
\label{eq:Moritaschematic}
\begin{tikzpicture}[baseline={([yshift=+.5ex]current bounding box.center)},vertex/.style={anchor=base,
    circle,fill=black!25,minimum size=18pt,inner sep=2pt},scale=0.6]
    \filldraw[grey] (-2,-2) rectangle ++(4,4);
    \draw[thick, dgrey] (-2,-2) -- (-2,+2);
    \draw[thick, dgrey] (-2,-2) -- (+2,-2);
    \draw[thick, dgrey] (+2,+2) -- (+2,-2);
    \draw[thick, dgrey] (+2,+2) -- (-2,+2);
    \draw[thick, red, -stealth] (0,-2) -- (0.354,-1.354);
    \draw[thick, red] (0,-2) -- (0.707,-0.707);
    \draw[thick, red, -stealth] (2,0) -- (1.354,-0.354);
    \draw[thick, red] (2,0) -- (0.707,-0.707);
    \draw[thick, red, -stealth] (-0.707,0.707) -- (-0.354,1.354);
    \draw[thick, red] (0,2) -- (-0.707,0.707);
    \draw[thick, red, -stealth] (-0.707,0.707) -- (-1.354,0.354);
    \draw[thick, red] (-2,0) -- (-0.707,0.707);
    \draw[thick, red, -stealth] (0.707,-0.707) -- (0,0);
    \draw[thick, red] (0.707,-0.707) -- (-0.707,0.707);

     \filldraw[red]  (-0.707,0.707) circle (3pt);
          \filldraw[red]  (0.707,-0.707) circle (3pt);
          
    \node[red, below] at (0,-2) {\scriptsize $ \cA_2$};
    \node[red, right] at (2,0) {\scriptsize $ \cA_2$};
    \node[red, above] at (0.2,0) {\scriptsize $ \cA_2$};
\end{tikzpicture} \hspace{0.1 in}=\hspace{0.1 in} \begin{tikzpicture}[baseline={([yshift=+.5ex]current bounding box.center)},vertex/.style={anchor=base,
    circle,fill=black!25,minimum size=18pt,inner sep=2pt},scale=0.6]
    \filldraw[grey] (-2,-2) rectangle ++(4,4);
    \draw[thick, dgrey] (-2,-2) -- (-2,+2);
    \draw[thick, dgrey] (-2,-2) -- (+2,-2);
    \draw[thick, dgrey] (+2,+2) -- (+2,-2);
    \draw[thick, dgrey] (+2,+2) -- (-2,+2);
    \draw[thick, red, -stealth] (0,-2) -- (0.354,-1.354);
    
    \draw[thick, red] (0,-2) -- (0.707,-0.707);

     \draw[thick, dgreen, ->-=0.8] (0.5,-2) -- (1.007,-1.007);
     \draw[thick, dgreen, ->-=0.8] (2,-0.5) -- (1.007,-1.007);
        
    \draw[thick, dgreen, ->-=0.8] (2,0.5) --  (0.707,-0.207);
     \draw[thick, dgreen, -<-=0.7] (0.5,2) -- (-0.207,0.707);
   \draw[thick, dgreen, ->-=0.7] (0.707,-0.207)-- (-0.207,0.707);
            
   \draw[thick, dgreen, ->-=0.8] (-0.5,-2) --(0.207,-0.707);
   \draw[thick, dgreen, ->-=0.8] (-0.707,0.207) --(-2,-0.5); 
   \draw[thick, dgreen, ->-=0.8] (0.207,-0.707)--(-0.707,0.207) ;    
     
     \draw[thick, dgreen, -<-=0.7] (-2,0.5) -- (-1.007,1.007);
       \draw[thick, dgreen, -<-=0.7] (-0.5,2) -- (-1.007,1.007);
     
    \draw[thick, red, -stealth] (2,0) -- (1.354,-0.354);
    \draw[thick, red] (2,0) -- (0.707,-0.707);
    \draw[thick, red, -stealth] (-0.707,0.707) -- (-0.354,1.354);
    \draw[thick, red] (0,2) -- (-0.707,0.707);
    \draw[thick, red, -stealth] (-0.707,0.707) -- (-1.354,0.354);
    \draw[thick, red] (-2,0) -- (-0.707,0.707);
    \draw[thick, red, -stealth] (0.707,-0.707) -- (0,0);
    \draw[thick, red] (0.707,-0.707) -- (-0.707,0.707);

     \filldraw[red]  (-0.707,0.707) circle (3pt);
          \filldraw[red]  (0.707,-0.707) circle (3pt);
          
            \filldraw[dgreen]  (-1.007,1.007) circle (3pt);
          \filldraw[dgreen]  (1.007,-1.007) circle (3pt);
          
          
    \node[red, below] at (0,-2) {\scriptsize $ \cA_1$};
    \node[red, right] at (2,0) {\scriptsize $ \cA_1$};
    
    \node[dgreen, below] at (-0.8,-2) {\scriptsize $ \cN$};
    \node[dgreen, below] at (0.8,-1.9) {\scriptsize $ \overline \cN $};
    
        \node[dgreen, right] at (2,-0.7) {\scriptsize $ \cN$};
    \node[dgreen, right] at (2,0.8) {\scriptsize $ \overline \cN $};
    
      \node[dgreen, left] at (-2,-0.7) {\scriptsize $ \cN$};
       \node[red, left] at (-2,0) {\scriptsize $ \cA_1$};
    \node[dgreen, left] at (-2,0.8) {\scriptsize $ \overline \cN $};
    
    
\end{tikzpicture} 
 \hspace{0.1 in}=\hspace{0.1 in} \begin{tikzpicture}[baseline={([yshift=+.5ex]current bounding box.center)},vertex/.style={anchor=base,
    circle,fill=black!25,minimum size=18pt,inner sep=2pt},scale=0.6]
    \filldraw[grey] (-2,-2) rectangle ++(4,4);
    \draw[thick, dgrey] (-2,-2) -- (-2,+2);
    \draw[thick, dgrey] (-2,-2) -- (+2,-2);
    \draw[thick, dgrey] (+2,+2) -- (+2,-2);
    \draw[thick, dgrey] (+2,+2) -- (-2,+2);
    \draw[thick, red, -stealth] (0,-2) -- (0.354,-1.354);
    
    \draw[thick, red] (0,-2) -- (0.707,-0.707);

    \draw[thick, red, -stealth] (2,0) -- (1.354,-0.354);
    \draw[thick, red] (2,0) -- (0.707,-0.707);
    \draw[thick, red, -stealth] (-0.707,0.707) -- (-0.354,1.354);
    \draw[thick, red] (0,2) -- (-0.707,0.707);
    \draw[thick, red, -stealth] (-0.707,0.707) -- (-1.354,0.354);
    \draw[thick, red] (-2,0) -- (-0.707,0.707);
    \draw[thick, red, -stealth] (0.707,-0.707) -- (0,0);
    \draw[thick, red] (0.707,-0.707) -- (-0.707,0.707);

     \filldraw[red]  (-0.707,0.707) circle (3pt);
          \filldraw[red]  (0.707,-0.707) circle (3pt);

          
    \node[red, below] at (0,-2) {\scriptsize $ \cA_1$};
    \node[red, right] at (2,0) {\scriptsize $ \cA_1$};
           \node[red, left] at (-2,0) {\scriptsize $ \cA_1$};

  \draw[dgreen,thick] (0.8,0.3) to [out = 180, in = 180,distance = 0.8 cm] node[rotate=90]{\midarrow} (0.8,1.6);
  \draw[dgreen,thick] (0.8,0.3)  to [out = 0, in = 0,distance=0.8 cm] node[rotate=90]{\midarrow} (0.8,1.6);
  
    \filldraw[dgreen]  (0.8,0.3) circle (3pt);
          \filldraw[dgreen]   (0.8,1.6)circle (3pt);

 \node[dgreen, left] at (0.3,0.95) {\scriptsize $\overline \cN$};
  \node[dgreen, right] at (1.3,0.95) {\scriptsize $\cN$};

     \end{tikzpicture} ~,
\eea
whereby we may shrink the bubble to produce a factor of $d_\cN$, which is then absorbed in the definition of the (co-)multiplication operations of $\cA_1$.

A particular case of this is that of Tambara-Yamagami categories, encountered in Section \ref{sec:TY}. 
In that case, defining $\cA = \bigoplus_g g$, we had 
\bea
 \cA = \cN \otimes \overline \cN =  \cN \otimes \mathds{1}\otimes \overline \cN~, 
\eea
and hence $\cA$ was Morita equivalent to the trivial algebra. In other words, gauging $\cA$ should give the same theory as gauging $\mathds{1}$, i.e. the original theory, and this is precisely the statement of Kramers-Wannier duality encountered above. As we saw in Section \ref{sec:TY}, the rigorous calculation is a bit more subtle than just (\ref{eq:Moritaschematic}).

\subsubsection{Example: gauging Fibonacci}

As a simple example, let us consider the Fibonacci category mentioned in Section \ref{sec:rank2}, with $F$-symbols given in (\ref{eq:FibonacciF}). 
We now try to gauge the object $\cA = 1 \oplus W$. The first step in doing so is choosing (co-)multiplications satisfying the necessary consistency conditions. Since no non-trivial multiplicities appear in $\cA$ or the fusion rules, we may drop the indices $\alpha, i,j,k$, and we obtain e.g. the following conditions,
\bea
&\mathrm{Separability:}& \hspace{0.2 in} \sum_{a,b} N_{ab}^c \sqrt{d_a d_b \over d_c} \mu_{ab}^c (\mu^c_{ab})^\vee = d_\cA \hspace{0.3 in} \forall \,c\in \cC
\no\\
&\mathrm{Associativity:}& \hspace{0.2 in} \ \sum_e \mu_{ab}^{e} \mu_{e c}^d (F_{a b c}^d)_{ef} = \mu_{b c}^f \mu_{a f}^d\hspace{0.4 in} \forall a,b,c,d,f \in \cC
\eea
where $N_{ab}^c$ are the fusion coefficients appearing in (\ref{eq:fusionrulesdef}), and $\mu_{ab}^c$ is non-zero only if $N_{ab}^c$ is. The second of these is the non-invertible analog of the relation in (\ref{eq:relomegadelmu}). 

One now attempts to find a solution to these equations. In general, this should be done with computer assistance, but in the current case things are simple enough to do by hand.  For example, setting $(a,b,c,d) = (W,1,W,1)$ in the associativity constraint gives $\mu_{W1}^W = \mu_{1W}^W$, and likewise setting  $(a,b,c,d) = (1,W,W,1)$ gives $\mu_{1W}^W = \mu_{11}^1$. Proceeding in this way, we ultimately find that
\bea
\mu_{11}^1 = \mu_{1W}^W = \mu_{W1}^W = \mu_{WW}^1 = \varphi^{3/4}  \mu_{WW}^W = 1~,
\eea
and similarly for the co-multiplications. In other words, as long as we define trivalent junctions as 
\bea
{\begin{tikzpicture}[baseline=0,scale=0.7]
\draw [red, thick, decoration={markings, mark=at position 0.5 with {\arrow{stealth}}}, postaction={decorate}] (-1,-1) -- (0,0);
\draw [red, thick, decoration={markings, mark=at position 0.5 with {\arrow{stealth}}}, postaction={decorate}] (1,-1) -- (0,0);
\draw [red, thick, decoration={markings, mark=at position 0.6 with {\arrow{stealth}}}, postaction={decorate}] (0,0) -- (0,1);

\filldraw[red] (0,0) circle (1.5pt);
 \node[above,red] at (0,-0.1) {\footnotesize$\times$};
\node[red,right] at (0,0) {\scriptsize$\mu$};

\node[below] at (-1.2,-1) {\footnotesize \color{red}$\cA$};
\node[below] at (1.1,-1) {\footnotesize \color{red}$\cA$};
\node[above] at (0,1) {\footnotesize \color{red}$\cA$};
\end{tikzpicture}}
\displaystyle &:=&{1\over \varphi}\left( 
 \begin{tikzpicture}[baseline={([yshift=-1ex]current bounding box.center)},vertex/.style={anchor=base,
    circle,fill=black!25,minimum size=18pt,inner sep=2pt},scale=0.3]
   \draw[->-=0.5,dashed] (-2,-2) -- (0,0);
   \draw[->-=0.5,dashed] (2,-2) -- (0,0);
   \draw[->-=0.7,dashed] (0,0)--(0,2) ;
   \filldraw[] (0,0) circle (3pt);
     \node[below] at (-2,-2.2) {$1$};
     \node[below] at (2,-2) {$1$};
     \node[above] at (0,2) {$1$};
    \end{tikzpicture}
    +
     \begin{tikzpicture}[baseline={([yshift=-1ex]current bounding box.center)},vertex/.style={anchor=base,
    circle,fill=black!25,minimum size=18pt,inner sep=2pt},scale=0.3]
   \draw[->-=0.5,dashed] (-2,-2) -- (0,0);
   \draw[->-=0.5,thick] (2,-2) -- (0,0);
   \draw[->-=0.7,thick] (0,0)--(0,2) ;
   \filldraw[] (0,0) circle (3pt);
     \node[below] at (-2,-2.2) {$1$};
     \node[below] at (2,-2) {$W$};
     \node[above] at (0,2) {$W$};
    \end{tikzpicture}
        +
     \begin{tikzpicture}[baseline={([yshift=-1ex]current bounding box.center)},vertex/.style={anchor=base,
    circle,fill=black!25,minimum size=18pt,inner sep=2pt},scale=0.3]
   \draw[->-=0.5,thick] (-2,-2) -- (0,0);
   \draw[->-=0.5,dashed] (2,-2) -- (0,0);
   \draw[->-=0.7,thick] (0,0)--(0,2) ;
   \filldraw[] (0,0) circle (3pt);
     \node[below] at (-2,-2.2) {$W$};
     \node[below] at (2,-2) {$1$};
     \node[above] at (0,2) {$W$};
    \end{tikzpicture}
    +
    \begin{tikzpicture}[baseline={([yshift=-1ex]current bounding box.center)},vertex/.style={anchor=base,
    circle,fill=black!25,minimum size=18pt,inner sep=2pt},scale=0.3]
   \draw[->-=0.5,thick] (-2,-2) -- (0,0);
   \draw[->-=0.5,thick] (2,-2) -- (0,0);
   \draw[->-=0.7,dashed] (0,0)--(0,2) ;
   \filldraw[] (0,0) circle (3pt);
     \node[below] at (-2,-2.2) {$W$};
     \node[below] at (2,-2) {$W$};
     \node[above] at (0,2) {$1$};
    \end{tikzpicture}
    +\varphi^{-3/4} \begin{tikzpicture}[baseline={([yshift=-1ex]current bounding box.center)},vertex/.style={anchor=base,
    circle,fill=black!25,minimum size=18pt,inner sep=2pt},scale=0.3]
   \draw[->-=0.5,thick] (-2,-2) -- (0,0);
   \draw[->-=0.5,thick] (2,-2) -- (0,0);
   \draw[->-=0.7,thick] (0,0)--(0,2) ;
   \filldraw[] (0,0) circle (3pt);
     \node[below] at (-2,-2.2) {$W$};
     \node[below] at (2,-2) {$W$};
     \node[above] at (0,2) {$W$};
     \node[above] at (0,-0.4) {\footnotesize$\times$};
    \end{tikzpicture}
    \right) ~,
    \no\\
\eea
then we can consistently insert a mesh of $\cA$ as in (\ref{eq:gaugingintermsofalgebra}), which implements gauging of the Fibonacci symmetry. 
Note that since $\cA = 1+W = W \otimes \overline W$, this algebra is Morita trivial, and hence we expect that gauging this algebra will give back the original theory. It is a good exercise to check this explicitly. 
\newline

\begin{tcolorbox}
\textbf{Exercise:} For those who are familiar with CFT,  use the action of the Verlinde line $W$ in $(G_2)_1$ and $(F_4)_1$ WZW models, together with the $F$-symbols above, to compute the various twisted partition functions, and then take the appropriate sum to explicitly show invariance under gauging.
\end{tcolorbox}

\subsection{Representation theory and the SymTFT}

From the point of view of Physics, one of the most important aspects of symmetry groups is the fact that fields/operators in the theory are  labelled by representations of the group. It is natural to ask what the analog of representation theory is for non-invertible symmetry. The answer is that irreducible representations of the tube algebra associated to a fusion category $\cC$ are given by simple objects of the so-called \textit{Drinfeld center} $\cZ(\cC)$  \cite{Lin:2022dhv,Bartsch:2022mpm,Bartsch:2022ytj,Bartsch:2023wvv}. More physically, irreducible representations of categorical symmetry in $(1+1)$d are in one-to-one correspondence with simple line operators in a certain $(2+1)$d topological theory known as the \textit{Symmetry TFT (SymTFT)}. In this subsection we first review the basic idea of the SymTFT (closely following \cite{Kaidi:2022cpf}), after which we illustrate the technique in the cases of $\cC = \mathrm{Vec}^\omega({\ZZ_2})$ and $\cC = \mathrm{Ising}$.

\subsubsection{SymTFT primer}
The SymTFT of a theory $\cX$ in $d$-dimensions with global symmetry $\cC$ is a topological theory SymTFT($\cC)$ in $(d+1)$-dimensions which, when compactified on an interval with appropriate boundary conditions, gives back the theory $\cX$ of interest. One of the many nice properties of the SymTFT is that it decouples the dynamics of $\cX$ from the symmetries. This makes the action of various topological manipulations $\phi$ (such as gauging a discrete symmetry or adding local counterterms in background fields) more transparent: indeed, by choosing appropriate topological boundary conditions for the SymTFT before compactification to $d$-dimensions, we can obtain not only the original $\cX$, but any theory of the form $\phi(\cX)$. 

\begin{figure}[!tbp]
	\centering
	\begin{tikzpicture}[scale=0.8]

	\shade[line width=2pt, top color=blue!30, bottom color=blue!5] 
	(0,0) to [out=90, in=-90]  (0,3)
	to [out=0,in=180] (6,3)
	to [out = -90, in =90] (6,0)
	to [out=180, in =0]  (0,0);
	
		\draw[very thick] (-7,0) -- (-7,3);
	\node[below] at (-7,0) {$Z_{\cX}[A]$};

	\draw[thick, snake it, <->] (-1.7,1.5) -- (-5, 1.5);
	
	\draw[thick] (0,0) -- (0,3);
	\draw[thick] (6,0) -- (6,3);
	\node at (3,1.5) {SymTFT($\cC$)};
	\node[below] at (0,0) {$\langle D(A)|$};
	\node[below] at (6,0) {$|\cX\rangle $}; 
	
	\end{tikzpicture}
	\\
	\vspace{0.5 in}
	\begin{tikzpicture}[scale=0.8]

	\shade[line width=2pt, top color=blue!30, bottom color=blue!5] 
	(0,0) to [out=90, in=-90]  (0,3)
	to [out=0,in=180] (6,3)
	to [out = -90, in =90] (6,0)
	to [out=180, in =0]  (0,0);
	
		\draw[very thick] (-7,0) -- (-7,3);
	\node[below] at (-7,0) {$Z_{\cX/G}[A]$};

	\draw[thick, snake it, <->] (-1.7,1.5) -- (-5, 1.5);
	
	\draw[thick] (0,0) -- (0,3);
	\draw[thick] (6,0) -- (6,3);
	\node at (3,1.5) {SymTFT($\cC$)};
	\node[below] at (0,0) {$\langle N(A)|$};
	\node[below] at (6,0) {$|\cX\rangle $}; 
	\end{tikzpicture} 
	
	\caption{Schematic picture of the SymTFT. By imposing  Dirichlet (resp. Neumann) boundary conditions on the left and the non-topological boundary condition (\ref{eq:dynamboundary}) on the right, we may compactify to obtain $\cX$ (resp. $\cX/G$). In general, there exist other topological boundary conditions as well.}
	\label{fig:symmTFTidea}
\end{figure}
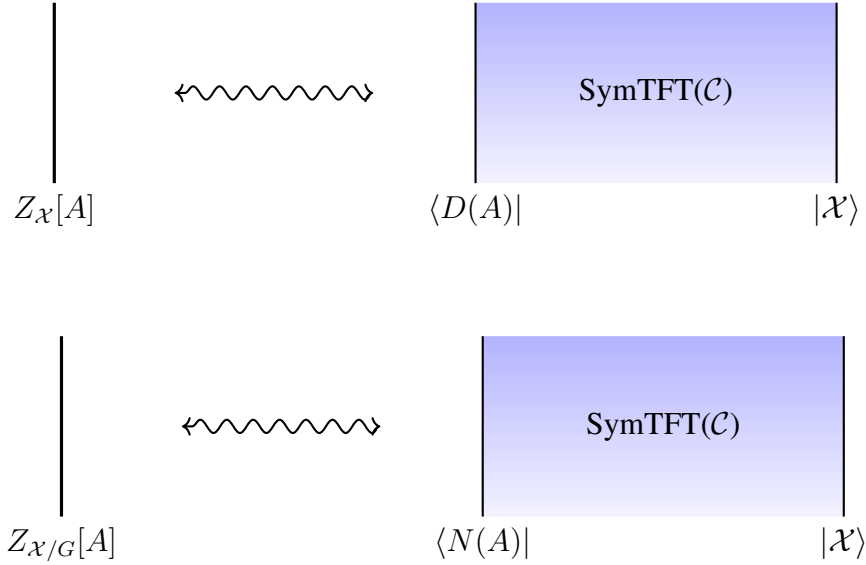

The basic idea is illustrated in Figure \ref{fig:symmTFTidea}. The $(d+1)$-dimensional SymTFT is placed on an interval, where the right boundary is endowed with non-topological boundary conditions capturing the dynamics of $\cX$, whereas the left boundary condition is topological. As described in Section \ref{sec:anomaly}, a theory in $(d+1)$ dimensions assigns a Hilbert space to $d$-dimensional subspaces, and hence the boundaries of SymTFT$(\cC)$ can be labeled by appropriate elements of its Hilbert space.
The dynamical boundary of SymTFT$(\cC)$ is taken to be 
\bea
\label{eq:dynamboundary}
| \cX\rangle = \sum_a Z_\cX[a] | a\rangle~,
\eea
where $|a \rangle$ is an appropriate basis of the state space.
When $\cC$ is a finite group $G$, the $a$ here represents the collective set of flat connections of $G$, and $Z_{\cX}[a]$ denotes the partition function of $\cX$ coupled to gauge fields $a$. When $\cC$ is a fusion category rather than a group, the label $a$ is an appropriate collective label for the topological defects in the theory. 

For the following formulas, we specialize to a finite Abelian, non-anomalous zero-form symmetry $G$. The topological boundary of SymTFT$(\cC)$ can take a number of forms, with common options being Dirichlet or Neumann boundary conditions. Dirichlet boundary conditions fix the fields $a$ to certain values $A$, whereas Neumann boundary conditions allow $a$ to fluctuate freely. In the state notation, these can be written as 
\bea\label{eq:bc}
\begin{split}
    \mathrm{Dirichlet:}& \hspace{0.5 in} |D(A)\rangle = \sum_a \delta(a-A) |a\rangle ~,
\\
\mathrm{Neumann:}& \hspace{0.5 in} |N(A)\rangle = \sum_a \exp\left(-i \int a \cup A\right) |a\rangle ~. 
\end{split}
\eea
The normalization factor in $\int a\cup A$ depends on the symmetry and is suppressed  here. For Dirichlet states, $A$ is a background gauge field for $G$; for Neumann states, it is a background gauge field for the quantum dual. Cup products use the natural pairing, and the sums and state normalizations include the suppressed gauge-volume factors.

Because the SymTFT is topological, the length of the interval is unimportant and can be shrunk to zero size, upon which one obtains a $d$-dimensional theory whose partition function can be computed by taking the inner product between the states on the left and right boundaries. In the case of Dirichlet boundary conditions, one obtains 
\bea
\langle D(A)| \cX \rangle = \sum_{a, a'} Z_{\cX}[a] \,\delta(a'-A) \langle a' | a \rangle = \sum_{a, a'} Z_{\cX}[a] \,\delta(a'-A) \delta(a-a') = Z_{\cX}[A]
\no\\
\eea
and hence reproduces the original $d$-dimensional theory $\cX$, coupled to background fields $A$. 
On the other hand, by putting Neumann boundary conditions on the left, we obtain 
\begin{equation}
    \langle N(A)| \cX \rangle =\sum_{a, a'} Z_{\cX}[a]\,\exp\left(i \int a' \cup A\right) \langle a' | a \rangle = \sum_{a} Z_{\cX}[a] \,\exp\left(i \int a \cup A\right)  = Z_{\cX/G}[A]~.
\end{equation}
In other words, in this case we obtain a $d$-dimensional theory $\phi(\cX)$, where $\phi$ represents gauging the discrete symmetry $G$ (which is indeed a topological operation).  Choosing mixed Dirichlet-Neumann boundary conditions also allows us to obtain $\cX$ with a subgroup of $G$ gauged. 

More generally, one can dress either boundary condition in \eqref{eq:bc} by a $d$-cocycle $\nu_d$.  The corresponding theories obtained by shrinking the slab are $\cX$ stacked with a counterterm, and $\cX/G$ with the gauging done with discrete torsion associated with the $d$-cocycle,
\begin{equation}
\begin{split}
    \ket{D(A)_{\nu_d}}= \sum_{a} \delta(a-A)\exp\left(-i \int \nu_d(a)\right)\ket{a}  &\longleftrightarrow  \langle D(A)_{\nu_d}| \cX \rangle= Z_{\cX}[A] \exp\left(i \int \nu_d(A)\right)~,\\
    \ket{N(A)_{\nu_d}}= \sum_{a} \exp\left(-i \int a\cup A -i \int \nu_d(a)\right)\ket{a}  &\longleftrightarrow  \langle N(A)_{\nu_d}| \cX \rangle= Z_{(\cX\times \nu_d)/G}[A]~. \\
\end{split}
\end{equation}
One therefore expects that any two theories $\cX$ and $\cX'$ that are related by a topological manipulation $\phi$, i.e. $\cX'=\phi(\cX)$, can be obtained by the slab construction with the \emph{same} SymTFT but with \emph{different} topological boundary conditions.

\subsubsection{Invertible symmetry and Dijkgraaf-Witten theories}
\label{sec:anomalyDW}

The SymTFT takes a particularly simple form when the symmetry in question is group-like. To see this, begin by considering a theory $\cX$ with background gauge fields $A$ and anomaly $\omega$. As discussed in Section \ref{sec:anomaly}, the anomaly inflow paradigm, in modern language, says that such a theory may be realized on the boundary of an appropriate invertible theory Inv$^\omega(A)$; see Figure \ref{fig:DWandanom}(a).

\hspace{-0.5 in}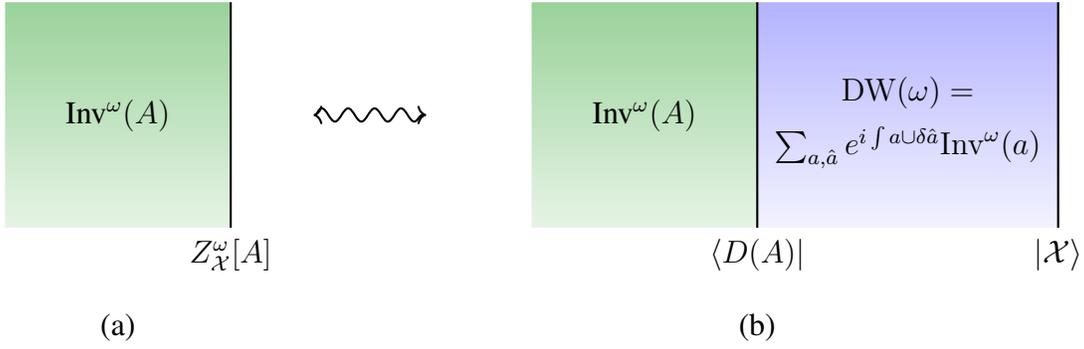
\begin{figure}[t]
	\centering
	\begin{tikzpicture}

	\shade[line width=2pt, top color=dgreen!40, bottom color=dgreen!10] 
	(0,0) to [out=90, in=-90]  (0,3)
	to [out=0,in=180] (3,3)
	to [out = -90, in =90] (3,0)
	to [out=180, in =0]  (0,0);

	\shade[line width=2pt, top color=blue!30, bottom color=blue!5] 
	(7+3,0) to [out=90, in=-90]  (7+3,3)
	to [out=0,in=180] (11+3,3)
	to [out = -90, in =90] (11+3,0)
	to [out=180, in =0]  (0,0);
	
	\shade[line width=2pt, top color=dgreen!40, bottom color=dgreen!10] 
	(11-4,0) to [out=90, in=-90]  (11-4,3)
	to [out=0,in=180] (14-4,3)
	to [out = -90, in =90] (14-4,0)
	to [out=180, in =0]  (0,0);
	
	\draw[thick] (3,0) -- (3,3);

	\node[below] at (1.5, -1) {(a)};
	
	\draw[thick, snake it, <->] (4.1,1.5) -- (5.6, 1.5);

	\draw[thick] (7+3,0) -- (7+3,3);
	\draw[thick] (11+3,0) -- (11+3,3);

	\node[below] at (10, -1) {(b)};
	
	\node at (1.5,1.5) {Inv$^\omega(A)$};
	\node[below] at (3,0) {$Z^\omega_{\cX}[A]$};
	\node at (9+3,1.8) {$\mathrm{DW}(\omega) =$};
	\node[below] at (9+3,1.5) {$ \sum_{a,\hat a }  e^{i \int a  \cup\delta \hat a}\mathrm{Inv}^\omega( a) $};
	\node[below] at (7+3,0) {$\langle D(A)| $};
	\node[below] at (11+3,0) {$\ket{\cX}$}; 
	\node at (12.5-4,1.5) {Inv$^\omega(A)$};
	
	\end{tikzpicture}
	
	\caption{A theory with anomaly $\omega$ can be realized on the boundary of an invertible theory Inv$^\omega(A)$. This is possible if and only if the Symmetry TFT is a (generalized) DW theory. }
	\label{fig:DWandanom}
\end{figure}

Given Figure \ref{fig:DWandanom}(a), one can promote the background gauge field $A$ to a dynamical gauge field $a$ both on the right boundary and in a slab in the bulk. The bulk theory within this slab is then a nontrivial TQFT, namely a
\textit{(generalized) Dijkgraaf-Witten (DW) theory}\footnote{The qualifier ``generalized" here refers to the fact that the relevant invertible theories are not restricted to the group cohomology elements studied in the original work of Dijkgraaf and Witten \cite{Dijkgraaf:1989pz}.}
\bea\label{eq:DW}
\mathrm{DW}(\omega) = \sum_{a,\hat a}  \exp\left(i \int a \cup\delta \hat a\right) \mathrm{Inv}^\omega(a)~,
\eea
where the displayed BF form assumes finite Abelian $G$, with $a$ a $G$-valued 1-cochain and $\hat a$ a $(d-1)$-cochain for the dual group, and as commented above we suppress the normalization in the BF term.   
On the left of the slab, one further imposes Dirichlet boundary conditions to pin the dynamical field $a$ to the background $A$.   
The theory $\cX$ is recovered by taking the thin slab limit where the non-topological boundary condition $\ket{\cX}$ and the Dirichlet boundary condition $\ket{D(A)}$ collide. 
We conclude that the SymTFT for an invertible symmetry with an anomaly is given by a (generalized) DW theory \eqref{eq:DW}, i.e. a discrete gauge theory, potentially with a non-trivial twist.

\subsubsection{Example: $\cC = \mathrm{Vec}(\ZZ_2)$}

We now illustrate the SymTFT technique for the simplest example of a $(1+1)$d theory with non-anomalous $\ZZ_2$ symmetry.
 By the above discussion, the SymTFT for such a theory is just $(2+1)$d $\ZZ_2$ gauge theory,
\bea
S = {\pi} \int_{M_3}  a\cup \delta \widehat a~.
\eea
This theory has four simple line operators, which can be written as 
\bea
L_{(e,m)}(\gamma) := e^{i \pi e \oint_\gamma a}  e^{i \pi m \oint_\gamma \widehat a}~, \hspace{0.5 in} e,m=0,1~. 
\eea
The fusions of these line operators are straightforward, 
\bea
L_{(e,m)}(\gamma) \times L_{(e',m')}(\gamma) = L_{(e+e',m+m')}(\gamma)~,
\eea
but the braidings between them are non-trivial---in particular, inserting $L_{(e,m)}$ on a loop $\gamma$ and $L_{(e',m')}$ on another loop $\gamma'$, one finds that\footnote{We work in $S^3$ with zero framing and assume that these loops are not linked with other insertions. } 
\bea
\label{eq:linkingZ2}
\langle L_{(e,m)}(\gamma) L_{(e',m')}(\gamma') \dots \rangle = (-1)^{(em'+ me') \mathrm{Link}(\gamma, \gamma') }\langle \dots \rangle~.
\eea
This may be derived from the path integral as follows. Inserting the line operators $ L_{(e,m)}$ and $L_{(e',m')}$ amounts to modifying the action to 
\bea
S &=& \pi \int_{M_3}  a \cup \delta \widehat a + \pi \int_\gamma (e a + m \widehat a) + \pi \int_{\gamma'}(e' a + m' \widehat a) 
\no\\
&=& \pi \int_{M_3}  a \cup \delta \widehat a + \pi \int (e \omega_\gamma + e' \omega_{\gamma'} )\cup  a+ \pi \int (m \omega_\gamma + m' \omega_{\gamma'} ) \cup \widehat a~,
\eea
where $\omega_\gamma$ is the Poincar{\'e} dual of $\gamma$. Integrating out $\widehat a$ enforces the following, 
\bea
\delta a = - (m \omega_\gamma + m' \omega_{\gamma'})~. 
\eea
This means that if we introduce a two-manifold $\Sigma$ such that $\p \Sigma  = m \gamma + m' \gamma'$, known as the \textit{Seifert surface} associated to the curve, then we have $ a= - \omega_\Sigma$, i.e. $a$ is the Poincar{\'e} dual of $\Sigma$. Plugging this back into the action gives 
\bea
-{\pi} \int (e \omega_\gamma + e' \omega_{\gamma'}) \cup \omega_\Sigma &=& -{\pi} \int \omega_{(e \gamma + e' {\gamma'}) \cap \Sigma}
\no\\
&=& -{\pi} \int \left[ e\, \omega_{\gamma \cap \Sigma} + e'\, \omega_{\gamma' \cap \Sigma} \right] ~.
\eea
Finally, we note that $\int \omega_{\gamma \cap \Sigma} =m'\, \mathrm{Link}(\gamma, \gamma')$ and likewise  $\int \omega_{\gamma' \cap \Sigma} =m\,\mathrm{Link}(\gamma, \gamma')$, so that in total we get 
\bea
 -{\pi}  \left( em'+ e' m\right) \mathrm{Link}(\gamma, \gamma') ~,
\eea
and hence the path integral gets a factor of $e^{i S } = (-1)^{ \left( em'+ e' m\right) \mathrm{Link}(\gamma, \gamma')}$, which is the factor that appears in (\ref{eq:linkingZ2}). 

Now in order to get the original 2d theory from the $\ZZ_2$ gauge theory described above, we need to choose a dynamical boundary condition for the  right boundary, and a topological boundary condition for the left boundary. For the topological boundary, let us first choose Dirichlet boundary conditions for the  gauge field $ a$. This means that $L_{(0,0)}$ and $L_{(1,0)}$ are trivialized on the boundary, whereas $L_{(0,1)}$ and $L_{(1,1)}$ both reduce to the same non-trivial defect on the boundary, which corresponds to the generator $\eta$ of the boundary $\ZZ_2$ symmetry. 

\begin{figure}[t]
    \centering
    {\begin{tikzpicture}[baseline=35]

  \shade[top color=red!40, bottom color=red!10,rotate=90]  (0,-0.7) -- (2,-0.7) -- (2.6,-0.2) -- (0.6,-0.2)-- (0,-0.7);
 \draw[thick,rotate=90] (0,-0.7) -- (2,-0.7);
\draw[thick,rotate=90] (0,-0.7) -- (0.6,-0.2);
\draw[thick,rotate=90]  (0.6,-0.2)--(2.6,-0.2);
\draw[thick,rotate=90]  (2.6,-0.2)-- (2,-0.7);

 \begin{scope}[xshift=-0.9in] \shade[top color=red!40, bottom color=red!10,rotate=90]  (0,-0.7) -- (2,-0.7) -- (2.6,-0.2) -- (0.6,-0.2)-- (0,-0.7);
 \draw[thick,rotate=90] (0,-0.7) -- (2,-0.7);
\draw[thick,rotate=90] (0,-0.7) -- (0.6,-0.2);
\draw[thick,rotate=90]  (0.6,-0.2)--(2.6,-0.2);
\draw[thick,rotate=90]  (2.6,-0.2)-- (2,-0.7);
\end{scope}

\node at (0.4,1.4) [circle,fill,dgreen, inner sep=1.2pt]{};
	
\draw[thick,blue,,dotted] (-1.85,1.4)--(0.4,1.4);
\node[right] at (-1.4,1.9) {$L_{(0,0)}$};

\begin{scope}[xshift=2.1in]

  \shade[top color=red!40, bottom color=red!10,rotate=90]  (0,-0.7) -- (2,-0.7) -- (2.6,-0.2) -- (0.6,-0.2)-- (0,-0.7);
 \draw[thick,rotate=90] (0,-0.7) -- (2,-0.7);
\draw[thick,rotate=90] (0,-0.7) -- (0.6,-0.2);
\draw[thick,rotate=90]  (0.6,-0.2)--(2.6,-0.2);
\draw[thick,rotate=90]  (2.6,-0.2)-- (2,-0.7);

 \begin{scope}[xshift=-0.9in] \shade[top color=red!40, bottom color=red!10,rotate=90]  (0,-0.7) -- (2,-0.7) -- (2.6,-0.2) -- (0.6,-0.2)-- (0,-0.7);
 \draw[thick,rotate=90] (0,-0.7) -- (2,-0.7);
\draw[thick,rotate=90] (0,-0.7) -- (0.6,-0.2);
\draw[thick,rotate=90]  (0.6,-0.2)--(2.6,-0.2);
\draw[thick,rotate=90]  (2.6,-0.2)-- (2,-0.7);
\end{scope}

\node at (0.4,1.4) [circle,fill,dgreen, inner sep=1.2pt]{};

\draw[thick,blue] (-1.6,1.4)--(0.4,1.4);
\draw[thick,blue,dotted] (-1.85,1.4)--(-1.6,1.4);
\node[right] at (-1.4,1.9) {$L_{(0,1)}$};
\draw[thick,violet] (-1.85,1.4) -- (-1.85,2.3);
\node[left] at (-1.5,2.5) {$\eta$};

\end{scope}

\begin{scope}[yshift=-1.5in]

  \shade[top color=red!40, bottom color=red!10,rotate=90]  (0,-0.7) -- (2,-0.7) -- (2.6,-0.2) -- (0.6,-0.2)-- (0,-0.7);
 \draw[thick,rotate=90] (0,-0.7) -- (2,-0.7);
\draw[thick,rotate=90] (0,-0.7) -- (0.6,-0.2);
\draw[thick,rotate=90]  (0.6,-0.2)--(2.6,-0.2);
\draw[thick,rotate=90]  (2.6,-0.2)-- (2,-0.7);

 \begin{scope}[xshift=-0.9in] \shade[top color=red!40, bottom color=red!10,rotate=90]  (0,-0.7) -- (2,-0.7) -- (2.6,-0.2) -- (0.6,-0.2)-- (0,-0.7);
 \draw[thick,rotate=90] (0,-0.7) -- (2,-0.7);
\draw[thick,rotate=90] (0,-0.7) -- (0.6,-0.2);
\draw[thick,rotate=90]  (0.6,-0.2)--(2.6,-0.2);
\draw[thick,rotate=90]  (2.6,-0.2)-- (2,-0.7);
\end{scope}

\node at (0.4,1.4) [circle,fill,dgreen, inner sep=1.2pt]{};

\draw[thick,blue] (-1.6,1.4)--(0.4,1.4);
\draw[thick,blue,dotted] (-1.85,1.4)--(-1.6,1.4);
\node[right] at (-1.4,1.9) {$L_{(1,0)}$};
\end{scope}

\begin{scope}[yshift=-1.5in,xshift=2.1in]

  \shade[top color=red!40, bottom color=red!10,rotate=90]  (0,-0.7) -- (2,-0.7) -- (2.6,-0.2) -- (0.6,-0.2)-- (0,-0.7);
 \draw[thick,rotate=90] (0,-0.7) -- (2,-0.7);
\draw[thick,rotate=90] (0,-0.7) -- (0.6,-0.2);
\draw[thick,rotate=90]  (0.6,-0.2)--(2.6,-0.2);
\draw[thick,rotate=90]  (2.6,-0.2)-- (2,-0.7);

 \begin{scope}[xshift=-0.9in] \shade[top color=red!40, bottom color=red!10,rotate=90]  (0,-0.7) -- (2,-0.7) -- (2.6,-0.2) -- (0.6,-0.2)-- (0,-0.7);
 \draw[thick,rotate=90] (0,-0.7) -- (2,-0.7);
\draw[thick,rotate=90] (0,-0.7) -- (0.6,-0.2);
\draw[thick,rotate=90]  (0.6,-0.2)--(2.6,-0.2);
\draw[thick,rotate=90]  (2.6,-0.2)-- (2,-0.7);
\end{scope}

\node at (0.4,1.4) [circle,fill,dgreen, inner sep=1.2pt]{};

\draw[thick,blue] (-1.6,1.4)--(0.4,1.4);
\draw[thick,blue,dotted] (-1.85,1.4)--(-1.6,1.4);
\node[right] at (-1.4,1.9) {$L_{(1,1)}$};
\draw[thick,violet] (-1.85,1.4) -- (-1.85,2.3);
\node[left] at (-1.5,2.5) {$\eta$};

\end{scope}

\end{tikzpicture}}

    \caption{The four sectors of a theory with $\ZZ_2$ global symmetry. The red planes represent the two boundaries of the SymTFT. The green dot represents a non-topological operator on the dynamical boundary.}
    \label{fig:Z2reps}
\end{figure}
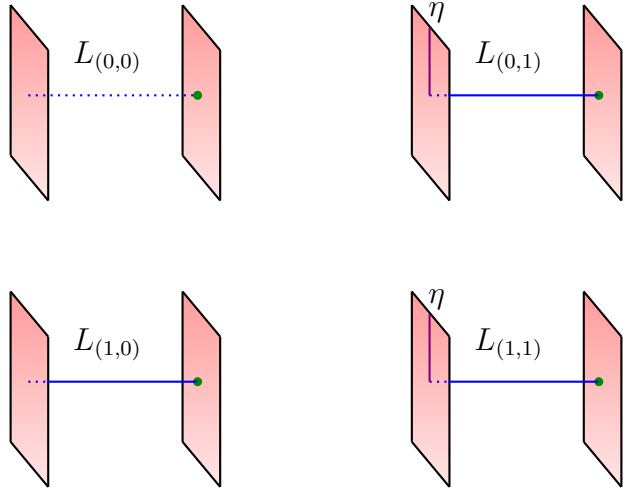

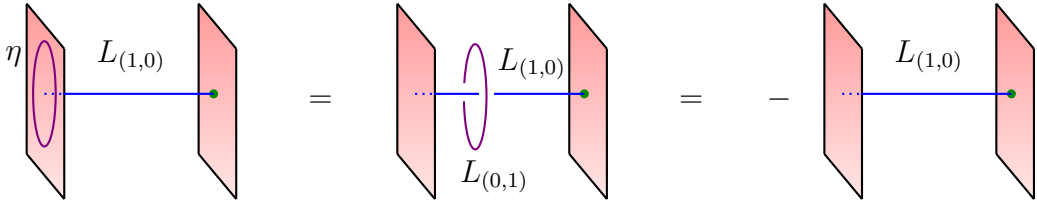
\begin{figure}[t]
    \centering
    {\begin{tikzpicture}[baseline=35]

  \shade[top color=red!40, bottom color=red!10,rotate=90]  (0,-0.7) -- (2,-0.7) -- (2.6,-0.2) -- (0.6,-0.2)-- (0,-0.7);
 \draw[thick,rotate=90] (0,-0.7) -- (2,-0.7);
\draw[thick,rotate=90] (0,-0.7) -- (0.6,-0.2);
\draw[thick,rotate=90]  (0.6,-0.2)--(2.6,-0.2);
\draw[thick,rotate=90]  (2.6,-0.2)-- (2,-0.7);

 \begin{scope}[xshift=-0.9in] \shade[top color=red!40, bottom color=red!10,rotate=90]  (0,-0.7) -- (2,-0.7) -- (2.6,-0.2) -- (0.6,-0.2)-- (0,-0.7);
 \draw[thick,rotate=90] (0,-0.7) -- (2,-0.7);
\draw[thick,rotate=90] (0,-0.7) -- (0.6,-0.2);
\draw[thick,rotate=90]  (0.6,-0.2)--(2.6,-0.2);
\draw[thick,rotate=90]  (2.6,-0.2)-- (2,-0.7);

\end{scope}

 \draw[thick,violet] (-1.7,1.4) arc(0:360:0.15cm and 0.7cm);

\node at (0.4,1.4) [circle,fill,dgreen, inner sep=1.2pt]{};

\draw[thick,blue] (-1.6,1.4)--(0.4,1.4);
\draw[thick,blue,dotted] (-1.85,1.4)--(-1.6,1.4);
\node[right] at (-1.3,1.9) {$L_{(1,0)}$};
\node[left] at (-2,1.9) {$\eta$};
\end{tikzpicture} 
\qquad$=$\qquad
\begin{tikzpicture}[baseline=35]

  \shade[top color=red!40, bottom color=red!10,rotate=90]  (0,-0.7) -- (2,-0.7) -- (2.6,-0.2) -- (0.6,-0.2)-- (0,-0.7);
 \draw[thick,rotate=90] (0,-0.7) -- (2,-0.7);
\draw[thick,rotate=90] (0,-0.7) -- (0.6,-0.2);
\draw[thick,rotate=90]  (0.6,-0.2)--(2.6,-0.2);
\draw[thick,rotate=90]  (2.6,-0.2)-- (2,-0.7);

 \begin{scope}[xshift=-0.9in] \shade[top color=red!40, bottom color=red!10,rotate=90]  (0,-0.7) -- (2,-0.7) -- (2.6,-0.2) -- (0.6,-0.2)-- (0,-0.7);
 \draw[thick,rotate=90] (0,-0.7) -- (2,-0.7);
\draw[thick,rotate=90] (0,-0.7) -- (0.6,-0.2);
\draw[thick,rotate=90]  (0.6,-0.2)--(2.6,-0.2);
\draw[thick,rotate=90]  (2.6,-0.2)-- (2,-0.7);

\end{scope}

 \draw[thick,violet] (-1.2,1.3) arc(-175:165:0.15cm and 0.7cm);

\node at (0.4,1.4) [circle,fill,dgreen, inner sep=1.2pt]{};

\draw[thick,blue] (-1.6,1.4)--(-1,1.4);
\draw[thick,blue] (-0.8,1.4)--(0.4,1.4);
\draw[thick,blue,dotted] (-1.85,1.4)--(-1.6,1.4);
\node[right] at (-0.9,1.8) {$L_{(1,0)}$};

\node[right] at (-1.4,0.3) {$L_{(0,1)}$};

\end{tikzpicture} 
\qquad$=$\qquad$-$\quad
\begin{tikzpicture}[baseline=35]

  \shade[top color=red!40, bottom color=red!10,rotate=90]  (0,-0.7) -- (2,-0.7) -- (2.6,-0.2) -- (0.6,-0.2)-- (0,-0.7);
 \draw[thick,rotate=90] (0,-0.7) -- (2,-0.7);
\draw[thick,rotate=90] (0,-0.7) -- (0.6,-0.2);
\draw[thick,rotate=90]  (0.6,-0.2)--(2.6,-0.2);
\draw[thick,rotate=90]  (2.6,-0.2)-- (2,-0.7);

 \begin{scope}[xshift=-0.9in] \shade[top color=red!40, bottom color=red!10,rotate=90]  (0,-0.7) -- (2,-0.7) -- (2.6,-0.2) -- (0.6,-0.2)-- (0,-0.7);
 \draw[thick,rotate=90] (0,-0.7) -- (2,-0.7);
\draw[thick,rotate=90] (0,-0.7) -- (0.6,-0.2);
\draw[thick,rotate=90]  (0.6,-0.2)--(2.6,-0.2);
\draw[thick,rotate=90]  (2.6,-0.2)-- (2,-0.7);

\end{scope}

\node at (0.4,1.4) [circle,fill,dgreen, inner sep=1.2pt]{};

\draw[thick,blue] (-1.6,1.4)--(0.4,1.4);
\draw[thick,blue,dotted] (-1.85,1.4)--(-1.6,1.4);
\node[right] at (-1.3,1.9) {$L_{(1,0)}$};

\end{tikzpicture}}

    \caption{Measuring the $\ZZ_2$ charge of the operators in the sector labelled by $L_{(1,0)}$. By making use of the bulk braidings (\ref{eq:linkingZ2}), we find that such operators are $\ZZ_2$ odd.  }
    \label{fig:Z2charge}
\end{figure}

We may now consider the four configurations shown in Figure \ref{fig:Z2reps}, in which various bulk lines stretch between the two boundaries. 
When ending on the non-topological boundary, the lines end on a potentially non-topological operator. Upon shrinking the bulk, these configurations define four sectors of operators in the original $(1+1)$d theory. Indeed, the two sectors in the first column define local operators, whereas the two sectors in the second column define operators in the $\eta$-twisted sector. We may measure the $\ZZ_2$ charge of each of these operators as follows. We begin by encircling the operators by the boundary line $\eta$. In the SymTFT picture, this encircling happens on the topological boundary. We may then pull the topological operator $\eta$ into the bulk, upon which it becomes $L_{(0,1)}$.  We then use the braiding of $L_{(0,1)}$  and the bulk line running perpendicular to the boundary to obtain the charge. This is shown for the sector labelled by $L_{(1,0)}$ in Figure \ref{fig:Z2charge}. In this way, we find that the sectors labelled by $L_{(0,0)}$ and $L_{(0,1)}$ are $\ZZ_2$ even, while the sectors labelled by $L_{(1,0)}$ and $L_{(1,1)}$ are $\ZZ_2$ odd.

In total then, the four topological lines of the SymTFT correspond to the following four sectors of a theory with $\ZZ_2$ global symmetry, 
\bea
\label{eq:Z2SymTFTcorrespondences}
L_{(0,0)} &\hspace{0.2 in}\leftrightarrow\hspace{0.2 in}& \ZZ_2\,\,\, \mathrm{even}, \,\, \mathrm{untwisted}~,
\no\\
L_{(1,0)} &\hspace{0.2 in}\leftrightarrow\hspace{0.2 in}& \ZZ_2\,\,\, \mathrm{odd}, \,\, \mathrm{untwisted}~,
\no\\
L_{(0,1)} &\hspace{0.2 in}\leftrightarrow\hspace{0.2 in}& \ZZ_2\,\,\, \mathrm{even}, \,\, \mathrm{twisted}~,
\no\\
L_{(1,1)} &\hspace{0.2 in}\leftrightarrow\hspace{0.2 in}& \ZZ_2\,\,\, \mathrm{odd}, \,\, \mathrm{twisted}~.
\eea
We see that the SymTFT knows about more than just usual representation theory (i.e. $\ZZ_2$ odd vs. even) of local operators---it also contains the data of twisted sectors. This will be crucial in the case of non-invertible symmetries, since non-invertible symmetries exchange untwisted and twisted sector operators (e.g. $\cN$ exchanging $\sigma$ and $\mu$ in the Ising model), and hence the representation theory of the non-invertible symmetry must accommodate both. 

Let us note here that we could have also chosen the topological boundary condition to be Dirichlet boundary conditions for $\widehat a$, instead of $a$. This would simply exchange $a$ and $\widehat a$, and consequently $L_{(1,0)}$ and $L_{(0,1)}$, throughout the above discussion. In particular, from the correspondences in (\ref{eq:Z2SymTFTcorrespondences}), this exchanges the $\ZZ_2$ odd untwisted sector with the $\ZZ_2$ even twisted sector, while leaving the other sectors unchanged.
Comparing this to the discussion around (\ref{tab:table1}) and (\ref{tab:table2}), we see that this is exactly the effect expected when gauging a $\ZZ_2$ symmetry in the boundary two-dimensional theory. 
We thus confirm that different discrete gaugings (which are examples of the topological manipulation $\phi$ discussed above) of a single theory correspond to different topological boundary conditions of a single SymTFT.\footnote{Note that there is no bosonic topological boundary condition for which $L_{(1,1)}$ is absorbed, since $L_{(1,1)}$ has non-trivial spin, and hence cannot be condensed.} The boundary gauging is implemented by a bulk zero-form symmetry, namely the electromagnetic duality symmetry $\ZZ_2^\mathrm{EM}$ exchanging $a \leftrightarrow \widehat a$. This will be important when we study the SymTFT for the Ising category below.

\subsubsection{Example: $\cC = \mathrm{Vec}^\omega(\ZZ_2)$}

We next consider the case of a $\ZZ_2$ symmetry with non-trivial `t Hooft anomaly. By the general discussion above, the SymTFT is a $\ZZ_2$ gauge theory with non-trivial Dijkgraaf-Witten twist, which may be written in the following form 
\bea
S = \pi \int_{M_3} a \cup \delta \widehat a + \widehat a \cup\widehat a\cup \widehat a ~. 
\eea
Alternatively, introducing the vector  $\mathbf{a} = (a, \widehat a)$ and switching to continuum notation, we may write this as
\bea
S = {1 \over 4 \pi} \int_{M_3} \mathbf{a} \wedge K d \mathbf{a}~, \hspace{0.5 in} K = \left( \begin{matrix} 0 & 2 \\ 2 & -2\end{matrix} \right) ~,
\eea
which is the so-called \textit{K-matrix} presentation. As usual, we are allowed to do field redefinitions of $a$ and $\widehat a$, so long as we are careful to preserve their quantization conditions.  In particular, we can change $a \rightarrow a$, $\widehat a \rightarrow \widehat a + a$, in which case we have  
\bea
S = {1 \over 4 \pi} \int_{M_3} \mathbf{a} \wedge K d \mathbf{a}~, \hspace{0.5 in} K = \left( \begin{matrix} 2 & 0 \\ 0 & -2\end{matrix} \right) ~.
\eea
In other words, the SymTFT for anomalous $\ZZ_2$ symmetry is simply $U(1)_2 \times U(1)_{-2}$ Chern-Simons theory.

In general, $U(1)_k$ Chern-Simons theory has $|k|$ lines $L_q$ labelled by the charge $q= 0 , \dots, |k|-1$ , with grouplike fusion rules 
\bea
L_{q_1} \times L_{q_2} = L_{q_1 + q_2}~,
\eea
with charges added modulo $|k|$, and braidings given by 
\bea
 \begin{tikzpicture}[baseline = -5]
\draw[thick] (0,-1.2)--(0,0);
\draw [white!30, line width=3pt, distance = 0.3 in] (0,0) ellipse (1cm and 0.4cm);
\draw[thick , -<-=0.7] (0,0) ellipse (1cm and 0.4cm);
\draw [white!30, line width=3pt, distance = 0.3 in] (0,0)-- (0,1.2);
\draw[thick,->-=0.6] (0,0)-- (0,1.2);

\node[below] at (0,-1.2) {$L_{q_1}$};
\node[left] at (-1,-0.2) {$L_{q_2}$};
    \end{tikzpicture} 
     \hspace{0.1 in}= e^{2 \pi i {q_1 q_2\over k}}     \hspace{0.1 in}\begin{tikzpicture}[baseline = -5]
\draw[thick] (0,-1.2)--(0,0);
\draw[thick,->-=0.1] (0,0)-- (0,1.2);

\node[below] at (0,-1.2) {$L_{q_1}$};
    \end{tikzpicture}~; 
    \eea
see e.g. \cite{wen1992classification}. In addition, the lines have the following spins
 \bea
{\begin{tikzpicture}[baseline = -5]
\draw[thick] (0.33,-1.2) to [out=90, in=225,distance=0.07in](0.35,-0.35);
\draw[thick] (0.35,0.35) to [out=135, in=-90,distance=0.07in] (0.33,1.2);
\draw [thick] (0,0) ++(-45:.5) to +(45:1);
\draw [thick, decoration = {markings, mark=at position .46 with {\arrowreversed[scale=1.5,rotate=10]{stealth}}}, postaction=decorate] (0,0) ++(-45:1) ++(45:.5) arc (-135:135:.5);
\draw [thick] (0,0) ++(45:.5) to +(-45:.3); \draw [thick] (0,0) ++(45:.5) ++ (-45:.7) to +(-45:.3);
\node[below] at (0.3,-1.2) {$L_{q}$};
\end{tikzpicture}}
\hspace{0.1in}= e^{2 \pi i {q^2\over 2k}}   \hspace{0.1 in}\begin{tikzpicture}[baseline = -5]
\draw[thick] (0,-1.2)--(0,0);
\draw[thick,->-=0.1] (0,0)-- (0,1.2);

\node[below] at (0,-1.2) {$L_{q}$};
    \end{tikzpicture}~. 
\eea
In the special case of $k=\pm 2$, this means that there is exactly one nontrivial line, with spin $\pm i$, referred to as a ``semion.''
Because of this, the theory $U(1)_2 \times U(1)_{-2}$ is known as the \textit{double semion} theory \cite{freedman2004class,levin2005string}. 
We denote the two semions by $s$ and $\overline s$, each of which has full braiding phase with another copy of itself $e^{\pm 2 \pi i {1 \over 2}} = -1$, and which have trivial linking with one another.  The theory $U(1)_2 \times U(1)_{-2}$ thus contains four topological lines, $1, s, \overline s$, and $b:=s \overline s$, with $b$ having trivial spin.

We now consider topological boundary conditions for this theory. 
It is impossible to have boundary conditions which absorb $s$ and $\overline s$, since these have non-trivial spin and cannot be condensed. 
Instead, we can only condense $b$, and from this we conclude the following correspondence between sectors, 
\bea
1 &\hspace{0.2 in}\leftrightarrow\hspace{0.2 in}& \ZZ_2\,\,\, \mathrm{even}, \,\, \mathrm{untwisted}~,
\no\\
b  &\hspace{0.2 in}\leftrightarrow\hspace{0.2 in}& \ZZ_2\,\,\, \mathrm{odd}, \,\, \mathrm{untwisted}~,
\no\\
s&\hspace{0.2 in}\leftrightarrow\hspace{0.2 in}& \ZZ_2\,\,\, \mathrm{even}, \,\, \mathrm{twisted}~,
\no\\
\overline s &\hspace{0.2 in}\leftrightarrow\hspace{0.2 in}& \ZZ_2\,\,\, \mathrm{odd}, \,\, \mathrm{twisted}~.
\eea
In the twisted sectors, ``even/odd'' are bookkeeping labels after rephasing the symmetry action; the canonical projective action has eigenvalues $\pm i$ and squares to $-1$.
In total then, we again have four sectors, just like for the non-anomalous $\ZZ_2$ symmetry studied before. 
 However, unlike the previous case, note that there is no longer a topological boundary condition that exchanges the $\ZZ_2$ odd untwisted sector ($b$) with the $\ZZ_2$ even twisted sector ($s$). This corresponds to the fact that it is no longer possible to gauge the boundary $\ZZ_2$ symmetry, as expected from the fact that it was anomalous.
 More generally, non-trivial spin or self-braiding of lines in the SymTFT is closely related to the presence of anomalies on the boundary, as discussed in e.g. \cite{Kaidi:2023maf}. 

\subsubsection{Example: $\cC = \mathrm{Ising}$}

We next consider the case of a theory with Ising symmetry; the Drinfeld center in this case was obtained in \cite{izumi2001structure,gelaki2009centers} in the mathematics literature and \cite{Barkeshli:2014cna,Teo:2015xla,Kaidi:2022cpf} in the physics literature. As described in \cite{Kaidi:2022cpf}, the SymTFT is obtained by gauging the $\ZZ_2$ EM duality symmetry of $(2+1)$d $\ZZ_2$ gauge theory.

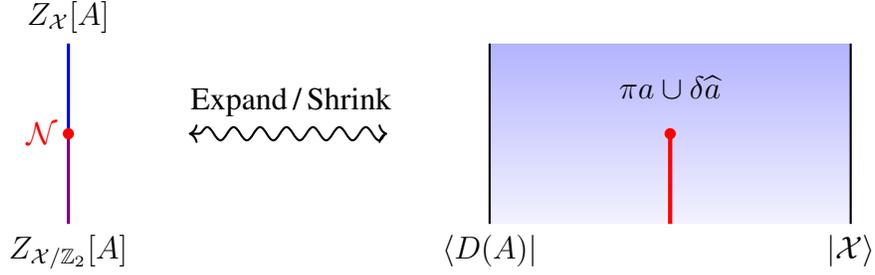
\begin{figure}[!tbp]
	\centering
	\begin{tikzpicture}[scale=0.8]
	
		\shade[line width=2pt, top color=blue!30, bottom color=blue!5] 
	(0,0) to [out=90, in=-90]  (0,3)
	to [out=0,in=180] (6,3)
	to [out = -90, in =90] (6,0)
	to [out=180, in =0]  (0,0);
	
	\draw[very thick, violet] (-7,0) -- (-7,1.5);
	\draw[very thick, blue] (-7,1.5) -- (-7,3);
	\node[above] at (-7,3) {$Z_{\cX}[A]$};
	\node[below] at (-7,0) {$Z_{\cX/\ZZ_2}[A]$};
	\node at (-7,1.5) [circle,fill,red, inner sep=1.5pt]{};
	\node[left,red] at (-7,1.5) {$\cN$};

	\draw[thick, snake it, <->] (-1.7,1.5) -- (-5, 1.5);
	\node[above] at (-3.3,1.6) {Expand\,/\,Shrink};
	
	\draw[thick] (0,0) -- (0,3);
	\draw[thick] (6,0) -- (6,3);
	\draw[ultra thick,red] (3,0) -- (3,1.5);
	\node[above] at (3,1.9) {$\pi {{a}}  \cup \delta \widehat{a}$};
	\node at (3,1.5) [circle,fill,red, inner sep=1.5pt]{};
	\node[below] at (0,0) {$\langle D(A)| $};
	\node[below] at (6,0) {$|\cX\rangle $}; 
	
	
	\end{tikzpicture}
	
	\caption{A $(1+1)$d QFT $\cX$ with $\ZZ_2$ symmetry and another $(1+1)$d QFT $\cX/\ZZ_2$  are separated by a topological interface $\cN$. This setup can be expanded into a $(2+1)$d slab, where the $(2+1)$d $\ZZ_2$ SymTFT has an insertion of a twist defect parallel to the Dirichlet boundary. 	}
	\label{fig:twistdefshrink}
\end{figure}

Indeed, in order to obtain a duality defect $\cN$ for $\ZZ_2$ on the boundary, one begins by considering a duality \textit{interface} separating a theory $\cX$ from the gauged theory  $\cX/\ZZ_2$. From the bulk perspective, the setup of interest is the insertion of a twist defect in the bulk, with the attached codimension-1 defect running parallel to the topological boundary; see Figure \ref{fig:twistdefshrink}. This defect should have the property that it exchanges the Dirichlet and Neumann boundary conditions for $a$, which is equivalent to exchanging the Dirichlet boundary condition for $a$ with the Dirichlet boundary condition for $\widehat a$. In other words, the defect in question is the generator of the $\ZZ_2^{EM}$ electromagnetic duality symmetry exchanging $a \leftrightarrow \widehat a$.  In order to turn $\cN$ from an interface into a defect, we must require that the theories  $\cX$ and  $\cX/\ZZ_2$ are equivalent, with a compatible action on the physical boundary. We then gauge the bulk $\ZZ_2^{EM}$ symmetry, so that the corresponding condensation defect becomes transparent, thereby making the twist defect into a genuine line operator in the bulk. 

\begin{figure}[t]
    \centering
    {\begin{tikzpicture}[baseline=35]

  \shade[top color=red!40, bottom color=red!10,rotate=90]  (0,-0.7) -- (2,-0.7) -- (2.6,-0.2) -- (0.6,-0.2)-- (0,-0.7);
 \draw[thick,rotate=90] (0,-0.7) -- (2,-0.7);
\draw[thick,rotate=90] (0,-0.7) -- (0.6,-0.2);
\draw[thick,rotate=90]  (0.6,-0.2)--(2.6,-0.2);
\draw[thick,rotate=90]  (2.6,-0.2)-- (2,-0.7);

 \begin{scope}[xshift=-0.9in] \shade[top color=red!40, bottom color=red!10,rotate=90]  (0,-0.7) -- (2,-0.7) -- (2.6,-0.2) -- (0.6,-0.2)-- (0,-0.7);
 \draw[thick,rotate=90] (0,-0.7) -- (2,-0.7);
\draw[thick,rotate=90] (0,-0.7) -- (0.6,-0.2);
\draw[thick,rotate=90]  (0.6,-0.2)--(2.6,-0.2);
\draw[thick,rotate=90]  (2.6,-0.2)-- (2,-0.7);

\end{scope}

 \draw[thick,violet] (-1.7,1.4) arc(0:360:0.15cm and 0.7cm);

\node at (0.4,1.4) [circle,fill,dgreen, inner sep=1.2pt]{};

\draw[thick,blue] (-1.6,1.4)--(0.4,1.4);
\draw[thick,blue,dotted] (-1.85,1.4)--(-1.6,1.4);
\node[right] at (-1.2,1.9) {$\widehat L_{(0)}^-$};
\node[left] at (-2,1.9) {$\eta$};
\end{tikzpicture} 
\qquad$=$\qquad
\begin{tikzpicture}[baseline=35]

  \shade[top color=red!40, bottom color=red!10,rotate=90]  (0,-0.7) -- (2,-0.7) -- (2.6,-0.2) -- (0.6,-0.2)-- (0,-0.7);
 \draw[thick,rotate=90] (0,-0.7) -- (2,-0.7);
\draw[thick,rotate=90] (0,-0.7) -- (0.6,-0.2);
\draw[thick,rotate=90]  (0.6,-0.2)--(2.6,-0.2);
\draw[thick,rotate=90]  (2.6,-0.2)-- (2,-0.7);

 \begin{scope}[xshift=-0.9in] \shade[top color=red!40, bottom color=red!10,rotate=90]  (0,-0.7) -- (2,-0.7) -- (2.6,-0.2) -- (0.6,-0.2)-- (0,-0.7);
 \draw[thick,rotate=90] (0,-0.7) -- (2,-0.7);
\draw[thick,rotate=90] (0,-0.7) -- (0.6,-0.2);
\draw[thick,rotate=90]  (0.6,-0.2)--(2.6,-0.2);
\draw[thick,rotate=90]  (2.6,-0.2)-- (2,-0.7);

\end{scope}

 \draw[thick,violet] (-1.2,1.3) arc(-175:165:0.15cm and 0.7cm);

\node at (0.4,1.4) [circle,fill,dgreen, inner sep=1.2pt]{};

\draw[thick,blue] (-1.6,1.4)--(-1,1.4);
\draw[thick,blue] (-0.8,1.4)--(0.4,1.4);
\draw[thick,blue,dotted] (-1.85,1.4)--(-1.6,1.4);
\node[right] at (-0.8,1.8) {$\widehat L_{(0)}^-$};

\node[right] at (-1.4,0.3) {$\widehat L_{(1)}^-$};

\end{tikzpicture} 
\qquad$=$\qquad$+$\quad
\begin{tikzpicture}[baseline=35]

  \shade[top color=red!40, bottom color=red!10,rotate=90]  (0,-0.7) -- (2,-0.7) -- (2.6,-0.2) -- (0.6,-0.2)-- (0,-0.7);
 \draw[thick,rotate=90] (0,-0.7) -- (2,-0.7);
\draw[thick,rotate=90] (0,-0.7) -- (0.6,-0.2);
\draw[thick,rotate=90]  (0.6,-0.2)--(2.6,-0.2);
\draw[thick,rotate=90]  (2.6,-0.2)-- (2,-0.7);

 \begin{scope}[xshift=-0.9in] \shade[top color=red!40, bottom color=red!10,rotate=90]  (0,-0.7) -- (2,-0.7) -- (2.6,-0.2) -- (0.6,-0.2)-- (0,-0.7);
 \draw[thick,rotate=90] (0,-0.7) -- (2,-0.7);
\draw[thick,rotate=90] (0,-0.7) -- (0.6,-0.2);
\draw[thick,rotate=90]  (0.6,-0.2)--(2.6,-0.2);
\draw[thick,rotate=90]  (2.6,-0.2)-- (2,-0.7);

\end{scope}

\node at (0.4,1.4) [circle,fill,dgreen, inner sep=1.2pt]{};

\draw[thick,blue] (-1.6,1.4)--(0.4,1.4);
\draw[thick,blue,dotted] (-1.85,1.4)--(-1.6,1.4);
\node[right] at (-1.2,1.9) {$\widehat L_{(0)}^-$};

\end{tikzpicture}

\vspace{0.2 in}
\begin{tikzpicture}[baseline=35]

  \shade[top color=red!40, bottom color=red!10,rotate=90]  (0,-0.7) -- (2,-0.7) -- (2.6,-0.2) -- (0.6,-0.2)-- (0,-0.7);
 \draw[thick,rotate=90] (0,-0.7) -- (2,-0.7);
\draw[thick,rotate=90] (0,-0.7) -- (0.6,-0.2);
\draw[thick,rotate=90]  (0.6,-0.2)--(2.6,-0.2);
\draw[thick,rotate=90]  (2.6,-0.2)-- (2,-0.7);

 \begin{scope}[xshift=-0.9in] \shade[top color=red!40, bottom color=red!10,rotate=90]  (0,-0.7) -- (2,-0.7) -- (2.6,-0.2) -- (0.6,-0.2)-- (0,-0.7);
 \draw[thick,rotate=90] (0,-0.7) -- (2,-0.7);
\draw[thick,rotate=90] (0,-0.7) -- (0.6,-0.2);
\draw[thick,rotate=90]  (0.6,-0.2)--(2.6,-0.2);
\draw[thick,rotate=90]  (2.6,-0.2)-- (2,-0.7);

\end{scope}

 \draw[thick,violet] (-1.7,1.4) arc(0:360:0.15cm and 0.7cm);

\node at (0.4,1.4) [circle,fill,dgreen, inner sep=1.2pt]{};

\draw[thick,blue] (-1.6,1.4)--(0.4,1.4);
\draw[thick,blue,dotted] (-1.85,1.4)--(-1.6,1.4);
\node[right] at (-1.2,1.9) {$\widehat L_{(0)}^-$};
\node[left] at (-2,1.9) {$\cN$};
\end{tikzpicture} 
\qquad$=$\qquad
\begin{tikzpicture}[baseline=35]

  \shade[top color=red!40, bottom color=red!10,rotate=90]  (0,-0.7) -- (2,-0.7) -- (2.6,-0.2) -- (0.6,-0.2)-- (0,-0.7);
 \draw[thick,rotate=90] (0,-0.7) -- (2,-0.7);
\draw[thick,rotate=90] (0,-0.7) -- (0.6,-0.2);
\draw[thick,rotate=90]  (0.6,-0.2)--(2.6,-0.2);
\draw[thick,rotate=90]  (2.6,-0.2)-- (2,-0.7);

 \begin{scope}[xshift=-0.9in] \shade[top color=red!40, bottom color=red!10,rotate=90]  (0,-0.7) -- (2,-0.7) -- (2.6,-0.2) -- (0.6,-0.2)-- (0,-0.7);
 \draw[thick,rotate=90] (0,-0.7) -- (2,-0.7);
\draw[thick,rotate=90] (0,-0.7) -- (0.6,-0.2);
\draw[thick,rotate=90]  (0.6,-0.2)--(2.6,-0.2);
\draw[thick,rotate=90]  (2.6,-0.2)-- (2,-0.7);

\end{scope}

 \draw[thick,violet] (-1.2,1.3) arc(-175:165:0.15cm and 0.7cm);

\node at (0.4,1.4) [circle,fill,dgreen, inner sep=1.2pt]{};

\draw[thick,blue] (-1.6,1.4)--(-1,1.4);
\draw[thick,blue] (-0.8,1.4)--(0.4,1.4);
\draw[thick,blue,dotted] (-1.85,1.4)--(-1.6,1.4);
\node[right] at (-0.8,1.8) {$\widehat L_{(0)}^-$};

\node[right] at (-1.4,0.3) {$\widehat \Sigma_{(i)}^\pm$};

\end{tikzpicture} 
\qquad$=$\qquad$-\sqrt{2}$\quad
\begin{tikzpicture}[baseline=35]

  \shade[top color=red!40, bottom color=red!10,rotate=90]  (0,-0.7) -- (2,-0.7) -- (2.6,-0.2) -- (0.6,-0.2)-- (0,-0.7);
 \draw[thick,rotate=90] (0,-0.7) -- (2,-0.7);
\draw[thick,rotate=90] (0,-0.7) -- (0.6,-0.2);
\draw[thick,rotate=90]  (0.6,-0.2)--(2.6,-0.2);
\draw[thick,rotate=90]  (2.6,-0.2)-- (2,-0.7);

 \begin{scope}[xshift=-0.9in] \shade[top color=red!40, bottom color=red!10,rotate=90]  (0,-0.7) -- (2,-0.7) -- (2.6,-0.2) -- (0.6,-0.2)-- (0,-0.7);
 \draw[thick,rotate=90] (0,-0.7) -- (2,-0.7);
\draw[thick,rotate=90] (0,-0.7) -- (0.6,-0.2);
\draw[thick,rotate=90]  (0.6,-0.2)--(2.6,-0.2);
\draw[thick,rotate=90]  (2.6,-0.2)-- (2,-0.7);

\end{scope}

\node at (0.4,1.4) [circle,fill,dgreen, inner sep=1.2pt]{};

\draw[thick,blue] (-1.6,1.4)--(0.4,1.4);
\draw[thick,blue,dotted] (-1.85,1.4)--(-1.6,1.4);
\node[right] at (-1.2,1.9) {$\widehat L_{(0)}^-$};

\end{tikzpicture}
}

    \caption{Measuring the charge of the operators in the sector labelled by $\widehat L_{(0)}^-$. By making use of the bulk braidings given in \cite{Kaidi:2022cpf}, we find that these states are $\ZZ_2$ even, but pick up a factor of $- \sqrt{2}$ when encircled by $\cN$.  }
    \label{fig:Isingcharge}
\end{figure}
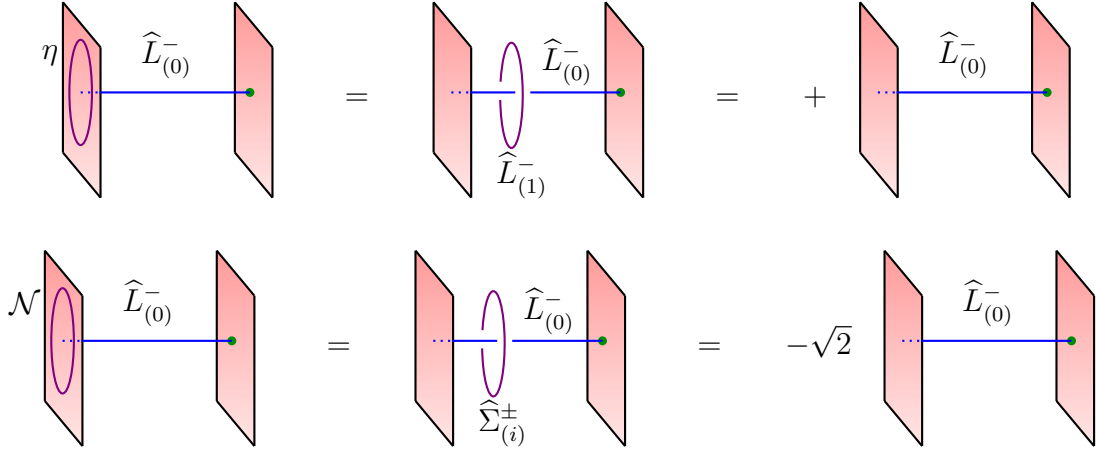

It is not hard to show that gauging the $\ZZ_2^{EM}$ symmetry of $\ZZ_2$ gauge theory gives a theory with a total of nine line operators, which we write as 
\bea
\widehat L_{(0)}^\pm~, \, \widehat L_{(1)}^\pm~, \, \widehat L_{[01]}~, \, \widehat \Sigma^\pm_{(0)}~, \, \widehat \Sigma^\pm_{(1)}~,
\eea
in the notation of \cite{Kaidi:2022cpf}; we refer the readers to that reference for more details. If we start as before with the Dirichlet-Neumann orbit of boundary conditions on the topological boundary, then after gauging $\ZZ_2^{EM}$, we find that the operators $\widehat L_{(0)}^\pm$ become trivial on the boundary, whereas $\widehat L_{(1)}^\pm$ become the non-trivial operator $\eta$ generating the $\ZZ_2$ symmetry of the boundary Ising category, and the operators $\widehat \Sigma_{(i)}^\pm$ all reduce to the non-invertible defect $\cN$. Finally, for $\widehat L_{[01]}$, we find that it reduces to a non-simple line $1+\eta$ on the boundary.

What the above analysis tells us is that we have nine irreducible representations with a total of ten components in the boundary theory---three of which contain local operators (namely $\widehat L_{(0)}^\pm$, $\widehat L_{[01]}$), three of which contain operators in the $\eta$-twisted sector (namely $\widehat L_{(1)}^\pm$, $\widehat L_{[01]}$), and four of which contain operators in the $\cN$-twisted sector (namely $\widehat \Sigma_{(0)}^\pm$, $\widehat \Sigma_{(1)}^\pm$). We may now measure the charges of operators in each sector under elements of the Ising category by encircling them by the appropriate defects, as is done for operators in sector $\widehat L_{(0)}^-$ in Figure \ref{fig:Isingcharge}. We see that operators in this sector are $\ZZ_2$ even, but pick up a factor of $-\sqrt{2}$ under the action of $\cN$. The local operator $\eps$ of the critical Ising model thus belongs to this sector. We may likewise label the other sectors by the operators that they contain in the critical Ising model,\footnote{We have not discussed $\psi$ and $\bar \psi$ in these lectures, but they are well-known operators in the $\eta$-twisted sector of the critical Ising model, and we include them here for completeness.} 
\bea
\widehat L_{(0)}^+ \hspace{0.2 in} &\leftrightarrow& \hspace{0.2 in} 1~, \hspace{0.53 in}\widehat L_{(0)}^- \hspace{0.2 in} \leftrightarrow \hspace{0.2 in} \eps~,
\no\\
\widehat L_{(1)}^+ \hspace{0.2 in} &\leftrightarrow& \hspace{0.2 in} \psi~, \hspace{0.5 in}\widehat L_{(1)}^- \hspace{0.2 in} \leftrightarrow \hspace{0.2 in} \overline \psi~,
\no\\
\widehat L_{[01]} \hspace{0.2 in} &\leftrightarrow& \hspace{0.2 in} (\sigma, \mu)~.
\eea
We see that the sector labelled by $\widehat L_{[01]}$ contains both $\sigma$ and $\mu$. This is because the corresponding representation is two-dimensional---indeed, the non-invertible symmetry $\cN$ maps between $\sigma$ and $\mu$.\footnote{More generally, the dimension of the representation is given, in the SymTFT description, by the dimension of the junction space of the bulk line on the topological boundary.}

 This may be seen in the SymTFT picture as follows. We begin by thinking of $\cN$ as a map from the untwisted Hilbert space to the untwisted Hilbert space, which corresponds to wrapping $\sigma$ by a loop of $\cN$, with no outgoing lines (see Figure \ref{fig:Nonsigma}). This is expected to give a vanishing result, and can be understood in the bulk as follows. We begin by working in the ungauged picture, in which $\cN$ lifts to a twist defect $\Sigma_{(i)}$ which is the boundary of the EM duality surface $D_\mathrm{EM}$, and the representation that $\sigma$ is in is labelled by the non-simple bulk line $L_{(0,1)}+L_{(1,0)}$. By shrinking the disk of $D_\mathrm{EM}$ in the bulk, we see that a non-zero value for this configuration requires the existence of a junction between $L_{(0,1)}$ and $L_{(1,0)}$, which by charge conservation cannot exist; see Figure \ref{fig:Nonsigma2}. On the other hand, if we treat $\cN$ as a map from the untwisted Hilbert space to the $\eta$-twisted Hilbert space, which corresponds to wrapping the operator in a loop of $\cN$ with an outgoing $\eta$ line (see Figure \ref{fig:Nonsigma}), then a non-zero result is indeed allowed by charge conservation (since $\eta$ is given by $L_{(1,1)}$ in the bulk), and gives $\sqrt{2}\, \mu$.

\begin{figure}[t]
    \centering
    {\begin{tikzpicture}[baseline=35]

\node at (0,0) [circle,fill,dgreen, inner sep=2pt]{};
\draw [red,thick] (0,0) circle[radius =0.8];

\node[right] at (-1.5,-0.4) {$\cN$};
\node[right] at (0.1,0) {$\sigma$};
\node[right] at (1.4,0) {$=$};
\node[right] at (2.2,0) {$0$};

\begin{scope}[xshift = 3 in]
\node at (0,0) [circle,fill,dgreen, inner sep=2pt]{};
\draw [red,thick] (0,0) circle[radius =0.8];

\draw [violet,thick] (0,0.8)--(0,1.5);

\node[right] at (-1.5,-0.4) {$\cN$};
\node[right] at (0.1,0) {$\sigma$};
\node[left] at (-0.1,1.2) {$\eta$};
\node[right] at (1.4,0) {$=$};

\node[right] at (1.80,0) {$\sqrt{2}\,\,\,\,\,\,\,$};
\draw [violet,thick] (2.7,0)--(2.7,1.5);
\node at (2.7,0) [circle,fill,dgreen, inner sep=2pt]{};

\node[right] at (2.9,0) {$\mu$};

\end{scope}

\end{tikzpicture}}

    \caption{When treating $\cN$ as a map from untwisted operators to untwisted operators, it annihilates $\sigma$, but when treated as a map from untwisted operators to the $\eta$-twisted sector, it produces $\sqrt{2}\,\mu$.  }
    \label{fig:Nonsigma}
\end{figure}

\begin{figure}[t]
    \centering
    {
\begin{tikzpicture}[baseline=35]

\begin{scope}[xshift=+0.2 in]
  \shade[top color=red!40, bottom color=red!10,rotate=90]  (0,-0.7) -- (2,-0.7) -- (2.6,-0.2) -- (0.6,-0.2)-- (0,-0.7);
 \draw[thick,rotate=90] (0,-0.7) -- (2,-0.7);
\draw[thick,rotate=90] (0,-0.7) -- (0.6,-0.2);
\draw[thick,rotate=90]  (0.6,-0.2)--(2.6,-0.2);
\draw[thick,rotate=90]  (2.6,-0.2)-- (2,-0.7);
\end{scope}

 \begin{scope}[xshift=-0.9in] \shade[top color=red!40, bottom color=red!10,rotate=90]  (0,-0.7) -- (2,-0.7) -- (2.6,-0.2) -- (0.6,-0.2)-- (0,-0.7);
 \draw[thick,rotate=90] (0,-0.7) -- (2,-0.7);
\draw[thick,rotate=90] (0,-0.7) -- (0.6,-0.2);
\draw[thick,rotate=90]  (0.6,-0.2)--(2.6,-0.2);
\draw[thick,rotate=90]  (2.6,-0.2)-- (2,-0.7);

\end{scope}

 \draw[thick,violet,fill={blue!30}] (-1.7,1.4) arc(0:360:0.15cm and 0.7cm);

\node at (0.9,1.4) [circle,fill,dgreen, inner sep=1.2pt]{};

\draw[thick,blue] (-1.6,1.4)--(0.9,1.4);
\draw[thick,blue,dotted] (-1.85,1.4)--(-1.6,1.4);
\node[right] at (-1.75,1.9) {$L_{(0,1)}+L_{(1,0)}$};
\node[left] at (-2,1.9) {$\cN$};
\end{tikzpicture} 
\qquad$=$\qquad
\begin{tikzpicture}[baseline=35]

\begin{scope}[xshift=+0.2 in]
  \shade[top color=red!40, bottom color=red!10,rotate=90]  (0,-0.7) -- (2,-0.7) -- (2.6,-0.2) -- (0.6,-0.2)-- (0,-0.7);
 \draw[thick,rotate=90] (0,-0.7) -- (2,-0.7);
\draw[thick,rotate=90] (0,-0.7) -- (0.6,-0.2);
\draw[thick,rotate=90]  (0.6,-0.2)--(2.6,-0.2);
\draw[thick,rotate=90]  (2.6,-0.2)-- (2,-0.7);
\end{scope}

 \begin{scope}[xshift=-0.9in] \shade[top color=red!40, bottom color=red!10,rotate=90]  (0,-0.7) -- (2,-0.7) -- (2.6,-0.2) -- (0.6,-0.2)-- (0,-0.7);
 \draw[thick,rotate=90] (0,-0.7) -- (2,-0.7);
\draw[thick,rotate=90] (0,-0.7) -- (0.6,-0.2);
\draw[thick,rotate=90]  (0.6,-0.2)--(2.6,-0.2);
\draw[thick,rotate=90]  (2.6,-0.2)-- (2,-0.7);

\end{scope}

 \shade[top color=violet!40, bottom color=violet!10] (-.45,1.35) circle(0.17cm and 0.7cm);
 
 \draw[thick,violet] (-0.6,1.3) arc(-175:165:0.15cm and 0.7cm);

\node at (0.9,1.4) [circle,fill,dgreen, inner sep=1.2pt]{};

\draw[thick,blue] (-1.6,1.4)--(-0.5,1.4);
\draw[thick,blue] (-0.2,1.4)--(0.9,1.4);
\draw[thick,blue,dotted] (-1.85,1.4)--(-1.6,1.4);
\node[right] at (-0.4,1.8) {$ L_{(1,0)}$};
\node[right] at (-1.7,1.8) {$ L_{(0,1)}$};

\node[right] at (-1,0.3) {$ \Sigma_{(i)}$};


\end{tikzpicture} 
\qquad$+$\qquad
\begin{tikzpicture}[baseline=35]

\begin{scope}[xshift=+0.2 in]
  \shade[top color=red!40, bottom color=red!10,rotate=90]  (0,-0.7) -- (2,-0.7) -- (2.6,-0.2) -- (0.6,-0.2)-- (0,-0.7);
 \draw[thick,rotate=90] (0,-0.7) -- (2,-0.7);
\draw[thick,rotate=90] (0,-0.7) -- (0.6,-0.2);
\draw[thick,rotate=90]  (0.6,-0.2)--(2.6,-0.2);
\draw[thick,rotate=90]  (2.6,-0.2)-- (2,-0.7);
\end{scope}

 \begin{scope}[xshift=-0.9in] \shade[top color=red!40, bottom color=red!10,rotate=90]  (0,-0.7) -- (2,-0.7) -- (2.6,-0.2) -- (0.6,-0.2)-- (0,-0.7);
 \draw[thick,rotate=90] (0,-0.7) -- (2,-0.7);
\draw[thick,rotate=90] (0,-0.7) -- (0.6,-0.2);
\draw[thick,rotate=90]  (0.6,-0.2)--(2.6,-0.2);
\draw[thick,rotate=90]  (2.6,-0.2)-- (2,-0.7);

\end{scope}

 \shade[top color=violet!40, bottom color=violet!10] (-.45,1.35) circle(0.17cm and 0.7cm);
 
 \draw[thick,violet] (-0.6,1.3) arc(-175:165:0.15cm and 0.7cm);

\node at (0.9,1.4) [circle,fill,dgreen, inner sep=1.2pt]{};

\draw[thick,blue] (-1.6,1.4)--(-0.5,1.4);
\draw[thick,blue] (-0.2,1.4)--(0.9,1.4);
\draw[thick,blue,dotted] (-1.85,1.4)--(-1.6,1.4);
\node[right] at (-0.4,1.8) {$ L_{(0,1)}$};
\node[right] at (-1.7,1.8) {$ L_{(1,0)}$};

\node[right] at (-1,0.3) {$ \Sigma_{(i)}$};


\end{tikzpicture} 

}

\vspace{0.2 in}
\hspace{1.6 in}\qquad$=$\qquad
\begin{tikzpicture}[baseline=35]

\begin{scope}[xshift=+0.2 in]
  \shade[top color=red!40, bottom color=red!10,rotate=90]  (0,-0.7) -- (2,-0.7) -- (2.6,-0.2) -- (0.6,-0.2)-- (0,-0.7);
 \draw[thick,rotate=90] (0,-0.7) -- (2,-0.7);
\draw[thick,rotate=90] (0,-0.7) -- (0.6,-0.2);
\draw[thick,rotate=90]  (0.6,-0.2)--(2.6,-0.2);
\draw[thick,rotate=90]  (2.6,-0.2)-- (2,-0.7);
\end{scope}

 \begin{scope}[xshift=-0.9in] \shade[top color=red!40, bottom color=red!10,rotate=90]  (0,-0.7) -- (2,-0.7) -- (2.6,-0.2) -- (0.6,-0.2)-- (0,-0.7);
 \draw[thick,rotate=90] (0,-0.7) -- (2,-0.7);
\draw[thick,rotate=90] (0,-0.7) -- (0.6,-0.2);
\draw[thick,rotate=90]  (0.6,-0.2)--(2.6,-0.2);
\draw[thick,rotate=90]  (2.6,-0.2)-- (2,-0.7);

\end{scope}

\node at (0.9,1.4) [circle,fill,dgreen, inner sep=1.2pt]{};

\draw[thick,blue] (-1.6,1.4)--(0.9,1.4);
\node at (-0.35,1.4) [circle,fill,violet, inner sep=1.2pt]{};

\draw[thick,blue,dotted] (-1.85,1.4)--(-1.6,1.4);
\node[right] at (-0.4,1.8) {$ L_{(1,0)}$};
\node[right] at (-1.7,1.8) {$ L_{(0,1)}$};



\end{tikzpicture} 
\qquad$+$\qquad
{\begin{tikzpicture}[baseline=35]

\begin{scope}[xshift=+0.2 in]
  \shade[top color=red!40, bottom color=red!10,rotate=90]  (0,-0.7) -- (2,-0.7) -- (2.6,-0.2) -- (0.6,-0.2)-- (0,-0.7);
 \draw[thick,rotate=90] (0,-0.7) -- (2,-0.7);
\draw[thick,rotate=90] (0,-0.7) -- (0.6,-0.2);
\draw[thick,rotate=90]  (0.6,-0.2)--(2.6,-0.2);
\draw[thick,rotate=90]  (2.6,-0.2)-- (2,-0.7);
\end{scope}

 \begin{scope}[xshift=-0.9in] \shade[top color=red!40, bottom color=red!10,rotate=90]  (0,-0.7) -- (2,-0.7) -- (2.6,-0.2) -- (0.6,-0.2)-- (0,-0.7);
 \draw[thick,rotate=90] (0,-0.7) -- (2,-0.7);
\draw[thick,rotate=90] (0,-0.7) -- (0.6,-0.2);
\draw[thick,rotate=90]  (0.6,-0.2)--(2.6,-0.2);
\draw[thick,rotate=90]  (2.6,-0.2)-- (2,-0.7);

\end{scope}

\node at (0.9,1.4) [circle,fill,dgreen, inner sep=1.2pt]{};

\draw[thick,blue] (-1.6,1.4)--(0.9,1.4);
\node at (-0.35,1.4) [circle,fill,violet, inner sep=1.2pt]{};

\draw[thick,blue,dotted] (-1.85,1.4)--(-1.6,1.4);
\node[right] at (-0.4,1.8) {$ L_{(0,1)}$};
\node[right] at (-1.7,1.8) {$ L_{(1,0)}$};



\end{tikzpicture} }

\vspace{0.2 in}
\hspace{-1.5 in}$=$\qquad $0$
    \caption{The bulk explanation for the fact that, when treated as a map between untwisted sector operators, $\cN$ maps $\sigma$ to 0. }
    \label{fig:Nonsigma2}
\end{figure}

\begin{tcolorbox}
\textbf{Exercise:} Fill in the details above. In particular, use the bulk braidings and quantum dimensions given in \cite{Kaidi:2022cpf} to reproduce the actions in (\ref{eq:Isingactions}). 
\end{tcolorbox}

\subsubsection*{Further references}

The first formal descriptions of categorical symmetries in $(1+1)$-dimensional systems were given in \cite{Bhardwaj:2017xup,Chang:2018iay}, and by now there have been a vast number of developments including \cite{Lin:2019hks,Thorngren:2019iar,Konechny:2019wff,Huang:2020lox,Gaiotto:2020iye,Komargodski:2020mxz,Chang:2020imq,Huang:2021ytb,Inamura:2021wuo,Thorngren:2021yso,Sharpe:2021srf,Huang:2021zvu,Huang:2021nvb,Vanhove:2021zop,Inamura:2021szw,Burbano:2021loy,Inamura:2022lun,Chang:2022hud,Lu:2022ver,Lin:2022dhv,Li:2023ani,Lin:2023uvm,Cao:2023doz,Jacobsen:2023isq,Choi:2023xjw,Haghighat:2023sax,Seiberg:2023cdc,Sun:2023xxv,Aksoy:2023hve,Duan:2023ykn,Chen:2023jht,Nagoya:2023zky,Bhardwaj:2023idu,Choi:2023vgk,Grover:2023loq,Damia:2024xju,Gutperle:2024vyp,Bharadwaj:2024gpj,Grimminger:2024mks,Lu:2024lzf,Bottini:2025hri,Albert:2025umy,Honda:2026bjy}.  
By far the most well-understood examples of categorical symmetries are line operators in RCFTs. The study of line operators in RCFT was initiated in \cite{Verlinde:1988sn,Moore:1988qv}---far predating the modern study of categorical symmetry---and these results have since been expanded on in a vast number of directions \cite{Petkova:2000ip,Fuchs:2002cm,Bhardwaj:2017xup,Chang:2018iay,Lin:2022dhv,Komargodski:2020mxz,Tachikawa:2017gyf, Frohlich:2004ef, Frohlich:2006ch, Frohlich:2009gb, Carqueville:2012dk, Brunner:2013xna, Huang:2021zvu, Thorngren:2019iar, Thorngren:2021yso, Lootens:2021tet, Huang:2021nvb, Inamura:2022lun}.

As we discussed above, the representation theory of categorical symmetries is captured by the SymTFT, which has now been studied in various contexts, and in various dimensions  \cite{Apruzzi:2023uma,Bhardwaj:2023bbf,DelZotto:2024tae,Franco:2024mxa,Hasan:2024aow,Copetti:2024onh,Antinucci:2024ltv,Bhardwaj:2024ydc,Bhardwaj:2024igy,Argurio:2024ewp,Duan:2024xbb,Chen:2024ulc,Bergman:2024its,Bonetti:2024etn,DelZotto:2025yoy,Schafer-Nameki:2025fiy,Kaidi:2025hyr}. Recent results have also shed light on anomalies of non-invertible symmetries \cite{Choi:2023xjw,Kaidi:2023maf,Zhang:2023wlu,Cordova:2023bja,Antinucci:2023ezl,Franco:2024mxa,Hsin:2025ria}, as well as their spontaneous breaking \cite{Damia:2023gtc,Bhardwaj:2024qrf}. 

An interesting topic which we have not had time to discuss here is the understanding of non-invertible symmetries on unorientable manifolds, and the relation to various structures in mathematics such as Jandl algebras. Some works in this direction include \cite{Fuchs:2003id,Kapustin:2015uma,Bhardwaj:2016dtk,Inamura:2021wuo,Harada:2025uhh}. 
\newpage
\section{Four dimensions and physical applications}
\label{sec:daythree}

Having discussed various aspects of non-invertible symmetries in $(1+1)$-dimensions, we now proceed to a similar discussion in $(3+1)$-dimensions. Fortunately, many of the conceptual tools developed previously, including the notion of half-space gauging, are immediately applicable. That being said, the space of non-invertible symmetries in higher dimensions remains less understood, and in fact there were no concrete examples of non-invertible symmetries in $(3+1)$-dimensions until the last few years \cite{Heidenreich:2021xpr,Koide:2021zxj,Kaidi:2021xfk,Choi:2021kmx}. As such, below we will not be able to give a comprehensive survey of non-invertible symmetries in $(3+1)$-dimensions, instead just offering some vignettes in particular models.

\subsection{Duality defects in $(3+1)$-dimensions}
\label{sec:4dduality}
Let us begin by discussing the particular case of duality defects in $(3+1)$-dimensions, 
which are the generalization of the TY defects in  $(1+1)$-dimensions studied in Section \ref{sec:TY}. 
There are two main constructions of such defects in $(3+1)$-dimensions, 
\begin{enumerate}
\item Start with a theory with a 0-form symmetry and 1-form symmetry with a certain anomaly, and gauge the 1-form symmetry. 
The 0-form symmetry will be broken due to the anomaly, but gives rise to a different, non-invertible symmetry. 
\item Gauge a 1-form symmetry in half of the space, possibly with discrete torsion.
\end{enumerate}
We now briefly review both of these constructions.

\subsubsection{Gauging a theory with an anomaly }
\label{sec:gaugingwithanom}

Say that we have a $(3+1)$-dimensional theory on a spin manifold $M_4$ with a zero-form symmetry $\ZZ_{2}^{(0)}$ and a one-form symmetry $\ZZ_2^{(1)}$. Let us denote the background gauge fields for these two symmetries by $A_1$ and $B_2$, and furthermore assume that they have a mixed `t Hooft anomaly given by the inflow action 
\bea
\widehat \cA[A_1, B_2]  = {{\pi \over 2}} \int_{M_5} A_1 \cup {\cP(B_2)}~, \hspace{0.5 in} \p M_5 = M_4~. 
\eea
Here $\cP$ is the Pontrjagin square, and the inflow expression is understood with its spin refinement: a zero-form symmetry transformation multiplies the partition function by $e^{{i\pi\over 2}\int_{M_4}\cP(B_2)}$. This is the same as the set-up in Section \ref{sec:highergroup}. 
Recall that in that case we gauged $\ZZ_2^{(0)}$, and as a result of the anomaly the symmetry  $\ZZ_2^{(1)}$ was extended by the 
dual $\widehat \ZZ_2^{(2)}$, in such a way that $\delta \widehat A_3 = - B_2 \cup B_2$. 
This gave a non-trivial 3-group structure. 

On the other hand, let us now consider what happens when we gauge $\ZZ_2^{(1)}$. In the current case, because the anomaly is quadratic in $b_2$, we can no longer  absorb the $M_5$ dependence by changing the cocycle condition on the dual background field $\widehat B_2$. Thus we cannot salvage $\ZZ_2^{(0)}$ as part of a higher-group. 

However, what we can do is the following. Although the generator of the $\ZZ_2^{(0)}$ symmetry is anomalous under $\ZZ_2^{(1)}$ transformations, we may stack it with an appropriate 3d TQFT such that the total anomaly is zero. This composite defect then becomes the generator of a non-invertible zero-form symmetry. 
Concretely, in the current case the inflow action for the defect is given by ${\pi \over 2} \int_{M_4} \cP(b_2)$, and it is known that the simplest 3d TQFT that can match this anomaly is $U(1)_2$ Chern-Simons theory \cite{Hsin:2018vcg}. Denoting the original defect by $D(M_3)$, we conclude that the composite defect 
\bea
\cN(M_3) \propto \int \cD c   \, D(M_3) \,e^{{i  \over 2\pi } \int_{M_3}c \wedge d  c + {i\over \pi} \int c \wedge b_2 }
\eea
is a non-anomalous topological defect. Note that we are writing $b_2$ as a $U(1)$ gauge field here---we may always add a BF term to the action that makes it into a $\ZZ_2$ gauge field later.

The fusion rules may be obtained via direct computation. First of all, denoting the one-form symmetry generator by $L(\Sigma_2) = e^{i \oint b_2}$, it is clear that we have,
\bea
L(\Sigma_2) \times \cN(M_3) &\propto& \int \cD c  \, D(M_3) \,e^{{i  \over 2 \pi} \int_{M_3}c\wedge  d  c + {i\over \pi}  \int c\wedge  b_2 + i  \int_{\Sigma_2}b_2 }
\no\\
&=&  \int \cD c   \, D(M_3) \,e^{{i  \over 2 \pi } \int_{M_3}c \wedge d c + {i\over \pi} \int (c+ \pi\omega_{\Sigma_2})\wedge b_2}~.
\eea
For the discrete phases, let $\omega_{\Sigma_2}$ denote an integer lift of the mod-2 Poincar{\'e} dual of $\Sigma_2$; the following manipulations are understood in the cochain formulation. Then changing variables via $c \rightarrow c -\pi \omega_{\Sigma_2}$, we have 
\bea
L(\Sigma_2) \times \cN(M_3) &=&e^{{i \pi \over 2 }\int \omega_{\Sigma_2} \cup \delta \omega_{\Sigma_2} }  \,  \cN(M_3)~.
\eea
The overall factor can be rewritten in terms of the \textit{triple intersection number},
\bea
\label{eq:tripleintdef}
Q(\Sigma_2) := {1\over 2} \int \omega_{\Sigma_2}  \cup \delta \omega_{\Sigma_2} ~, 
\eea
which finally gives
\bea
L(\Sigma_2) \times \cN(M_3) &=& (-1)^{Q(\Sigma_2) }\cN(M_3)~. 
\eea

As for the fusion of $\cN(M_3) $ with itself (or its orientation reversal), we have,
\bea
\cN(M_3) \times \cN (M_3) ^\dagger &\propto& \int \cD c \cD \bar c \, e^{{i  \over  2 \pi} \int_{M_3} (c \wedge d c - \bar c \wedge d \bar c) + {i\over \pi}  \int_{M_3} (c- \bar c) \wedge b_2 }
\no\\
&=&  \int \cD c \cD \widehat c \, e^{{ i  \over \pi } \int c \wedge d \widehat c - {i  \over 2 \pi} \int \widehat c\wedge d \widehat c + {i\over \pi}  \int \widehat c \wedge b_2}~,  
\eea
where in the second equality we have changed variables to $\widehat c := c - \bar c$. Note that the first term is simply a BF term, and we can easily integrate $c$ out. Returning to $\ZZ_2$-valued fields (so that the preceding $U(1)$ two-form is $\pi b_2$), we are then left with 
\bea
\cN(M_3) \times \cN (M_3) ^\dagger\propto \sum_{\widehat c \,\in\, H^1(M_3, \ZZ_2)} e^{- {i \pi \over 2 } \int \widehat c \cup \delta  \widehat c + i \pi \int \widehat c \cup b_2}~,
\eea
or in homology notation,
\bea
\cN(M_3) \times \cN (M_3) ^\dagger\propto \sum_{\Sigma_2 \in H_2(M_3, \ZZ_2)} (-1)^{Q(\Sigma_2)} L(\Sigma_2)~.
\eea
Note that we have been somewhat cavalier with the overall normalization---as shown in e.g. Appendix B of \cite{Kaidi:2021xfk}, the correct normalization is 
\bea
\cN(M_3) \times \cN (M_3) ^\dagger= {1 \over |H^0(M_3, \ZZ_2)|} \sum_{\Sigma_2 \in H_2(M_3, \ZZ_2)} (-1)^{Q(\Sigma_2)} L(\Sigma_2)~. 
\eea
This can be understood by noting that, from the point of view of $M_3$, the submanifold $\Sigma_2$ is codimension-1, so the right-hand side is analogous to the gauging of a zero-form symmetry in $(2+1)$-dimensions, whose coefficient is given in (\ref{eq:zeroformcoeff}). 
Using the equivalence of $\cN^\dagger$ and $\cN$ in the spin theory, up to purely gravitational invertible factors, we obtain the following fusion rules, 
\bea
&\vphantom{.}& L(\Sigma_2)^2 = 1~, \hspace{0.5 in }L(\Sigma_2) \times \cN(M_3) = (-1)^{Q(\Sigma_2)} \cN(M_3) ~, 
\no\\
&\vphantom{.}&\cN(M_3) \times \cN (M_3) = {1 \over |H^0(M_3, \ZZ_2)|} \sum_{\Sigma_2 \in H_2(M_3, \ZZ_2)} (-1)^{Q(\Sigma_2)} L(\Sigma_2)~. 
\eea
Let us note, incidentally, that the right-hand side of the final equation is often referred to as a \textit{condensation defect}, 
\bea
\label{eq:condesatedef}
\cC(M_3) :=  {1 \over |H^0(M_3, \ZZ_2)|} \sum_{\Sigma_2 \in H_2(M_3, \ZZ_2)} (-1)^{Q(\Sigma_2)} L(\Sigma_2)~.
\eea

Let us now compare these fusion rules to the Ising fusion rules in (\ref{eq:Isingfusionrules}). We see that they are extremely similar, with two main differences. First, in two dimensions all topological defects appearing were lines, but in the current case the duality defect is codimension-1, while the generators of the one-form symmetry are codimension-2. In particular, the fusion of two codimension-1 surfaces gives rise to a sum of codimension-2 surfaces. Because of this interplay between objects of various dimensions, the mathematical structure describing these symmetries is no longer a category, but rather a \textit{higher-category}, though we will not give details of this mathematical structure.\footnote{See e.g. \cite{Decoppet:2023juy} for some details in the fusion 2-category case. } Second, there are factors of $(-1)^{Q(\Sigma_2)}$ appearing above which do not appear in the Ising fusion rules---indeed, in the Ising case, all coefficients were required to be non-negative integers. More generally, the coefficients for $d$-dimensional defects can be $d$-dimensional TQFTs \cite{Choi:2022,Copetti:2023mcq}. In the case of lines, these were $1$-dimensional topological quantum mechanical theories, i.e. theories of free $n$-dimensional qunits, with the resulting coefficient being the non-negative integer $n$ itself.

 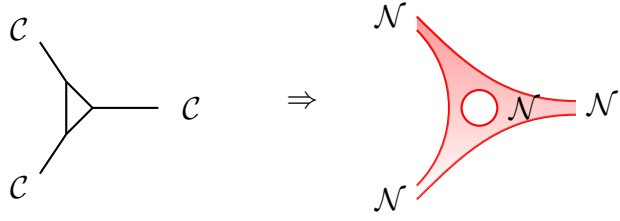
\begin{figure}[!tbp]
\begin{center}
\begin{tikzpicture}[baseline=0,scale=0.35]
\draw[ thick] (-2,2.5)--(-1,1);
\draw[ thick] (-1,1)--(0,0);
\draw[ thick] (-2,-2.5)--(-1,-1);
\draw[ thick] (-1,-1)--(0,0);
\draw[thick] (-1,1)--(-1,-1);
\draw[thick] (0,0)--(2.5,0);
  \node[left] at (-2,3) {$\cC$};
    \node[left] at (-2,-3) {$\cC$};
     \node[right] at (3,0) {$\cC$};
   \end{tikzpicture}
\hspace{0.35 in}$\Rightarrow$\hspace{0.2 in}
\begin{tikzpicture}[baseline=0,scale=0.45]
  \draw[thick,red] (-2.2,2.7) to[out=-45,in=180] (2.5,0.2);
   \draw[thick,red] (-2.2,-2.7) to[out=45,in=180] (2.5,-0.2);
    \draw[thick,red] (-2.2,2.3) to[out=-45,in=45](-2.2,-2.3);
    
 \shade[line width=2pt, top color=red,opacity=0.4, red] 
   (-2.2,2.7) to[out=-45,in=180] (2.5,0.2)
  to[out=-90,in=90] (2.5,-0.2)
 to[out=180, in=45] (-2.2,-2.7)
  to[out=90, in=-90] (-2.2,-2.3)
  to[out=45,in=-45](-2.2,2.3)
   to[out=90,in=-90]  (-2.2,2.7);
\draw[red,fill = white,thick]  (-0.35,0) circle (15pt);
\node[right] at (0.2,0){$\cN$};
\node[left] at (-2.2,2.7){$\cN$};
\node[left] at (-2.2,-2.7){$\cN$};
\node[right] at (2.5,0.1){$\cN$};
 \end{tikzpicture}
\caption{A theory with a Kramers-Wannier defect has a self-duality since the mesh of $\cC$ can be replaced by a set of topologically trivial loops of $\cN$. }
\label{fig:reversefig}
\end{center}
\end{figure}

As in $(1+1)$-dimensions, the presence of a duality defect tells us that the theory has a self-duality under gauging, with the role of the algebra object described in Section \ref{sec:non-invgauging} now being played by the condensation defect. This may be argued in a manner similar to Section \ref{sec:TY}. In particular, we begin by inserting a fine mesh of the condensate $\cC(M_3)$ (which is itself a fine mesh of 2-surfaces), and then replacing it with pairs of $\cN(M_3)$ as shown schematically in Figure \ref{fig:reversefig}. But assuming that we
started with a fine enough mesh, each loop of $\cN(M_3)$ is now contractible, and may be evaluated to a number. Thus with appropriate normalization we reobtain the original theory. Note that the factor of $(-1)^{Q(\Sigma_2)}$ in the expression for $\cC(M_3)$ tells us that the gauging of $\ZZ_2^{(1)}$ must be performed with a certain discrete torsion in order to get a self-duality. In Section \ref{sec:masslessQED} we will see this in a bit more detail in a concrete example.

\subsubsection{Half-space gauging}

We may also generate non-invertible symmetries in $(3+1)$-dimensions by performing half-space gauging of non-anomalous symmetries.
In particular, say that we gauge a $\ZZ_2^{(1)}$ one-form symmetry in half of the space, with Dirichlet boundary conditions at the interface, 
\bea
\begin{tikzpicture}[baseline={([yshift=-1ex]current bounding box.center)},vertex/.style={anchor=base,
    circle,fill=black!25,minimum size=18pt,inner sep=2pt},scale=0.5]
  
   \draw[thick] (0,-1.5) -- (0,3.5); 
   
   \shade[top color=blue, bottom color=white,opacity = 0.1] (0,-1.5) --(0,3.5) --(-5,3.5) -- (-5, -1.5) -- (0,-1.5) ;
      \shade[top color=red, bottom color=white,opacity = 0.1] (0,-1.5) --(0,3.5) --(10,3.5) -- (10, -1.5) -- (0,-1.5) ;
   
    \node[left] at (-1,1) {$\cL[B_2]$}; 
     \node[right] at (1,1) {$\cL[b_2] + \pi b_2 \cup B_2$}; 
     \node[below ] at (0,-1.5) {$b_2 |_0 = 0$};
     
      \node[left] at (-2,2.7) {\footnotesize$x<0$}; 
     \node[right] at (7,2.7) {\footnotesize$x>0$};

\end{tikzpicture}~~~.
\eea
If the theories on the left and the right are the same, then this topological interface gives a topological defect in a single theory,
and it is in fact non-invertible. The non-invertible nature can be verified by directly checking the fusion rules. 
Indeed, consider fusing the defect with its orientation reversal by first inserting $\cN$ at $x=0$ and $\cN^\dagger$ at $x = \eps$, 
\bea
\begin{tikzpicture}[baseline={([yshift=-1ex]current bounding box.center)},vertex/.style={anchor=base,
    circle,fill=black!25,minimum size=18pt,inner sep=2pt},scale=0.5]
  
   \draw[thick] (0,-1.5) -- (0,3.5); 
   
   \shade[top color=blue, bottom color=white,opacity = 0.1] (0,-1.5) --(0,3.5) --(-5,3.5) -- (-5, -1.5) -- (0,-1.5) ;
      \shade[top color=red, bottom color=white,opacity = 0.1] (0,-1.5) --(0,3.5) --(10,3.5) -- (10, -1.5) -- (0,-1.5) ;
        \shade[top color=violet, bottom color=white,opacity = 0.1] (10,-1.5) --(10,3.5) --(25,3.5) -- (25, -1.5) -- (10,-1.5) ;
   
    \node[left] at (-1,1) {$\cL[B_2]$}; 
     \node[right] at (1,1) {$\cL[b_2] + \pi b_2 \cup B_2$}; 
      \node[right] at (11,1) {$\cL[b_2] +  \pi (b_2 \cup \widehat b_2 - \widehat b_2 \cup B_2)$}; 
     \node[below ] at (0,-1.5) {$b_2|_0 = 0$};
        \node[below ] at (10,-1.5) {$\widehat b_2|_\eps = 0$};
     
      \node[left] at (-2,2.7) {\footnotesize$x<0$}; 
     \node[right] at (6,2.7) {\footnotesize$0<x<\eps$}; 
      \node[right] at (22,2.7) {\footnotesize$x>\eps$}; 
     
      \draw[thick] (10,-1.5) -- (10,3.5);

\end{tikzpicture}~~~.
\eea
and then taking the $\eps \rightarrow 0$ limit, 
\bea
\begin{tikzpicture}[baseline={([yshift=-1ex]current bounding box.center)},vertex/.style={anchor=base,
    circle,fill=black!25,minimum size=18pt,inner sep=2pt},scale=0.5]
  
   \draw[thick] (0,-1.5) -- (0,3.5); 
   
   \shade[top color=blue, bottom color=white,opacity = 0.1] (0,-1.5) --(0,3.5) --(-5,3.5) -- (-5, -1.5) -- (0,-1.5) ;
      \shade[top color=violet, bottom color=white,opacity = 0.1] (0,-1.5) --(0,3.5) --(15,3.5) -- (15, -1.5) -- (0,-1.5) ;
   
    \node[left] at (-1,1) {$\cL[B_2]$}; 
  \node[right] at (1,1) {$\cL[b_2] + \pi (b_2\cup \widehat b_2 - \widehat b_2 \cup B_2)$}; 
     \node[below ] at (0,-1.5) {$b_2|_0, \widehat b_2|_0= 0$};
     
      \node[left] at (-2,2.7) {\footnotesize$x<0$}; 
     \node[right] at (7,2.7) {\footnotesize$x>0$};

\end{tikzpicture}~~~.
\eea
Note that away from the interface, we may integrate out $\widehat b_2$, which imposes $b_2 = B_2$. 
Thus away from the interface, we obtain $\cL[B_2]$ everywhere. 
On the other hand, at the location of the interface, the Dirichlet boundary condition $\widehat b_2=0$ 
means that we no longer have  $b_2= B_2$. Instead, we are left with a copy of $\ZZ_2^{(1)}$ gauge theory in the tubular neighborhood of the interface at $M_3$, which gives precisely a condensate $\cC(M_3)$ of the type in (\ref{eq:condesatedef}); see \cite{Choi:2022} for details.\footnote{In the current case, the condensate does not contain $(-1)^{Q(\Sigma_2)}$ though. This is because the current gauging does not involve discrete torsion. } The upshot is that we reobtain fusion rules 
\bea
\cN(M_3 ) \times \cN(M_3)^\dagger = \cC(M_3)
\eea
of the familiar type. We likewise have $L(\Sigma_2) \times \cN(M_3) = \cN(M_3)$, as follows immediately from the Dirichlet boundary conditions at the location of the interface. Hence this is again a defect of duality type. 

We should note that the two constructions of non-invertible symmetries discussed above, i.e. via gauging in a theory with an anomaly or via half-space gauging, may give the same defects in some cases, but are not in general interchangeable. For example, there exist non-invertible defects which can be constructed via half-space gauging, but which do not arise from any theory with an anomaly.\footnote{Such non-invertible symmetries are sometimes referred to as \textit{intrinsically non-invertible} \cite{Kaidi:2022uux}.} There could also be non-invertible defects which cannot be constructed via either of these pictures.

\subsection{Maxwell theory}
We will now switch gears and discuss non-invertible symmetries in $(3+1)$d in a variety of concrete models. Our first example will be Maxwell theory \cite{Choi:2021kmx}, which we had discussed before in Sections \ref{sec:oneform} and \ref{sec:anomaly}.  Unlike before, in the current case we will also allow for a theta angle, 
\bea
\cL = {1\over 2 e^2} f\wedge * f + { \theta \over 8 \pi^2} f \wedge f~. 
\eea
The theta term is a total derivative and hence at the classical level seems to do nothing, but at the quantum level it has important effects.
The resulting theory is labelled by two parameters $e$ and $\theta$, which we group into a single complex parameter 
\bea
\tau =  {\theta \over 2 \pi }+  {2 \pi i \over e^2}~. 
\eea
Let us denote the theory labelled by the complex coupling $\tau$ by $\cX[\tau]$. An important property of Maxwell theory (on spin manifolds) is that it has a Montonen-Olive duality symmetry $\mathrm{SL}(2, \ZZ)$ \cite{Montonen:1977sn}, which says (suppressing the standard gravitational factors) that\footnote{Recall that $\mathrm{SL}(2, \ZZ)$ is the group of $2 \times 2$ integer matrices with unit determinant.}
\bea
\cX [\tau] \cong \cX \left[{a \tau + b \over c \tau + d} \right] ~, \hspace{0.5 in} \forall \left(\begin{matrix} a & b \\ c & d \end{matrix}  \right) \in \mathrm{SL}(2, \ZZ)~. 
\eea
For example, taking $\left(\begin{smallmatrix} a & b \\ c & d \end{smallmatrix}  \right) = \left(\begin{smallmatrix} 0 & 1 \\ -1 & 0 \end{smallmatrix}  \right) $, we find that $\cX [\tau] \cong \cX \left[-1/\tau \right]$. This is an extremely non-trivial identity, which relates the theory at strong coupling to the theory at weak coupling.

The presence of the theta angle modifies the currents in (\ref{eq:Maxwellcurrents}) to
\bea
j^E =  { 1 \over e^2} f - { \theta \over 4 \pi^2} * f~, \hspace{0.5 in} j^M = {1\over 2 \pi} * f~,
\eea
but as before they are both conserved, and they correspond to two one-form symmetries $U(1)_E^{(1)}$ and $U(1)_M^{(1)}$. 
In order to construct non-invertible defects, we now consider half-space gauging subgroups of these two symmetry groups. For example, consider gauging a $\ZZ_N^{(1)} \subset U(1)_E^{(1)}$ with $N>1$ in half-space with Dirichlet boundary conditions on the boundary. 
Gauging $\ZZ_N^{(1)} $ amounts to shifting $a \rightarrow {1\over N}a$,\footnote{This follows since gauging $\ZZ_N^{(1)} $ keeps only Wilson lines of charge $q \in N \ZZ$. Thus the minimal charge in the gauged theory is $N$ times the minimal charge of the original theory, and this can alternatively be understood as rescaling $a \rightarrow {1\over N} a$. } and hence maps $\cX[\tau]$ to $\cX[\tau/N^2]$. We thus get an interface, 
\bea
\begin{tikzpicture}[baseline={([yshift=-1ex]current bounding box.center)},vertex/.style={anchor=base,
    circle,fill=black!25,minimum size=18pt,inner sep=2pt},scale=0.5]
  
   \draw[thick] (0,-1.5) -- (0,3.5); 
   
   \shade[top color=blue, bottom color=white,opacity = 0.1] (0,-1.5) --(0,3.5) --(-5,3.5) -- (-5, -1.5) -- (0,-1.5) ;
      \shade[top color=red, bottom color=white,opacity = 0.1] (0,-1.5) --(0,3.5) --(5,3.5) -- (5, -1.5) -- (0,-1.5) ;
   
    \node[left] at (-1,1.3) {$\cX[\tau]$}; 
     \node[right] at (1,1.3) {$\cX[\tau/N^2]$}; 
\end{tikzpicture}~~~.
\eea
In general, the theories on the left and right of the interface are distinct. 
However, recall that we have the duality $\cX[\tau] \cong \cX[-1/\tau]$, so this interface can alternatively be written as an interface between $\cX[\tau]$ and $\cX[-N^2 / \tau]$,
\bea
\begin{tikzpicture}[baseline={([yshift=-1ex]current bounding box.center)},vertex/.style={anchor=base,
    circle,fill=black!25,minimum size=18pt,inner sep=2pt},scale=0.5]
  
   \draw[thick] (0,-1.5) -- (0,3.5); 
   
   \shade[top color=blue, bottom color=white,opacity = 0.1] (0,-1.5) --(0,3.5) --(-5,3.5) -- (-5, -1.5) -- (0,-1.5) ;
      \shade[top color=red, bottom color=white,opacity = 0.1] (0,-1.5) --(0,3.5) --(10,3.5) -- (10, -1.5) -- (0,-1.5) ;
   
    \node[left] at (-1,1.3) {$\cX[\tau]$}; 
     \node[right] at (1,1.3) {$\cX[\tau/N^2] \cong \cX[-N^2/ \tau]$}; 
\end{tikzpicture}~~~.
\eea
In particular, when $\tau = i N$ we see that the theories on the left and the right are the same, and hence we get a defect in a single theory.
By the same arguments as in Section \ref{sec:4dduality}, this defect is topological, and gives a duality defect $\cN$. 

In the case of the Ising model, moving the local operator $\sigma$ past the non-invertible defect $\cN$ gave rise to a non-local operator $\mu$, as shown in (\ref{eq:sigmatomu})---this was one of the hallmarks of non-invertibility.  In the current case, a similar phenomenon occurs when we try to move a Wilson line through $\cN(M_3)$.\footnote{In the following formulas and diagrams, the right-hand operators are written in the gauged electric frame, before the S-duality identification.} In a common electric frame, the outgoing line is instead a fractional `t Hooft line. Indeed, since the gauging step maps $a \rightarrow {1\over N} a$, the charge-1 Wilson line becomes a charge ${1\over N}$ Wilson line upon passing it, but such a Wilson line is not well-defined as a line---it must be thought of as attached to a surface,\footnote{It is not well-defined since it is not gauge invariant. }
\bea
e^{i \oint a} \rightarrow ``\,\,{e^{{i \over N} \oint a}}\,\,"= e^{{i \over N} \int f}~,
\eea
or pictorially, 
\bea
\begin{tikzpicture}[scale = 1.2,baseline = 40]
   
   \shade[top color=red!40, bottom color=red!10,xshift=-0.6in, rotate=90]  (0,-0.7) -- (2,-0.7) -- (2.6,-0.2) -- (0.6,-0.2)-- (0,-0.7);
\draw[thick,xshift=-0.6in,rotate=90] (0,-0.7) -- (2,-0.7);
\draw[thick,xshift=-0.6in,rotate=90] (0,-0.7) -- (0.6,-0.2);
\draw[thick,xshift=-0.6in,rotate=90]  (0.6,-0.2)--(2.6,-0.2);
\draw[thick,xshift=-0.6in,rotate=90]  (2.6,-0.2)-- (2,-0.7);

\node[below] at (-1, 0){$\cN$};

 \draw[thick,blue] (-2,1.3) ellipse (0.125cm and 0.25cm);
  
    \node[left] at (-2.1,1.25) {$W$};
    \end{tikzpicture} 
\hspace{0.6 in}\Longrightarrow \hspace{0.6 in} 
\begin{tikzpicture}[scale = 1.2,baseline = 40]
  
\shade[top color=red!40, bottom color=red!10,xshift=-0.6in, rotate=90]  (0,-0.7) -- (2,-0.7) -- (2.6,-0.2) -- (0.6,-0.2)-- (0,-0.7);
  
\draw[thick,xshift=-0.6in,rotate=90] (0,-0.7) -- (2,-0.7);
\draw[thick,xshift=-0.6in,rotate=90] (0,-0.7) -- (0.6,-0.2);
\draw[thick,xshift=-0.6in,rotate=90]  (0.6,-0.2)--(2.6,-0.2);
\draw[thick,xshift=-0.6in,rotate=90]  (2.6,-0.2)-- (2,-0.7);

  \shade[top color=dgreen!40, bottom color=dgreen!10] (0.3,1.05) -- (-0.8,1.05)-- (-0.8,1.55) -- (0.3,1.55)-- (0.3,1.05) ;
    \draw[thick] (-1.1,1.3) ellipse (0.125cm and 0.25cm);
         \draw[thick,blue,fill=dgreen!20] (0.3,1.3) ellipse (0.125cm and 0.25cm);


\node[below] at (-1, 0){$\cN$};

     \draw[thick,dgreen] (0.3,1.55)--(-0.8,1.55);
  \draw[thick,dgreen] (0.3,1.05)--(-0.8,1.05);
  \draw[thick,dotted,dgreen] (-0.8,1.55)--(-1.1,1.55);
  \draw[thick,dotted,dgreen] (-0.8,1.05)--(-1.1,1.05);

     \node[right] at (0.4,1.25) {$W^{1\over N}$};

\end{tikzpicture}~~~.
\eea
If the cycle on which $W$ was defined is contractible, we can further deform the right-hand side to obtain a droplet, 
\bea
\begin{tikzpicture}[scale = 1.2,baseline = 40]
   
   \shade[top color=red!40, bottom color=red!10,xshift=-0.6in, rotate=90]  (0,-0.7) -- (2,-0.7) -- (2.6,-0.2) -- (0.6,-0.2)-- (0,-0.7);
\draw[thick,xshift=-0.6in,rotate=90] (0,-0.7) -- (2,-0.7);
\draw[thick,xshift=-0.6in,rotate=90] (0,-0.7) -- (0.6,-0.2);
\draw[thick,xshift=-0.6in,rotate=90]  (0.6,-0.2)--(2.6,-0.2);
\draw[thick,xshift=-0.6in,rotate=90]  (2.6,-0.2)-- (2,-0.7);

\node[below] at (-1, 0){$\cN$};

 \draw[thick,blue] (-2,1.3) ellipse (0.125cm and 0.25cm);
  
    \node[left] at (-2.1,1.25) {$W$};
    \end{tikzpicture} 
\hspace{0.6 in}\Longrightarrow \hspace{0.6 in} 
\begin{tikzpicture}[scale = 1.2,baseline = 40]
  
\shade[top color=red!40, bottom color=red!10,xshift=-0.6in, rotate=90]  (0,-0.7) -- (2,-0.7) -- (2.6,-0.2) -- (0.6,-0.2)-- (0,-0.7);
  
\draw[thick,xshift=-0.6in,rotate=90] (0,-0.7) -- (2,-0.7);
\draw[thick,xshift=-0.6in,rotate=90] (0,-0.7) -- (0.6,-0.2);
\draw[thick,xshift=-0.6in,rotate=90]  (0.6,-0.2)--(2.6,-0.2);
\draw[thick,xshift=-0.6in,rotate=90]  (2.6,-0.2)-- (2,-0.7);

         \draw[thick,blue,fill=dgreen!20] (0.3,1.3) ellipse (0.125cm and 0.25cm);

\node[below] at (-1, 0){$\cN$};
     \node[right] at (0.4,1.25) {$W^{1\over N}$};

\end{tikzpicture}~~~.
\eea
Either way, we see that the Wilson line is mapped to a non-genuine line operator, i.e. a line attached to a surface. 

Thus far we have discussed the gauging of $\ZZ_N^{(1)} \subset U(1)_E^{(1)}$. But it is also possible to gauge a subset $\ZZ_N^{(1)} \subset U(1)_M^{(1)}$ of the magnetic one-form symmetry. In this case one can show that $a \rightarrow N a$, so that $\cX[\tau]$ is mapped to $\cX[N^2 \tau]$, 
\bea
\begin{tikzpicture}[baseline={([yshift=-1ex]current bounding box.center)},vertex/.style={anchor=base,
    circle,fill=black!25,minimum size=18pt,inner sep=2pt},scale=0.5]
  
   \draw[thick] (0,-1.5) -- (0,3.5); 
   
   \shade[top color=blue, bottom color=white,opacity = 0.1] (0,-1.5) --(0,3.5) --(-5,3.5) -- (-5, -1.5) -- (0,-1.5) ;
      \shade[top color=red, bottom color=white,opacity = 0.1] (0,-1.5) --(0,3.5) --(5,3.5) -- (5, -1.5) -- (0,-1.5) ;
   
    \node[left] at (-1,1.3) {$\cX[\tau]$}; 
     \node[right] at (1,1.3) {$\cX[N^2 \tau]$}; 
\end{tikzpicture}~~~.
\eea
Using Montonen-Olive duality to rewrite $\cX[N^2 \tau] \cong \cX[- {1\over N^2 \tau}]$, we see that for $\tau = {i / N}$ we obtain a new defect, which is topological and non-invertible by the usual arguments. Hence at both $\tau = i N$ and $\tau = {i / N}$, we have non-invertible duality defects, which are constructed by gauging $\ZZ_N^{(1)}$ subgroups of the electric and magnetic one-form symmetries, respectively. 

Even more generally, we could consider simultaneously gauging $\ZZ_p^{(1)} \times \ZZ_q^{(1)} \subset U(1)_E^{(1)} \times U(1)_M^{(1)}$. 
However, in this case we must be careful about the values $p$ and $q$. As discussed in Section \ref{sec:anomaly},  the electric and magnetic one-form symmetries have a mixed anomaly and hence cannot be gauged simultaneously, but it is still possible for there to be non-anomalous subgroups. To analyze this, let us restrict $B_E$ and $B_M$ in (\ref{eq:Maxwellinflowact}) to be $\ZZ_p^{(1)}$ and  $\ZZ_q^{(1)}$ gauge fields by including appropriate BF terms, in which case the anomaly inflow action becomes 
\bea 
\widehat \cA [B_E, B_M] =\int_{M_5} {1\over 2 \pi}  B_E \wedge d B_M + {p \over 2 \pi} B_E \wedge d a_E + {q \over 2 \pi} B_M \wedge d a_M~. 
\eea
Here $a_E$ and $a_M$ are two-form gauge fields on $M_5$. Integrating out $a_E$ and $a_M$ allows us to replace $B_{E} \rightarrow {2 \pi \over p} B_E$, $B_{M} \rightarrow {2 \pi \over q} B_M$, where $B_{E}$ (resp. $B_M$) now has periods in $\ZZ_{p}$ (resp. $\ZZ_q$),  
\bea
\widehat \cA [B_E, B_M] ={2 \pi \over p q} \int_{M_5} B_E \cup \delta B_M~. 
\eea
Since $B_M$ can be shifted by $q$ times any cochain, we may replace $\delta B_M$ by $(1- q y) \delta B_M$ for any integer $y$. We now make use of B{\'e}zout's identity, which claims that there exist integers $x,y$ such that $px + qy = 1$ if and only if $p$ and $q$ are coprime.\footnote{More generally,  B{\'e}zout's identity says that for integers $p$ and $q$ with greatest common divisor $d = \mathrm{gcd}(p,q)$, there exist integers $x$ and $y$ such that $p x + q y = d$. } What this means is that when  $\mathrm{gcd}(p,q)=1$, we may rewrite the inflow action as 
\bea
\widehat \cA [B_E, B_M] ={2 \pi \over p q} \int_{M_5} (1- q y) B_E \cup \delta B_M = {2 \pi x \over  q} \int_{M_5}  B_E \cup \delta B_M ~.
\eea
Imposing that $B_M$ is flat mod $q$, we then conclude that $\widehat \cA [B_E, B_M]  \in 2 \pi \ZZ$, and hence that the anomaly is trivialized.

The conclusion is thus that we are allowed to gauge $\ZZ_p^{(1)} \times \ZZ_q^{(1)} \subset U(1)_E^{(1)} \times U(1)_M^{(1)}$ as long as $\mathrm{gcd}(p,q)=1$. Doing so sends $\tau \rightarrow \left( {q/ p}\right)^2 \tau$, and coupling this with Montonen-Olive duality, we conclude that there is a non-invertible duality symmetry at every point $\tau = i {p / q}$ with positive coprime $p,q$ and $pq>1$. For $p=q=1$, this construction gives the ordinary invertible S-duality defect.

\begin{tcolorbox}
\textbf{Exercise:}  Consider the $(1+1)$-dimensional theory of a single compact boson at radius $R$. This theory has two  $U(1)$ zero-form symmetries, namely the momentum and winding symmetries $U(1)_\text{mom}$ and $U(1)_\text{wind}$. Argue that gauging a $\ZZ_p \times \ZZ_q$ subgroup of $U(1)_\text{mom} \times U(1)_\text{wind}$ leads to a change in the radius by $R \rightarrow {q \over p}R$. 
(Note that we must  take $\mathrm{gcd}(p,q) = 1$ in order to avoid a mixed anomaly). 
Then use this to argue that there are non-invertible defects at every radius $R^2\in\mathbb{Q}$ with $R^2>0$ and $R^2\ne 1$ (\textit{Hint: follow the same steps as for Maxwell theory, replacing S-duality with T-duality}). At $R^2=1$, this construction reduces to ordinary invertible T-duality.
\end{tcolorbox}

\subsection{Massless QED}
\label{sec:masslessQED}

Let us next analyze the example of massless QED \cite{Cordova:2022ieu,Choi:2022jqy}, i.e. Maxwell theory coupled to a unit charge Dirac fermion,
\bea
\cL = {1\over 4 e^2 } f_{\m\n} f^{\m\n} + i \overline \Psi (\p_\m - i a_\m) \g^\m \Psi~.
\eea
We work on spin manifolds, without dynamical magnetic monopoles, and suppress gravitational background terms. In the above action we have not included a $\theta$ angle, but this does not lead to a loss of generality---indeed, in 
massless QED the theta angle is not physical. The reason for this is the ABJ anomaly discussed in Section \ref{sec:anomaly}. 
In particular, doing an axial rotation of $\Psi$ of the form $\Psi \rightarrow e^{i \alpha \gamma_5/2} \Psi$ produces a theta angle $\alpha$, 
and conversely any theta angle can be removed by simply doing a field redefinition of $\Psi$.

In the presence of the fermion $\Psi$, the current $j^E$ defined in (\ref{eq:Maxwellcurrents}) is no longer conserved, since the conservation of $j^E$ followed from the equation of motion $d * f = 0$, which now picks up a source term.\footnote{If the fermion had non-unit charge $q = N$, we would not break $U(1)_E^{(1)}$ completely, but rather to a $\ZZ_N^{(1)}$ subgroup. } As a result, the defects that we obtained by gauging subgroups of $U(1)_E^{(1)}$ are no longer present. On the other hand, the $U(1)_M^{(1)}$ symmetry is not affected by the charged fermion, and hence we can still consider half-space gauging of $\ZZ_N^{(1)}\subset U(1)_M^{(1)}$. However, since we can no longer use Montonen-Olive duality in QED (electromagnetic duality is broken by the presence of the electrically charged fermion), in order to obtain a non-invertible defect we will have to perform our gauging in a somewhat clever way.

Our starting point is the action in (\ref{eq:Maxwellcoupledtobkg}) without the background field $B_E$, i.e.
\bea
S[B_M] &=& \int {1\over 2e^2 } f\wedge * f + {1 \over 2 \pi} f \wedge B_M ~. 
\eea  
Restricting $B_M$ to a $\ZZ_N^{(1)}$ background, we are free to add local counterterms in $B_M$, and we choose to do this as follows, 
\bea
S[ B_M] &=& \int {1\over 2e^2 } f\wedge * f + {1 \over 2 \pi} f \wedge B_M +  { N k  \over 4 \pi} B_M \wedge B_M~,
\eea  
where we assume that $\mathrm{gcd}(k,N)=1$. We now gauge $\ZZ_N^{(1)}\subset U(1)_M^{(1)}$, which is done by including a BF term and making $B_M$ dynamical, i.e. $B_M \rightarrow b_M$, 
\bea
S[b_M] &=& \int {1\over 2e^2 } f\wedge * f + {1 \over 2 \pi} f \wedge b_M +  {N \over 2 \pi} b_M \wedge da_M + {N k  \over 4 \pi} b_M \wedge b_M~. 
\eea  
 Including local counterterms before gauging is known as including {discrete torsion}; it is sometimes also referred to as \textit{stacking with an SPT} before gauging. 

What is the effect of such a gauging? Note that because $\mathrm{gcd}(k,N)=1$, there exists an integer $p$ such that $pk = 1$ mod $N$. We may then rewrite the action $S[b_M]$ as follows, 
\bea
S[b_M] &=& \int {1\over 2e^2 } f\wedge * f + {1 \over 8 \pi^2} \left(-{2 \pi p \over N}\right) f \wedge f  + { N k  \over 4 \pi} \left( b_M + {p \over N}f \right)^2 ~, 
\eea
where we have dropped the BF term and terms that evaluate to $2 \pi  \ZZ$. Integrating out $b_M$ then gives, up to an overall gravitational counterterm, the following, 
\bea
S[b_M] &=& \int {1\over 2e^2 } f\wedge * f + {1 \over 8 \pi^2} \left(-{2 \pi p \over N}\right) f \wedge f~. 
\eea
Hence we see that the gauging of $\ZZ_N^{(1)}\subset U(1)_M^{(1)}$ with discrete torsion labelled by $k$ shifts $\theta \rightarrow \theta - {2 \pi p \over N}$, where $p$ is the mod $N$ inverse of $k$. On the other hand, this gauging does not affect the coupling $e$ at all. 

The importance of this is that, as we discussed above, the $\theta$ angle in QED can be removed by a redefinition of the fermion $\Psi$. As such, the theory obtained by half-space gauging of $\ZZ_N^{(1)}\subset U(1)_M^{(1)}$  with discrete torsion can be brought back to the original theory by chiral rotation,
\bea
\begin{tikzpicture}[baseline={([yshift=-1ex]current bounding box.center)},vertex/.style={anchor=base,
    circle,fill=black!25,minimum size=18pt,inner sep=2pt},scale=0.5]
  
   \draw[thick] (0,-1.5) -- (0,3.5); 
   
   \shade[top color=blue, bottom color=white,opacity = 0.1] (0,-1.5) --(0,3.5) --(-5,3.5) -- (-5, -1.5) -- (0,-1.5) ;
      \shade[top color=red, bottom color=white,opacity = 0.1] (0,-1.5) --(0,3.5) --(10,3.5) -- (10, -1.5) -- (0,-1.5) ;
   
    \node[left] at (-1,1.3) {$\cX[\tau]$}; 
     \node[right] at (1,1.3) {$\cX[\tau - { p \over N}] \cong \cX[\tau]$}; 
\end{tikzpicture}~~~,
\eea
and hence half-space gauging gives rise to a topological defect $\cN_{p/ N}$ in a single theory.

How does this defect act on operators of the theory? On the elementary field $\Psi$, it acts as a standard axial rotation with angle $\alpha = {2 \pi p \over N}$. In this sense, we see that the axial symmetry survives as a non-anomalous symmetry for angles $\alpha \in 2 \pi \QQ$, at the cost of becoming non-invertible.  The non-invertible nature can be seen by analyzing the action of the symmetry on extended operators of the theory---in particular, when acting on `t Hooft lines it gives them fractional electric charge via the Witten effect, and hence `t Hooft lines are mapped to non-genuine lines attached to surfaces,
\bea
\begin{tikzpicture}[scale = 1.2,baseline = 40]
   
   \shade[top color=red!40, bottom color=red!10,xshift=-0.6in, rotate=90]  (0,-0.7) -- (2,-0.7) -- (2.6,-0.2) -- (0.6,-0.2)-- (0,-0.7);
\draw[thick,xshift=-0.6in,rotate=90] (0,-0.7) -- (2,-0.7);
\draw[thick,xshift=-0.6in,rotate=90] (0,-0.7) -- (0.6,-0.2);
\draw[thick,xshift=-0.6in,rotate=90]  (0.6,-0.2)--(2.6,-0.2);
\draw[thick,xshift=-0.6in,rotate=90]  (2.6,-0.2)-- (2,-0.7);

\node[below] at (-1, 0){$\cN_{p/N}$};

 \draw[thick,blue] (-2,1.3) ellipse (0.125cm and 0.25cm);
  
    \node[left] at (-2.1,1.25) {$T$};
    \end{tikzpicture} 
\hspace{0.6 in}\Longrightarrow \hspace{0.6 in} 
\begin{tikzpicture}[scale = 1.2,baseline = 40]
  
\shade[top color=red!40, bottom color=red!10,xshift=-0.6in, rotate=90]  (0,-0.7) -- (2,-0.7) -- (2.6,-0.2) -- (0.6,-0.2)-- (0,-0.7);
  
\draw[thick,xshift=-0.6in,rotate=90] (0,-0.7) -- (2,-0.7);
\draw[thick,xshift=-0.6in,rotate=90] (0,-0.7) -- (0.6,-0.2);
\draw[thick,xshift=-0.6in,rotate=90]  (0.6,-0.2)--(2.6,-0.2);
\draw[thick,xshift=-0.6in,rotate=90]  (2.6,-0.2)-- (2,-0.7);

  \shade[top color=dgreen!40, bottom color=dgreen!10] (0.3,1.05) -- (-0.8,1.05)-- (-0.8,1.55) -- (0.3,1.55)-- (0.3,1.05) ;
    \draw[thick] (-1.1,1.3) ellipse (0.125cm and 0.25cm);
         \draw[thick,blue,fill=dgreen!20] (0.3,1.3) ellipse (0.125cm and 0.25cm);


\node[below] at (-1, 0){$\cN_{p/N}$};

     \draw[thick,dgreen] (0.3,1.55)--(-0.8,1.55);
  \draw[thick,dgreen] (0.3,1.05)--(-0.8,1.05);
  \draw[thick,dotted,dgreen] (-0.8,1.55)--(-1.1,1.55);
  \draw[thick,dotted,dgreen] (-0.8,1.05)--(-1.1,1.05);

     \node[right] at (0.4,1.25) {$T W^{-{p\over N}}$};

\end{tikzpicture}~~~.
\eea
Note that the existence of these defects is independent of the coupling $e$---i.e. at any coupling, we have an infinite number of non-invertible defects labelled by  $\QQ$.\footnote{Contrast this with the case of Maxwell theory, where in each theory labelled by $\tau \in i \QQ$ with $\mathrm{Im}\,\tau>0$ and $\tau\ne i$, we had constructed one non-invertible defect. }

One important implication of this defect is that massless QED is \textit{natural}, i.e. we do not expect $m \overline \Psi \Psi$ to be produced radiatively. Indeed, if we add a mass $m \overline \Psi \Psi$, then we explicitly break the non-invertible symmetry.\footnote{Previous arguments for this relied on the invertible $U(1)_A$, which is itself anomalous, and no longer a symmetry in topologically non-trivial spacetimes.}  Another application of the non-invertible symmetry is that it can be used to prove helicity conservation in electron-positron scattering---for more on this, see \cite{Choi:2022jqy}. 

Before moving on to more realistic models, let us discuss a slightly more explicit construction of the above non-invertible defect. As we have already said, the defect involves an axial rotation by ${2 \pi p\over N}$, so we might imagine that it is constructed out of the axial current $\cN_{p\over N} \stackrel{?}{\sim} e^{{2 \pi i p\over N} \half \int * j_A}$.\footnote{The extra factor of $\half$ follows from the comments in footnote \ref{footnote:axial}.} But because of the ABJ anomaly, 
\bea 
d * j_A = {1\over 4 \pi^2 }f \wedge f~,
\eea
we see that $j_A$ is not conserved, and hence that $\cN$ defined in this manner would not be topological. 
On the other hand, moving the right-hand side to the left, we see that 
\bea 
d \left ( * j_A - {1\over 4 \pi^2 }a \wedge da \right) = 0~, 
\eea
and hence the operator 
\bea
\cN_{p \over N}(M_3) \stackrel{?}{\sim}  e^{{2 \pi i p\over N} \half  \int_{M_3} \left( * j_A - {1\over 4 \pi^2 }a \wedge da \right) }
\eea
is topological. In flat space, this operator is also gauge-invariant, and hence it is a well-defined generator for the invertible axial symmetry. But for more general spacetimes (or in the presence of magnetic monopoles), it is not gauge-invariant, due to the fractional Chern-Simons level $k = {p\over N}$. 

However, there is a well-known way to make this term gauge-invariant. In particular, for  $p=1$, we replace it with a \textit{fractional quantum Hall system},
\bea
\label{eq:fQHs}
{ N \over 4 \pi} c d c + {1 \over 2 \pi} c d a~,
\eea
where $c$ and $a$ are properly quantized gauge fields, which in particular means that $\oint dc , \oint da \in 2 \pi \ZZ$. Integrating out $c$ would naively give $c \stackrel{?}{=} - {a \over N}$, and plugging this back into (\ref{eq:fQHs}) would then reproduce $-{1/N \over 4 \pi} a da$. But in reality the equation $c \stackrel{?}{=} - {a \over N}$ is not quite correct, since it would imply that $c$ is not properly quantized. The correct, gauge invariant expression is the one in (\ref{eq:fQHs}). 

Thus, at the cost of introducing a new degree of freedom $c$ on $M_3$, we are able to construct a topological, gauge-invariant defect, 
\bea
\label{eq:masslesQEDdefectwv}
\cN_{1 \over N}(M_3) := \int \cD c \,\, e^{i \int_{M_3} \left( { \pi  \over N}  * j_A + { N \over 4 \pi} c d c + {1 \over 2 \pi} c d a  \right)}~.
\eea
For $N>1$, the new degree of freedom $c$ is the origin of the non-invertible nature of $\cN_{1 \over N}(M_3)$. 

To see that this is non-invertible, we may compute the fusion rules of this defect with its orientation reversal, following the steps in Section \ref{sec:gaugingwithanom}. Starting with,
\bea
\cN_{1 \over N}(M_3)  \times \cN_{1 \over N}(M_3)^\dagger = \int \cD c \cD \bar c\,\, e^{i \int_{M_3} { N \over 4 \pi} c dc - { N \over 4 \pi }\bar c d \bar c + {1 \over 2 \pi} (c - \bar c) d a}~
\eea
we change variables via $\bar c \rightarrow c - \widehat c$, integrate out $c$,
switch to homology notation, and note that the generators of the magnetic one-form symmetry take the form $L(\Sigma_2) = e^{{i \over N} \oint_{\Sigma_2} da}$ to obtain for even $N$
\bea
\cN_{1 \over N}(M_3)  \times \cN_{1 \over N}(M_3)^\dagger = {{1\over |H^0(M_3,\ZZ_N)|}} \sum_{\Sigma_2\, \in\, H_2(M_3, \ZZ_N)} e^{{2 \pi i \over N }Q(\Sigma_2)} L(\Sigma_2)~,
\eea
For even $N$, $Q(\Sigma_2)$ is defined by the integer-lift expression in (\ref{eq:tripleintdef}), now using $\omega_{\Sigma_2}$ as an integer lift of the mod-$N$ Poincar{\'e} dual. For odd $N$, the spin-theory condensate is untwisted and the exponential phase in this equation is omitted. The even-$N$ phase comes from the discrete torsion in gauging $\ZZ_N^{(1)}\subset U(1)_M^{(1)}$.

For $p=1$, we used a standard fractional quantum Hall system to define the defect $\cN_{1/N}$. A similar procedure may be carried out for more general $p$ such that $\mathrm{gcd}(p,N) = 1$. In that case, we couple to the so-called \textit{minimal $\ZZ_N^{(1)}$ TQFT} $\cA^{N,p}$, defined in \cite{Hsin:2018vcg}. 

\subsection{Massless QCD below the electroweak scale}

We now make our example slightly more realistic, and consider massless QCD below the electroweak scale, where $SU(2) \times U(1)_Y$ is broken to $U(1)_{EM}$ \cite{Cordova:2022ieu,Choi:2022jqy}.\footnote{We take the direct-product color and electromagnetic gauge group; in electron-charge units, $a/3$ is the properly quantized $U(1)$ connection. Thus the defects below need not be minimal.} The IR chiral Lagrangian takes the following form,
\bea
\cL_{IR} = {1\over 2} d \pi^0 \wedge * d \pi^0 +  g\, \pi^0 f \wedge f + \dots~,
\eea
where we have focused on the part that involves the neutral pion. What we will now show is that, using a non-invertible symmetry, we may predict the value of the $\pi^0 \rightarrow \gamma \gamma$ coupling $g$, and in particular prove that it must be non-zero. 

To see this, note that for massless quarks, the QCD Lagrangian is known to have the following symmetry,
\bea
U(1)_{A_3}: \hspace{0.2 in} \binom{u}{d} \rightarrow e^{{i \alpha \gamma_5 \sigma_3/2}} \binom{u}{d}~, 
\eea
whose corresponding classically conserved current is 
\bea
j_\mu^{A_3} = {1\over 2} \bar u \gamma_5 \gamma_\m u - {1\over 2} \bar d \gamma_5 \gamma_\mu d~. 
\eea
In fact, this symmetry has an ABJ anomaly of exactly the same type as before, and hence we may use precisely the same construction to obtain a non-invertible symmetry for every element of $U(1)_{A_3}$ labelled by $\alpha = 2 \pi p /N$. 
We now take $p=1$ and consider an insertion of this non-invertible defect at $x = 0$, giving
\bea
&\vphantom{.}& \int_{x < 0} \left(\half d \pi^0 \wedge * d \pi^0 + g \pi^0 f \wedge f \right) + \int_{x > 0} \left(\half d \pi^0 \wedge * d \pi^0 + g \pi^0 f \wedge f \right) 
\no\\
&\vphantom{.}& \hspace{0.8 in}+ \int_{x = 0} \left( {2\pi  \over N} * j^{A_3} + { N \over 4 \pi} c d c + {1 \over 2 \pi } c da\right) ~. 
\eea
Note that in the chiral Lagrangian, the axial current becomes simply
\bea
j^{A_3} =  -f_\pi d \pi^0 + \dots ~,
\eea
where $f_\pi \sim 92.4$ MeV is the pion decay constant, and corresponds to the symmetry shifting the neutral pion by $\pi^0 \rightarrow \pi^0 -\alpha f_\pi$.  We may then compute the following equations of motion,
\bea
&\pi^0:& \hspace{0.3 in} \pi^0 |_{x = 0^+} - \pi^0 |_{x = 0^-} = - {2 \pi \over N} f_\pi~,
\no\\
&c:& \hspace{0.3 in} N dc + f = 0~,
\no\\
&a:& \hspace{0.3 in} 2 g \left(\pi^0 |_{x = 0^+} - \pi^0 |_{x = 0^-}  \right) f = {1 \over 2 \pi } d c~. 
\eea
Plugging the first equation into the third gives 
\bea
- {4 \pi g f_\pi\over N}\,  f = {1 \over 2 \pi} dc~,
\eea
and then further using the second equation gives 
\bea
4 \pi g f_\pi = {1 \over 2 \pi} \hspace{0.2 in}\Rightarrow \hspace{0.2 in} g = {1 \over 8 \pi^2 f_\pi}~.
\eea
Thus, as promised, the presence of the non-invertible symmetry $\cN_{1\over N}$ in QCD tells us the coupling of the $\pi^0 f\wedge f$ term in the chiral Lagrangian, and hence the leading chiral prediction for the rate of $\pi^0 \rightarrow \gamma\gamma$, once small quark masses are restored. 

\subsection{Axion models} 

Let us now move on to some beyond the Standard Model (BSM) models. In particular, we consider axion-Maxwell theory \cite{Choi:2022fgx,Yokokura:2022alv,Choi:2023pdp}, 
\bea
\cL = {v^2 \over 2} d \phi \wedge * d \phi + {1\over 2 e^2 } f \wedge * f - { k \over 8 \pi^2} \phi f \wedge f~, 
\eea
where $\phi \sim \phi + 2 \pi$ is the axion field. We take $k\in\ZZ$ and work on spin manifolds. The theory of a compact scalar was studied previously in Section \ref{sec:higherform}, where it was shown that it has a 0-form and 2-form symmetry. In total then, we can try to define the following four currents \bea
j_\mathrm{shift}^{(1)} = v^2 d\phi~, \hspace{0.2 in} j_\mathrm{winding}^{(3)} = {1\over 2 \pi} * d \phi~, \hspace{0.2 in} j_E^{(2)} = {1 \over e^2} f ~, \hspace{0.2 in} j_M^{(2)} = {1\over 2 \pi} * f~, 
\eea
which satisfy the following 
\bea
d * j_\mathrm{shift}^{(1)} = -{k \over 8 \pi^2} f \wedge f~, \hspace{0.2 in} d * j_\mathrm{winding}^{(3)}= 0~,  \hspace{0.2 in} d *  j_E^{(2)} = {k \over 4 \pi^2} d \phi \wedge f~, \hspace{0.2in} d * j_M^{(2)} = 0~
\eea
using the equations of motion and the Bianchi identities. 
When $k=0$, we see that there are four conserved currents, and hence four symmetries,
\bea
U(1)_\mathrm{shift}^{(0)}~, \hspace{0.2 in} U(1)_\mathrm{wind.}^{(2)} ~, \hspace{0.2 in} U(1)_E^{(1)}~, \hspace{0.2 in}U(1)_M^{(1)}~,
\eea
which are the symmetries of the decoupled axion and Maxwell theories, but for non-zero $k$ the shift symmetry and electric one-form symmetries $U(1)_\mathrm{shift}^{(0)}$ and $U(1)_E^{(1)}$ are broken. 

For the shift symmetry $U(1)_\mathrm{shift}^{(0)}$, we see that the symmetry is broken by an ABJ-like term of the same form as the $U(1)_A$ symmetry in QED. As such, we can resurrect a $\QQ$ subset as non-invertible symmetries by stacking the naive defect constructed out of $* j_\mathrm{shift}^{(1)}$ with an appropriate fractional quantum Hall system. Alternatively, we can gauge a subgroup of $U(1)_M^{(1)}$ in half-space. Either way, we get a non-invertible zero-form symmetry whose defect takes the form analogous to (\ref{eq:masslesQEDdefectwv}), with the opposite Chern-Simons sign for $k>0$. 

Next we consider $U(1)_E^{(1)}$. As in the previous case, non-conservation of $*j_E^{(2)}$ means that $\cN^{(1)}_\alpha(M_2) \stackrel{?}{\sim} e^{i \alpha \oint_{M_2} * j_E^{(2)}}$ is no longer topological. We may attempt to obtain a topological defect by instead using the modified current,
 \bea
 \label{eq:DM2firstform}
 \cN^{(1)}_\alpha(M_2) \stackrel{?}{\sim} e^{i \alpha \oint_{M_2} \left( * j_E^{(2)} -{k \over 4 \pi^2} \phi \wedge f \right) }~,
 \eea
 but this term does not, in general, respect the identification $\phi\sim\phi+2\pi$. The solution, as before, is to couple the defect to new degrees of freedom which, when naively integrated out, reproduce the above non-gauge-invariant expression. 
The final result is that, for $\alpha = {2\pi p \over N}$ with $p$ and $N$ coprime, we may write
\bea
\cN^{(1)}_{p/N}(M_2) = \int \cD \lambda \cD c \, e^{i \oint_{M_2} \left({2 \pi p \over N e^2} * f + {N \over 2 \pi} \lambda dc + {pk \over 2 \pi} \phi d c + {1 \over 2 \pi } \lambda f \right) }~. 
\eea
Here $\lambda\sim\lambda+2\pi$ is a scalar and $c$ is a $U(1)$ gauge field on $M_2$.
\begin{tcolorbox}
\textbf{Exercise:} Check that naively integrating out $\lambda$ and $c$ reproduces the ill-defined defect in (\ref{eq:DM2firstform}). 
\end{tcolorbox}

\noindent
This is our first example of a non-invertible one-form symmetry in $(3+1)$d,\footnote{In $(2+1)$-dimensions, non-Abelian anyons give a wealth of examples. } combining two generalizations of the notion of symmetry. 
Note that for $k=1$, this defect may also be obtained by \textit{higher half-space gauging} of a $\ZZ_N^{(1)} \times \ZZ_N^{(2)}\subset U(1)_M^{(1)} \times U(1)_\mathrm{wind}^{(2)}$ on a codimension-1 slice of spacetime, as described in \cite{Choi:2022fgx}. 

We will now use this non-invertible symmetry to make some non-trivial physical claims.  As we have just said, the non-invertible symmetry $\cN^{(1)}_{p/N}(M_2)$ is constructed by gauging a subgroup of $U(1)_M^{(1)} \times U(1)_\mathrm{wind}^{(2)}$, and as such it forms an algebra with these symmetries. Thus if we break the $U(1)_M^{(1)} \times U(1)_\mathrm{wind}^{(2)}$ symmetry, we must also break the full non-invertible one-form symmetry. Likewise, breaking $U(1)_M^{(1)}$ breaks the full non-invertible zero-form symmetry. 

Now say that we try to UV complete axion-Maxwell theory, and upon doing so break some of the symmetries. If we call the scales of symmetry breaking $E_\mathrm{shift}, E_\mathrm{wind}, E_E,$ and $E_M$ (i.e. these are the scales below which the symmetries are emergent), then from the condensation structure explained above, we must have 
\bea
E_\mathrm{shift} \lesssim E_M~, \hspace{0.5 in} E_E \lesssim \mathrm{min} \{ E_\mathrm{wind}, E_M\}~.
\eea
In particular, one typically expects that $E_E \approx m_E$ and $E_M \approx m_M$ with $m_{E,M}$ the masses of the lightest electrically/magnetically charged particles and $E_\mathrm{wind} {\lesssim} \sqrt{T}$ with $T$ the tension of a dynamical axion string, from which we conclude that
\bea
m_E \lesssim m_M~, \hspace{0.5 in} m_E \lesssim \sqrt{T}~.
\eea
Under these assumptions, the lightest electrically charged excitation cannot be parametrically heavier than the lightest monopole. 

For a semiclassical illustration, consider the $SU(2)$ Georgi-Glashow model coupled to an axion. In this model, we have gauge bosons $g_{\m}^a$, an adjoint Higgs field $H$, and an axion $\phi$ coupled to the $SU(2)$ instanton density, with the Higgs being used to break down to the axion-Maxwell theory described above. In this case we have 
\bea
m_E \sim m_W \sim g v~, 
\eea
where $v^2 = \langle \mathrm{Tr}H^2 \rangle$ and $g$ is the $SU(2)$ coupling at the scale $v$. On the other hand, since the winding symmetry is preserved in this model, we have $E_\mathrm{wind} = \infty$. Finally, $m_M \sim {v \over g}$. Thus the above inequalities imply that
\bea
g v \lesssim { v \over g }\hspace{0.2 in}\Rightarrow \hspace{0.2 in} g \lesssim 1~,
\eea 
i.e. the semiclassical Higgs description is consistent with this hierarchy in the weak-coupling regime. This is sensible: when the theory is strongly coupled at the Higgsing scale, it is believed to confine instead of flowing to axion-Maxwell \cite{Brennan:2020ehu}. 

\subsection{Neutrino mass models}

One sharp signal of BSM physics is non-zero neutrino masses, 
\bea
|m_2^2 - m_1^2| \sim (10\,\mathrm{meV})^2~, \hspace{0.3 in} |m_3^2 - m_2^2| \sim (50\,\mathrm{meV})^2 ~, \hspace{0.3in} \sum_\nu m_\nu \leq 150\,\mathrm{meV}~. 
\eea
The bound on the sum is cosmology dependent. There are two standard ways to account for these masses,
\begin{itemize}
\item \textbf{Majorana mass:} In this case we use an effective dimension-5 operator (the \textit{Weinberg operator}), 
\bea
\cL \supset {1\over \Lambda} Y_{ij}^N ( H L_i)( H L_j) \hspace{0.3 in} \stackrel{\langle H \rangle \sim v}{\longrightarrow} \hspace{0.3 in}{v^2 \over \Lambda} Y_{ij}^N \nu_i \nu_j~.
\eea
\item \textbf{Dirac mass:} In this case we add a right-handed neutrino (whose left-handed conjugate we denote by $N$) and use the renormalizable interaction,
\bea
\cL \supset Y_{ij}^N H L_i N_j\hspace{0.3 in} \stackrel{\langle H \rangle \sim v}{\longrightarrow} \hspace{0.3 in} v\, Y_{ij}^N \nu_i N_j~. 
\eea
\end{itemize}
We will focus mainly on the latter case here. A troublesome feature is the tiny Yukawa couplings needed to reproduce experiment, i.e. $Y_\nu / Y_\tau \lesssim 10^{-11}$. 
In the current section, we briefly explain how non-invertible symmetries can be used to make this result natural \cite{Cordova:2022fhg}. 

The key fact is that, in order to break a higher-form symmetry in an EFT, we cannot just modify the action by adding higher-derivative terms in the original fields, since local operators are not charged under the higher-form symmetry. 
Instead, the breaking of higher-form symmetries must involve the introduction of new degrees of freedom. 

Now say that we have a $U(1)^{(1)}_M$ symmetry in our theory. Breaking this symmetry requires finite action configurations carrying magnetic charge. Let us assume that this happens via monopoles. The mass and cutoff in this case are given by 
\bea
m_M \sim {v \over g}~, \hspace{0.5 in} \Lambda \sim g v~, 
\eea
with $v$ some e.g. Higgsing scale that determines the monopole mass. 
The quantum vacuum contains such monopoles which propagate for proper time $\delta t \sim 1/\Lambda$, and hence give rise to terms in the quantum effective action of the scale 
\bea
\delta \cL \sim e^{- S_\mathrm{mon}}\sim e^{- m_M \delta t}  \sim e^{- \# / g^2}~. 
\eea
We thus see that there is a natural way to violate magnetic one-form symmetries non-perturbatively, i.e. by exponentially small amounts. 

In situations in which there is a non-invertible zero-form symmetry coming from half-space gauging of the magnetic one-form symmetry, there is consequently also a natural way to break the non-invertible zero-form symmetry by exponentially small amounts, since breaking the one-form symmetry also breaks the non-invertible symmetry. The goal is then to build models in which the neutrino masses explicitly break such a non-invertible symmetry. This would mean that having $m_\nu= 0$ would be natural as long as the non-invertible symmetry is preserved, but if the non-invertible symmetry is violated by $e^{- \# / g^2}$, as we have described above, then we would expect $m_\nu \sim e^{- \# / g^2}$ as well. This gives a mechanism for having exponentially small neutrino masses. 

We now describe one particular model with this behavior, in the context of the Dirac mass case; other models can be found in  \cite{Cordova:2022fhg} and follow-up works. 
We begin by coupling the Standard Model to right-handed neutrinos and gauging $U(1)_{\mu - \tau}$. The charges of the lepton and Higgs sectors under the gauge groups (ignoring color) are shown in Table \ref{table:neutrinos}. Note that we have also included the charge under the global symmetry $U(1)_{\widetilde L}$, corresponding to Standard Model lepton number. 

 \begin{table}[tp]

\begin{center}
\begin{tabular}{c|ccccc}
& $SU(2)_L$ & $U(1)_Y$ & $U(1)_{\mu - \tau}$ & $U(1)_{\widetilde L}$
\\\hline
$L_e$ & $2$  & $-1$ & $0 $& $1$
\\
$L_ \mu$ & $2$ & $-1$ & $1 $& $1$
\\
$L_\tau$ & $2$ & $-1$ & $-1$ & $1$
\\\hline
$\bar e_e$  & $1 $& $2$ & $0 $& $-1 $
\\
$\bar e_\mu$  & $1$ &$ 2$ &${-1}$ & $-1$ 
\\
$\bar e_\tau$  & $1$ & $2$ & ${+1}$ & $-1 $
\\\hline
$N_e$ & $1$ & $0$  & $0$&$ 0$  
\\
$N_\m$ &$1$ & $0$  & ${-1}$& $0$ 
\\
$N_\tau$ &$1$ & $0$  & ${+1}$& $0 $
\\\hline
$H$&$ 2 $& ${+1}$ &$ 0$ & $0$
\end{tabular}
\end{center}
\caption{Charges of lepton and Higgs sectors under the gauge groups in this model, as well as the global $U(1)_{\widetilde L}$ symmetry corresponding to Standard Model lepton number, which is not conserved quantum mechanically.}
\label{table:neutrinos}
\end{table}
 
 In fact, it is known that in the Standard Model, an anomaly with $SU(2)_L$ breaks $U(1)_{\widetilde L}$  to a non-anomalous $\ZZ_3^{\widetilde L}$ subgroup. This subgroup prevents generation of the Dirac mass term $H L N$, and hence the presence of $\ZZ_3^{\widetilde L}$ ensures that having massless neutrinos is natural.

 However, in the current model we are also gauging $U(1)_{\mu - \tau}$. It turns out that $\ZZ_3^{\widetilde L}$ has yet another ABJ anomaly with $U(1)_{\mu - \tau}$, which this time breaks it altogether. However, as described in previous subsections, it is still possible to salvage $\ZZ_3^{\widetilde L}$ as a \textit{non-invertible symmetry}, and once again, it turns out that this non-invertible symmetry can be obtained by half-space gauging of a magnetic one-form symmetry. In a UV completion where monopole effects break the magnetic one-form symmetry and the associated instanton effects are the only source of violation of the protecting non-invertible symmetry, exponentially suppressed Dirac masses can therefore be generated.

\subsection*{Further references}

Early works on non-invertible symmetries in higher dimensions include \cite{Nguyen:2021yld,Heidenreich:2021xpr,Koide:2021zxj,Kaidi:2021xfk,Choi:2021kmx}, and there have now  been countless follow-ups, including but not limited to \cite{Choi:2022zal,Apruzzi:2021nmk,Arias-Tamargo:2022nlf,Hayashi:2022fkw,Roumpedakis:2022aik,Bhardwaj:2022yxj,Kaidi:2022uux,Choi:2022jqy,Cordova:2022ieu,Antinucci:2022eat,Bashmakov:2022jtl,Damia:2022seq,Damia:2022bcd,Choi:2022rfe,Lu:2022ver,Bhardwaj:2022lsg,Bartsch:2022mpm,Lin:2022xod,Apruzzi:2022rei,GarciaEtxebarria:2022vzq, Benini:2022hzx, Wang:2021vki, Chen:2021xuc, DelZotto:2022ras,Bhardwaj:2022dyt,Brennan:2022tyl,Delmastro:2022pfo,Heckman:2022muc,Freed:2022qnc,Niro:2022ctq,Kaidi:2022cpf,Mekareeya:2022spm,vanBeest:2022fss,Antinucci:2022vyk,Chen:2022cyw,Damia:2023ses,Copetti:2023mcq,Argurio:2023lwl,vanBeest:2023dbu,vanBeest:2023mbs,Lawrie:2023tdz,Chen:2023czk,Nardoni:2024sos,Kan:2024fuu}.

Relations between higher-dimensional non-invertible symmetries and M-theory/string theory have been explored in \cite{Apruzzi:2021nmk,GarciaEtxebarria:2022vzq,Apruzzi:2022rei,Heckman:2022muc,vanBeest:2022fss,Antinucci:2022vyk,Fernandez-Melgarejo:2024ffg,DeMarco:2025pza}, while the role played by non-invertible symmetries on string worldsheets has been studied in \cite{Kaidi:2024wio,Heckman:2024obe}, with some recent phenomenological applications discussed in \cite{Kobayashi:2024yqq,Kobayashi:2024cvp,Kobayashi:2025znw,Suzuki:2025oov,Kobayashi:2025cwx,Kobayashi:2025lar,Dong:2025jra,Suzuki:2025bxg,Kobayashi:2025ocp,Suzuki:2025kxz,Nakai:2025thw}.  
Other applications to phenomenology or amplitudes include \cite{Cordova:2022ieu,Choi:2022jqy,Choi:2022fgx,Yokokura:2022alv,Cordova:2022fhg,Choi:2023pdp,Cordova:2024ypu,Copetti:2024dcz,Chen:2024tsx,Liang:2025dkm,Choi:2025vxr}. 
Finally, lattice realizations in various dimensions have been discussed in e.g. \cite{Koide:2021zxj,Honda:2024yte, Ando:2024nlk,Honda:2024sdz,Bhardwaj:2024kvy,Honda:2024xmk,Choi:2024rjm,Lu:2024ytl,Cao:2024qjj,Honda:2024kvf,Katayama:2025pmz,Kawana:2025vbi,Fidkowski:2025rsq,Su:2025aag}.

\newpage

\bibliographystyle{JHEP}
\bibliography{bib.bib}

\providecommand{\href}[2]{#2}\begingroup\raggedright\begin{thebibliography}{100}

\bibitem{McGreevy:2022oyu}
J.~McGreevy, \emph{{Generalized Symmetries in Condensed Matter}},
  \href{http://dx.doi.org/10.1146/annurev-conmatphys-040721-021029}{\emph{Ann.
  Rev. Condensed Matter Phys.} {\bf 14} (2023) 57--82},
  [\href{http://arxiv.org/abs/2204.03045}{{\tt 2204.03045}}].

\bibitem{Freed:2022iao}
D.~S. Freed, \emph{{Introduction to topological symmetry in QFT.}},
  \href{http://dx.doi.org/10.1090/pspum/107/01946}{\emph{Proc. Symp. Pure
  Math.} {\bf 107} (2024) 93--106},
  [\href{http://arxiv.org/abs/2212.00195}{{\tt 2212.00195}}].

\bibitem{Shao:2023gho}
S.-H. Shao, \emph{{What's Done Cannot Be Undone: TASI Lectures on
  Non-Invertible Symmetries}},  \href{http://arxiv.org/abs/2308.00747}{{\tt
  2308.00747}}.

\bibitem{Brennan:2023mmt}
T.~D. Brennan and S.~Hong, \emph{{Introduction to Generalized Global Symmetries
  in QFT and Particle Physics}},  \href{http://arxiv.org/abs/2306.00912}{{\tt
  2306.00912}}.

\bibitem{Bhardwaj:2023kri}
L.~Bhardwaj, L.~E. Bottini, L.~Fraser-Taliente, L.~Gladden, D.~S.~W. Gould,
  A.~Platschorre et~al., \emph{{Lectures on generalized symmetries}},
  \href{http://dx.doi.org/10.1016/j.physrep.2023.11.002}{\emph{Phys. Rept.}
  {\bf 1051} (2024) 1--87}, [\href{http://arxiv.org/abs/2307.07547}{{\tt
  2307.07547}}].

\bibitem{Gomes:2023ahz}
P.~R.~S. Gomes, \emph{{An introduction to higher-form symmetries}},
  \href{http://dx.doi.org/10.21468/SciPostPhysLectNotes.74}{\emph{SciPost Phys.
  Lect. Notes} {\bf 74} (2023) 1}, [\href{http://arxiv.org/abs/2303.01817}{{\tt
  2303.01817}}].

\bibitem{Schafer-Nameki:2023jdn}
S.~Schafer-Nameki, \emph{{ICTP lectures on (non-)invertible generalized
  symmetries}},
  \href{http://dx.doi.org/10.1016/j.physrep.2024.01.007}{\emph{Phys. Rept.}
  {\bf 1063} (2024) 1--55}, [\href{http://arxiv.org/abs/2305.18296}{{\tt
  2305.18296}}].

\bibitem{Luo:2023ive}
R.~Luo, Q.-R. Wang and Y.-N. Wang, \emph{{Lecture notes on generalized
  symmetries and applications}},
  \href{http://dx.doi.org/10.1016/j.physrep.2024.02.002}{\emph{Phys. Rept.}
  {\bf 1065} (2024) 1--43}, [\href{http://arxiv.org/abs/2307.09215}{{\tt
  2307.09215}}].

\bibitem{Seiberg:2019vrp}
N.~Seiberg, \emph{{Field Theories With a Vector Global Symmetry}},
  \href{http://dx.doi.org/10.21468/SciPostPhys.8.4.050}{\emph{SciPost Phys.}
  {\bf 8} (2020) 050}, [\href{http://arxiv.org/abs/1909.10544}{{\tt
  1909.10544}}].

\bibitem{TachikawaTASI}
Y.~Tachikawa, ``{TASI lectures on anomalies and topological phases}.''
  \url{https://member.ipmu.jp/yuji.tachikawa/lectures/2019-top-anom/}.

\bibitem{Freed:2014iua}
D.~S. Freed, \emph{{Anomalies and Invertible Field Theories}},
  \href{http://dx.doi.org/10.1090/pspum/088/01462}{\emph{Proc. Symp. Pure
  Math.} {\bf 88} (2014) 25--46}, [\href{http://arxiv.org/abs/1404.7224}{{\tt
  1404.7224}}].

\bibitem{Freed:2016rqq}
D.~S. Freed and M.~J. Hopkins, \emph{{Reflection positivity and invertible
  topological phases}},
  \href{http://dx.doi.org/10.2140/gt.2021.25.1165}{\emph{Geom. Topol.} {\bf 25}
  (2021) 1165--1330}, [\href{http://arxiv.org/abs/1604.06527}{{\tt
  1604.06527}}].

\bibitem{Monnier:2019ytc}
S.~Monnier, \emph{{A Modern Point of View on Anomalies}},
  \href{http://dx.doi.org/10.1002/prop.201910012}{\emph{Fortsch. Phys.} {\bf
  67} (2019) 1910012}, [\href{http://arxiv.org/abs/1903.02828}{{\tt
  1903.02828}}].

\bibitem{Freed:2012bs}
D.~S. Freed and C.~Teleman, \emph{{Relative quantum field theory}},
  \href{http://dx.doi.org/10.1007/s00220-013-1880-1}{\emph{Commun. Math. Phys.}
  {\bf 326} (2014) 459--476}, [\href{http://arxiv.org/abs/1212.1692}{{\tt
  1212.1692}}].

\bibitem{Faddeev:1984ung}
L.~D. Faddeev and S.~L. Shatashvili, \emph{{Algebraic and Hamiltonian Methods
  in the Theory of Nonabelian Anomalies}},
  \href{http://dx.doi.org/10.1007/BF01018976}{\emph{Teor. Mat. Fiz.} {\bf 60}
  (1984) 206--217}.

\bibitem{Callan:1984sa}
C.~G. Callan, Jr. and J.~A. Harvey, \emph{{Anomalies and Fermion Zero Modes on
  Strings and Domain Walls}},
  \href{http://dx.doi.org/10.1016/0550-3213(85)90489-4}{\emph{Nucl. Phys. B}
  {\bf 250} (1985) 427--436}.

\bibitem{Kapustin:2014gua}
A.~Kapustin and N.~Seiberg, \emph{{Coupling a QFT to a TQFT and Duality}},
  \href{http://dx.doi.org/10.1007/JHEP04(2014)001}{\emph{JHEP} {\bf 04} (2014)
  001}, [\href{http://arxiv.org/abs/1401.0740}{{\tt 1401.0740}}].

\bibitem{Vafa:1989ih}
C.~Vafa, \emph{{Quantum Symmetries of String Vacua}},
  \href{http://dx.doi.org/10.1142/S0217732389001842}{\emph{Mod. Phys. Lett. A}
  {\bf 4} (1989) 1615}.

\bibitem{Baez:2010ya}
J.~C. Baez and J.~Huerta, \emph{{An Invitation to Higher Gauge Theory}},
  \href{http://dx.doi.org/10.1007/s10714-010-1070-9}{\emph{Gen. Rel. Grav.}
  {\bf 43} (2011) 2335--2392}, [\href{http://arxiv.org/abs/1003.4485}{{\tt
  1003.4485}}].

\bibitem{Cordova:2018cvg}
C.~C\'ordova, T.~T. Dumitrescu and K.~Intriligator, \emph{{Exploring 2-Group
  Global Symmetries}},
  \href{http://dx.doi.org/10.1007/JHEP02(2019)184}{\emph{JHEP} {\bf 02} (2019)
  184}, [\href{http://arxiv.org/abs/1802.04790}{{\tt 1802.04790}}].

\bibitem{Benini:2018reh}
F.~Benini, C.~C\'ordova and P.-S. Hsin, \emph{{On 2-Group Global Symmetries and
  their Anomalies}},
  \href{http://dx.doi.org/10.1007/JHEP03(2019)118}{\emph{JHEP} {\bf 03} (2019)
  118}, [\href{http://arxiv.org/abs/1803.09336}{{\tt 1803.09336}}].

\bibitem{Frohlich:2009gb}
J.~Frohlich, J.~Fuchs, I.~Runkel and C.~Schweigert, \emph{{Defect lines,
  dualities, and generalised orbifolds}},  in \emph{{16th International
  Congress on Mathematical Physics}}, 9, 2009.
\newblock \href{http://arxiv.org/abs/0909.5013}{{\tt 0909.5013}}.
\newblock \href{http://dx.doi.org/10.1142/9789814304634_0056}{DOI}.

\bibitem{Gaiotto:2014kfa}
D.~Gaiotto, A.~Kapustin, N.~Seiberg and B.~Willett, \emph{{Generalized Global
  Symmetries}}, \href{http://dx.doi.org/10.1007/JHEP02(2015)172}{\emph{JHEP}
  {\bf 02} (2015) 172}, [\href{http://arxiv.org/abs/1412.5148}{{\tt
  1412.5148}}].

\bibitem{Batista:2004sc}
C.~D. Batista and Z.~Nussinov, \emph{{Generalized Elitzur's theorem and
  dimensional reduction}},
  \href{http://dx.doi.org/10.1103/PhysRevB.72.045137}{\emph{Phys. Rev. B} {\bf
  72} (2005) 045137}, [\href{http://arxiv.org/abs/cond-mat/0410599}{{\tt
  cond-mat/0410599}}].

\bibitem{Nussinov:2006iva}
Z.~Nussinov and G.~Ortiz, \emph{{Sufficient symmetry conditions for Topological
  Quantum Order}}, \href{http://dx.doi.org/10.1073/pnas.0803726105}{\emph{Proc.
  Nat. Acad. Sci.} {\bf 106} (2009) 16944--16949},
  [\href{http://arxiv.org/abs/cond-mat/0605316}{{\tt cond-mat/0605316}}].

\bibitem{Nussinov:2009zz}
Z.~Nussinov and G.~Ortiz, \emph{{A symmetry principle for topological quantum
  order}}, \href{http://dx.doi.org/10.1016/j.aop.2008.11.002}{\emph{Annals
  Phys.} {\bf 324} (2009) 977--1057},
  [\href{http://arxiv.org/abs/cond-mat/0702377}{{\tt cond-mat/0702377}}].

\bibitem{Gaiotto:2017yup}
D.~Gaiotto, A.~Kapustin, Z.~Komargodski and N.~Seiberg, \emph{{Theta, Time
  Reversal, and Temperature}},
  \href{http://dx.doi.org/10.1007/JHEP05(2017)091}{\emph{JHEP} {\bf 05} (2017)
  091}, [\href{http://arxiv.org/abs/1703.00501}{{\tt 1703.00501}}].

\bibitem{Cobanera:2011wn}
E.~Cobanera, G.~Ortiz and Z.~Nussinov, \emph{{The Bond-Algebraic Approach to
  Dualities}}, \href{http://dx.doi.org/10.1080/00018732.2011.619814}{\emph{Adv.
  Phys.} {\bf 60} (2011) 679--798}, [\href{http://arxiv.org/abs/1103.2776}{{\tt
  1103.2776}}].

\bibitem{DelZotto:2015isa}
M.~Del~Zotto, J.~J. Heckman, D.~S. Park and T.~Rudelius, \emph{{On the Defect
  Group of a 6D SCFT}},
  \href{http://dx.doi.org/10.1007/s11005-016-0839-5}{\emph{Lett. Math. Phys.}
  {\bf 106} (2016) 765--786}, [\href{http://arxiv.org/abs/1503.04806}{{\tt
  1503.04806}}].

\bibitem{Eckhard:2019jgg}
J.~Eckhard, H.~Kim, S.~Schafer-Nameki and B.~Willett, \emph{{Higher-Form
  Symmetries, Bethe Vacua, and the 3d-3d Correspondence}},
  \href{http://dx.doi.org/10.1007/JHEP01(2020)101}{\emph{JHEP} {\bf 01} (2020)
  101}, [\href{http://arxiv.org/abs/1910.14086}{{\tt 1910.14086}}].

\bibitem{Bergman:2020ifi}
O.~Bergman, Y.~Tachikawa and G.~Zafrir, \emph{{Generalized symmetries and
  holography in ABJM-type theories}},
  \href{http://dx.doi.org/10.1007/JHEP07(2020)077}{\emph{JHEP} {\bf 07} (2020)
  077}, [\href{http://arxiv.org/abs/2004.05350}{{\tt 2004.05350}}].

\bibitem{Morrison:2020ool}
D.~R. Morrison, S.~Schafer-Nameki and B.~Willett, \emph{{Higher-Form Symmetries
  in 5d}}, \href{http://dx.doi.org/10.1007/JHEP09(2020)024}{\emph{JHEP} {\bf
  09} (2020) 024}, [\href{http://arxiv.org/abs/2005.12296}{{\tt 2005.12296}}].

\bibitem{DelZotto:2020sop}
M.~Del~Zotto and K.~Ohmori, \emph{{2-Group Symmetries of 6D Little String
  Theories and T-Duality}},
  \href{http://dx.doi.org/10.1007/s00023-021-01018-3}{\emph{Annales Henri
  Poincare} {\bf 22} (2021) 2451--2474},
  [\href{http://arxiv.org/abs/2009.03489}{{\tt 2009.03489}}].

\bibitem{Albertini:2020mdx}
F.~Albertini, M.~Del~Zotto, I.~n. Garc\'\i{}a~Etxebarria and S.~S. Hosseini,
  \emph{{Higher Form Symmetries and M-theory}},
  \href{http://dx.doi.org/10.1007/JHEP12(2020)203}{\emph{JHEP} {\bf 12} (2020)
  203}, [\href{http://arxiv.org/abs/2005.12831}{{\tt 2005.12831}}].

\bibitem{Bah:2020uev}
I.~Bah, F.~Bonetti and R.~Minasian, \emph{{Discrete and higher-form symmetries
  in SCFTs from wrapped M5-branes}},
  \href{http://dx.doi.org/10.1007/JHEP03(2021)196}{\emph{JHEP} {\bf 03} (2021)
  196}, [\href{http://arxiv.org/abs/2007.15003}{{\tt 2007.15003}}].

\bibitem{DelZotto:2020esg}
M.~Del~Zotto, I.~n. Garc\'\i{}a~Etxebarria and S.~S. Hosseini, \emph{{Higher
  form symmetries of Argyres-Douglas theories}},
  \href{http://dx.doi.org/10.1007/JHEP10(2020)056}{\emph{JHEP} {\bf 10} (2020)
  056}, [\href{http://arxiv.org/abs/2007.15603}{{\tt 2007.15603}}].

\bibitem{Bhardwaj:2020phs}
L.~Bhardwaj and S.~Sch\"afer-Nameki, \emph{{Higher-form symmetries of 6d and 5d
  theories}}, \href{http://dx.doi.org/10.1007/JHEP02(2021)159}{\emph{JHEP} {\bf
  02} (2021) 159}, [\href{http://arxiv.org/abs/2008.09600}{{\tt 2008.09600}}].

\bibitem{Apruzzi:2020zot}
F.~Apruzzi, M.~Dierigl and L.~Lin, \emph{{The Fate of Discrete 1-Form
  Symmetries in 6d}},
  \href{http://dx.doi.org/10.21468/SciPostPhys.12.2.047}{\emph{SciPost Phys.}
  {\bf 12} (2022) 047}, [\href{http://arxiv.org/abs/2008.09117}{{\tt
  2008.09117}}].

\bibitem{BenettiGenolini:2020doj}
P.~Benetti~Genolini and L.~Tizzano, \emph{{Instantons, symmetries and anomalies
  in five dimensions}},
  \href{http://dx.doi.org/10.1007/JHEP04(2021)188}{\emph{JHEP} {\bf 04} (2021)
  188}, [\href{http://arxiv.org/abs/2009.07873}{{\tt 2009.07873}}].

\bibitem{Gukov:2020btk}
S.~Gukov, P.-S. Hsin and D.~Pei, \emph{{Generalized global symmetries of $T[M]$
  theories. Part I}},
  \href{http://dx.doi.org/10.1007/JHEP04(2021)232}{\emph{JHEP} {\bf 04} (2021)
  232}, [\href{http://arxiv.org/abs/2010.15890}{{\tt 2010.15890}}].

\bibitem{Closset:2020scj}
C.~Closset, S.~Schafer-Nameki and Y.-N. Wang, \emph{{Coulomb and Higgs Branches
  from Canonical Singularities: Part 0}},
  \href{http://dx.doi.org/10.1007/JHEP02(2021)003}{\emph{JHEP} {\bf 02} (2021)
  003}, [\href{http://arxiv.org/abs/2007.15600}{{\tt 2007.15600}}].

\bibitem{Closset:2020afy}
C.~Closset, S.~Giacomelli, S.~Schafer-Nameki and Y.-N. Wang, \emph{{5d and 4d
  SCFTs: Canonical Singularities, Trinions and S-Dualities}},
  \href{http://dx.doi.org/10.1007/JHEP05(2021)274}{\emph{JHEP} {\bf 05} (2021)
  274}, [\href{http://arxiv.org/abs/2012.12827}{{\tt 2012.12827}}].

\bibitem{Apruzzi:2021phx}
F.~Apruzzi, M.~van Beest, D.~S.~W. Gould and S.~Sch\"afer-Nameki,
  \emph{{Holography, 1-form symmetries, and confinement}},
  \href{http://dx.doi.org/10.1103/PhysRevD.104.066005}{\emph{Phys. Rev. D} {\bf
  104} (2021) 066005}, [\href{http://arxiv.org/abs/2104.12764}{{\tt
  2104.12764}}].

\bibitem{Apruzzi:2021vcu}
F.~Apruzzi, S.~Schafer-Nameki, L.~Bhardwaj and J.~Oh, \emph{{The Global Form of
  Flavor Symmetries and 2-Group Symmetries in 5d SCFTs}},
  \href{http://dx.doi.org/10.21468/SciPostPhys.13.2.024}{\emph{SciPost Phys.}
  {\bf 13} (2022) 024}, [\href{http://arxiv.org/abs/2105.08724}{{\tt
  2105.08724}}].

\bibitem{Hosseini:2021ged}
S.~S. Hosseini and R.~Moscrop, \emph{{Maruyoshi-Song flows and defect groups of
  $ {\mathrm{D}}_{\mathrm{p}}^{\mathrm{b}} $(G) theories}},
  \href{http://dx.doi.org/10.1007/JHEP10(2021)119}{\emph{JHEP} {\bf 10} (2021)
  119}, [\href{http://arxiv.org/abs/2106.03878}{{\tt 2106.03878}}].

\bibitem{Cvetic:2021sxm}
M.~Cvetic, M.~Dierigl, L.~Lin and H.~Y. Zhang, \emph{{Higher-form symmetries
  and their anomalies in M-/F-theory duality}},
  \href{http://dx.doi.org/10.1103/PhysRevD.104.126019}{\emph{Phys. Rev. D} {\bf
  104} (2021) 126019}, [\href{http://arxiv.org/abs/2106.07654}{{\tt
  2106.07654}}].

\bibitem{Buican:2021xhs}
M.~Buican and H.~Jiang, \emph{{1-form symmetry, isolated $ \mathcal{N} $ = 2
  SCFTs, and Calabi-Yau threefolds}},
  \href{http://dx.doi.org/10.1007/JHEP12(2021)024}{\emph{JHEP} {\bf 12} (2021)
  024}, [\href{http://arxiv.org/abs/2106.09807}{{\tt 2106.09807}}].

\bibitem{Iqbal:2021rkn}
N.~Iqbal and J.~McGreevy, \emph{{Mean string field theory: Landau-Ginzburg
  theory for 1-form symmetries}},
  \href{http://dx.doi.org/10.21468/SciPostPhys.13.5.114}{\emph{SciPost Phys.}
  {\bf 13} (2022) 114}, [\href{http://arxiv.org/abs/2106.12610}{{\tt
  2106.12610}}].

\bibitem{Braun:2021sex}
A.~P. Braun, M.~Larfors and P.-K. Oehlmann, \emph{{Gauged 2-form symmetries in
  6D SCFTs coupled to gravity}},
  \href{http://dx.doi.org/10.1007/JHEP12(2021)132}{\emph{JHEP} {\bf 12} (2021)
  132}, [\href{http://arxiv.org/abs/2106.13198}{{\tt 2106.13198}}].

\bibitem{Cvetic:2021maf}
M.~Cvetic, J.~J. Heckman, E.~Torres and G.~Zoccarato, \emph{{Reflections on the
  matter of 3D N=1 vacua and local Spin(7) compactifications}},
  \href{http://dx.doi.org/10.1103/PhysRevD.105.026008}{\emph{Phys. Rev. D} {\bf
  105} (2022) 026008}, [\href{http://arxiv.org/abs/2107.00025}{{\tt
  2107.00025}}].

\bibitem{Closset:2021lhd}
C.~Closset and H.~Magureanu, \emph{{The $U$-plane of rank-one 4d
  $\mathcal{N}=2$ KK theories}},
  \href{http://dx.doi.org/10.21468/SciPostPhys.12.2.065}{\emph{SciPost Phys.}
  {\bf 12} (2022) 065}, [\href{http://arxiv.org/abs/2107.03509}{{\tt
  2107.03509}}].

\bibitem{Lee:2021obi}
Y.~Lee and Y.~Zheng, \emph{{Remarks on compatibility between conformal symmetry
  and continuous higher-form symmetries}},
  \href{http://dx.doi.org/10.1103/PhysRevD.104.085005}{\emph{Phys. Rev. D} {\bf
  104} (2021) 085005}, [\href{http://arxiv.org/abs/2108.00732}{{\tt
  2108.00732}}].

\bibitem{Lee:2021crt}
Y.~Lee, K.~Ohmori and Y.~Tachikawa, \emph{{Matching higher symmetries across
  Intriligator-Seiberg duality}},
  \href{http://dx.doi.org/10.1007/JHEP10(2021)114}{\emph{JHEP} {\bf 10} (2021)
  114}, [\href{http://arxiv.org/abs/2108.05369}{{\tt 2108.05369}}].

\bibitem{Bah:2021brs}
I.~Bah, F.~Bonetti, E.~Leung and P.~Weck, \emph{{M5-branes probing flux
  backgrounds}}, \href{http://dx.doi.org/10.1007/JHEP10(2022)122}{\emph{JHEP}
  {\bf 10} (2022) 122}, [\href{http://arxiv.org/abs/2111.01790}{{\tt
  2111.01790}}].

\bibitem{Closset:2021lwy}
C.~Closset, S.~Sch{\"a}fer-Nameki and Y.-N. Wang, \emph{{Coulomb and Higgs
  branches from canonical singularities. Part I. Hypersurfaces with smooth
  Calabi-Yau resolutions}},
  \href{http://dx.doi.org/10.1007/JHEP04(2022)061}{\emph{JHEP} {\bf 04} (2022)
  061}, [\href{http://arxiv.org/abs/2111.13564}{{\tt 2111.13564}}].

\bibitem{DelZotto:2022fnw}
M.~Del~Zotto, J.~J. Heckman, S.~N. Meynet, R.~Moscrop and H.~Y. Zhang,
  \emph{{Higher symmetries of 5D orbifold SCFTs}},
  \href{http://dx.doi.org/10.1103/PhysRevD.106.046010}{\emph{Phys. Rev. D} {\bf
  106} (2022) 046010}, [\href{http://arxiv.org/abs/2201.08372}{{\tt
  2201.08372}}].

\bibitem{Cvetic:2022uuu}
M.~Cveti{\v{c}}, M.~Dierigl, L.~Lin and H.~Y. Zhang, \emph{{All eight- and
  nine-dimensional string vacua from junctions}},
  \href{http://dx.doi.org/10.1103/PhysRevD.106.026007}{\emph{Phys. Rev. D} {\bf
  106} (2022) 026007}, [\href{http://arxiv.org/abs/2203.03644}{{\tt
  2203.03644}}].

\bibitem{Beratto:2021xmn}
E.~Beratto, N.~Mekareeya and M.~Sacchi, \emph{{Zero-form and one-form
  symmetries of the ABJ and related theories}},
  \href{http://dx.doi.org/10.1007/JHEP04(2022)126}{\emph{JHEP} {\bf 04} (2022)
  126}, [\href{http://arxiv.org/abs/2112.09531}{{\tt 2112.09531}}].

\bibitem{Cvetic:2023pgm}
M.~Cveti{\v{c}}, J.~J. Heckman, M.~H{\"u}bner and E.~Torres, \emph{{Generalized
  symmetries, gravity, and the swampland}},
  \href{http://dx.doi.org/10.1103/PhysRevD.109.026012}{\emph{Phys. Rev. D} {\bf
  109} (2024) 026012}, [\href{http://arxiv.org/abs/2307.13027}{{\tt
  2307.13027}}].

\bibitem{Anber:2024gis}
M.~M. Anber and S.~Y.~L. Chan, \emph{{Global aspects of 3-form gauge theory:
  implications for axion-Yang-Mills systems}},
  \href{http://dx.doi.org/10.1007/JHEP10(2024)113}{\emph{JHEP} {\bf 10} (2024)
  113}, [\href{http://arxiv.org/abs/2407.03416}{{\tt 2407.03416}}].

\bibitem{Balasubramanian:2024nei}
M.~Balasubramanian, M.~Buican and R.~Radhakrishnan, \emph{{On the
  Classification of Bosonic and Fermionic One-Form Symmetries in $2+1$d and
  {\textquoteright}t Hooft Anomaly Matching}},
  \href{http://dx.doi.org/10.1007/s00220-025-05494-0}{\emph{Commun. Math.
  Phys.} {\bf 406} (2025) 319}, [\href{http://arxiv.org/abs/2408.00866}{{\tt
  2408.00866}}].

\bibitem{Berean-Dutcher:2025ohp}
J.~Berean-Dutcher, M.~Derda and J.~Parra-Martinez, \emph{{Soft Theorems from
  Higher Symmetries}},  \href{http://arxiv.org/abs/2505.03566}{{\tt
  2505.03566}}.

\bibitem{Cordova:2020tij}
C.~C\'ordova, T.~T. Dumitrescu and K.~Intriligator, \emph{{2-Group Global
  Symmetries and Anomalies in Six-Dimensional Quantum Field Theories}},
  \href{http://dx.doi.org/10.1007/JHEP04(2021)252}{\emph{JHEP} {\bf 04} (2021)
  252}, [\href{http://arxiv.org/abs/2009.00138}{{\tt 2009.00138}}].

\bibitem{DeWolfe:2020uzb}
O.~DeWolfe and K.~Higginbotham, \emph{{Generalized symmetries and 2-groups via
  electromagnetic duality in $AdS/CFT$}},
  \href{http://dx.doi.org/10.1103/PhysRevD.103.026011}{\emph{Phys. Rev. D} {\bf
  103} (2021) 026011}, [\href{http://arxiv.org/abs/2010.06594}{{\tt
  2010.06594}}].

\bibitem{Iqbal:2020lrt}
N.~Iqbal and N.~Poovuttikul, \emph{{2-group global symmetries, hydrodynamics
  and holography}},
  \href{http://dx.doi.org/10.21468/SciPostPhys.15.2.063}{\emph{SciPost Phys.}
  {\bf 15} (2023) 063}, [\href{http://arxiv.org/abs/2010.00320}{{\tt
  2010.00320}}].

\bibitem{Hidaka:2020izy}
Y.~Hidaka, M.~Nitta and R.~Yokokura, \emph{{Global 3-group symmetry and 't
  Hooft anomalies in axion electrodynamics}},
  \href{http://dx.doi.org/10.1007/JHEP01(2021)173}{\emph{JHEP} {\bf 01} (2021)
  173}, [\href{http://arxiv.org/abs/2009.14368}{{\tt 2009.14368}}].

\bibitem{Brennan:2020ehu}
T.~D. Brennan and C.~Cordova, \emph{{Axions, higher-groups, and emergent
  symmetry}}, \href{http://dx.doi.org/10.1007/JHEP02(2022)145}{\emph{JHEP} {\bf
  02} (2022) 145}, [\href{http://arxiv.org/abs/2011.09600}{{\tt 2011.09600}}].

\bibitem{Bhardwaj:2021wif}
L.~Bhardwaj, \emph{{2-Group symmetries in class S}},
  \href{http://dx.doi.org/10.21468/SciPostPhys.12.5.152}{\emph{SciPost Phys.}
  {\bf 12} (2022) 152}, [\href{http://arxiv.org/abs/2107.06816}{{\tt
  2107.06816}}].

\bibitem{Hidaka:2021mml}
Y.~Hidaka, M.~Nitta and R.~Yokokura, \emph{{Topological axion electrodynamics
  and 4-group symmetry}},
  \href{http://dx.doi.org/10.1016/j.physletb.2021.136762}{\emph{Phys. Lett. B}
  {\bf 823} (2021) 136762}, [\href{http://arxiv.org/abs/2107.08753}{{\tt
  2107.08753}}].

\bibitem{Hidaka:2021kkf}
Y.~Hidaka, M.~Nitta and R.~Yokokura, \emph{{Global 4-group symmetry and
  {\textquoteright}t Hooft anomalies in topological axion electrodynamics}},
  \href{http://dx.doi.org/10.1093/ptep/ptab150}{\emph{PTEP} {\bf 2022} (2022)
  04A109}, [\href{http://arxiv.org/abs/2108.12564}{{\tt 2108.12564}}].

\bibitem{Apruzzi:2021mlh}
F.~Apruzzi, L.~Bhardwaj, D.~S.~W. Gould and S.~Schafer-Nameki, \emph{{2-Group
  symmetries and their classification in 6d}},
  \href{http://dx.doi.org/10.21468/SciPostPhys.12.3.098}{\emph{SciPost Phys.}
  {\bf 12} (2022) 098}, [\href{http://arxiv.org/abs/2110.14647}{{\tt
  2110.14647}}].

\bibitem{DelZotto:2022joo}
M.~Del~Zotto, I.~Garc{\'\i}a~Etxebarria and S.~Schafer-Nameki, \emph{{2-Group
  Symmetries and M-Theory}},
  \href{http://dx.doi.org/10.21468/SciPostPhys.13.5.105}{\emph{SciPost Phys.}
  {\bf 13} (2022) 105}, [\href{http://arxiv.org/abs/2203.10097}{{\tt
  2203.10097}}].

\bibitem{Kitaev:2005hzj}
A.~Kitaev, \emph{{Anyons in an exactly solved model and beyond}},
  \href{http://dx.doi.org/10.1016/j.aop.2005.10.005}{\emph{Annals Phys.} {\bf
  321} (2006) 2--111}, [\href{http://arxiv.org/abs/cond-mat/0506438}{{\tt
  cond-mat/0506438}}].

\bibitem{Bhardwaj:2017xup}
L.~Bhardwaj and Y.~Tachikawa, \emph{{On Finite Symmetries and Their Gauging in
  Two Dimensions}},
  \href{http://dx.doi.org/10.1007/JHEP03(2018)189}{\emph{JHEP} {\bf 03} (2018)
  189}, [\href{http://arxiv.org/abs/1704.02330}{{\tt 1704.02330}}].

\bibitem{Chang:2018iay}
C.-M. Chang, Y.-H. Lin, S.-H. Shao, Y.~Wang and X.~Yin, \emph{{Topological
  Defect Lines and Renormalization Group Flows in Two Dimensions}},
  \href{http://dx.doi.org/10.1007/JHEP01(2019)026}{\emph{JHEP} {\bf 01} (2019)
  026}, [\href{http://arxiv.org/abs/1802.04445}{{\tt 1802.04445}}].

\bibitem{Simon:2022ohj}
S.~H. Simon and J.~K. Slingerland, \emph{{Straightening Out the Frobenius-Schur
  Indicator}},  \href{http://arxiv.org/abs/2208.14500}{{\tt 2208.14500}}.

\bibitem{Choi:2024tri}
Y.~Choi, B.~C. Rayhaun and Y.~Zheng, \emph{{Generalized Tube Algebras,
  Symmetry-Resolved Partition Functions, and Twisted Boundary States}},
  \href{http://arxiv.org/abs/2409.02159}{{\tt 2409.02159}}.

\bibitem{etingof2005fusion}
P.~Etingof, D.~Nikshych and V.~Ostrik, \emph{On fusion categories},
  {\emph{Annals of Mathematics} (2005) 581--642}.

\bibitem{gainutdinov2023davydov}
A.~M. Gainutdinov, J.~Haferkamp and C.~Schweigert, \emph{Davydov-yetter
  cohomology, comonads and ocneanu rigidity}, {\emph{Advances in Mathematics}
  {\bf 414} (2023) 108853}.

\bibitem{ostrik2002fusion}
V.~Ostrik, \emph{Fusion categories of rank 2}, {\emph{arXiv preprint
  math/0203255} (2002) }.

\bibitem{ostrik2013pivotal}
V.~Ostrik, \emph{Pivotal fusion categories of rank 3 (with an appendix written
  jointly with dmitri nikshych)}, {\emph{arXiv preprint arXiv:1309.4822} (2013)
  }.

\bibitem{bruillard2016classification}
P.~Bruillard, S.-H. Ng, E.~C. Rowell and Z.~Wang, \emph{On classification of
  modular categories by rank. preprint}, {\emph{arXiv preprint
  arXiv:1507.05139} (2016) }.

\bibitem{vercleyen2024low}
G.~Vercleyen, \emph{On Low-Rank Multiplicity-Free Fusion Categories}.
\newblock National University of Ireland, Maynooth (Ireland), 2024.

\bibitem{Kramers:1941kn}
H.~A. Kramers and G.~H. Wannier, \emph{{Statistics of the two-dimensional
  ferromagnet. Part 1.}},
  \href{http://dx.doi.org/10.1103/PhysRev.60.252}{\emph{Phys. Rev.} {\bf 60}
  (1941) 252--262}.

\bibitem{Kramers:1941zz}
H.~A. Kramers and G.~H. Wannier, \emph{{Statistics of the Two-Dimensional
  Ferromagnet. Part II}},
  \href{http://dx.doi.org/10.1103/PhysRev.60.263}{\emph{Phys. Rev.} {\bf 60}
  (1941) 263--276}.

\bibitem{Ising}
S.~Yamaguchi, ``{Kramers-Wannier duality of the 2d Ising model (Japanese)}.''
  \url{https://www-het.phys.sci.osaka-u.ac.jp/~yamaguch/j/pdf/ising.pdf}.

\bibitem{tambara1998tensor}
D.~Tambara and S.~Yamagami, \emph{Tensor categories with fusion rules of
  self-duality for finite abelian groups}, {\emph{Journal of Algebra} {\bf 209}
  (1998) 692--707}.

\bibitem{Kaidi:2025hyr}
J.~Kaidi, X.~Shi, S.~Shimamori and Z.~Sun, \emph{{The SymTFT for $N$-ality
  defects: Part I}},  \href{http://arxiv.org/abs/2509.19429}{{\tt 2509.19429}}.

\bibitem{Diatlyk:2023fwf}
O.~Diatlyk, C.~Luo, Y.~Wang and Q.~Weller, \emph{{Gauging non-invertible
  symmetries: topological interfaces and generalized orbifold groupoid in 2d
  QFT}}, \href{http://dx.doi.org/10.1007/JHEP03(2024)127}{\emph{JHEP} {\bf 03}
  (2024) 127}, [\href{http://arxiv.org/abs/2311.17044}{{\tt 2311.17044}}].

\bibitem{Perez-Lona:2023djo}
A.~Perez-Lona, D.~Robbins, E.~Sharpe, T.~Vandermeulen and X.~Yu, \emph{{Notes
  on gauging noninvertible symmetries. Part I. Multiplicity-free cases}},
  \href{http://dx.doi.org/10.1007/JHEP02(2024)154}{\emph{JHEP} {\bf 02} (2024)
  154}, [\href{http://arxiv.org/abs/2311.16230}{{\tt 2311.16230}}].

\bibitem{Perez-Lona:2024sds}
A.~Perez-Lona, D.~Robbins, E.~Sharpe, T.~Vandermeulen and X.~Yu, \emph{{Notes
  on gauging noninvertible symmetries. Part II. Higher multiplicity cases}},
  \href{http://dx.doi.org/10.1007/JHEP05(2025)066}{\emph{JHEP} {\bf 05} (2025)
  066}, [\href{http://arxiv.org/abs/2408.16811}{{\tt 2408.16811}}].

\bibitem{Lin:2022dhv}
Y.-H. Lin, M.~Okada, S.~Seifnashri and Y.~Tachikawa, \emph{{Asymptotic Density
  of States in 2D CFTs with Non-Invertible Symmetries}},
  \href{http://dx.doi.org/10.1007/JHEP03(2023)094}{\emph{JHEP} {\bf 03} (2023)
  094}, [\href{http://arxiv.org/abs/2208.05495}{{\tt 2208.05495}}].

\bibitem{Bartsch:2022mpm}
T.~Bartsch, M.~Bullimore, A.~E.~V. Ferrari and J.~Pearson,
  \emph{{Non-invertible symmetries and higher representation theory I}},
  \href{http://dx.doi.org/10.21468/SciPostPhys.17.1.015}{\emph{SciPost Phys.}
  {\bf 17} (2024) 015}, [\href{http://arxiv.org/abs/2208.05993}{{\tt
  2208.05993}}].

\bibitem{Bartsch:2022ytj}
T.~Bartsch, M.~Bullimore, A.~E.~V. Ferrari and J.~Pearson,
  \emph{{Non-invertible symmetries and higher representation theory II}},
  \href{http://dx.doi.org/10.21468/SciPostPhys.17.2.067}{\emph{SciPost Phys.}
  {\bf 17} (2024) 067}, [\href{http://arxiv.org/abs/2212.07393}{{\tt
  2212.07393}}].

\bibitem{Bartsch:2023wvv}
T.~Bartsch, M.~Bullimore and A.~Grigoletto, \emph{{Representation theory for
  categorical symmetries}},  \href{http://arxiv.org/abs/2305.17165}{{\tt
  2305.17165}}.

\bibitem{Kaidi:2022cpf}
J.~Kaidi, K.~Ohmori and Y.~Zheng, \emph{{Symmetry TFTs for Non-invertible
  Defects}}, \href{http://dx.doi.org/10.1007/s00220-023-04859-7}{\emph{Commun.
  Math. Phys.} {\bf 404} (2023) 1021--1124},
  [\href{http://arxiv.org/abs/2209.11062}{{\tt 2209.11062}}].

\bibitem{Dijkgraaf:1989pz}
R.~Dijkgraaf and E.~Witten, \emph{{Topological Gauge Theories and Group
  Cohomology}}, \href{http://dx.doi.org/10.1007/BF02096988}{\emph{Commun. Math.
  Phys.} {\bf 129} (1990) 393}.

\bibitem{wen1992classification}
X.-G. Wen and A.~Zee, \emph{Classification of abelian quantum hall states and
  matrix formulation of topological fluids}, {\emph{Physical Review B} {\bf 46}
  (1992) 2290}.

\bibitem{freedman2004class}
M.~Freedman, C.~Nayak, K.~Shtengel, K.~Walker and Z.~Wang, \emph{A class of p,
  t-invariant topological phases of interacting electrons}, {\emph{Annals of
  Physics} {\bf 310} (2004) 428--492}.

\bibitem{levin2005string}
M.~A. Levin and X.-G. Wen, \emph{String-net condensation: A physical mechanism
  for topological phases}, {\emph{Physical Review B?Condensed Matter and
  Materials Physics} {\bf 71} (2005) 045110}.

\bibitem{Kaidi:2023maf}
J.~Kaidi, E.~Nardoni, G.~Zafrir and Y.~Zheng, \emph{{Symmetry TFTs and
  anomalies of non-invertible symmetries}},
  \href{http://dx.doi.org/10.1007/JHEP10(2023)053}{\emph{JHEP} {\bf 10} (2023)
  053}, [\href{http://arxiv.org/abs/2301.07112}{{\tt 2301.07112}}].

\bibitem{izumi2001structure}
M.~Izumi, \emph{The structure of sectors associated with {L}ongo--{R}ehren
  inclusions {II}: examples},
  \href{http://dx.doi.org/10.1142/S0129055X01000818}{\emph{Reviews in
  Mathematical Physics} {\bf 13} (2001) 603--674}.

\bibitem{gelaki2009centers}
S.~Gelaki, D.~Naidu and D.~Nikshych, \emph{Centers of graded fusion
  categories}, \href{http://dx.doi.org/10.2140/ant.2009.3.959}{\emph{Algebra
  Number Theory} {\bf 3} (2009) 959--990},
  [\href{http://arxiv.org/abs/0905.3117}{{\tt 0905.3117}}].

\bibitem{Barkeshli:2014cna}
M.~Barkeshli, P.~Bonderson, M.~Cheng and Z.~Wang, \emph{{Symmetry
  Fractionalization, Defects, and Gauging of Topological Phases}},
  \href{http://dx.doi.org/10.1103/PhysRevB.100.115147}{\emph{Phys. Rev. B} {\bf
  100} (2019) 115147}, [\href{http://arxiv.org/abs/1410.4540}{{\tt
  1410.4540}}].

\bibitem{Teo:2015xla}
J.~C.~Y. Teo, T.~L. Hughes and E.~Fradkin, \emph{{Theory of Twist Liquids:
  Gauging an Anyonic Symmetry}},
  \href{http://dx.doi.org/10.1016/j.aop.2015.05.012}{\emph{Annals Phys.} {\bf
  360} (2015) 349--445}, [\href{http://arxiv.org/abs/1503.06812}{{\tt
  1503.06812}}].

\bibitem{Lin:2019hks}
Y.-H. Lin and S.-H. Shao, \emph{{Duality Defect of the Monster CFT}},
  \href{http://dx.doi.org/10.1088/1751-8121/abd69e}{\emph{J. Phys. A} {\bf 54}
  (2021) 065201}, [\href{http://arxiv.org/abs/1911.00042}{{\tt 1911.00042}}].

\bibitem{Thorngren:2019iar}
R.~Thorngren and Y.~Wang, \emph{{Fusion category symmetry. Part I. Anomaly
  in-flow and gapped phases}},
  \href{http://dx.doi.org/10.1007/JHEP04(2024)132}{\emph{JHEP} {\bf 04} (2024)
  132}, [\href{http://arxiv.org/abs/1912.02817}{{\tt 1912.02817}}].

\bibitem{Konechny:2019wff}
A.~Konechny, \emph{{Open topological defects and boundary RG flows}},
  \href{http://dx.doi.org/10.1088/1751-8121/ab7c8b}{\emph{J. Phys. A} {\bf 53}
  (2020) 155401}, [\href{http://arxiv.org/abs/1911.06041}{{\tt 1911.06041}}].

\bibitem{Huang:2020lox}
T.-C. Huang and Y.-H. Lin, \emph{{The $F$-Symbols for Transparent
  Haagerup-Izumi Categories with $G = \mathbb{Z}_{2n+1}$}},
  \href{http://arxiv.org/abs/2007.00670}{{\tt 2007.00670}}.

\bibitem{Gaiotto:2020iye}
D.~Gaiotto and J.~Kulp, \emph{{Orbifold groupoids}},
  \href{http://dx.doi.org/10.1007/JHEP02(2021)132}{\emph{JHEP} {\bf 02} (2021)
  132}, [\href{http://arxiv.org/abs/2008.05960}{{\tt 2008.05960}}].

\bibitem{Komargodski:2020mxz}
Z.~Komargodski, K.~Ohmori, K.~Roumpedakis and S.~Seifnashri, \emph{{Symmetries
  and strings of adjoint QCD$_{2}$}},
  \href{http://dx.doi.org/10.1007/JHEP03(2021)103}{\emph{JHEP} {\bf 03} (2021)
  103}, [\href{http://arxiv.org/abs/2008.07567}{{\tt 2008.07567}}].

\bibitem{Chang:2020imq}
C.-M. Chang and Y.-H. Lin, \emph{{Lorentzian dynamics and factorization beyond
  rationality}}, \href{http://dx.doi.org/10.1007/JHEP10(2021)125}{\emph{JHEP}
  {\bf 10} (2021) 125}, [\href{http://arxiv.org/abs/2012.01429}{{\tt
  2012.01429}}].

\bibitem{Huang:2021ytb}
T.-C. Huang and Y.-H. Lin, \emph{{Topological field theory with Haagerup
  symmetry}}, \href{http://dx.doi.org/10.1063/5.0079062}{\emph{J. Math. Phys.}
  {\bf 63} (2022) 042306}, [\href{http://arxiv.org/abs/2102.05664}{{\tt
  2102.05664}}].

\bibitem{Inamura:2021wuo}
K.~Inamura, \emph{{Topological field theories and symmetry protected
  topological phases with fusion category symmetries}},
  \href{http://dx.doi.org/10.1007/JHEP05(2021)204}{\emph{JHEP} {\bf 05} (2021)
  204}, [\href{http://arxiv.org/abs/2103.15588}{{\tt 2103.15588}}].

\bibitem{Thorngren:2021yso}
R.~Thorngren and Y.~Wang, \emph{{Fusion category symmetry. Part II.
  Categoriosities at c = 1 and beyond}},
  \href{http://dx.doi.org/10.1007/JHEP07(2024)051}{\emph{JHEP} {\bf 07} (2024)
  051}, [\href{http://arxiv.org/abs/2106.12577}{{\tt 2106.12577}}].

\bibitem{Sharpe:2021srf}
E.~Sharpe, \emph{{Topological operators, noninvertible symmetries and
  decomposition}},
  \href{http://dx.doi.org/10.4310/ATMP.2023.v27.n8.a2}{\emph{Adv. Theor. Math.
  Phys.} {\bf 27} (2023) 2319--2407},
  [\href{http://arxiv.org/abs/2108.13423}{{\tt 2108.13423}}].

\bibitem{Huang:2021zvu}
T.-C. Huang, Y.-H. Lin and S.~Seifnashri, \emph{{Construction of
  two-dimensional topological field theories with non-invertible symmetries}},
  \href{http://dx.doi.org/10.1007/JHEP12(2021)028}{\emph{JHEP} {\bf 12} (2021)
  028}, [\href{http://arxiv.org/abs/2110.02958}{{\tt 2110.02958}}].

\bibitem{Huang:2021nvb}
T.-C. Huang, Y.-H. Lin, K.~Ohmori, Y.~Tachikawa and M.~Tezuka, \emph{{Numerical
  Evidence for a Haagerup Conformal Field Theory}},
  \href{http://dx.doi.org/10.1103/PhysRevLett.128.231603}{\emph{Phys. Rev.
  Lett.} {\bf 128} (2022) 231603}, [\href{http://arxiv.org/abs/2110.03008}{{\tt
  2110.03008}}].

\bibitem{Vanhove:2021zop}
R.~Vanhove, L.~Lootens, M.~Van~Damme, R.~Wolf, T.~J. Osborne, J.~Haegeman
  et~al., \emph{{Critical Lattice Model for a Haagerup Conformal Field
  Theory}}, \href{http://dx.doi.org/10.1103/PhysRevLett.128.231602}{\emph{Phys.
  Rev. Lett.} {\bf 128} (2022) 231602},
  [\href{http://arxiv.org/abs/2110.03532}{{\tt 2110.03532}}].

\bibitem{Inamura:2021szw}
K.~Inamura, \emph{{On lattice models of gapped phases with fusion category
  symmetries}}, \href{http://dx.doi.org/10.1007/JHEP03(2022)036}{\emph{JHEP}
  {\bf 03} (2022) 036}, [\href{http://arxiv.org/abs/2110.12882}{{\tt
  2110.12882}}].

\bibitem{Burbano:2021loy}
I.~M. Burbano, J.~Kulp and J.~Neuser, \emph{{Duality defects in E$_{8}$}},
  \href{http://dx.doi.org/10.1007/JHEP10(2022)187}{\emph{JHEP} {\bf 10} (2022)
  186}, [\href{http://arxiv.org/abs/2112.14323}{{\tt 2112.14323}}].

\bibitem{Inamura:2022lun}
K.~Inamura, \emph{{Fermionization of fusion category symmetries in 1+1
  dimensions}}, \href{http://dx.doi.org/10.1007/JHEP10(2023)101}{\emph{JHEP}
  {\bf 10} (2023) 101}, [\href{http://arxiv.org/abs/2206.13159}{{\tt
  2206.13159}}].

\bibitem{Chang:2022hud}
C.-M. Chang, J.~Chen and F.~Xu, \emph{{Topological defect lines in two
  dimensional fermionic CFTs}},
  \href{http://dx.doi.org/10.21468/SciPostPhys.15.5.216}{\emph{SciPost Phys.}
  {\bf 15} (2023) 216}, [\href{http://arxiv.org/abs/2208.02757}{{\tt
  2208.02757}}].

\bibitem{Lu:2022ver}
D.-C. Lu and Z.~Sun, \emph{{On triality defects in 2d CFT}},
  \href{http://dx.doi.org/10.1007/JHEP02(2023)173}{\emph{JHEP} {\bf 02} (2023)
  173}, [\href{http://arxiv.org/abs/2208.06077}{{\tt 2208.06077}}].

\bibitem{Li:2023ani}
L.~Li, M.~Oshikawa and Y.~Zheng, \emph{{Noninvertible duality transformation
  between symmetry-protected topological and spontaneous symmetry breaking
  phases}}, \href{http://dx.doi.org/10.1103/PhysRevB.108.214429}{\emph{Phys.
  Rev. B} {\bf 108} (2023) 214429},
  [\href{http://arxiv.org/abs/2301.07899}{{\tt 2301.07899}}].

\bibitem{Lin:2023uvm}
Y.-H. Lin and S.-H. Shao, \emph{{Bootstrapping noninvertible symmetries}},
  \href{http://dx.doi.org/10.1103/PhysRevD.107.125025}{\emph{Phys. Rev. D} {\bf
  107} (2023) 125025}, [\href{http://arxiv.org/abs/2302.13900}{{\tt
  2302.13900}}].

\bibitem{Cao:2023doz}
W.~Cao, L.~Li, M.~Yamazaki and Y.~Zheng, \emph{{Subsystem non-invertible
  symmetry operators and defects}},
  \href{http://dx.doi.org/10.21468/SciPostPhys.15.4.155}{\emph{SciPost Phys.}
  {\bf 15} (2023) 155}, [\href{http://arxiv.org/abs/2304.09886}{{\tt
  2304.09886}}].

\bibitem{Jacobsen:2023isq}
J.~L. Jacobsen and H.~Saleur, \emph{{Non-invertible symmetries and RG flows in
  the two-dimensional O(n) loop model}},
  \href{http://dx.doi.org/10.1007/JHEP12(2023)090}{\emph{JHEP} {\bf 12} (2023)
  090}, [\href{http://arxiv.org/abs/2305.05746}{{\tt 2305.05746}}].

\bibitem{Choi:2023xjw}
Y.~Choi, B.~C. Rayhaun, Y.~Sanghavi and S.-H. Shao, \emph{{Remarks on
  boundaries, anomalies, and noninvertible symmetries}},
  \href{http://dx.doi.org/10.1103/PhysRevD.108.125005}{\emph{Phys. Rev. D} {\bf
  108} (2023) 125005}, [\href{http://arxiv.org/abs/2305.09713}{{\tt
  2305.09713}}].

\bibitem{Haghighat:2023sax}
B.~Haghighat and Y.~Sun, \emph{{Topological Defect Lines in bosonized
  Parafermionic CFTs}},
  \href{http://dx.doi.org/10.4310/ATMP.241031012317}{\emph{Adv. Theor. Math.
  Phys.} {\bf 28} (2024) 1987--2023},
  [\href{http://arxiv.org/abs/2306.16555}{{\tt 2306.16555}}].

\bibitem{Seiberg:2023cdc}
N.~Seiberg and S.-H. Shao, \emph{{Majorana chain and Ising model -
  (non-invertible) translations, anomalies, and emanant symmetries}},
  \href{http://dx.doi.org/10.21468/SciPostPhys.16.3.064}{\emph{SciPost Phys.}
  {\bf 16} (2024) 064}, [\href{http://arxiv.org/abs/2307.02534}{{\tt
  2307.02534}}].

\bibitem{Sun:2023xxv}
Z.~Sun and Y.~Zheng, \emph{{When are duality defects group-theoretical?}},
  \href{http://dx.doi.org/10.1007/JHEP10(2025)104}{\emph{JHEP} {\bf 10} (2025)
  104}, [\href{http://arxiv.org/abs/2307.14428}{{\tt 2307.14428}}].

\bibitem{Aksoy:2023hve}
{\"O}.~M. Aksoy, C.~Mudry, A.~Furusaki and A.~Tiwari,
  \emph{{Lieb-Schultz-Mattis anomalies and web of dualities induced by gauging
  in quantum spin chains}},
  \href{http://dx.doi.org/10.21468/SciPostPhys.16.1.022}{\emph{SciPost Phys.}
  {\bf 16} (2024) 022}, [\href{http://arxiv.org/abs/2308.00743}{{\tt
  2308.00743}}].

\bibitem{Duan:2023ykn}
Z.~Duan, Q.~Jia and S.~Lee, \emph{{{\ensuremath{\mathbb{Z}}}$_{N}$ duality and
  parafermions revisited}},
  \href{http://dx.doi.org/10.1007/JHEP11(2023)206}{\emph{JHEP} {\bf 11} (2023)
  206}, [\href{http://arxiv.org/abs/2309.01913}{{\tt 2309.01913}}].

\bibitem{Chen:2023jht}
J.~Chen, B.~Haghighat and Q.-R. Wang, \emph{{Para-fusion Category and
  Topological Defect Lines in $\mathbb Z_N$-parafermionic CFTs}},
  \href{http://arxiv.org/abs/2309.01914}{{\tt 2309.01914}}.

\bibitem{Nagoya:2023zky}
Y.~Nagoya and S.~Shimamori, \emph{{Non-invertible duality defect and
  non-commutative fusion algebra}},
  \href{http://dx.doi.org/10.1007/JHEP12(2023)062}{\emph{JHEP} {\bf 12} (2023)
  062}, [\href{http://arxiv.org/abs/2309.05294}{{\tt 2309.05294}}].

\bibitem{Bhardwaj:2023idu}
L.~Bhardwaj, L.~E. Bottini, D.~Pajer and S.~Sch{\"a}fer-Nameki, \emph{{Gapped
  phases with non-invertible symmetries: (1+1)d}},
  \href{http://dx.doi.org/10.21468/SciPostPhys.18.1.032}{\emph{SciPost Phys.}
  {\bf 18} (2025) 032}, [\href{http://arxiv.org/abs/2310.03784}{{\tt
  2310.03784}}].

\bibitem{Choi:2023vgk}
Y.~Choi, D.-C. Lu and Z.~Sun, \emph{{Self-duality under gauging a
  non-invertible symmetry}},
  \href{http://dx.doi.org/10.1007/JHEP01(2024)142}{\emph{JHEP} {\bf 01} (2024)
  142}, [\href{http://arxiv.org/abs/2310.19867}{{\tt 2310.19867}}].

\bibitem{Grover:2023loq}
S.~Grover, S.~Hegde and D.~P. Jatkar, \emph{{Duality defects in D$_{n}$-type
  Niemeier lattice CFTs}},
  \href{http://dx.doi.org/10.1007/JHEP05(2024)057}{\emph{JHEP} {\bf 05} (2024)
  057}, [\href{http://arxiv.org/abs/2312.17165}{{\tt 2312.17165}}].

\bibitem{Damia:2024xju}
J.~A. Damia, G.~Galati, O.~Hulik and S.~Mancani, \emph{{Exploring duality
  symmetries, multicriticality and RG flows at c = 2}},
  \href{http://dx.doi.org/10.1007/JHEP04(2024)028}{\emph{JHEP} {\bf 04} (2024)
  028}, [\href{http://arxiv.org/abs/2401.04166}{{\tt 2401.04166}}].

\bibitem{Gutperle:2024vyp}
M.~Gutperle, Y.-Y. Li, D.~Rathore and K.~Roumpedakis, \emph{{Non-invertible
  symmetries in S$_{N}$ orbifold CFTs and holography}},
  \href{http://dx.doi.org/10.1007/JHEP09(2024)110}{\emph{JHEP} {\bf 09} (2024)
  110}, [\href{http://arxiv.org/abs/2405.15693}{{\tt 2405.15693}}].

\bibitem{Bharadwaj:2024gpj}
S.~Bharadwaj, P.~Niro and K.~Roumpedakis, \emph{{Non-invertible defects on the
  worldsheet}}, \href{http://dx.doi.org/10.1007/JHEP03(2025)164}{\emph{JHEP}
  {\bf 03} (2025) 164}, [\href{http://arxiv.org/abs/2408.14556}{{\tt
  2408.14556}}].

\bibitem{Grimminger:2024mks}
J.~F. Grimminger, W.~Harding and N.~Mekareeya, \emph{{Wreathing, discrete
  gauging, and non-invertible symmetries}},
  \href{http://dx.doi.org/10.1007/JHEP01(2025)124}{\emph{JHEP} {\bf 01} (2025)
  124}, [\href{http://arxiv.org/abs/2410.12906}{{\tt 2410.12906}}].

\bibitem{Lu:2024lzf}
D.-C. Lu, Z.~Sun and Z.~Zhang, \emph{{Exploring G-ality defects in 2-dim
  QFTs}}, \href{http://dx.doi.org/10.1007/JHEP11(2025)081}{\emph{JHEP} {\bf 11}
  (2025) 081}, [\href{http://arxiv.org/abs/2406.12151}{{\tt 2406.12151}}].

\bibitem{Bottini:2025hri}
L.~E. Bottini and S.~Schafer-Nameki, \emph{{Construction of a Gapless Phase
  with Haagerup Symmetry}},
  \href{http://dx.doi.org/10.1103/PhysRevLett.134.191602}{\emph{Phys. Rev.
  Lett.} {\bf 134} (2025) 191602}, [\href{http://arxiv.org/abs/2410.19040}{{\tt
  2410.19040}}].

\bibitem{Albert:2025umy}
J.~Albert, Y.~Honda, J.~Kaidi and Y.~Zheng, \emph{{Haagerup Symmetry in
  $(E_8)_1$?}},  \href{http://arxiv.org/abs/2512.08225}{{\tt 2512.08225}}.

\bibitem{Honda:2026bjy}
Y.~Honda, J.~Kaidi and I.~Orii, \emph{{Parafermionizing the Monster}},
  \href{http://arxiv.org/abs/2605.10902}{{\tt 2605.10902}}.

\bibitem{Verlinde:1988sn}
E.~P. Verlinde, \emph{{Fusion Rules and Modular Transformations in 2D Conformal
  Field Theory}},
  \href{http://dx.doi.org/10.1016/0550-3213(88)90603-7}{\emph{Nucl. Phys. B}
  {\bf 300} (1988) 360--376}.

\bibitem{Moore:1988qv}
G.~W. Moore and N.~Seiberg, \emph{{Classical and Quantum Conformal Field
  Theory}}, \href{http://dx.doi.org/10.1007/BF01238857}{\emph{Commun. Math.
  Phys.} {\bf 123} (1989) 177}.

\bibitem{Petkova:2000ip}
V.~B. Petkova and J.~B. Zuber, \emph{{Generalized twisted partition
  functions}},
  \href{http://dx.doi.org/10.1016/S0370-2693(01)00276-3}{\emph{Phys. Lett. B}
  {\bf 504} (2001) 157--164}, [\href{http://arxiv.org/abs/hep-th/0011021}{{\tt
  hep-th/0011021}}].

\bibitem{Fuchs:2002cm}
J.~Fuchs, I.~Runkel and C.~Schweigert, \emph{{TFT construction of RCFT
  correlators 1. Partition functions}},
  \href{http://dx.doi.org/10.1016/S0550-3213(02)00744-7}{\emph{Nucl. Phys. B}
  {\bf 646} (2002) 353--497}, [\href{http://arxiv.org/abs/hep-th/0204148}{{\tt
  hep-th/0204148}}].

\bibitem{Tachikawa:2017gyf}
Y.~Tachikawa, \emph{{On gauging finite subgroups}},
  \href{http://dx.doi.org/10.21468/SciPostPhys.8.1.015}{\emph{SciPost Phys.}
  {\bf 8} (2020) 015}, [\href{http://arxiv.org/abs/1712.09542}{{\tt
  1712.09542}}].

\bibitem{Frohlich:2004ef}
J.~Frohlich, J.~Fuchs, I.~Runkel and C.~Schweigert, \emph{{Kramers-Wannier
  duality from conformal defects}},
  \href{http://dx.doi.org/10.1103/PhysRevLett.93.070601}{\emph{Phys. Rev.
  Lett.} {\bf 93} (2004) 070601},
  [\href{http://arxiv.org/abs/cond-mat/0404051}{{\tt cond-mat/0404051}}].

\bibitem{Frohlich:2006ch}
J.~Frohlich, J.~Fuchs, I.~Runkel and C.~Schweigert, \emph{{Duality and defects
  in rational conformal field theory}},
  \href{http://dx.doi.org/10.1016/j.nuclphysb.2006.11.017}{\emph{Nucl. Phys. B}
  {\bf 763} (2007) 354--430}, [\href{http://arxiv.org/abs/hep-th/0607247}{{\tt
  hep-th/0607247}}].

\bibitem{Carqueville:2012dk}
N.~Carqueville and I.~Runkel, \emph{{Orbifold completion of defect
  bicategories}}, \href{http://dx.doi.org/10.4171/qt/76}{\emph{Quantum Topol.}
  {\bf 7} (2016) 203--279}, [\href{http://arxiv.org/abs/1210.6363}{{\tt
  1210.6363}}].

\bibitem{Brunner:2013xna}
I.~Brunner, N.~Carqueville and D.~Plencner, \emph{{A quick guide to defect
  orbifolds}}, \href{http://dx.doi.org/10.1090/pspum/088/01456}{\emph{Proc.
  Symp. Pure Math.} {\bf 88} (2014) 231--242},
  [\href{http://arxiv.org/abs/1310.0062}{{\tt 1310.0062}}].

\bibitem{Lootens:2021tet}
L.~Lootens, C.~Delcamp, G.~Ortiz and F.~Verstraete, \emph{{Dualities in
  One-Dimensional Quantum Lattice Models: Symmetric Hamiltonians and Matrix
  Product Operator Intertwiners}},
  \href{http://dx.doi.org/10.1103/PRXQuantum.4.020357}{\emph{PRX Quantum} {\bf
  4} (2023) 020357}, [\href{http://arxiv.org/abs/2112.09091}{{\tt
  2112.09091}}].

\bibitem{Apruzzi:2023uma}
F.~Apruzzi, F.~Bonetti, D.~S.~W. Gould and S.~Schafer-Nameki, \emph{{Aspects of
  categorical symmetries from branes: SymTFTs and generalized charges}},
  \href{http://dx.doi.org/10.21468/SciPostPhys.17.1.025}{\emph{SciPost Phys.}
  {\bf 17} (2024) 025}, [\href{http://arxiv.org/abs/2306.16405}{{\tt
  2306.16405}}].

\bibitem{Bhardwaj:2023bbf}
L.~Bhardwaj, L.~E. Bottini, D.~Pajer and S.~Schafer-Nameki, \emph{{The club
  sandwich: Gapless phases and phase transitions with non-invertible
  symmetries}},
  \href{http://dx.doi.org/10.21468/SciPostPhys.18.5.156}{\emph{SciPost Phys.}
  {\bf 18} (2025) 156}, [\href{http://arxiv.org/abs/2312.17322}{{\tt
  2312.17322}}].

\bibitem{DelZotto:2024tae}
M.~Del~Zotto, S.~N. Meynet and R.~Moscrop, \emph{{Remarks on geometric
  engineering, symmetry TFTs and anomalies}},
  \href{http://dx.doi.org/10.1007/JHEP07(2024)220}{\emph{JHEP} {\bf 07} (2024)
  220}, [\href{http://arxiv.org/abs/2402.18646}{{\tt 2402.18646}}].

\bibitem{Franco:2024mxa}
S.~Franco and X.~Yu, \emph{{Generalized symmetries in 2D from string theory:
  SymTFTs, intrinsic relativeness, and anomalies of non-invertible
  symmetries}}, \href{http://dx.doi.org/10.1007/JHEP11(2024)004}{\emph{JHEP}
  {\bf 11} (2024) 004}, [\href{http://arxiv.org/abs/2404.19761}{{\tt
  2404.19761}}].

\bibitem{Hasan:2024aow}
A.~Hasan, S.~Meynet and D.~Migliorati,
  \emph{{SL$_{2}$({\ensuremath{\mathbb{R}}}) symmetries of SymTFT and
  non-invertible U(1) symmetries of Maxwell theory}},
  \href{http://dx.doi.org/10.1007/JHEP12(2024)131}{\emph{JHEP} {\bf 12} (2024)
  131}, [\href{http://arxiv.org/abs/2405.19218}{{\tt 2405.19218}}].

\bibitem{Copetti:2024onh}
C.~Copetti, \emph{{Defect Charges, Gapped Boundary Conditions, and the Symmetry
  TFT}},  \href{http://arxiv.org/abs/2408.01490}{{\tt 2408.01490}}.

\bibitem{Antinucci:2024ltv}
A.~Antinucci, C.~Copetti and S.~Schafer-Nameki, \emph{{SymTFT for (3+1)d
  Gapless SPTs and Obstructions to Confinement}},
  \href{http://dx.doi.org/10.21468/SciPostPhys.18.3.114}{\emph{SciPost Phys.}
  {\bf 18} (2025) 114}, [\href{http://arxiv.org/abs/2408.05585}{{\tt
  2408.05585}}].

\bibitem{Bhardwaj:2024ydc}
L.~Bhardwaj, K.~Inamura and A.~Tiwari, \emph{{Fermionic non-invertible
  symmetries in (1+1)d: Gapped and gapless phases, transitions, and symmetry
  TFTs}}, \href{http://dx.doi.org/10.21468/SciPostPhys.18.6.194}{\emph{SciPost
  Phys.} {\bf 18} (2025) 194}, [\href{http://arxiv.org/abs/2405.09754}{{\tt
  2405.09754}}].

\bibitem{Bhardwaj:2024igy}
L.~Bhardwaj, C.~Copetti, D.~Pajer and S.~Schafer-Nameki, \emph{{Boundary
  SymTFT}},
  \href{http://dx.doi.org/10.21468/SciPostPhys.19.2.061}{\emph{SciPost Phys.}
  {\bf 19} (2025) 061}, [\href{http://arxiv.org/abs/2409.02166}{{\tt
  2409.02166}}].

\bibitem{Argurio:2024ewp}
R.~Argurio, A.~Collinucci, G.~Galati, O.~Hulik and E.~Paznokas,
  \emph{{Non-Invertible T-duality at Any Radius via Non-Compact SymTFT}},
  \href{http://dx.doi.org/10.21468/SciPostPhys.18.3.089}{\emph{SciPost Phys.}
  {\bf 18} (2025) 089}, [\href{http://arxiv.org/abs/2409.11822}{{\tt
  2409.11822}}].

\bibitem{Duan:2024xbb}
Z.~Duan, Q.~Jia and S.~Lee, \emph{{Web of 4D dualities, supersymmetric
  partition functions and SymTFT}},
  \href{http://dx.doi.org/10.1007/JHEP01(2025)161}{\emph{JHEP} {\bf 01} (2025)
  161}, [\href{http://arxiv.org/abs/2410.10036}{{\tt 2410.10036}}].

\bibitem{Chen:2024ulc}
J.~Chen and Q.~Jia, \emph{{SymTFT approach to 2D orbifold groupoids:
  {\textquoteright}t Hooft anomalies, gauging, and partition functions}},
  \href{http://dx.doi.org/10.1007/JHEP04(2025)043}{\emph{JHEP} {\bf 04} (2025)
  043}, [\href{http://arxiv.org/abs/2411.18056}{{\tt 2411.18056}}].

\bibitem{Bergman:2024its}
O.~Bergman and F.~Mignosa, \emph{{String theory and the SymTFT of 3d
  orthosymplectic Chern-Simons theory}},
  \href{http://dx.doi.org/10.1007/JHEP04(2025)047}{\emph{JHEP} {\bf 04} (2025)
  047}, [\href{http://arxiv.org/abs/2412.00184}{{\tt 2412.00184}}].

\bibitem{Bonetti:2024etn}
F.~Bonetti, M.~Del~Zotto and R.~Minasian, \emph{{SymTFTs and non-invertible
  symmetries of 6d (2,0) SCFTs of type D from M-theory}},
  \href{http://dx.doi.org/10.1007/JHEP02(2025)156}{\emph{JHEP} {\bf 02} (2025)
  156}, [\href{http://arxiv.org/abs/2412.07842}{{\tt 2412.07842}}].

\bibitem{DelZotto:2025yoy}
M.~Del~Zotto, A.~Hasan and E.~Riedel~G{\r{a}}rding, \emph{{SymTFT, Protected
  Gaplessness, and Spontaneous Breaking of Non-invertible Symmetries}},
  \href{http://arxiv.org/abs/2504.18501}{{\tt 2504.18501}}.

\bibitem{Schafer-Nameki:2025fiy}
S.~Schafer-Nameki, A.~Tiwari, A.~Warman and C.~Zhang, \emph{{SymTFT Approach
  for Mixed States with Non-Invertible Symmetries}},
  \href{http://arxiv.org/abs/2507.05350}{{\tt 2507.05350}}.

\bibitem{Zhang:2023wlu}
C.~Zhang and C.~C{\'o}rdova, \emph{{Anomalies of (1+1)-dimensional categorical
  symmetries}},
  \href{http://dx.doi.org/10.1103/PhysRevB.110.035155}{\emph{Phys. Rev. B} {\bf
  110} (2024) 035155}, [\href{http://arxiv.org/abs/2304.01262}{{\tt
  2304.01262}}].

\bibitem{Cordova:2023bja}
C.~Cordova, P.-S. Hsin and C.~Zhang, \emph{{Anomalies of non-invertible
  symmetries in (3+1)d}},
  \href{http://dx.doi.org/10.21468/SciPostPhys.17.5.131}{\emph{SciPost Phys.}
  {\bf 17} (2024) 131}, [\href{http://arxiv.org/abs/2308.11706}{{\tt
  2308.11706}}].

\bibitem{Antinucci:2023ezl}
A.~Antinucci, F.~Benini, C.~Copetti, G.~Galati and G.~Rizi, \emph{{Anomalies of
  non-invertible self-duality symmetries: fractionalization and gauging}},
  \href{http://arxiv.org/abs/2308.11707}{{\tt 2308.11707}}.

\bibitem{Hsin:2025ria}
P.-S. Hsin, R.~Kobayashi and C.~Zhang, \emph{{Anomalies of Coset Non-Invertible
  Symmetries}},
  \href{http://dx.doi.org/10.21468/SciPostPhys.20.1.006}{\emph{SciPost Phys.}
  {\bf 20} (2026) 006}, [\href{http://arxiv.org/abs/2503.00105}{{\tt
  2503.00105}}].

\bibitem{Damia:2023gtc}
J.~A. Damia, R.~Argurio and S.~Chaudhuri, \emph{{When the moduli space is an
  orbifold: spontaneous breaking of continuous non-invertible symmetries}},
  \href{http://dx.doi.org/10.1007/JHEP03(2024)042}{\emph{JHEP} {\bf 03} (2024)
  042}, [\href{http://arxiv.org/abs/2309.06491}{{\tt 2309.06491}}].

\bibitem{Bhardwaj:2024qrf}
L.~Bhardwaj, D.~Pajer, S.~Schafer-Nameki and A.~Warman, \emph{{Hasse diagrams
  for gapless SPT and SSB phases with non-invertible symmetries}},
  \href{http://dx.doi.org/10.21468/SciPostPhys.19.4.113}{\emph{SciPost Phys.}
  {\bf 19} (2025) 113}, [\href{http://arxiv.org/abs/2403.00905}{{\tt
  2403.00905}}].

\bibitem{Fuchs:2003id}
J.~Fuchs, I.~Runkel and C.~Schweigert, \emph{{TFT construction of RCFT
  correlators. 2. Unoriented world sheets}},
  \href{http://dx.doi.org/10.1016/j.nuclphysb.2003.11.026}{\emph{Nucl. Phys. B}
  {\bf 678} (2004) 511--637}, [\href{http://arxiv.org/abs/hep-th/0306164}{{\tt
  hep-th/0306164}}].

\bibitem{Kapustin:2015uma}
A.~Kapustin and A.~Turzillo, \emph{{Equivariant Topological Quantum Field
  Theory and Symmetry Protected Topological Phases}},
  \href{http://dx.doi.org/10.1007/JHEP03(2017)006}{\emph{JHEP} {\bf 03} (2017)
  006}, [\href{http://arxiv.org/abs/1504.01830}{{\tt 1504.01830}}].

\bibitem{Bhardwaj:2016dtk}
L.~Bhardwaj, \emph{{Unoriented 3d TFTs}},
  \href{http://dx.doi.org/10.1007/JHEP05(2017)048}{\emph{JHEP} {\bf 05} (2017)
  048}, [\href{http://arxiv.org/abs/1611.02728}{{\tt 1611.02728}}].

\bibitem{Harada:2025uhh}
W.~Harada, J.~Kaidi, Y.~Kusuki and Y.~Liu, \emph{{New Crosscap States}},
  \href{http://arxiv.org/abs/2508.18357}{{\tt 2508.18357}}.

\bibitem{Heidenreich:2021xpr}
B.~Heidenreich, J.~McNamara, M.~Montero, M.~Reece, T.~Rudelius and
  I.~Valenzuela, \emph{{Non-invertible global symmetries and completeness of
  the spectrum}}, \href{http://dx.doi.org/10.1007/JHEP09(2021)203}{\emph{JHEP}
  {\bf 09} (2021) 203}, [\href{http://arxiv.org/abs/2104.07036}{{\tt
  2104.07036}}].

\bibitem{Koide:2021zxj}
M.~Koide, Y.~Nagoya and S.~Yamaguchi, \emph{{Non-invertible topological defects
  in 4-dimensional $\mathbb {Z}_2$ pure lattice gauge theory}},
  \href{http://dx.doi.org/10.1093/ptep/ptab145}{\emph{PTEP} {\bf 2022} (2022)
  013B03}, [\href{http://arxiv.org/abs/2109.05992}{{\tt 2109.05992}}].

\bibitem{Kaidi:2021xfk}
J.~Kaidi, K.~Ohmori and Y.~Zheng, \emph{{Kramers-Wannier-like Duality Defects
  in (3+1)D Gauge Theories}},
  \href{http://dx.doi.org/10.1103/PhysRevLett.128.111601}{\emph{Phys. Rev.
  Lett.} {\bf 128} (2022) 111601}, [\href{http://arxiv.org/abs/2111.01141}{{\tt
  2111.01141}}].

\bibitem{Choi:2021kmx}
Y.~Choi, C.~Cordova, P.-S. Hsin, H.~T. Lam and S.-H. Shao, \emph{{Noninvertible
  duality defects in 3+1 dimensions}},
  \href{http://dx.doi.org/10.1103/PhysRevD.105.125016}{\emph{Phys. Rev. D} {\bf
  105} (2022) 125016}, [\href{http://arxiv.org/abs/2111.01139}{{\tt
  2111.01139}}].

\bibitem{Hsin:2018vcg}
P.-S. Hsin, H.~T. Lam and N.~Seiberg, \emph{{Comments on One-Form Global
  Symmetries and Their Gauging in 3d and 4d}},
  \href{http://dx.doi.org/10.21468/SciPostPhys.6.3.039}{\emph{SciPost Phys.}
  {\bf 6} (2019) 039}, [\href{http://arxiv.org/abs/1812.04716}{{\tt
  1812.04716}}].

\bibitem{Decoppet:2023juy}
T.~D{\'e}coppet, \emph{{On fusion 2-categories}}.
\newblock PhD thesis, Oxford University, 2023.
\newblock 10.5287/ora-jnnrdekyg.

\bibitem{Choi:2022}
Y.~Choi, C.~Cordova, P.-S. Hsin, H.~T. Lam and S.-H. Shao, \emph{Non-invertible
  {Condensation}, {Duality}, and {Triality} {Defects} in 3+1 {Dimensions}},
  \href{http://dx.doi.org/10.1007/s00220-023-04727-4}{\emph{arXiv.org} (Apr.,
  2022) }.

\bibitem{Copetti:2023mcq}
C.~Copetti, M.~Del~Zotto, K.~Ohmori and Y.~Wang, \emph{{Higher Structure of
  Chiral Symmetry}},
  \href{http://dx.doi.org/10.1007/s00220-024-05227-9}{\emph{Commun. Math.
  Phys.} {\bf 406} (2025) 73}, [\href{http://arxiv.org/abs/2305.18282}{{\tt
  2305.18282}}].

\bibitem{Kaidi:2022uux}
J.~Kaidi, G.~Zafrir and Y.~Zheng, \emph{{Non-invertible symmetries of $
  \mathcal{N} $ = 4 SYM and twisted compactification}},
  \href{http://dx.doi.org/10.1007/JHEP08(2022)053}{\emph{JHEP} {\bf 08} (2022)
  053}, [\href{http://arxiv.org/abs/2205.01104}{{\tt 2205.01104}}].

\bibitem{Montonen:1977sn}
C.~Montonen and D.~I. Olive, \emph{{Magnetic Monopoles as Gauge Particles?}},
  \href{http://dx.doi.org/10.1016/0370-2693(77)90076-4}{\emph{Phys. Lett. B}
  {\bf 72} (1977) 117--120}.

\bibitem{Cordova:2022ieu}
C.~Cordova and K.~Ohmori, \emph{{Noninvertible Chiral Symmetry and Exponential
  Hierarchies}},
  \href{http://dx.doi.org/10.1103/PhysRevX.13.011034}{\emph{Phys. Rev. X} {\bf
  13} (2023) 011034}, [\href{http://arxiv.org/abs/2205.06243}{{\tt
  2205.06243}}].

\bibitem{Choi:2022jqy}
Y.~Choi, H.~T. Lam and S.-H. Shao, \emph{{Noninvertible Global Symmetries in
  the Standard Model}},
  \href{http://dx.doi.org/10.1103/PhysRevLett.129.161601}{\emph{Phys. Rev.
  Lett.} {\bf 129} (2022) 161601}, [\href{http://arxiv.org/abs/2205.05086}{{\tt
  2205.05086}}].

\bibitem{Choi:2022fgx}
Y.~Choi, H.~T. Lam and S.-H. Shao, \emph{{Non-invertible Gauss law and
  axions}}, \href{http://dx.doi.org/10.1007/JHEP09(2023)067}{\emph{JHEP} {\bf
  09} (2023) 067}, [\href{http://arxiv.org/abs/2212.04499}{{\tt 2212.04499}}].

\bibitem{Yokokura:2022alv}
R.~Yokokura, \emph{{Non-invertible symmetries in axion electrodynamics}},
  \href{http://arxiv.org/abs/2212.05001}{{\tt 2212.05001}}.

\bibitem{Choi:2023pdp}
Y.~Choi, M.~Forslund, H.~T. Lam and S.-H. Shao, \emph{{Quantization of
  Axion-Gauge Couplings and Noninvertible Higher Symmetries}},
  \href{http://dx.doi.org/10.1103/PhysRevLett.132.121601}{\emph{Phys. Rev.
  Lett.} {\bf 132} (2024) 121601}, [\href{http://arxiv.org/abs/2309.03937}{{\tt
  2309.03937}}].

\bibitem{Cordova:2022fhg}
C.~Cordova, S.~Hong, S.~Koren and K.~Ohmori, \emph{{Neutrino Masses from
  Generalized Symmetry Breaking}},
  \href{http://dx.doi.org/10.1103/PhysRevX.14.031033}{\emph{Phys. Rev. X} {\bf
  14} (2024) 031033}, [\href{http://arxiv.org/abs/2211.07639}{{\tt
  2211.07639}}].

\bibitem{Nguyen:2021yld}
M.~Nguyen, Y.~Tanizaki and M.~\"Unsal, \emph{{Semi-Abelian gauge theories,
  non-invertible symmetries, and string tensions beyond $N$-ality}},
  \href{http://dx.doi.org/10.1007/JHEP03(2021)238}{\emph{JHEP} {\bf 03} (2021)
  238}, [\href{http://arxiv.org/abs/2101.02227}{{\tt 2101.02227}}].

\bibitem{Choi:2022zal}
Y.~Choi, C.~Cordova, P.-S. Hsin, H.~T. Lam and S.-H. Shao,
  \emph{{Non-invertible Condensation, Duality, and Triality Defects in 3+1
  Dimensions}},
  \href{http://dx.doi.org/10.1007/s00220-023-04727-4}{\emph{Commun. Math.
  Phys.} {\bf 402} (2023) 489--542},
  [\href{http://arxiv.org/abs/2204.09025}{{\tt 2204.09025}}].

\bibitem{Apruzzi:2021nmk}
F.~Apruzzi, F.~Bonetti, I.~Garc{\'\i}a~Etxebarria, S.~S. Hosseini and
  S.~Schafer-Nameki, \emph{{Symmetry TFTs from String Theory}},
  \href{http://dx.doi.org/10.1007/s00220-023-04737-2}{\emph{Commun. Math.
  Phys.} {\bf 402} (2023) 895--949},
  [\href{http://arxiv.org/abs/2112.02092}{{\tt 2112.02092}}].

\bibitem{Arias-Tamargo:2022nlf}
G.~Arias-Tamargo and D.~Rodriguez-Gomez, \emph{{Non-invertible symmetries from
  discrete gauging and completeness of the spectrum}},
  \href{http://dx.doi.org/10.1007/JHEP04(2023)093}{\emph{JHEP} {\bf 04} (2023)
  093}, [\href{http://arxiv.org/abs/2204.07523}{{\tt 2204.07523}}].

\bibitem{Hayashi:2022fkw}
Y.~Hayashi and Y.~Tanizaki, \emph{{Non-invertible self-duality defects of
  Cardy-Rabinovici model and mixed gravitational anomaly}},
  \href{http://dx.doi.org/10.1007/JHEP08(2022)036}{\emph{JHEP} {\bf 08} (2022)
  036}, [\href{http://arxiv.org/abs/2204.07440}{{\tt 2204.07440}}].

\bibitem{Roumpedakis:2022aik}
K.~Roumpedakis, S.~Seifnashri and S.-H. Shao, \emph{{Higher Gauging and
  Non-invertible Condensation Defects}},
  \href{http://dx.doi.org/10.1007/s00220-023-04706-9}{\emph{Commun. Math.
  Phys.} {\bf 401} (2023) 3043--3107},
  [\href{http://arxiv.org/abs/2204.02407}{{\tt 2204.02407}}].

\bibitem{Bhardwaj:2022yxj}
L.~Bhardwaj, L.~E. Bottini, S.~Schafer-Nameki and A.~Tiwari,
  \emph{{Non-invertible higher-categorical symmetries}},
  \href{http://dx.doi.org/10.21468/SciPostPhys.14.1.007}{\emph{SciPost Phys.}
  {\bf 14} (2023) 007}, [\href{http://arxiv.org/abs/2204.06564}{{\tt
  2204.06564}}].

\bibitem{Antinucci:2022eat}
A.~Antinucci, G.~Galati and G.~Rizi, \emph{{On continuous 2-category symmetries
  and Yang-Mills theory}},
  \href{http://dx.doi.org/10.1007/JHEP12(2022)061}{\emph{JHEP} {\bf 12} (2022)
  061}, [\href{http://arxiv.org/abs/2206.05646}{{\tt 2206.05646}}].

\bibitem{Bashmakov:2022jtl}
V.~Bashmakov, M.~Del~Zotto and A.~Hasan, \emph{{On the 6d origin of
  non-invertible symmetries in 4d}},
  \href{http://dx.doi.org/10.1007/JHEP09(2023)161}{\emph{JHEP} {\bf 09} (2023)
  161}, [\href{http://arxiv.org/abs/2206.07073}{{\tt 2206.07073}}].

\bibitem{Damia:2022seq}
J.~A. Damia, R.~Argurio and L.~Tizzano, \emph{{Continuous Generalized
  Symmetries in Three Dimensions}},
  \href{http://dx.doi.org/10.1007/JHEP05(2023)164}{\emph{JHEP} {\bf 05} (2023)
  164}, [\href{http://arxiv.org/abs/2206.14093}{{\tt 2206.14093}}].

\bibitem{Damia:2022bcd}
J.~A. Damia, R.~Argurio and E.~Garcia-Valdecasas, \emph{{Non-invertible defects
  in 5d, boundaries and holography}},
  \href{http://dx.doi.org/10.21468/SciPostPhys.14.4.067}{\emph{SciPost Phys.}
  {\bf 14} (2023) 067}, [\href{http://arxiv.org/abs/2207.02831}{{\tt
  2207.02831}}].

\bibitem{Choi:2022rfe}
Y.~Choi, H.~T. Lam and S.-H. Shao, \emph{{Noninvertible Time-Reversal
  Symmetry}},
  \href{http://dx.doi.org/10.1103/PhysRevLett.130.131602}{\emph{Phys. Rev.
  Lett.} {\bf 130} (2023) 131602}, [\href{http://arxiv.org/abs/2208.04331}{{\tt
  2208.04331}}].

\bibitem{Bhardwaj:2022lsg}
L.~Bhardwaj, S.~Schafer-Nameki and J.~Wu, \emph{{Universal Non-Invertible
  Symmetries}}, \href{http://dx.doi.org/10.1002/prop.202200143}{\emph{Fortsch.
  Phys.} {\bf 70} (2022) 2200143}, [\href{http://arxiv.org/abs/2208.05973}{{\tt
  2208.05973}}].

\bibitem{Lin:2022xod}
L.~Lin, D.~G. Robbins and E.~Sharpe, \emph{{Decomposition, Condensation
  Defects, and Fusion}},
  \href{http://dx.doi.org/10.1002/prop.202200130}{\emph{Fortsch. Phys.} {\bf
  70} (2022) 2200130}, [\href{http://arxiv.org/abs/2208.05982}{{\tt
  2208.05982}}].

\bibitem{Apruzzi:2022rei}
F.~Apruzzi, I.~Bah, F.~Bonetti and S.~Schafer-Nameki, \emph{{Noninvertible
  Symmetries from Holography and Branes}},
  \href{http://dx.doi.org/10.1103/PhysRevLett.130.121601}{\emph{Phys. Rev.
  Lett.} {\bf 130} (2023) 121601}, [\href{http://arxiv.org/abs/2208.07373}{{\tt
  2208.07373}}].

\bibitem{GarciaEtxebarria:2022vzq}
I.~Garc{\'\i}a~Etxebarria, \emph{{Branes and Non-Invertible Symmetries}},
  \href{http://dx.doi.org/10.1002/prop.202200154}{\emph{Fortsch. Phys.} {\bf
  70} (2022) 2200154}, [\href{http://arxiv.org/abs/2208.07508}{{\tt
  2208.07508}}].

\bibitem{Benini:2022hzx}
F.~Benini, C.~Copetti and L.~Di~Pietro, \emph{{Factorization and global
  symmetries in holography}},
  \href{http://dx.doi.org/10.21468/SciPostPhys.14.2.019}{\emph{SciPost Phys.}
  {\bf 14} (2023) 019}, [\href{http://arxiv.org/abs/2203.09537}{{\tt
  2203.09537}}].

\bibitem{Wang:2021vki}
J.~Wang and Y.-Z. You, \emph{{Gauge Enhanced Quantum Criticality Between Grand
  Unifications: Categorical Higher Symmetry Retraction}},
  \href{http://arxiv.org/abs/2111.10369}{{\tt 2111.10369}}.

\bibitem{Chen:2021xuc}
X.~Chen, A.~Dua, P.-S. Hsin, C.-M. Jian, W.~Shirley and C.~Xu, \emph{{Loops in
  4+1d topological phases}},
  \href{http://dx.doi.org/10.21468/SciPostPhys.15.1.001}{\emph{SciPost Phys.}
  {\bf 15} (2023) 001}, [\href{http://arxiv.org/abs/2112.02137}{{\tt
  2112.02137}}].

\bibitem{DelZotto:2022ras}
M.~Del~Zotto and I.~Garc{\'\i}a~Etxebarria, \emph{{Global structures from the
  infrared}}, \href{http://dx.doi.org/10.1007/JHEP11(2023)058}{\emph{JHEP} {\bf
  11} (2023) 058}, [\href{http://arxiv.org/abs/2204.06495}{{\tt 2204.06495}}].

\bibitem{Bhardwaj:2022dyt}
L.~Bhardwaj, M.~Bullimore, A.~E.~V. Ferrari and S.~Schafer-Nameki,
  \emph{{Anomalies of generalized symmetries from solitonic defects}},
  \href{http://dx.doi.org/10.21468/SciPostPhys.16.3.087}{\emph{SciPost Phys.}
  {\bf 16} (2024) 087}, [\href{http://arxiv.org/abs/2205.15330}{{\tt
  2205.15330}}].

\bibitem{Brennan:2022tyl}
T.~D. Brennan, C.~Cordova and T.~T. Dumitrescu, \emph{{Line Defect Quantum
  Numbers {\&} Anomalies}},  \href{http://arxiv.org/abs/2206.15401}{{\tt
  2206.15401}}.

\bibitem{Delmastro:2022pfo}
D.~G. Delmastro, J.~Gomis, P.-S. Hsin and Z.~Komargodski, \emph{{Anomalies and
  symmetry fractionalization}},
  \href{http://dx.doi.org/10.21468/SciPostPhys.15.3.079}{\emph{SciPost Phys.}
  {\bf 15} (2023) 079}, [\href{http://arxiv.org/abs/2206.15118}{{\tt
  2206.15118}}].

\bibitem{Heckman:2022muc}
J.~J. Heckman, M.~H{\"u}bner, E.~Torres and H.~Y. Zhang, \emph{{The Branes
  Behind Generalized Symmetry Operators}},
  \href{http://dx.doi.org/10.1002/prop.202200180}{\emph{Fortsch. Phys.} {\bf
  71} (2023) 2200180}, [\href{http://arxiv.org/abs/2209.03343}{{\tt
  2209.03343}}].

\bibitem{Freed:2022qnc}
D.~S. Freed, G.~W. Moore and C.~Teleman, \emph{{Topological symmetry in quantum
  field theory}},  \href{http://arxiv.org/abs/2209.07471}{{\tt 2209.07471}}.

\bibitem{Niro:2022ctq}
P.~Niro, K.~Roumpedakis and O.~Sela, \emph{{Exploring non-invertible symmetries
  in free theories}},
  \href{http://dx.doi.org/10.1007/JHEP03(2023)005}{\emph{JHEP} {\bf 03} (2023)
  005}, [\href{http://arxiv.org/abs/2209.11166}{{\tt 2209.11166}}].

\bibitem{Mekareeya:2022spm}
N.~Mekareeya and M.~Sacchi, \emph{{Mixed anomalies, two-groups, non-invertible
  symmetries, and 3d superconformal indices}},
  \href{http://dx.doi.org/10.1007/JHEP01(2023)115}{\emph{JHEP} {\bf 01} (2023)
  115}, [\href{http://arxiv.org/abs/2210.02466}{{\tt 2210.02466}}].

\bibitem{vanBeest:2022fss}
M.~van Beest, D.~S.~W. Gould, S.~Schafer-Nameki and Y.-N. Wang, \emph{{Symmetry
  TFTs for 3d QFTs from M-theory}},
  \href{http://dx.doi.org/10.1007/JHEP02(2023)226}{\emph{JHEP} {\bf 02} (2023)
  226}, [\href{http://arxiv.org/abs/2210.03703}{{\tt 2210.03703}}].

\bibitem{Antinucci:2022vyk}
A.~Antinucci, F.~Benini, C.~Copetti, G.~Galati and G.~Rizi, \emph{{The
  holography of non-invertible self-duality symmetries}},
  \href{http://dx.doi.org/10.1007/JHEP03(2025)052}{\emph{JHEP} {\bf 03} (2025)
  052}, [\href{http://arxiv.org/abs/2210.09146}{{\tt 2210.09146}}].

\bibitem{Chen:2022cyw}
S.~Chen and Y.~Tanizaki, \emph{{Solitonic Symmetry beyond Homotopy:
  Invertibility from Bordism and Noninvertibility from Topological Quantum
  Field Theory}},
  \href{http://dx.doi.org/10.1103/PhysRevLett.131.011602}{\emph{Phys. Rev.
  Lett.} {\bf 131} (2023) 011602}, [\href{http://arxiv.org/abs/2210.13780}{{\tt
  2210.13780}}].

\bibitem{Damia:2023ses}
J.~A. Damia, R.~Argurio, F.~Benini, S.~Benvenuti, C.~Copetti and L.~Tizzano,
  \emph{{Non-invertible symmetries along 4d RG flows}},
  \href{http://dx.doi.org/10.1007/JHEP02(2024)084}{\emph{JHEP} {\bf 02} (2024)
  084}, [\href{http://arxiv.org/abs/2305.17084}{{\tt 2305.17084}}].

\bibitem{Argurio:2023lwl}
R.~Argurio and R.~Vandepopeliere, \emph{{When {\ensuremath{\mathbb{Z}}}$_{2}$
  one-form symmetry leads to non-invertible axial symmetries}},
  \href{http://dx.doi.org/10.1007/JHEP08(2023)205}{\emph{JHEP} {\bf 08} (2023)
  205}, [\href{http://arxiv.org/abs/2306.01414}{{\tt 2306.01414}}].

\bibitem{vanBeest:2023dbu}
M.~van Beest, P.~Boyle~Smith, D.~Delmastro, Z.~Komargodski and D.~Tong,
  \emph{{Monopoles, scattering, and generalized symmetries}},
  \href{http://dx.doi.org/10.1007/JHEP03(2025)014}{\emph{JHEP} {\bf 03} (2025)
  014}, [\href{http://arxiv.org/abs/2306.07318}{{\tt 2306.07318}}].

\bibitem{vanBeest:2023mbs}
M.~van Beest, P.~Boyle~Smith, D.~Delmastro, R.~Mouland and D.~Tong,
  \emph{{Fermion-monopole scattering in the Standard Model}},
  \href{http://dx.doi.org/10.1007/JHEP08(2024)004}{\emph{JHEP} {\bf 08} (2024)
  004}, [\href{http://arxiv.org/abs/2312.17746}{{\tt 2312.17746}}].

\bibitem{Lawrie:2023tdz}
C.~Lawrie, X.~Yu and H.~Y. Zhang, \emph{{Intermediate defect groups,
  polarization pairs, and noninvertible duality defects}},
  \href{http://dx.doi.org/10.1103/PhysRevD.109.026005}{\emph{Phys. Rev. D} {\bf
  109} (2024) 026005}, [\href{http://arxiv.org/abs/2306.11783}{{\tt
  2306.11783}}].

\bibitem{Chen:2023czk}
S.~Chen and Y.~Tanizaki, \emph{{Solitonic symmetry as non-invertible symmetry:
  cohomology theories with TQFT coefficients}},
  \href{http://arxiv.org/abs/2307.00939}{{\tt 2307.00939}}.

\bibitem{Nardoni:2024sos}
E.~Nardoni, M.~Sacchi, O.~Sela, G.~Zafrir and Y.~Zheng, \emph{{Dimensionally
  reducing generalized symmetries from (3+1)-dimensions}},
  \href{http://dx.doi.org/10.1007/JHEP07(2024)110}{\emph{JHEP} {\bf 07} (2024)
  110}, [\href{http://arxiv.org/abs/2403.15995}{{\tt 2403.15995}}].

\bibitem{Kan:2024fuu}
N.~Kan, K.~Kawabata and H.~Wada, \emph{{Symmetry fractionalization and duality
  defects in Maxwell theory}},
  \href{http://dx.doi.org/10.1007/JHEP10(2024)238}{\emph{JHEP} {\bf 10} (2024)
  238}, [\href{http://arxiv.org/abs/2404.14481}{{\tt 2404.14481}}].

\bibitem{Fernandez-Melgarejo:2024ffg}
J.~J. Fernandez-Melgarejo, G.~Giorgi, D.~Marques and J.~A. Rosabal,
  \emph{{Noninvertible symmetries in type IIB supergravity}},
  \href{http://dx.doi.org/10.1103/PhysRevD.111.066024}{\emph{Phys. Rev. D} {\bf
  111} (2025) 066024}, [\href{http://arxiv.org/abs/2407.09402}{{\tt
  2407.09402}}].

\bibitem{DeMarco:2025pza}
M.~De~Marco and S.~N. Meynet, \emph{{Symmetries beyond branes: geometric
  engineering and isometries}},
  \href{http://dx.doi.org/10.1007/JHEP08(2025)082}{\emph{JHEP} {\bf 08} (2025)
  082}, [\href{http://arxiv.org/abs/2503.19022}{{\tt 2503.19022}}].

\bibitem{Kaidi:2024wio}
J.~Kaidi, Y.~Tachikawa and H.~Y. Zhang, \emph{{On a class of selection rules
  without group actions in field theory and string theory}},
  \href{http://arxiv.org/abs/2402.00105}{{\tt 2402.00105}}.

\bibitem{Heckman:2024obe}
J.~J. Heckman, J.~McNamara, M.~Montero, A.~Sharon, C.~Vafa and I.~Valenzuela,
  \emph{{On the Fate of Stringy Non-Invertible Symmetries}},
  \href{http://arxiv.org/abs/2402.00118}{{\tt 2402.00118}}.

\bibitem{Kobayashi:2024yqq}
T.~Kobayashi and H.~Otsuka, \emph{{Non-invertible flavor symmetries in
  magnetized extra dimensions}},
  \href{http://dx.doi.org/10.1007/JHEP11(2024)120}{\emph{JHEP} {\bf 11} (2024)
  120}, [\href{http://arxiv.org/abs/2408.13984}{{\tt 2408.13984}}].

\bibitem{Kobayashi:2024cvp}
T.~Kobayashi, H.~Otsuka and M.~Tanimoto, \emph{{Yukawa textures from
  non-invertible symmetries}},
  \href{http://dx.doi.org/10.1007/JHEP12(2024)117}{\emph{JHEP} {\bf 12} (2024)
  117}, [\href{http://arxiv.org/abs/2409.05270}{{\tt 2409.05270}}].

\bibitem{Kobayashi:2025znw}
T.~Kobayashi, Y.~Nishioka, H.~Otsuka and M.~Tanimoto, \emph{{More about quark
  Yukawa textures from selection rules without group actions}},
  \href{http://dx.doi.org/10.1007/JHEP05(2025)177}{\emph{JHEP} {\bf 05} (2025)
  177}, [\href{http://arxiv.org/abs/2503.09966}{{\tt 2503.09966}}].

\bibitem{Suzuki:2025oov}
M.~Suzuki and L.-X. Xu, \emph{{Phenomenological implications of a class of
  non-invertible selection rules}},
  \href{http://arxiv.org/abs/2503.19964}{{\tt 2503.19964}}.

\bibitem{Kobayashi:2025cwx}
T.~Kobayashi, H.~Okada and H.~Otsuka, \emph{{Radiative neutrino mass models
  from non-invertible selection rules}},
  \href{http://dx.doi.org/10.1007/JHEP12(2025)111}{\emph{JHEP} {\bf 12} (2025)
  111}, [\href{http://arxiv.org/abs/2505.14878}{{\tt 2505.14878}}].

\bibitem{Kobayashi:2025lar}
T.~Kobayashi, H.~Mita, H.~Otsuka and R.~Sakuma, \emph{{Matter symmetries in
  supersymmetric standard models from non-invertible selection rules}},
  \href{http://arxiv.org/abs/2506.10241}{{\tt 2506.10241}}.

\bibitem{Dong:2025jra}
J.~Dong, T.~Jeric, T.~Kobayashi, R.~Nishida and H.~Otsuka, \emph{{On discrete
  gauging and non-invertible selection rules}},
  \href{http://arxiv.org/abs/2507.02375}{{\tt 2507.02375}}.

\bibitem{Suzuki:2025bxg}
M.~Suzuki, L.-X. Xu and H.~Y. Zhang, \emph{{Spurion Analysis for Non-Invertible
  Selection Rules from Near-Group Fusions}},
  \href{http://arxiv.org/abs/2508.14970}{{\tt 2508.14970}}.

\bibitem{Kobayashi:2025ocp}
T.~Kobayashi, R.~Nishida and H.~Otsuka, \emph{{Non-Invertible Selection Rules
  on Heterotic Non-Abelian Orbifolds}},
  \href{http://arxiv.org/abs/2509.10019}{{\tt 2509.10019}}.

\bibitem{Suzuki:2025kxz}
M.~Suzuki and L.-X. Xu, \emph{{Spurion Analysis of $\mathbb{Z}_M/\mathbb{Z}_2$
  Non-Invertible Selection Rules: Low-Order versus All-Order Zeros}},
  \href{http://arxiv.org/abs/2510.18972}{{\tt 2510.18972}}.

\bibitem{Nakai:2025thw}
Y.~Nakai, H.~Otsuka, Y.~Shigekami and Z.~Zhang, \emph{{The Minimal
  Supersymmetric Standard Model with Non-Invertible Selection Rules}},
  \href{http://arxiv.org/abs/2512.21509}{{\tt 2512.21509}}.

\bibitem{Cordova:2024ypu}
C.~Cordova, S.~Hong and S.~Koren, \emph{{Noninvertible Peccei-Quinn Symmetry
  and the Massless Quark Solution to the Strong CP Problem}},
  \href{http://dx.doi.org/10.1103/PhysRevX.15.031011}{\emph{Phys. Rev. X} {\bf
  15} (2025) 031011}, [\href{http://arxiv.org/abs/2402.12453}{{\tt
  2402.12453}}].

\bibitem{Copetti:2024dcz}
C.~Copetti, L.~Cordova and S.~Komatsu, \emph{{S-matrix bootstrap and
  non-invertible symmetries}},
  \href{http://dx.doi.org/10.1007/JHEP03(2025)204}{\emph{JHEP} {\bf 03} (2025)
  204}, [\href{http://arxiv.org/abs/2408.13132}{{\tt 2408.13132}}].

\bibitem{Chen:2024tsx}
S.~Chen, A.~Cherman, G.~Choi and M.~Neuzil, \emph{{Cheshire
  {\ensuremath{\theta}} terms, Aharonov-Bohm effects, and axions}},
  \href{http://dx.doi.org/10.1103/jnxt-m1dk}{\emph{Phys. Rev. D} {\bf 112}
  (2025) 045012}, [\href{http://arxiv.org/abs/2410.23355}{{\tt 2410.23355}}].

\bibitem{Liang:2025dkm}
Q.~Liang and T.~T. Yanagida, \emph{{Non-invertible symmetry as an axion-less
  solution to the strong CP problem}},
  \href{http://dx.doi.org/10.1016/j.physletb.2025.139706}{\emph{Phys. Lett. B}
  {\bf 868} (2025) 139706}, [\href{http://arxiv.org/abs/2505.05142}{{\tt
  2505.05142}}].

\bibitem{Choi:2025vxr}
G.~Choi, T.~Gherghetta and J.~Terning, \emph{{Noninvertible chiral symmetry and
  axions under electromagnetic duality}},
  \href{http://dx.doi.org/10.1103/vdcp-k462}{\emph{Phys. Rev. D} {\bf 112}
  (2025) 095023}, [\href{http://arxiv.org/abs/2509.14395}{{\tt 2509.14395}}].

\bibitem{Honda:2024yte}
Y.~Honda, O.~Morikawa, S.~Onoda and H.~Suzuki, \emph{{Lattice Realization of
  the Axial U(1) Noninvertible Symmetry}},
  \href{http://dx.doi.org/10.1093/ptep/ptae040}{\emph{PTEP} {\bf 2024} (2024)
  043B04}, [\href{http://arxiv.org/abs/2401.01331}{{\tt 2401.01331}}].

\bibitem{Ando:2024nlk}
T.~Ando, \emph{{Gauging on the lattice and gapped/gapless topological phases}},
   \href{http://arxiv.org/abs/2402.03566}{{\tt 2402.03566}}.

\bibitem{Honda:2024sdz}
Y.~Honda, S.~Onoda and H.~Suzuki, \emph{{Action of the Axial U(1)
  Non-Invertible Symmetry on the {\textquoteright}t~Hooft Line Operator: A
  Lattice Gauge Theory Study}},
  \href{http://dx.doi.org/10.1093/ptep/ptae093}{\emph{PTEP} {\bf 2024} (2024)
  073B04}, [\href{http://arxiv.org/abs/2403.16752}{{\tt 2403.16752}}].

\bibitem{Bhardwaj:2024kvy}
L.~Bhardwaj, L.~E. Bottini, S.~Schafer-Nameki and A.~Tiwari, \emph{{Lattice
  Models for Phases and Transitions with Non-Invertible Symmetries}},
  \href{http://arxiv.org/abs/2405.05964}{{\tt 2405.05964}}.

\bibitem{Honda:2024xmk}
Y.~Honda, S.~Onoda and H.~Suzuki, \emph{{Action of the Axial U(1) Noninvertible
  Symmetry on the {\textquoteright}t Hooft Line Operator: A Simple Argument}},
  \href{http://dx.doi.org/10.1093/ptep/ptae167}{\emph{PTEP} {\bf 2024} (2024)
  113B02}, [\href{http://arxiv.org/abs/2405.07669}{{\tt 2405.07669}}].

\bibitem{Choi:2024rjm}
Y.~Choi, Y.~Sanghavi, S.-H. Shao and Y.~Zheng, \emph{{Non-invertible and
  higher-form symmetries in 2+1d lattice gauge theories}},
  \href{http://dx.doi.org/10.21468/SciPostPhys.18.1.008}{\emph{SciPost Phys.}
  {\bf 18} (2025) 008}, [\href{http://arxiv.org/abs/2405.13105}{{\tt
  2405.13105}}].

\bibitem{Lu:2024ytl}
D.-C. Lu, Z.~Sun and Y.-Z. You, \emph{{Realizing triality and $p$-ality by
  lattice twisted gauging in (1+1)d quantum spin systems}},
  \href{http://dx.doi.org/10.21468/SciPostPhys.17.5.136}{\emph{SciPost Phys.}
  {\bf 17} (2024) 136}, [\href{http://arxiv.org/abs/2405.14939}{{\tt
  2405.14939}}].

\bibitem{Cao:2024qjj}
W.~Cao, L.~Li and M.~Yamazaki, \emph{{Generating lattice non-invertible
  symmetries}},
  \href{http://dx.doi.org/10.21468/SciPostPhys.17.4.104}{\emph{SciPost Phys.}
  {\bf 17} (2024) 104}, [\href{http://arxiv.org/abs/2406.05454}{{\tt
  2406.05454}}].

\bibitem{Honda:2024kvf}
Y.~Honda, S.~Onoda and H.~Suzuki, \emph{{Axion QED as a Lattice Gauge Theory
  and Non-Invertible Symmetry}},
  \href{http://dx.doi.org/10.22323/1.466.0358}{\emph{PoS} {\bf LATTICE2024}
  (2025) 358}, [\href{http://arxiv.org/abs/2412.08142}{{\tt 2412.08142}}].

\bibitem{Katayama:2025pmz}
N.~Katayama and Y.~Tanizaki, \emph{{2d Cardy-Rabinovici model with the modified
  Villain lattice: exact dualities and symmetries}},
  \href{http://dx.doi.org/10.1007/JHEP11(2025)004}{\emph{JHEP} {\bf 11} (2025)
  004}, [\href{http://arxiv.org/abs/2505.19412}{{\tt 2505.19412}}].

\bibitem{Kawana:2025vbi}
K.~Kawana, \emph{{Landau theory for lattice higher-form gauge theories and the
  Kramers-Wannier duality}},
  \href{http://dx.doi.org/10.1103/ftcr-m4mr}{\emph{Phys. Rev. D} {\bf 112}
  (2025) 074514}, [\href{http://arxiv.org/abs/2507.06555}{{\tt 2507.06555}}].

\bibitem{Fidkowski:2025rsq}
L.~Fidkowski, C.~Xu and C.~Zhang, \emph{{Non-invertible bosonic chiral symmetry
  on the lattice}},  \href{http://arxiv.org/abs/2510.17969}{{\tt 2510.17969}}.

\bibitem{Su:2025aag}
L.~Su, \emph{{$\mathbb{Z}_2$ lattice gauge theories: fermionic gauging,
  transmutation, and Kramers-Wannier dualities}},
  \href{http://arxiv.org/abs/2510.20893}{{\tt 2510.20893}}.

\end{thebibliography}\endgroup

\end{document}